\documentclass[11pt,a4paper,oneside]{book}

\pdfoutput=1
\usepackage{dstyle}
\usepackage{geometry}
\usepackage{times}
\usepackage{graphics}
\usepackage{wrapfig}
\usepackage{pdfpages}
\usepackage{doublespace}
\usepackage{amssymb}
\usepackage{bm}
\usepackage{multirow}
\usepackage{tabularx,booktabs}
\usepackage[pdfencoding=auto,psdextra]{hyperref}
\usepackage{threeparttable}
\usepackage{setspace}
\usepackage{subfig}
\usepackage{amsmath}
\usepackage{txfonts}
\usepackage{wasysym}
\usepackage{fancyhdr}
\usepackage{rotating}
\usepackage{natbib}
\usepackage[titletoc]{appendix}
\def\laeq{\raise.2ex\hbox{$<$}\kern-.75em\lower.9ex\hbox{$\sim$}\,}
\def\gaeq{\raise.2ex\hbox{$>$}\kern-.75em\lower.9ex\hbox{$\sim$}\,}

\begin{document}

\includepdf[fitpaper]{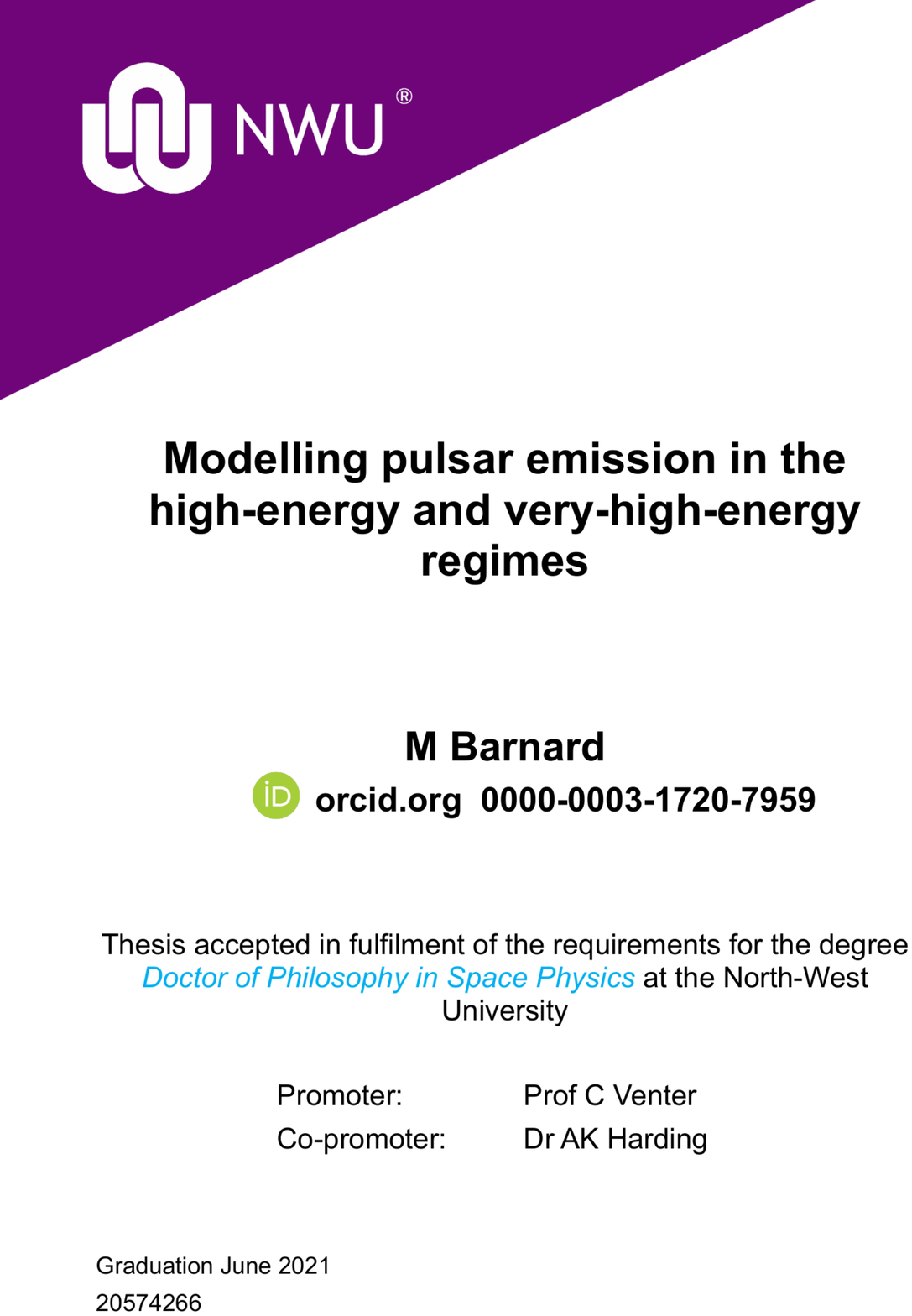}

\onehalfspacing
\frontmatter
\chapter*{Acknowledgements}

I dedicate this work to my loving Father for his grace, guidance, and love. Special thanks to Him for
providing me with the love and support from family, friends, and work colleagues at the Physics Department.
Special thanks to my husband for his unwavering faith in me, and encouraging me to follow my dreams. It
gives me great pleasure in acknowledging the support of my supervisor (Prof. Christo Venter) and co-supervisor (Dr. Alice Harding) for their guidance, patience, invaluable advice, and constantly challenging me. Your perseverance and
ideas are truly inspirational. Special thanks goes to other collaborators, Constantinos Kalapotharakos, and Tyrel Johnson for their valuable contributions made to my PhD research.
\\
\\
\noindent \textit{``I would have lost heart, unless I had believed that I would see the goodness of the Lord in the land of the living. Wait on the Lord; be of good courage, and He shall strengthen your heart. Wait, I say, on the Lord!''}{ --- \textup{Psalm 27:13$-$14, NKJV}}\\

\noindent \textit{``He counts the number of the stars; \\
He calls them all by name. \\
Great is our Lord, and mighty in power; \\
His understanding is infinite.''}{ --- \textup{Psalm 147:4$-$5, NKJV}} \\

\noindent \textit{``When I consider Your heavens, the work of Your fingers,
The moon and the stars, which You have ordained, \\
What is man that You are mindful of him, And the son of man that You visit him?''}{ --- \textup{Psalm 8:3$-$4, NKJV}}  \\

\noindent \textit{``Now faith is the substance of things hoped for, \\ the evidence of things not seen.''}{ --- \textup{Hebrews 11:1, NKJV}} \\

\noindent \textit{``Call to Me, and I will answer you, and show you great and mighty \\ things, which you do not know.''}{ --- \textup{Jeremiah 33:3, NKJV}} \\

\noindent \textit{``Fear not, for I am with you; \\
Be not dismayed, for I am your God. \\
I will strengthen you, \\
Yes, I will help you, \\
I will uphold you with My righteous right hand.''}{ --- \textup{Isaiah 41:10, NKJV}} \\

\newgeometry{top=0mm,bottom=20mm,left=25mm,right=20mm}
\chapter*{\huge \centering Abstract}
\vspace*{-1cm}
\section*{\centering \bf Modelling pulsar emission in the high-energy and very-high-energy regimes} %Maximum 500 words and five to ten descriptive key terms. Include aims, methods, en main findings

The {\it Fermi} Large Area Telescope has revolutionised the $\gamma$-ray pulsar field, increasing the population to over 250 detected pulsars. The majority display spectra with exponential cutoffs in a narrow range around a few GeV. Models predicted cutoffs up to 100 GeV; it was therefore not expected that pulsars would be visible in the very-high-energy ($>$100~GeV) regime. Subsequent surprise discoveries by ground-based telescopes of pulsed emission from four pulsars above tens of GeV have marked the beginning of a new era, raising important questions about the electrodynamics and local environment of pulsar magnetospheres. I have performed geometric light curve modelling using static, retarded vacuum, and offset polar cap dipole $B$-fields, in conjunction with standard two-pole caustic and outer gap geometries. I also considered a slot gap $E$-field associated with the offset polar cap $B$-field and found that its inclusion leads to qualitatively different light curves. Solving the particle transport equation shows that the particle energy only becomes large enough to yield significant curvature radiation at large altitudes above the stellar surface, given this relatively low $E$-field. Therefore, particles do not always attain the radiation-reaction limit. Increasing the slot gap $E$-field by a factor of 100 led to improved light curve fits, as well as curvature radiation reaction at lower altitudes. The overall optimal light curve fit was for the retarded vacuum dipole field and outer gap model. Recent kinetic simulations sparked a debate regarding the emission mechanism of pulsed $\gamma$-ray emission from pulsars. Some models invoke curvature radiation, while others assume synchrotron radiation in the current sheet. Detection of the Vela pulsar by H.E.S.S.\ ($20-120$~GeV) and \emph{Fermi} provides evidence for a curved spectrum. We posit this to result from curvature radiation via primary particles in the pulsar magnetosphere and current sheet. We present energy-dependent light curves using an extended slot gap and current sheet model and invoking a two-step accelerating $E$-field as motivated by kinetic simulations. I include a refined calculation of the curvature radius of particle trajectories, impacting the particle transport, predicted light curves, and spectra. The model reproduces the decrease of flux of the first light-curve peak relative to the second one, evolution of the bridge emission, near constant phase positions of peaks, and narrowing of pulses with increasing energy. We can fundamentally explain the first of these trends, since I found that the curvature radii of the particle trajectories in regions where the second $\gamma$-ray light curve peak originates are systematically larger than those associated with the first peak, implying a correspondingly larger cutoff for the second peak. An unknown azimuthal dependence of the $E$-field as well as uncertainty in the precise emission locale preclude a simplistic discrimination of emission mechanisms. Finally, H.E.S.S.\ recently announced the detection of pulsed emission from the Vela pulsar up to 7~TeV, constraining particle energies to exceed several TeV. I contributed to a paper invoking synchrotron self-Compton emission to model this new radiation component, thus providing a consistent framework to describe the TeV emission from Vela.

{\bf Keywords:} Gamma rays --- Pulsars --- Vela pulsar (PSR J0835$-$4510) --- Magnetic fields --- \textit{Fermi} Large Area Telescope.

\restoregeometry
\tableofcontents
\listoffigures
\listoftables
\mainmatter
\chapter{Introduction}

\section{Recent developments in  $\gamma$-ray pulsar astronomy} \label{sec:DevGamPSRAstro}% What is going on in the field and limitations

\subsection{Historical perspective of $\gamma$-ray pulsar detections and theoretical expectations} \label{subsec:HistVHE}
Since the launch in June~2008 of the \textit{Fermi} Large Area Telescope (LAT; \citealp{Atwood2009}), a high-energy (HE) satellite measuring $\gamma$-rays in the 20~MeV to $>300$~GeV range, there has been a consistent discovery rate of new pulsars. The \textit{Fermi} LAT Collaboration has already released two pulsar catalogues (1PC, \citealp{Abdo2010FirstCat}; 2PC, \citealp{Abdo2013SecondCat}) discussing the light curve and spectral properties of these (117 in 2PC) pulsars. Prior to \textit{Fermi}, only~7 $\gamma$-ray pulsars were known \citep{Thompson1997}. The bulk of the \textit{Fermi}-detected pulsars display exponentially cutoff spectra with cutoffs falling in a narrow range around a few GeV. During this time (early 2000s), there was no detection of TeV pulsed emission.

Earlier pulsar models mostly expected HE emission at tens of GeV, while some made predictions of TeV emission, but this was rather uncertain. For example, pulsar models (see Chapter~\ref{chap:PSRastro} for more details), assuming the standard outer gap (OG) scenario, predicted spectral components in the very-high-energy (VHE; $>100$~GeV) regime when estimating the inverse Compton scattering (ICS) flux of primary electrons on synchrotron radiation (SR) or other soft photons (\citealt{Cheng1986,Romani1996,Hirotani2001}). This resulted in a natural bump around a few TeV (involving $\sim10$~TeV particles) in the extreme Klein-Nishina limit. However, these components may not survive up to the light cylinder\footnote{The radius $R_{\rm LC}$ where the co-rotation speed equals the speed of light $c$.} and beyond, since magnetic pair creation leads to absorption of the TeV $\gamma$-ray flux \citep{Hirotani2001}. Other studies assumed standard pulsar models and curvature radiation (CR) to be the dominant radiation mechanism producing $\gamma$-ray emission and found spectral cutoffs of up to 100~GeV. For example, \citet{Bulik2000} modelled the cutoffs of millisecond pulsars (MSPs) that possess relatively low $B$-fields and short periods. Their model assumed a static dipole $B$-field and a polar cap (PC) geometry, and predicted CR from the primary electrons that are released from the PC and accelerated along curved $B$-field lines. Their predicted spectrum cut off at $\sim100$~GeV. The CR photons may undergo magnetic pair production in the intense low-altitude $B$-fields, and the newly formed electron-positron secondaries will emit SR in the optical and X-ray band. \citet{Harding2002} also found CR spectral cutoffs at energies between 50$-$100~GeV. %(see Figure~\ref{fig:IACTs}).
\citet{Harding2005c} investigated the X-ray and $\gamma$-ray spectrum of rotation-powered MSPs using a pair-starved polar cap (PSPC) model, and found CR cutoffs of $\sim$10$-$50~GeV (see also \citealp{Frackowiak2005,Venter2005}). \citet{Harding2008} modelled the optical to $\gamma$-ray emission from a slot gap (SG) accelerator and applied it to the Crab pulsar (assuming a retarded vacuum dipole (RVD) $B$-field), finding spectral cutoffs of up to a few GeV. \citet{Hirotani2008a} modelled phase-resolved spectra of the Crab pulsar using the OG and SG models, and found HE cutoffs of up to $\sim$25~GeV (see also \citet{Tang2008} who used the RVD $B$-field to model phase-resolved spectra of the Crab, finding HE cutoffs around $\sim$10~GeV). \textit{Therefore, it was more or less the consensus of the field prior to 2008 that the HE emission from pulsars occurred in an energy band that was perhaps above the detection range of satellite detectors like the Energetic Gamma-Ray Experiment Telescope (EGRET) in some cases, and below that of ground-based Cherenkov detectors that had energy thresholds above 100 GeV, unless there were TeV spectral components (but the older OG predictions of the latter had been scaled down based on available upper limits at the time, so this was not a strong expectation).}

\subsection{Observational revolution} \label{subsec:obsRev}
\textit{In view of the above, it was not strongly expected that pulsars should be visible in the VHE regime.} It was therefore surprising when the Major Atmospheric Gamma-ray Imaging Cherenkov Telescope (MAGIC) detected pulsed emission from the Crab pulsar at energies up to $\sim$25~GeV~\citep{Aliu2008,Aleksic2011,Aleksic2012}, and even more surprising when the Very Energetic Radiation Imaging Telescope Array System (VERITAS) announced the detection of the same, but up to $\sim$400~GeV \citep{Aliu2011}. The Crab pulsar is thus the first source from which pulsations have been detected over almost all energies ranging from radio to VHE $\gamma$-rays.

The detection of the Crab pulsar above several GeV prompted \textit{Fermi} to search for pulsed emission at HEs. They detected significant pulsations above 10~GeV from 20 pulsars and above 25~GeV from 12 pulsars~\citep{Ackermann2013}. A stacking analysis involving 115 \textit{Fermi}-detected pulsars (excluding the Crab pulsar) was performed by \citet{McCann2015}. However, no emission above 50~GeV was detected, implying that VHE pulsar detections may be rare, given current telescope sensitivities. Notably, pulsed emission was also detected from the Vela pulsar up to ~80~GeV with the \textit{Fermi} LAT \citep{Leung2014}.

Ground-based Cherenkov telescopes are now searching for more examples of VHE pulsars, and they have had some success in recent years. In the VHE band, MAGIC detected pulsations from the Crab pulsar at energies up to 1~TeV \citep{Ansoldi2016}. Pulsed emission from the Vela pulsar was detected in the sub-20 GeV to 100~GeV range with H.E.S.S.\ \citep{Abdalla2018}. New observations by H.E.S.S.\ reveal pulsed emission from Vela up to several TeV (H.E.S.S.\ Collaboration, in preparation). VERITAS furthermore detected no emission from Geminga above 100~GeV \citep{Aliu2015}. However, pulsed emission from the Geminga pulsar between 15~GeV and 75~GeV at a significance of 6.3$\sigma$ was recently announced by MAGIC, although only the second light curve peak is visible at these energies. The MAGIC spectrum is an extension of the \emph{Fermi} LAT spectrum, ruling out the possibility of a sub-exponential cutoff in the same energy range at the $3.6\sigma$ level \citep{Acciari2020}. H.E.S.S.~II furthermore detected pulsed emission from PSR~B1706$-$44 in the sub-100~GeV energy range \citep{SpirJacob2019}. 

From these VHE observations, four trends in the energy-dependent pulse profiles seem to emerge: as the photon energy $E_\gamma$ is increased (above several GeV), the main light curve peaks of Crab, Vela and Geminga seem to remain at the same phase positions, the intensity ratio of the first to second peak (P1/P2) decreases with an increase of $E_{\gamma}$ for Vela and Geminga, the inter-peak ``bridge'' emission evolves for Vela, and the peak widths decrease for Crab \citep{Aliu2011}, Vela \citep{Abdo2010Vela} and Geminga \citep{Abdo2010Geminga}. The second peak of Crab for MAGIC is harder and extends to a bit higher energy ($\sim$2~TeV) than the first peak. The P1/P2 vs.\ $E_{\gamma}$ effect was also seen by \emph{Fermi} for a number of pulsars \citep{Abdo2010FirstCat,Abdo2013SecondCat}. 

\subsection{Debate regarding high-energy radiation mechanisms}  \label{subsec:debate_radmech}

In general, multi-wavelength pulsar light curves exhibit an intricate structure that evolves with $E_{\gamma}$ \citep[e.g.,][]{Buehler2014}, reflecting the various underlying emitting particle populations and spectral radiation components that contribute to this emission, as well as the local $B$-field geometry and $E$-field spatial distribution. In addition, Special Relativistic effects modify the emission beam, given the fact that the co-rotation speeds may reach close to the speed of light $c$ in the outer magnetosphere.

Some traditional physical emission models invoke CR from extended regions within the magnetosphere to explain the HE spectra and light curves. These include the SG (\citealp{Arons1983,Harding2003}) and OG (\citealt{Romani1995,Cheng1986}) models. However, they fall short of fully addressing global magnetospheric characteristics, e.g., the particle acceleration and pair production, current closure, and radiation of a complex multi-wavelength spectrum. Geometric light curve modelling \citep{Dyks2004b,Venter2009,Watters2009,Johnson2014,Pierbattista2015} presented an important interim avenue for probing the pulsar magnetosphere in the context of traditional pulsar models, focusing on the spatial rather than physical origin of HE photons. More recent developments include global magnetospheric models such as the force-free (FF) inside and dissipative outside (FIDO) model \citep{Brambilla2015,Kalapotharakos2009,Kalapotharakos2014}, equatorial current sheet models (e.g., \citealt{Bai2010b,Petri2012}), the striped-wind models (e.g., \citealt{Petri2011}), and kinetic / particle-in-cell simulations (PIC; \citealt{Brambilla2018,Cerutti2016current,Cerutti2016,Cerutti2020,Kalapotharakos2018,Philippov2018}). Some studies using the FIDO models assume that particles are accelerated by induced $E$-fields in dissipative magnetospheres and produce GeV emission via CR (e.g., \citealt{Kalapotharakos2014}). Conversely, in some of the wind or current-sheet models, HE emission originates beyond the light cylinder via SR by relativistic, hot particles that have been accelerated via magnetic reconnection inside the current sheet \citep[e.g.,][]{Petri2011,Philippov2018}. Other studies assume ICS to be the dominant emission mechanism of
HE $\gamma$-rays in an OG scenario (see \citealt{Lyutikov2012,Lyutikov2013} who modelled the broadband spectrum of the Crab pulsar). There is thus an ongoing debate regarding pulsar emission mechanisms, and it is hoped that future observations will help discriminate between models.

\subsection{Latest \emph{NICER} results} \label{subsec:NICER}

The Neutron star Interior Composition Explorer (\emph{NICER}; \citealt{Gendreau2016})\footnote{\url{https://heasarc.gsfc.nasa.gov/docs/nicer/}} is an instrument that is dedicated to study thermal and non-thermal emission from neutron stars (NS) in the soft X-ray band ($0.2-12$~keV) through soft X-ray timing and a spectroscopy instrument on-board the International Space Station, with exceptional sensitivity. \emph{NICER} has a star-tracker-based pointing system that allows the X-ray timing instrument to target and track celestial objects over nearly the full hemisphere.

Earlier modelling have long expected multipolar $B$-fields in MSPs (\citealt{Ruderman1975,Arons1983,Asseo2002}). For example, \citet{Harding2011a,Harding2011c} used a generalised solution of an offset dipole of which the PCs are assumed to be offset from the dipole axis and applied it to MSPs. Since MSPs such as PSR J0437$-$4715 and PSR J0030+0451 are too old to suffer significant cooling, their thermal X-ray emission is believed to be from hot spots on the PCs. These hotspots may not be strictly antipodal, given a non-dipolar $B$-field structure.

Since the launch of \emph{NICER} in June~2017, the rotation-powered MSPs such as PSR~J0030+0451 and PSR~J0437$-$4715, have been studied in much detail. Modelling of the observed thermal X-ray pulsations from these sources gave valuable insight into the global $B$-field structures associated with MSPs. These studies support the existence of a multi-polar $B$-field, including offset-dipole plus quadrupole components, that deviates from a centred dipole (e.g., \citealp{Miller2019,Riley2019}; see Section~\ref{subsec:StatDip}) after modelling \emph{NICER} X-ray waveforms from PSR~J0030+0451. \citet{Kalapotharakos2020} investigated the $B$-field structure that includes offset dipole plus
quadrupole components using a static vacuum field and FF global magnetosphere models. They modelled the $\gamma$-ray and X-ray emission and compared it to the \emph{Fermi} data (see also \citealt{Chen2020}). These observations thus confirm earlier expectations of more complicated, multi-polar $B$-field structures in pulsars, the effect of which are particularly evident near the stellar surface.

\subsection{Upcoming developments} 
% What are the outcome and what will be the significance of your research in the future
% Patrizia Caraveo and Isabelle Grenier
The population of pulsars detected by the \emph{Fermi} LAT has increased to over 250, leading to the preparation of the \emph{Fermi}'s Third Pulsar Catalogue (3PC). This catalogue builds on the 2PC, and will include updated timing solutions, pulse profiles, spectra, and ancillary data. In addition to an increase in the number of pulsars, the 3PC also includes novel pulsars, e.g., the first radio-quiet MSP and first extra-Galactic $\gamma$-ray pulsar \citep{Limyansky2019}. The All-sky Medium Energy Gamma-ray Observatory (\emph{AMEGO}; \citealt{McEnery2019}) is a proposed MeV $\gamma$-ray surveyor probe that fills the gap between hard X-ray instruments, e.g., NuSTAR and the HE $\gamma$-ray telescopes, e.g., \emph{Fermi} LAT and the Astro-Rivelatore Gamma a Immagini Leggero (\emph{AGILE}), and is planned to launch in 2029. Current and future missions (including the Square Kilometre Array, SKA) are dedicated to search for more pulsars over the entire electromagnetic spectrum. More pulsars emitting VHE emission may be found by present and future ground-based telescopes, e.g., the Cherenkov Telescope Array (CTA), which will have a ten-fold increase in sensitivity compared to present-day Cherenkov telescopes.

\section{Problem identification and research aims} \label{subsec:Problem}
% What are the shortcomings you want to address by finding a solution
%  MOTIVERING een-tot-een. Se iets oor elke punt (paragraaf oor elke punt, `link' met jou motivering).}

The \emph{NICER} mission shows evidence of MSPs possessing offset-dipole structures. These studies pave the way for investigating new $B$-field structures, similar to our study done in Chapter~\ref{chap:OffsetPC}. As a first approach, we will study the effect of the $B$-field structure on the predicted GeV light curves of the Vela pulsar by developing a geometric modelling code \citep{Dyks2004b} based on different $B$-field solutions, i.e., static dipole, RVD, and an offset-PC dipole (the latter is additionally implemented; \citealt{Harding2011a,Harding2011c}), assuming constant emissivity $\epsilon_\nu$. Also, we implement an SG $E$-field to modulate $\epsilon_\nu$ for such an offset-PC dipole and examine the effect thereof on the GeV light curves. Since this $E$-field is relatively low, we will multiply it by a factor 100 and illustrate the effective change in the light curves as well as the best fits of the \emph{Fermi} data to the model, and compare our fits to multi-wavelength fits from independent studies.

As we have seen, a few major developments have shaped the field of pulsar science over the past decade. One of these include the increase in pulsar detections by the \emph{Fermi} LAT. The light curves and phase-resolved spectra exhibit unique trends and different cutoffs for each emission peak in the different energy bands. The phase-resolved spectral cutoff for the second peak appears larger than that for the first peak in many cases, as well as the trends pointed out earlier. Our main goal for this study is to explain these trends and the spectral cutoffs for the Vela pulsar in the GeV range (see Chapter~\ref{chap:CREdepLCmod}). Additionally, the more advanced kinetic and global models led to the debate between emission mechanisms. Given this ongoing debate between the emission mechanisms of HE emission, our motivation in this study is to explain the curved GeV spectrum and light curves of Vela as measured by \emph{Fermi} and H.E.S.S.\ that result from primary particles emitting CR. Specifically, by modelling the $E_\gamma$-dependent light curves (and P1/P2 signature) and phase-resolved spectra in the CR regime of synchro-curvature (SC) radiation, we hope to probe whether this effect can serve as a potential discriminator between emission mechanisms and models (see also the reviews of~\citealt{Harding2016,Venter2016,Venter2017} on using pulsar light curves to scrutinise magnetospheric structure and emission distribution). 

To date, four VHE pulsars have been detected, i.e., Crab, Vela, Geminga and PSR~B1706$-$44, being some of the brightest $\gamma$-ray sources. To explain the measured VHE pulsed emission as seen from these pulsars leads to motivation for updating existing or implementing new spectral components. \citet{Harding2015} implemented a synchrotron self-Compton (SSC) radiation component that can explain the VHE emission seen from the Crab pulsar. A follow up paper \citep{Harding2018} extended the SSC emission code and modelled the emission for Vela in this same energy range, and will be discussed in greater detail in Chapter~\ref{chap:VelaTeV}. The immense rise in the number of pulsars detected makes population studies possible in order to better understand pulsar physics. 

We have access to an SSC emission code that predicts light curves and spectra, and already includes the SG current sheet model, FF $B$-field solution, a constant $E_\parallel$ (as motivated by the kinetic models), standard radiation processes including CR, SR, ICS, and SSC, as well as pair cascades (associated with magnetic pair production and calculated in a separate code) that originate near the PC \citep{Harding2015}. In order to model emission pulse profiles as a function of energy, as well as predicting phase-resolved spectra for Vela, we will apply this SSC emission code assuming emission from primary particles that emit only CR. Since particles that emit CR radiation mostly follow the curved $B$-field lines in the rotating frame, our proposed project involves implementing a refined calculation of the curvature radius $\rho_{\rm c}$ of the particle trajectory. We will investigate the behaviour of the light curve peaks as well as the light curve trends as a function of $\rho_{\rm c}$. For the optimal light curve and spectral fits, we will study the local environment of the peaks' emission regions, finding a systematic difference in $\rho_{\rm c}$, particle Lorentz factor $\gamma$, and spectral cutoff energy $E_{\gamma,\rm CR}$ for the two peaks. Lastly we will compare our results to measurements of \emph{Fermi} and H.E.S.S.\ for the Vela pulsar (see Chapter~\ref{chap:CREdepLCmod}). Our improved $\rho_{\rm c}$ was also one of the adaptions made by \citet{Harding2018}, therefore the results we obtained in this study accompany theirs (see Chapter~\ref{chap:VelaTeV}).

\section{Publications} % Own research papers / proceedings published

The publications that emanated directly from this study are listed below.

\subsection{Peer-reviewed conference proceedings}
\begin{enumerate}
%\item Breed, M., Venter, C., Harding, A.K., \& Johnson, T.J., 2014, \textit{Implementation of an offset-dipole magnetic field in a pulsar modelling code}, in Conf.\ Proc.\ of SAIP2013: the 58$^{th}$ Ann.\ Conf. of the South African Inst.\ of Phys., ed.\ by R.\ Botha \& T.\ Jili, pp.\ 350$-$355.
%\item Breed, M., Venter, C., \& Harding, A. K., 2015, \textit{Pulsar emission in the very high energy regime}, in Conf.\ Proc.\ of HEASA2015: the 3$^{rd}$ Ann.\ Conf.\ on High Energy Astrophys.\ in Southern Africa, ed.\ by M. Boettcher, D. Buckley, S. Colafrancesco, P. Meintjes \& S. Razzaque, id.\ 27.
%\item Breed, M., Venter, C., Harding, A. K., \& Johnson, T. J., 2015, \textit{The Effect of Different Magnetospheric Structures on Predictions of $\gamma$-ray Pulsar Light Curves}, in Conf. Proc. of SAIP2012: the 57$^{th}$ Ann.\ Conf.\ of the South African Inst.\ of Phys., ed.\ by J. Janse van Rensburg, pp.\ 316$-$321.
%\item Breed, M., Venter, C., Harding, A. K., \& Johnson, T. J., 2015, \textit{The effect of an offset-dipole magnetic field on the Vela pulsar's $\gamma$-ray light curves}, in Conf.\ Proc.\ of SAIP2014: the 59$^{th}$ Ann.\ Conf.\ of the South African Inst.\ of Phys., ed.\ by C. Engelbrecht \& S.\ Karataglidis, pp.\ 311$-$316.
\item Breed, M.; Venter, C.; Harding, A. K., 2016, \textit{Very-high energy emission from pulsars}, in Conf.\ Proc.\ of SAIP2015: Proc.\ of the 60$^{th}$ Ann.\ Conf.\ of the South African Inst.\ of Phys., ed.\ by M. Chithambo \& A. Venter, pp. 278$-$283.
\item Barnard, M., Venter, C., \& Harding, A. K., 2017, \textit{High-energy pulsar light curves in an offset polar cap $B$-field geometry}, in Conf.\ Proc.\ of HEASA2016: the 4$^{th}$ Ann.\ Conf.\ on High Energy Astrophys.\ in Southern Africa, ed.\ by M. Boettcher, D. Buckley, S. Colafrancesco, P. Meintjes, \& S. Razzaque, id.\ 42. 
\item Barnard, M., Venter, C., Harding, A.~K., \& Kalapotharakos, C., 2017, \textit{Modelling energy-dependent pulsar light curves due to curvature radiation}, in Conf.\ Proc.\ of HEASA2017: the 5$^{th}$ Ann.\ Conf.\ on High Energy Astrophys.\ in Southern Africa, ed.\ by M. Boettcher, D. Buckley, S. Colafrancesco, P. Meintjes, \& S. Razzaque, id.\ 22.
\item Venter, C., Barnard, M., Harding, A.~K., \& Kalapotharakos, C., 2018, \textit{Modelling energy-dependent pulsar light curves}, in Conf.\ Proc.\ of IAUS No.\ 337: Pulsar Astrophysics - The Next 50 Years, ed.\ Weltevrede, P., Perera, B.~B.~P., Preston, L.~L., \& Sanidas, S., 337, 120$-$123.
\end{enumerate}

\subsection{Journal articles}
\begin{enumerate}
\item Barnard, M., Venter, C., \& Harding, A.~K., 2016, \textit{The Effect of an Offset Polar Cap Dipolar Magnetic Field on the Modeling of the Vela Pulsar's $\gamma$-Ray Light Curves}, ApJ, 832, 107.
\item Harding, A.~K., Kalapotharakos, C., Barnard, M., \& Venter, C., 2018, \textit{Multi-TeV Emission from the Vela Pulsar}, ApJ, 869, L18. \\
My contribution is the calculation of a refined curvature radius $\rho_{\rm c}$ as discussed in Chapter~\ref{chap:ch4}.
\item Barnard, M., Venter, C., Harding, A.~K., \& Kalapotharakos, C., 2020, \textit{Probing the $\gamma$-ray Pulsar Emission Mechanism via Energy-dependent Light Curve Modeling}, in preparation.
\end{enumerate}

\section{Thesis outline}

\textbf{Chapter~\ref{chap:PSRastro}}: This Chapter gives an overview of various topics related to pulsar science, and more specifically, those that are relevant to this study on pulsar emission modelling, e.g., the history of pulsars, their formation, different pulsar classes, standard models of pulsar electrodynamics, important radiation mechanisms, later pulsar emission models, and models of pulsar magnetospheres. \\

\noindent\textbf{Chapter~\ref{chap:OffsetPC}}: 
A summary of a published journal article investigating the implication of magnetospheric structures on pulsar model light curves. Additionally, I studied an offset-PC dipole $B$-field structure and an SG $E$-field solution, and the effect on light curve predictions when I increased such an $E$-field. \\

\noindent\textbf{Chapter~\ref{chap:ch4}}: In this Chapter, I describe the emission modelling code I used to study pulsar emission and explain the implementation of a more refined $\rho_{\rm c}$ calculation. I also discuss additional technical details that this study entails, such as the calibration of the code and getting the parallelised version thereof running on the local cluster. \\

\noindent\textbf{Chapter~\ref{chap:CREdepLCmod}}: This Chapter describes first results that followed from the implementations discussed in Chapter~\ref{chap:ch4} for the Vela pulsar, and include the energy-dependent light curve and spectral modelling (Barnard et al., in prep.). This Chapter highlights the main results of my PhD thesis work, in accordance with the aims set earlier. The results also accompany those discussed in Chapter~\ref{chap:VelaTeV}. \\

\noindent\textbf{Chapter~\ref{chap:VelaTeV}}: Here I emphasise my main contribution to the accompanying VHE paper for the Vela pulsar \citep{Harding2018}. \\

\noindent\textbf{Chapter~\ref{chap:conclusion}}: Summarises the conclusions drawn from this study. \\

% Motivation, publications, outline
\chapter{Pulsar astrophysics}\label{chap:PSRastro}

% Write short introduction
I give an overview of several relevant pulsar topics in order to provide context for the present study. I briefly describe the historical development of the pulsar field (Section~\ref{subsec:PSRhistory}), the mechanism of pulsar formation (Section~\ref{subsec:PSRformation}), different classes of pulsars (Section~\ref{subsec:PSRcat}), the standard braking model that explains the conversion of rotational energy of pulsars into radiation and particle acceleration (Section~\ref{sec:BrakingMod}), the traditional Goldreich-Julian model (Section~\ref{sec:GJmodel}), some relevant radiation mechanisms and pair production (Section~\ref{sec:RadMechanisms}),
and pulsar emission models (Section~\ref{sec:PhysPSRmodels}). Given the fact that this project mainly deals with pulsar magnetospheres and the HE and VHE $\gamma$-ray light curves of the Vela pulsar as measured by the \textit{Fermi} and ground-based telescopes, I lastly describe developments in $B$-field structures and models (Sections~\ref{sec:BandEfields} and~\ref{sec:bmodels}). This Chapter represents an update on what was presented in \citet{Breed2015MSc}. 

%----------------------------------------------------------------------
\section{Pulsar discovery, formation and classes} \label{PSRs}
\subsection{A survey of pulsar history}\label{subsec:PSRhistory}

The neutron was discovered by James Chadwick in 1932 \citep{Chadwick1932}. The concept of a neutron star (NS) originated more or less at the same time. Chandrasekhar studied stellar evolution and discovered that a collapsing stellar core consisting of a mass larger than 1.4 $M_{\odot}$ (the well-known Chandrasekhar limit, applicable to white dwarf stars) should continue collapsing, since it can not balance its own gravity after all its nuclear fuel has been exhausted \citep{Chandrasekhar1931}. \citet{Landau1932} also studied white dwarf stars and speculated on the existence of a star that could be more dense than white dwarf stars, and is described as a gigantic atom. Walter Baade and Fritz Zwicky analysed observations of supernova explosions and discovered that supernovae appeared to be less frequent than common novae, and to emit enormous amounts of energy during each explosion \citep{Baade1934a}. They also observed that supernovae explode faster than novae. Their calculations implied that a supernova remnant can not have a larger radius than a nova. Baade and Zwicky proposed that NSs could form in supernova explosions, since a supernova represents a transition from an ordinary star into a very dense object with a small radius and mass \citep{Baade1934b}. In 1939, Oppenheimer and Volkoff constructed the first models that could describe the structure of an NS, also incorporating general relativity. They stated that NSs are so dense that spacetime is curved around and within them, motivating the importance of general relativistic effects \citep{Haensel2007}. They calculated that stars reaching a mass larger than 3 $M_{\odot}$ (known as the Oppenheimer-Volkoff limit) would undergo gravitational collapse to form a black hole. The concept of NSs was not taken too seriously until the late 1960s when new discoveries were made in high-energy (HE) and radio astronomy \citep{Becker2002}.

Results from HE cosmic-ray experiments implied that there could be astrophysical objects, e.g., supernova remnants, which could produce high-energy cosmic rays as well as X-rays and $\gamma$-rays \citep{Morrison1954,Morrison1958}. In 1962, Rossi and Giaconni confirmed these notions when they detected X-rays from Sco X-1 (a source located in the constellation Scorpio), the brightest X-ray source in the sky \citep{Giacconi1962}. These X-rays were believed to be the result of SR by cosmic electrons carrying energies of the order of tens of keV. \citet{Bowyer1964} detected a second X-ray source Tau X-1, situated in the constellation Taurus. This source coincided with the Crab supernova remnant. Among all the different theories and processes proposed for the origin of these X-rays, \citet{Chiu1964} proposed that this was due to thermal radiation emitted from the surface of a hot NS. Since NSs are expected to appear as point sources and the X-radiation from the Crab supernova remnant had a finite angular size of $\sim1^{\prime}$, the existence of an actual NS still remained uncertain. \citet{Hoyle1964} made the visionary prediction that there could be an NS with a strong $B$-field of $\sim10^{10}$ G at the centre of the Crab Nebula. 

In 1967, Anthony Hewish directed the construction of a radio telescope at the Mullard Radio Astronomy Observatory, which was designed to detect interplanetary scintillation from cosmic sources \citep{Hewish1968}. The first discovery made with this new radio telescope was by Jocelyn Bell, a graduate student from Cambridge University supervised by Hewish. She detected a weak, variable radio source displaying a series of stable periodic pulses \citep{Hewish1968,Hewish1975}. These radio pulses arrived at a precise period of 1.3373012 s. They jestingly called this source ``Little Green Man 1''. After three more similar pulsating radio sources were detected (PSR B1133$+$16, PSR B0834$+$06, PSR B0950$+$08), it became clear that a new kind of natural phenomenon was discovered. Another faster pulsar -- the Vela pulsar -- was discovered in 1968 by the Molonglo group, possessing a pulse period of 0.089 s and situated near the centre of the Vela X supernova remnant \citep{Large1968}. Staelin and Reifenstein discovered two more pulsars in 1968, one of which (the Crab pulsar) was located within $5^{\prime}$ from the centre of the famous Crab Nebula, having a period of 33 ms \citep{Staelin1968}. In the same year that the first known pulsar (PSR B1919+21) was discovered, over 100 theoretical papers were published proposing interpretations or models for pulsars \citep{Will1994}. During this time, \citet{Wheeler1966} and \citet{Pacini1967} proposed that the energy source in the Crab Nebula could possibly be a rapidly rotating, and highly magnetised NS. Gold \citeyearpar{Gold1968,Gold1969} suggested that since supernova remnants are associated with fast rotating NSs, a pulsar is none other than a rotating NS. Therefore, it is believed that NSs are born in core-collapsed supernovae of highly evolved massive stars. \citet{Cocke1969} next discovered strong optical pulses from the Crab pulsar. This important discovery that the ``remnant star'' that survived the Crab supernova explosion \citep{Minkowski1942} was in fact a pulsar, a rapidly rotating NS, therefore solidified the link between supernovae, NSs, and pulsars. Soon after, \citet{Bradt1969} and \citet{Fritz1969} detected X-ray pulsations from the Crab pulsar in the $1.5-10$ keV range, and \citet{Hillier1970} detected $\gamma$-ray pulsations at energies $>0.6$ MeV with a significance of $\sim3.5\sigma$. 

During the mid-seventies $\gamma$-ray astronomy expanded with the launch of two satellites: {\it Small Astronomy Satellite 2 (SAS-2)} in 1972 \citep{Fichtel1975}, which confirmed the existence of $\gamma$-ray emission from the Crab pulsar \citep{Kniffen1974} and the Vela pulsar \citep{Thompson1975}, and \textit{Cosmic Ray Satellite-B (COS-B)} in 1975, which provided a complete detailed map of the $\gamma$-ray sky \citep{Schonfelder2001}. The number of detected radio pulsars also increased rapidly in this era. The idea that pulsars have high $B$-fields ($\sim10^{12}$~G) was confirmed by the \emph{Uhuru} (i.e., {\it Small Astronomy Satellite 1 (SAS-1)}) observation of an accreting X-ray binary pulsar Her X-1 in the constellation Hercules \citep{Tananbaum1972}. A spectral feature at $58$ keV was interpreted as resonant electron cyclotron emission or absorption in the hot polar plasma of the NS, implying a $B$-field of $\sim6\times10^{12}$ G \citep{Truemper1978}. 

The launch of other satellite missions that made important contributions to HE astrophysics, especially isolated NSs, include \textit{High Energy Astrophysical Observatories} (\textit{HEAO 1}, \textit{HEAO 2}, and \textit{HEAO 3}), \textit{Chandra X-ray Observatory}, and \textit{X-ray Multi-Mirror Mission} (\emph{XMM-Newton}; \citealp{Rudak2002}). The field of $\gamma$-ray pulsars has been revolutionised by the launch of \textit{Astro-rivelatore Gamma a Immagini LEggero (AGILE)} and the \textit{Fermi} LAT, which is much more sensitive than its predecessor, \textit{EGRET} \citep{Atwood2009}. Very recently, the ground-based imaging atmospheric Cherenkov telescopes, \textit{Major Atmospheric Gamma-Ray Imaging Cherenkov} ({\it MAGIC}; \citealp{Aleksic2011,Aleksic2012,Aleksic2015,Aliu2008}) and \textit{Very Energetic Radiation Imaging Telescope Array System} ({\it VERITAS}; \citealp{Aliu2011}) detected $\gamma$-ray pulsations from the Crab pulsar up to sev-eral hundred GeV. Furthermore, the H.E.S.S.-II has earlier detected pulsed emission from the Vela pulsar above 20 GeV \citep{Abdalla2018}, and recently up to 7~TeV \citep{Djannati2017}. Pulsed emission was also detected from Geminga between 15~GeV and 75~GeV by MAGIC \citep{Acciari2020}, and PSR~B1706$-$44 in the sub-100~GeV energy range by H.E.S.S.~II \citep{SpirJacob2019}. 

\begin{figure}
\begin{center}
\includegraphics[scale=0.8]{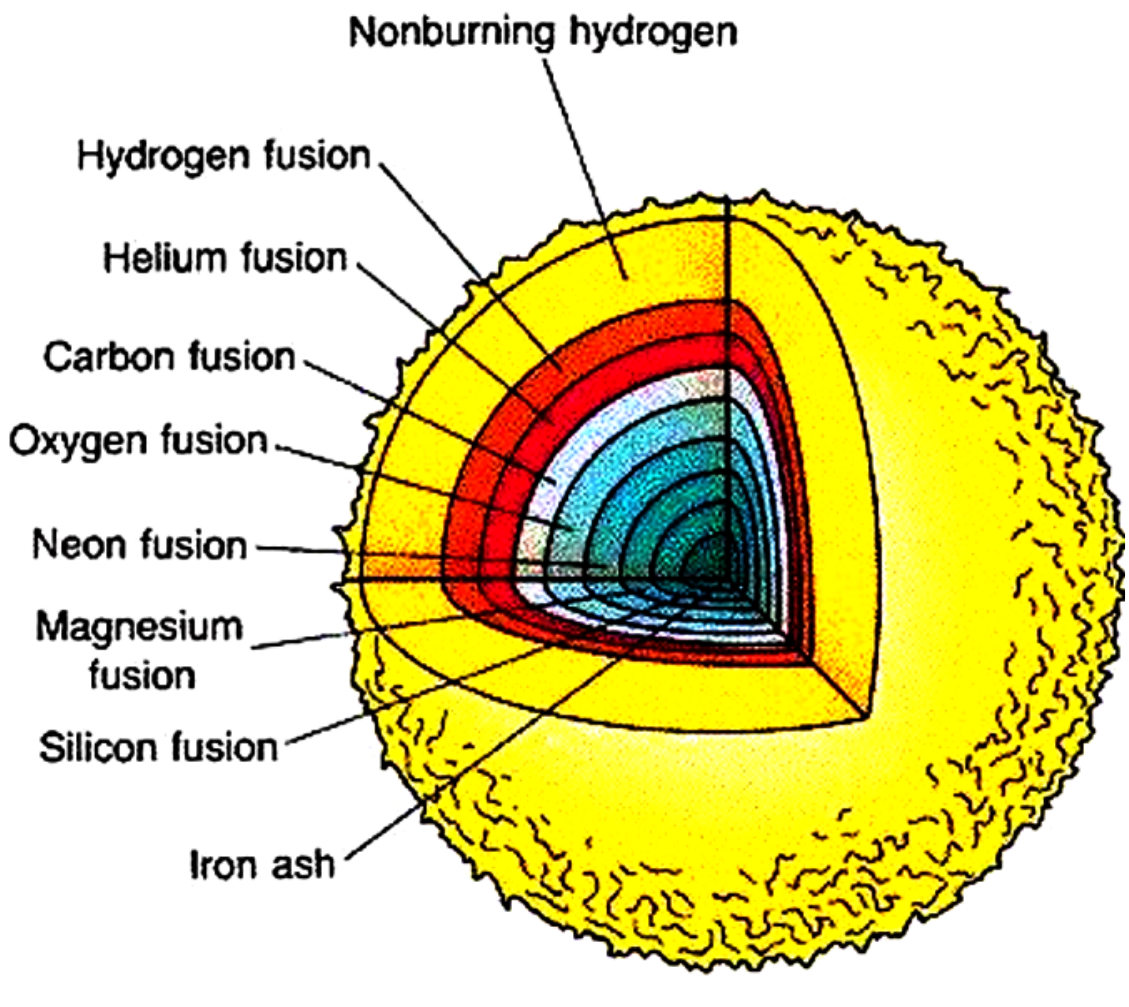}
\end{center}
\caption[Schematic representation of a massive star's interior]{\label{fig:massive_star} Illustration of the chemical composition of a highly evolved massive star, with each layer representing a different element, and an iron core at the centre. From \citet{Chaisson2002}.}
\end{figure}

\subsection{Pulsar formation}\label{subsec:PSRformation}

\begin{figure}
\begin{center}
\includegraphics[scale=0.6]{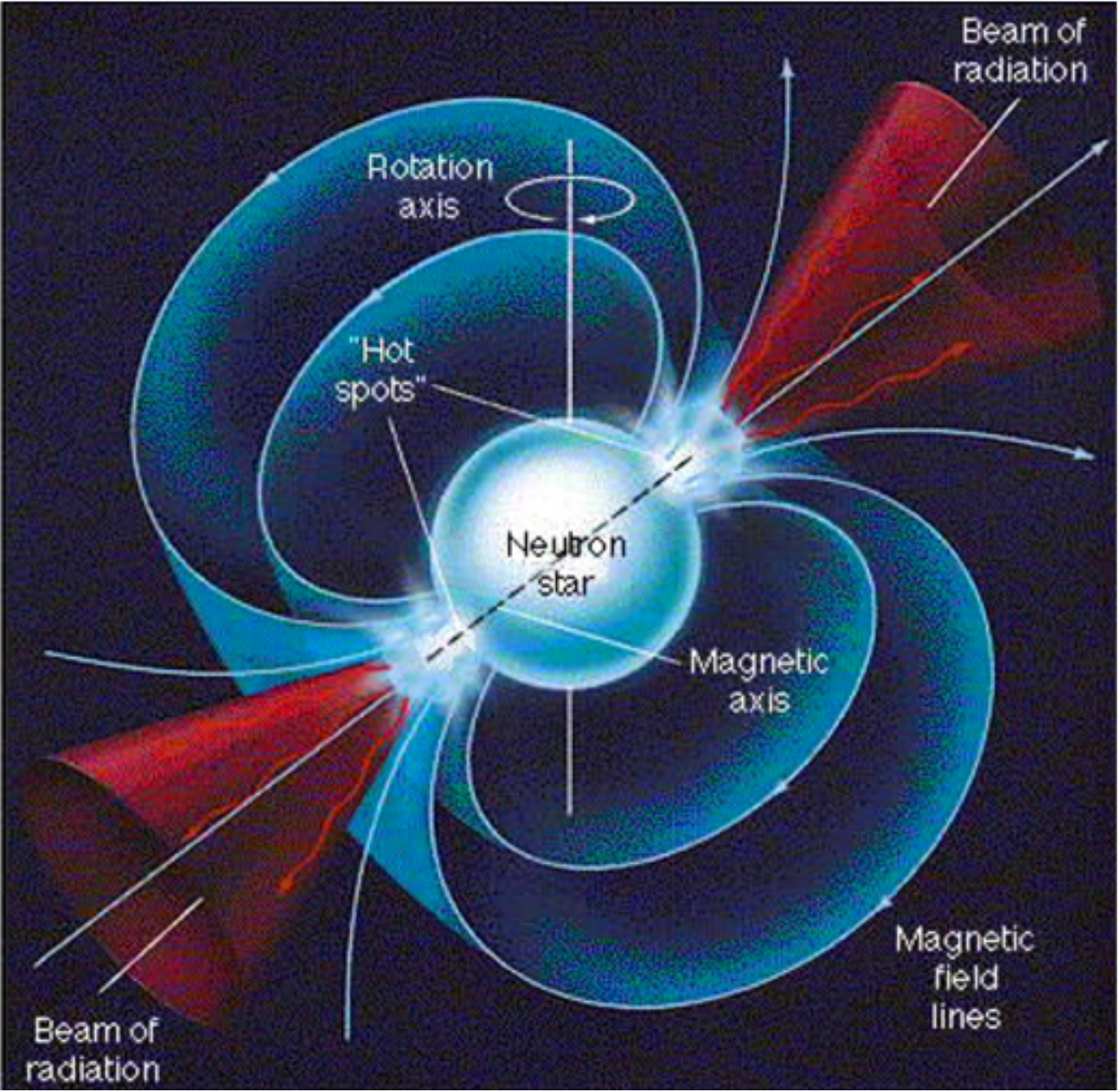}
\end{center}
\caption[The ``lighthouse'' pulsar model]{\label{fig:Lighthouse} A pulsar may be compared to a lighthouse. The charged particles are accelerated along the $B$-field lines of the rotating NS, producing radiation in the form of beams. Figure from \citet{Chaisson2002}.}
\end{figure}

The formation of pulsars is initiated by the death of high-mass ($M>8M_{\odot}$) stars \citep{Chaisson2002}. A high-mass star is made up of various layers of elements, starting with the hydrogen surface, then helium, carbon, oxygen, and other heavier elements at the core, as illustrated in Figure~\ref{fig:massive_star}. There are two mechanisms operating during the burning and evolutionary stages of such stars, namely fusion and fission. Fusion takes place during the burning process. Each element (from the outer layers down to the inner layers) burns its nuclei, causing an increase in temperature with depth. The released nuclear energy produces gas and radiation pressure which counteracts the star's gravity. Once a particular element is exhausted, the burning of a heavier one is initiated by gravitational contraction \citep{Chaisson2002}.

The burning process continues until an iron core is established. Since iron is the most stable element, it serves as the division between operation of the fusion and fission processes. The iron core becomes unstable when the star attempts to contract again and the nuclear reactions (which have been supplying energy) cease, so that all equilibrium is destroyed \citep{Tayler1994}. The gravity exceeds the gas pressure and the core collapses in on itself, causing the central regions to reach high densities and extremely high temperatures. After the collapse, fission takes place and the thermal energy from the core is absorbed to enable the photons to break the iron up into lighter nuclei, which in turn dissociate into protons and neutrons (a process known as photo-disintegration, \citealp{Chaisson2002}). As the temperature and pressure of the core (now consisting of elementary particles) decrease, the gravitational force becomes stronger and the density increases even more, allowing the collapse to continue. The compression inside the core causes the protons and electrons to combine, producing neutrons and neutrinos (the process is known as neutronisation, \citealp{Tayler1994}). These neutrinos escape from the star, carrying energy with them. The pressure decreases again, so that the core collapses to a point were the neutrons make contact with each other, reaching stellar core densities of $\sim10^{15}$ kg m$^{-3}$. Neutron degeneracy pressure now opposes further gravitational collapse, slowing it down. The core contracts, exceeding the equilibrium point, and is accompanied by the release of gravitational binding energy and emission of neutrinos and gravitational waves \citep{Bowers1984}. A ``hydrodynamic bounce'' may occur as the core rebounds and a shock wave will sweep through the star at high speed, outward into the mantle, and may lead to a spectacular supernova explosion \citep{Bowers1984,Tayler1994}. 

Historically supernovae have been divided into two classes, i.e., core-collapse and thermal runaway supernova, from a physical point of view of their mechanism of explosion. Type I supernovae occur in binary systems \citep{Palen2002} involving white dwarfs, and Type II supernovae involve isolated, highly evolved massive stars. When a massive star explodes as a Type II supernova, the remains of the star are carried outward into space by the shock wave. These remains may form a nebula, sometimes observed as being surrounded by a supernova remnant shell. Nebulae are regions of glowing, ionised gas with the brightness of these clouds depending on the brightness of the central degenerate NS \citep{Chaisson2002}.

The maximum predicted mass of an NS is between $1.5M_{\odot}$ and $2.7M_{\odot}$ \citep{Palen2002}. The highest mass observed so far is $2.01\pm0.04M_{\odot}$, for PSR J0348$+$0431 \citep{Antoniadis2013}\footnote{\url{https://greenbankobservatory.org/most-massive-neutron-star-ever-detected/}}. For somewhat higher stellar masses, it is believed that a black hole will be formed after gravitational collapse \citep{Kanbach2001}. NSs are small, very dense objects. According to the law of conservation of angular momentum, a rigidly rotating object will spin faster as it shrinks, implying that the NS rotates very rapidly, with millisecond to subsecond periods, and having strong $B$-fields, e.g., $B\sim10^{8-13}$ G. Such a rapidly, highly magnetised NS is known as a (rotation-powered) pulsar that radiates energy into space. The simplest analogy of a pulsar is a lighthouse, as shown in Figure~\ref{fig:Lighthouse}.

The magnetic poles of the pulsar are known as polar caps (PCs), from where charged particles may be accelerated more or less steadily along the $B$-field lines to very high energies (although newer models prefer the dominant site of acceleration to be the equatorial current sheet - see Section~\ref{sec:bmodels}). The radio radiation is emitted in a searchlight pattern, and as the radio beam sweeps past Earth, a pulse is observed. All pulsars are NSs but not all NSs are (observable as) pulsars, for two reasons. First, an NS only pulses because of a strong $B$-field and rapid rotation, which diminish with time, causing the radio pulses to weaken and occur less frequently. Second, young pulsars are not always visible from Earth because the radio beam is very narrow, and may miss Earth \citep{Chaisson2002}.

\subsection{Pulsar classes}\label{subsec:PSRcat}
%Pulsar populations: young and old, 
%Millisecond, Anomalous X-ray Pulsar (AXPs), Soft gamma-ray repeaters (SGRs or magnetars), Rotating radio transients (RRATs), 
%Binaries (x-ray / blackwidow systems)
%figures: PPdot diagram or from Alice's Zoo article

Pulsars are generally divided into two categories according to the $B$-field and age. Canonical pulsars are young ($\tau\sim10^{3}-10^{6}$ yr) and have high $B$-fields ($B\sim10^{12}-10^{13}$ G), while MSPs are old ($\tau\sim10^{8}-10^{9}$ yr) and are characterised by low $B$-fields ($B\sim10^{8}-10^{9}$~G). Since the launch of several satellite observatories, for instance \textit{R\"{o}ntgensatellit (ROSAT)}, \textit{Extreme Ultraviolet Explorer (EUVE)}, \textit{Advanced Satellite for Cosmology and Astrophysics (ASCA)}, \textit{Rossi X-ray Timing Explorer (RXTE)}, \textit{Chandra}, \emph{XMM-Newton}, and the \textit{Fermi} LAT, the number of detections of rotation-powered pulsars (RPPs, pulsars driven by the rotational energy of the NS) has increased dramatically \citep{Becker2002}. These RPPs have been detected in various energy bands including radio, X-ray, $\gamma$-ray, and optical, enabling the study of multi-wavelength pulsar emission. 

The Crab pulsar is a famous canonical pulsar. Its light curves have been detected in radio, optical, X-ray, and $\gamma$-ray bands, all being phase-aligned \citep{Abdo2010Crab}. Several other pulsars have similar emission properties as those of the Crab pulsar, including B0540$-$69, J0537$-$6909, and B1509$-$58 \citep{Becker2002}. Another well-known example is the Vela pulsar (PSR B0833$-$45), the brightest persistent GeV source in the sky \citep{Abdo2009Vela}. It has a period $P=0.089$ s, period derivative $\dot{P}=1.24\times10^{-13}$ s s$^{-1}$, a characteristic age $\tau=P/2\dot{P}=1.2\times10^{4}$ yr, and it is also one of the closet pulsars to Earth, lying at a distance of $d=287_{-17}^{+19}$ pc \citep{Dodson2003}. Vela was first detected emitting HE pulses by \textit{SAS-2} \citep{Thompson1975}, followed by phase-resolved studies with \textit{COS-B} \citep{Grenier1988} and \textit{EGRET} \citep{Kanbach1994,Fierro1998}. Vela was the first source investigated by \textit{AGILE} \citep{Pellizzoni2009}, and the \textit{Fermi} LAT used the Vela pulsar as a calibration source. Vela-like pulsars (e.g., PSR B0833$-$45, PSR B1706$-$44, PSR B1046$-$58, and PSR B1951$+$32) possess spin-down ages in the range $\sim10^{4-5}$ years and are detected in various wavebands. Another source detected by \textit{SAS-2} and \textit{COS-B} was Geminga, which was identified as a radio-quiet pulsar when the \textit{ROSAT} satellite detected pulsed X-ray emission from it \citep{Halpern1992}. \textit{SAS-2} and \textit{COS-B} confirmed, using a timing solution from \textit{ROSAT} data, that Geminga is also a bright $\gamma$-ray pulsar \citep{Mattox1992}.

A new class of radio pulsars was discovered in 1981 by Backer and his colleagues, following the detection of PSR B1937$+$21, which has a period of 1.56 ms \citep{Backer1982}. MSPs originate from ordinary pulsars that are in binary systems. These normal pulsars ``switch off'' due to continued rotational energy loss, but following angular momentum and mass transfer via accretion from their companion star, they ``switch on'' again and become visible as MSPs \citep{Alpar1982}. MSPs have relatively short spin periods ($P\lesssim10$ ms), small period derivatives ($\dot{P}\sim10^{-21}-10^{-19}$, i.e., they are very stable rotators), large spin-down ages, and low $B$-field strengths compared to those of normal pulsars and magnetars \citep{Alpar1982}. 

An interesting new class of pulsars has recently been discovered. These so-called rotating radio transients (RRATs) are associated with single, dispersed bursts of emission having durations in the range of $2-30$ ms, with the average time interval between bursts ranging from a few minutes to hours. It is suggested that these sources originate from rotating NSs, since radio emission from these objects is usually detectable for $<1$ s per day, with their periodicities ranging between $0.4-7.0$ s \citep{McLaughlin2006}. RRATs may be examples of pulsars whose magnetospheres switch between several stable configurations \citep{Keane2011}.

Magnetars, including anomalous X-ray pulsars (AXPs) and soft $\gamma$-ray repeaters (SGRs), are NSs that have extremely strong surface $B$-fields of $B\sim10^{14-15}$ G, increasing in strength from the surface down to the core \citep{Duncan1992}. These sources are also characterised by burst-like emission. They exhibit very strong X-ray emission, which is too high and variable to be explained by conversion of rotational energy alone, but possibly involve the decay and instability of their enormous $B$-fields \citep{Rea2011}. They have long rotation periods that range from $2-12$ s (exceeding those of radio pulsars), as well as large period derivatives ($\dot{P}\sim10^{-13}-10^{-9}$~s~s$^{-1}$; \citealp{Mereghetti2008}).

%----------------------------------------------------------------------
\section{Standard braking model for rotation-powered pulsars} \label{sec:BrakingMod}

Let us consider the NS to be a rapidly rotating object possessing a dipolar $B$-field. This NS has an angular momentum $J\approx M_{i}R_{i}^{2}{\mathrm{\Omega}}_{i}$, which is assumed to be conserved during the collapse of the progenitor,
with $M_{i}$, $R_{i}$, and $\Omega_{i}=2\pi/P_{i}$ the initial mass, radius, angular velocity, and $P_{i}$ the initial rotational period. The relation between the initial and final angular velocity is therefore (since $M_{i}\approx M_{f}$)
\begin{equation}
\Omega_{f}\sim\Omega_{i}\biggl(\frac{R_{i}}{R_{f}}\biggr)^{2}.
\end{equation}
This relation states that for values $R_{i}>R_{f}$ the angular velocity increases so that the rotational period $P_{f}$ becomes much shorter, ranging from milliseconds up to seconds. The interior of the NS is assumed to be fully conductive, implying conservation of the magnetic flux $\Phi=\oint{\mathbf{B}}\cdot{d{\rm {\mathbf{a}}}}\sim B_{i}R_{i}^{2}$ during the collapse of the core. The magnitude of the final $B$-field is then given by

\begin{equation}
B_{f}\sim B_{i}\biggl(\frac{R_{i}}{R_{f}}\biggr)^{2}.
\end{equation}
From this relation it follows that for $R_{i}>R_{f}$ the $B$-field will increase, yielding high values of $B_{f}\sim10^{12}$ G. The collapse of a compact neutron core therefore leads to high magnetic strengths and short periods. The rotational energy of the pulsar will be converted into electromagnetic and particle energy, leading to a slower rotational rate. The basic outcome of this rotation-powered pulsar model is to predict the rate at which this slow-down occurs. The angular kinetic energy of the rotating NS is given by
\begin{equation}
E_{{\rm rot}}=\frac{1}{2}I\Omega^{2},
\end{equation}
with $I\sim MR^{2}$ the moment of inertia. In this model the polar $B$-field strength at the stellar surface can be estimated by equating the rotational energy loss rate to the magnetic dipole radiation loss rate $L_{{\rm md}}$ \citep{Ostriker1969} 
\begin{equation}\label{eq:Edot}
\dot{E}_{{\rm rot}}=\frac{d}{dt}\Bigl(\frac{1}{2}I\Omega^{2}\Bigr)=I\Omega\dot{\Omega}=-\frac{4\pi^{2}I}{P^{3}}\dot{P}{\approx}L_{{\rm md}}=-\frac{2}{3c^{3}}\mu^{2}\Omega^{4}\sin^{2}\alpha,
\end{equation}
with $\mu\equiv B_{0}R^{3}/2$ the magnetic moment of the dipole, $\dot{P}$ the time derivative of the period in s$\,$s$^{-1}$, $B_{{\rm 0}}$ the surface $B$-field strength (polar $B$-field strength in Gaussian units), $R$ the stellar radius, $\alpha$ the inclination angle between the magnetic and spin axes of the NS, and $c$ the speed of light. The magnitude of $B_{{\rm 0}}$ can now be estimated by inserting typical values of $I=10^{45}$ g cm$^{2}$, $R=10^{6}$ cm and $\alpha\sim90^{\circ}$, giving %\citep{Manchester1977}
%\begin{equation}\label{B0}
%B_{0}\approx4.5\times10^{19}\sqrt{P\dot{P}}.
%\end{equation}
%For Eq.~\eqref{B0}, we neglected the term associated with radiation ($L_{\gamma}$), since this is thought to be much smaller than $L_{{\rm pf}}$ (particle acceleration) and $L_{{\rm md}}$. For the vacuum case one may set $\dot{E}_{\rm rot}=L_{\rm md}$ which yields the more familiar expression
\begin{equation}\label{eq:B0}
B_{0}\approx6.4\times10^{19}\sqrt{P\dot{P}}.
\end{equation}
Later calculations by, e.g., \citet{Spitkovsky2006,Li2012} resulted in $L_{\rm pf}\propto\left({1+\sin^{2}\alpha}\right)$, the Poynting flux. By equating this $\dot{E}_{{\rm rot}}$ yields a similar value for $B_{0}$.

We can estimate the pulsar rotational (characteristic) age as follows. Assume that the change in $\dot{\Omega}=-K\Omega^{(n-1)}$ is due to magnetic dipole radiation losses (Bowers \& Deeming 1984), where $K$ is a positive constant, and the parameter $n=\ddot{\Omega}\Omega/\dot{\Omega}^{2}$ is the braking index, which comes from differentiating the equation for $\dot{\Omega}$. This expression for $\dot{\Omega}$ is motivated by Eq.~\eqref{eq:Edot}, assuming that $\mu_{\perp}\equiv\sin\alpha$ stays constant. Next, integrate this expression and substitute $\Omega^{2}=-{\dot{\Omega}}/{k_{1}\Omega}$ where $k_{1}$ is a constant (see Eq.~[\ref{eq:Edot}]). The characteristic age is then given by \citep{Manchester1977}
\begin{equation}\label{eq:pulsar-age}
\tau=-\frac{\Omega}{(n-1)\dot{\Omega}}\biggl[1-\Bigl(\frac{\Omega}{\Omega_{0}}\Bigr)^{n-1}\biggr]\approx-\frac{\Omega}{(n-1)\dot{\Omega}}\equiv\frac{P}{(n-1)\dot{P}},
\end{equation}
with the assumptions $n\neq1$ and $\Omega\ll\Omega_{{\rm 0}}$, with $\Omega_{{\rm 0}}$ the angular velocity at time $t=0$. This is approximately equal to
\begin{equation}\label{eq:pulsar-age-1}
\tau\approx-\frac{\Omega}{2\dot{\Omega}}\equiv\frac{P}{2\dot{P}},
\end{equation}
when setting $n=3$ for the case of magneto-dipole braking \citep{Becker2002}. This characteristic age serves as an upper limit for the true age of the pulsar, since the value for $n$ is chosen to be a constant. However, when $\Omega\lesssim\Omega_{{\rm 0}}$, the true age of the pulsar will be smaller than $\tau$. 

\begin{figure}
\begin{center}
\includegraphics[scale=0.4]{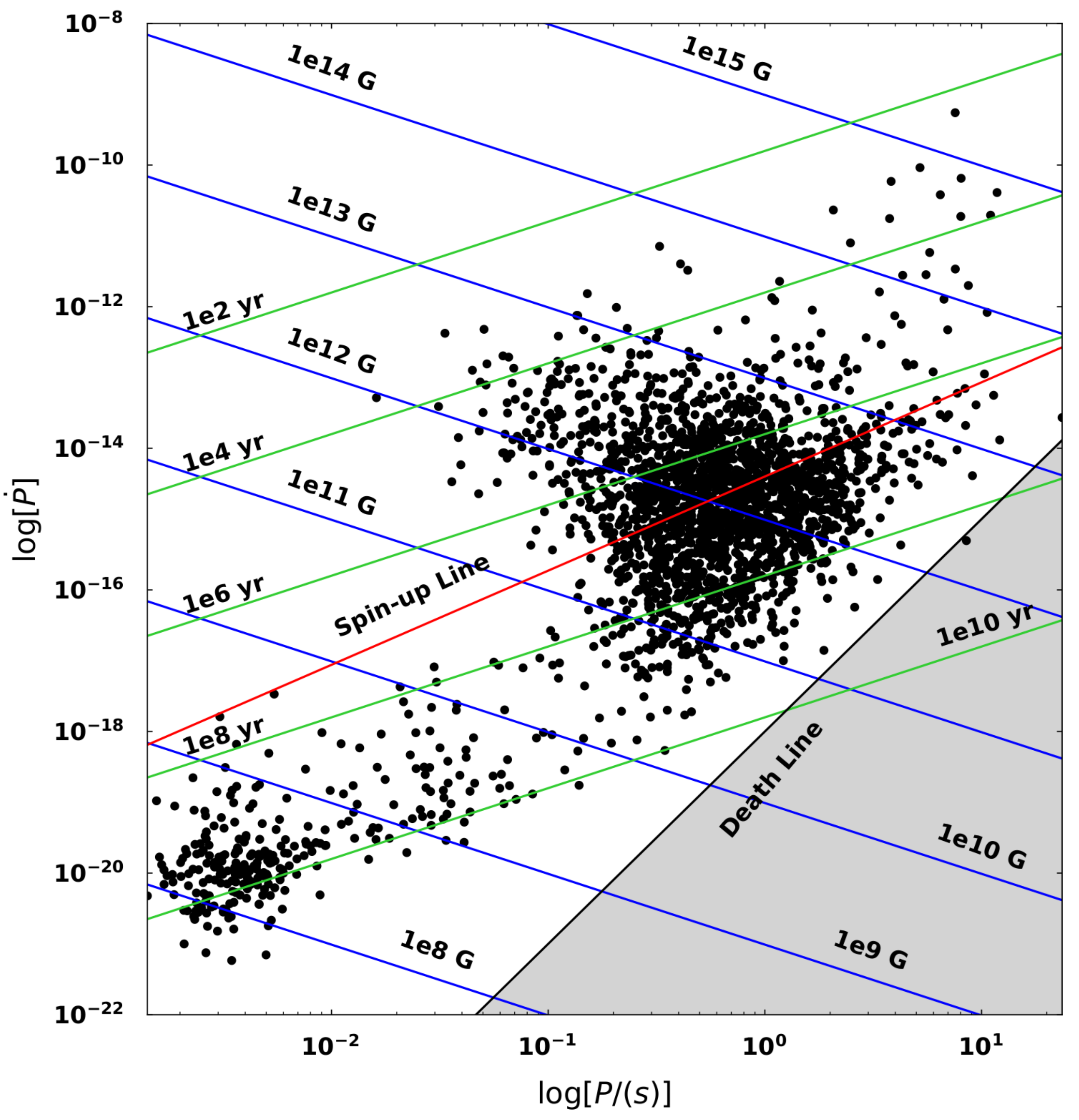}
\end{center}
\caption[A $P\dot{P}$-diagram representing the canonical and millisecond pulsar populations]{\label{fig:PPdot} A $P\dot{P}$-diagram indicating the two pulsar populations including the canonical pulsars (in the centre) and the MSPs (in the lower left corner). The black dots are radio pulsars from the Parkes Observatory \textit{ATNF} Pulsar Catalogue for $\dot{P}>0$ \citep{Manchester2005}. The blue solid lines represent constant surface magnetic field $B_0$ contours, while the green solid lines represent characteristic pulsar ages $\tau$. The grey area is the ``graveyard'' where the canonical pulsars turn off and are spun up again so that they eventually enter the MSP region. The spin-up line (red line) is the equilibrium period of spin-up by accretion, which is the Keplerian orbital period at the Alfv\'{e}n radius \citep{Alpar1982}.}
\end{figure}

The evolution and properties of different pulsar populations are best described by drawing a $P\dot{P}$-diagram (Figure~\ref{fig:PPdot}, the time derivative of the period $\dot{{\it P}}$ versus ${\it P}$, using the pulsars from the Parkes Observatory \textit{ATNF} Pulsar Catalogue for $\dot{P}>0$; \citealp{Manchester2005}). Rotation-powered pulsars could also have $\dot{P}<0$, e.g., when there is acceleration along the line of sight for such objects embedded in a globular cluster. As mentioned in Section~\ref{subsec:PSRcat}, one can distinguish two pulsar populations: the canonical pulsars and MSPs. The canonical radio pulsar population is identified with the younger pulsars and is situated at the centre of the $P\dot{P}$-diagram. The canonical pulsars typically have high surface magnetic fields of $B_{0}\sim10^{12}-10^{13}$~G and rotational ages of $\tau\sim10^{3}-10^{6}$ yr (as indicated by the contours of constant $B_{0}$ and $\tau$). During the evolution of pulsars as they age, three things happen. First, the magnetic dipole field drops (although the timescale for this process is uncertain), second, the pulsar slows down due to energy losses (mostly by dipole radiation and particle loss), causing the pulse period to increase, and lastly the particles emitted by the pulsar form a pulsar wind. On the $P\dot{P}$-diagram there is a ``death valley'' where the canonical pulsars turn off \citep{Chen1993}. This turn-off is due to the fact that the PC potential responsible for electron-positron ($e^{\pm}$) pair creation and subsequent radio emission becomes too low, inhibiting pair production (see Section~\ref{subsec:PairProduct}), and leading to the ``death'' of canonical radio pulsars (i.e., they become invisible). Some pulsars inside the death valley are spun up again by the transfer of mass and angular momentum from a binary companion \citep{Alpar1982}, so that they enter the MSP region (lower left corner). These MSPs have relatively short periods ($P\lesssim{10}$ ms) and lower surface $B$-fields ($B_{0}\sim{10^{8}-10^{9}}$ G) compared to the canonical pulsars. The spin-up line, representing the spin-up upper limit of MSPs (via accretion), is also indicated. 

%----------------------------------------------------------------------
\section{The Goldreich-Julian model}\label{sec:GJmodel}

\begin{figure}
\begin{center}
\includegraphics[scale=0.9]{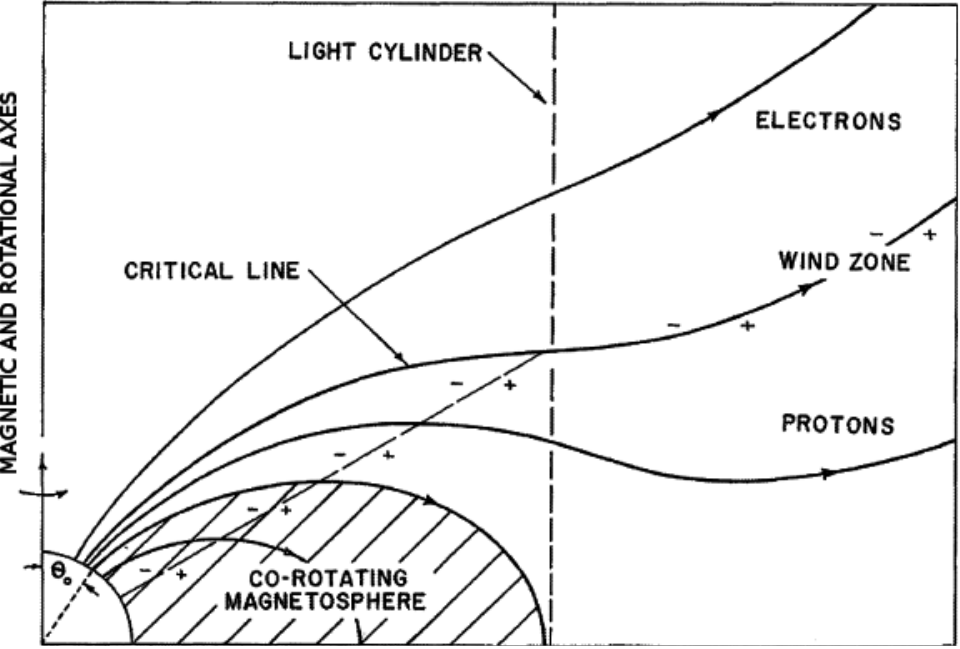}
\end{center}
\caption[The pulsar magnetosphere as envisioned by \citet{GJ1969}]{\label{fig:GJMagneto} The pulsar magnetosphere as envisioned by \citet{GJ1969}. The corotating zone is represented by the shaded region within the light cylinder, where the particles corotate with the closed $B$-field lines. The $B$-field lines which go beyond the light cylinder are forced open and the particles escape along them. The electrons flow out near the PC along the higher-latitude lines, whereas the protons (or possibly iron nuclei) flow out near the PC angle ($\theta_{\rm PC}$) along the lower-latitude lines. These two magnetospheric regions are separated by the critical field line (which is at the same potential as the interstellar medium). The dashed line represents the condition where the charge density $\rho_{{\rm GJ}}\propto-{\boldsymbol{\Omega}}\cdot{\bf B}=0$. Above this dashed line ${\boldsymbol{\Omega}}\cdot{\bf B}>0$ (negative $\rho_{{\rm GJ}}$), and below it ${\boldsymbol{\Omega}}\cdot{\bf B}<0$ (positive $\rho_{{\rm GJ}}$).}
\end{figure}

In 1969, Peter Goldreich and William Julian studied a simple model describing the properties of the magnetosphere around a highly magnetised, rotating pulsar. In this model, they considered an NS to be a uniformly magnetised, perfectly conducting sphere, with an internal magnetic field ${\bf{B}}_{\rm in}=B_{0}\vec{e}_{z}\parallel{\boldsymbol\mu}$, and with an external dipole $B$-field (e.g., \citealp{Padmanabhan2001}). They considered an aligned rotator, i.e., the rotation axis being aligned with a magnetic dipole vector ($\boldsymbol{\Omega}\parallel\boldsymbol{\mu}$; see Figure~\ref{fig:GJMagneto}). Another assumption is that there are initially no charges filling the surrounding magnetosphere \citep{Meszaros1992}.

As the pulsar rotates with a velocity ${\bf v}={\boldsymbol{\Omega}}\times{\bf r}$, the charged particles at the stellar surface will experience a Lorentz force $(q/c)({\bf v}\times{\bf B}_{{\rm in}})$, with $q$ the particle charge. Since the NS is a perfect conductor (${\bf E}_{{\rm in}}\cdot{\bf B}_{{\rm in}}=0$, implying that the $B$-field lines are equipotentials) the charges will be redistributed in order for the electric force to counter balance the magnetic force, leading to charge separation. This implies 
\begin{equation}\label{eq:E-in-field}
{\mathbf{E}}_{\rm in}=-\frac{({\boldsymbol{\Omega}}\times{\bf r})\times{\bf B}_{{\rm in}}}{c}=-\frac{\Omega B_{0}r\sin\theta}{c}(\sin\theta\vec{e_{r}}+\cos\theta\vec{e_{\theta}}).
\end{equation}
Since ${\rm \nabla}\times{\rm {\bf E}_{in}}=0$, we can write
\begin{equation}
{\rm {\bf{E}}_{in}}=-\nabla\Phi_{{\rm in}}(r,\theta),
\end{equation}
with $\Phi_{{\rm in}}(r,\theta)$ the electric potential. After integration, we find 
\begin{equation}
{\Phi}_{\rm in}(r,\theta)=\frac{\Omega B_{0}r^{2}}{2c}\sin^{2}\theta+\Phi_{0},
\end{equation}
with ${\Phi}_{0}$ a constant. This implies a potential difference between the magnetic axis and PC angle (colatitude of the PC rim, $\theta_{\rm PC}\sim\sqrt{\Omega{R}/c}$) of
\begin{equation}
\Delta{\Phi_{\rm in}}=\frac{1}{2}\left(\frac{\Omega R}{c}\right)^{2}B_{0}R.
\end{equation}
The external $E$-field now follows by solving the Laplace equation and requiring a continuous electric potential at the stellar surface:
\begin{eqnarray}
E_{r{\rm ,out}} & = & -\frac{9Q}{r^{4}}\left(\cos^{2}\theta-\frac{1}{3}\right), \\
E_{\theta{\rm ,out}} & = & -\frac{6Q}{r^{4}}\cos\theta\sin\theta,
\end{eqnarray}
with $Q=B_{0}\Omega R^{5}/6c$. Using these expressions for $E_{{\rm out}}$ it follows that the electric force on surface charges vastly exceeds the gravitational force (by a factor of $\sim5\times10^8B_{12}P$ for a proton and $\sim8\times10^{11}B_{12}P$ for an electron with $B_{12}=B_{0}/10^{12}$~G; \citealp{GJ1969}). This constitutes an existence proof for a plasma-filled pulsar magnetosphere, since the accelerating $E$-field parallel to the local $B$-field ($E_\parallel$) will extract particles from the stellar surface to fill the magnetosphere. 

An expression for the charge density in the corotating magnetosphere follows from Eq.~\eqref{eq:E-in-field}
\begin{equation}\label{eq:rhoGJ}
\rho_{{\rm GJ}}=\frac{\nabla\cdot{\bf E}}{4\pi}\approx-\frac{\boldsymbol{\Omega}\cdot{\bf B}}{2\pi c}.
\end{equation}
This implies a number density of 
\begin{equation}\label{eq:numberdensity}
n_{\rm e}=7\times10^{-2}\frac{B_{0}}{P}{\rm \, cm^{-3}},
\end{equation}
at the stellar surface. Despite its success, the model has a few problems, most notably the question of the return current (charge neutrality) and its inherent instability, as well as the charge supply (which cannot be only from the NS surface). 

%----------------------------------------------------------------------

\section{High-energy radiation mechanisms and pair creation processes}\label{sec:RadMechanisms}

\subsection{Particle acceleration}\label{subsec:pacc}

Charged particles that are accelerated will emit electromagnetic radiation. If the speed $v$ of the charged particle is much less than the speed of light in vacuum $c$, i.e., ${v}{\ll}{c}$, the particle is non-relativistic. The power radiated by such charged particles in the non-relativistic regime is calculated by using the Larmor formula \citep{Jackson1999}
\begin{equation}
{P_{\rm total}}=\frac{2q^{2}a^{2}}{3c^{3}},
\end{equation}
with $q$ the charge of the particle and $a$ its acceleration. 

However, when charged particles are accelerated to extremely high energies (GeV$-$TeV), they will emit HE $\gamma$-ray photons, e.g., those detected by \textit{Fermi}. At these HEs the particle's speed becomes relativistic ($\beta\equiv v/c\approx1$) with a Lorentz factor of $\gamma{\equiv}1/\sqrt{1-\beta^{2}}\gg1$. The relativistic Larmor formula (or Li\'{e}nard formula, see \citealp{Jackson1999}) for these HE particles is as follows 
\begin{equation}\label{eq:Lienard-formula}
P_{\rm total}=\frac{2q^{2}}{3c^{3}}\gamma^{4}(a_{\perp}^{2}+\gamma^{2}a_{\parallel}^{2}),
\end{equation}
with $a_{\perp}$ the perpendicular acceleration component and $a_{\parallel}$ the parallel acceleration component (with respect to the particle's velocity direction). In the following subsections, radiation mechanisms including synchrotron radiation (SR), curvature radiation (CR), and inverse Compton scattering (ICS), which are relevant for HE pulsar emission models, are discussed. The first two are due to relativistic particles that are accelerated along curved paths inside the magnetosphere, whereas the latter occurs due to the interaction between photons and the relativistic particles. In the last subsection we discuss pair production, where an HE photon converts into an electron and positron pair.

\subsection{Synchrotron radiation}\label{subsec:SR}
 
SR (magneto-bremsstrahlung) occurs when relativistic charged particles gyrate about a $B$-field line. For non-relativistic particles, this is known as cyclotron radiation. When the particle's perpendicular momentum becomes relativistic, it is known as SR \citep{Rybicki1979}. Neglecting radiation losses, the equation of motion for a relativistic particle reveals that the particle travels at a constant speed parallel to the $B$-field with an acceleration perpendicular to the $B$-field. This implies that the particle will follow a helical path as it gyrates along a $B$-field. The gyration angular frequency (rotation around a field line) is given by \citep{Rybicki1979}
\begin{equation}
\omega_{{\rm B}}=\frac{qB}{\gamma mc},
\end{equation}
with $m$ the particle's mass, and $B$ the magnitude of the $B$-field. If ${\bf v}\cdot{\bf B}=0$ the gyroradius is 
\begin{equation}
r_{{\rm B}}=\frac{v}{\omega_{{\rm B}}}.
\end{equation}
Since the particle is accelerated it will emit radiation and the assumption of no radiation losses will no longer be valid. The total SR energy loss rate is given by
\begin{equation}
\dot{E}_{{\rm SR}}=\frac{2}{3c}(r_{0}\gamma Bv_{\perp})^{2},
\end{equation}
with $v_{\perp}$ the charged particle's speed perpendicular to the $B$-field \citep{Blumenthal1970} and $r_{0}\equiv e^{2}/m_{e}c^{2}$ the classical electron radius (with $m_{e}$ the electron mass and $m_{e}c^{2}$ its rest-mass energy). For the gyrating component we assume $a_{\perp}=\omega_{{\rm B}}v_{\perp}$ and $a_{\parallel}=0$, then Eq.~\eqref{eq:Lienard-formula} is the total emitted radiation
\begin{equation}\label{eq:P-total}
P_{\rm total}=\frac{2q^{2}}{3c^{3}}\gamma^{4}\left(\frac{qB}{\gamma mc}\right)^{2}v_{\perp}^{2}.
\end{equation}
When Eq.~\eqref{eq:P-total} is averaged over all angles, for an isotropic distribution of velocities, the SR power emitted is \citep{Padmanabhan2000} 
\begin{equation}\label{eq:P-SR}
P_{\rm SR,total}=\frac{4}{3}\sigma_{{\rm T}}(c\beta^{2}\gamma^{2})U_{{\rm B}}\propto E_{\rm e}^{2}B^{2},
\end{equation}
with $\sigma_{{\rm T}}\equiv8\pi r_{0}^{2}/3$ the Thomson cross section, $E_{\rm e}$ the particle energy, and $U_{{\rm B}}=B^{2}/8\pi$ the magnetic energy density.

The radiation emitted by these relativistic particles will be beamed into a cone with an angular width $\sim1/\gamma$ around the velocity direction. Since the particle's acceleration and velocity are perpendicular for SR, the observed pulses are a factor of $\gamma^{3}$ shorter in time than the gyration period, leading to a broader spectrum with a maximum characterised by a critical frequency
\begin{equation}
\omega_{{\rm c}}=\frac{3}{2}\gamma^{3}\omega_{{\rm B}}\sin\alpha^{{\rm P}},
\end{equation}
with $\alpha^{{\rm P}}=\arctan(v_{\perp}/v_{\parallel})$ the pitch angle \citep{Rybicki1979}. The total SR power per unit frequency emitted by a single electron is
\begin{equation}
P_{\rm SR}(\omega)=\frac{\sqrt{3}}{2\pi}\frac{q^{3}B}{mec^{2}}\sin\alpha^{{\rm P}}F\left(\frac{\omega}{\omega_{{\rm c}}}\right),
\end{equation}
with
\begin{equation} \label{eq:SRBesselfunction}
F(x)\equiv x\int_{x}^{\infty}K_{5/3}(\xi)d\xi,
\end{equation}
where $K_{5/3}$ is the modified Bessel function of order 5/3, and
\begin{equation}\label{eq:SRBesselfn}
F(x)\sim\Biggl\{\begin{array}{ll}
\frac{4\pi}{\sqrt{3}\Gamma(\frac{1}{3})}\left(\frac{x}{2}\right)^{1/3} & \: x\ll1\\
(\frac{\pi}{2})^{1/2}e^{-x}x^{1/2} & \: x\gg1,
\end{array}
\end{equation}
with $x=\omega/\omega_{c}$. For $\omega\ll\omega_{c}$, $F\propto\omega^{1/3}$, while for $\omega\gg\omega_{c}$, $F\propto e^{-(\omega/\omega_{c})}\omega^{1/2}$.

In many astrophysical sources, the photon spectra reveal a power law distribution of energies. Assume that the number density $N(E_{\gamma})$ of particles over some energy range $(E_{\rm e},E_{\rm e}+dE_{\rm e})$ can be described by a power law $N(E_{\rm e})dE_{\rm e}=CE_{\rm e}^{-p}dE_{\rm e}$, with $C$ a constant and $p$ the power-law index of the emitting particles. Following \citet{Rybicki1979}, the total SR power radiated per unit volume per unit frequency can be shown to be a power-law spectrum
\begin{eqnarray}\label{eq:SR-spectrum}
P_{\rm SR}(\omega) & \propto & \omega^{-s},
\end{eqnarray}
and is only valid between the minimum and the maximum cutoff frequencies depending on the minimum and maximum values for $\gamma$, and with $s=(p-1)/2$ the index of the energy spectrum. The latter relation implies that the injection and radiation spectral indices are related in this case. 

SR is an important process for pulsars. For example, in PC and SG models primary photons are emitted via CR and undergo magnetic photon absorption (see Section~\ref{subsec:PairProduct}) to create $e^{\pm}$ pairs. The perpendicular energy from these secondary pairs is converted to HE radiation via SR. It is possible that radio photons are absorbed by charged particles present in the $B$-field via the process of synchrotron self-absorption \citep{Harding2008}. The above discussion is only valid for $B$-field strengths $B<4\times10^{12}$ G. For larger $B$-fields, a quantum SR approach is necessary (e.g., \citealp{Sokolov1968,Harding1987,Harding2006}).

\subsection{Curvature radiation}\label{subsec:CR}

CR is the radiation process associated with relativistic particles that are constrained to move along a curved $B$-field line. This implies that its perpendicular velocity component $v_{\perp}=0$, and $\alpha^{{\rm P}}=0$ (see above Sections for definitions). CR is therefore linked to a change in longitudinal kinetic energy with respect to the $B$-field, as opposed to SR, where there is change in transverse energy (see Figure~\ref{fig:Pair-production}). These two processes in fact represent two limits of the more general synchro-curvature (SC) process \citep{Torres2018}. In some pulsar models, primary particles are accelerated from the stellar surface along the open field lines. The kinetic energy longitudinal to the $B$-field will exceed the transverse energy (which will be radiated away very rapidly via SR), and therefore CR will be more important than SR regarding energy loss of primary particles \citep{Sturrock1971}. The curvature radius is the instantaneous radius of curvature of the particle trajectory, i.e., $\rho=\rho_{\rm c}$. The critical frequency is then defined as \citep{Daugherty1982,Story2007,Venter2009}
\begin{equation}\label{eq:CR-critfreq}
\omega_{\rm CR}=\frac{3c}{2\rho_{\rm c}}\gamma^{3},
\end{equation}
and the critical energy
\begin{equation}\label{eq:energy-CR}
E_{{\rm CR}}=\hbar\omega_{{\rm CR}}=\frac{3\hbar c\gamma^{3}}{2\rho_{\rm c}}=\frac{3\lambdabar_{\rm c}\gamma^{3}}{2\rho_{\rm c}}m_{e}c^{2},
\end{equation}
where $h=6.626\times10^{-27}$ erg s$^{-1}$ is Planck's constant, $\lambdabar_{\rm c}\equiv\hbar/m_{e}c$ (with $\hbar=h/2\pi$ and $\lambdabar_{\rm c}=\lambda_{\rm c}/2\pi$), and $\lambda_{\rm c}$ the Compton wavelength. The instantaneous power spectrum (in units of erg s$^{-1}$ erg$^{-1}$) is given by (e.g., \citealp{Venter2010})
\begin{equation}\label{eq:P-total-CR}
\left(\frac{dP}{dE}\right)_{{\rm CR}}=\frac{\sqrt{3}\alpha_{f}\gamma c}{2\pi\rho_{\rm c}}F\left(\frac{E_{\gamma}}{E_{{\rm CR}}}\right),
\end{equation}
with $\alpha_{f}$ the fine structure constant, $K_{5/3}$ the modified Bessel function of order 5/3, $x=E_{\gamma}/E_{{\rm CR}}$, with $E_{\gamma}$ the photon energy and $F$ given by Eq.~(\ref{eq:SRBesselfunction}). Similar to SR, for $E_{\gamma}\ll E_{{\rm CR}}$, $F\propto E_{\gamma}^{1/3}$, while for $E_{\gamma}\gg E_{{\rm CR}}$, $F\propto e^{-(E_{\gamma}/E_{{\rm CR}})}E_{\gamma}^{1/2}$ (see Eq.~[\ref{eq:SRBesselfn}], \citealp{Erber1966}). The total power radiated by the electron primary can be determined by integrating Eq.~\eqref{eq:P-total-CR} over energy. The latter is equal to the total CR loss rate of electrons,
\begin{equation}\label{eq:CR-lossrate}
\dot{E}_{{\rm CR}}=\frac{2e^{2}c\gamma^{4}}{3\rho_{\rm c}^{2}},
\end{equation}
with $e$ the electron charge. 

Traditionally, HE emission in standard pulsar models is believed to be from CR of primary electrons accelerated tangentially to the $B$-field in the radiation-reaction regime. The curvature radiation reaction (CRR) limit is reached when the energy gained via acceleration of relativistic electrons (by an $E$-field parallel to the $B$-field) is equal to the energy loss via radiation, and can be expressed as follows (e.g., \citealp{Harding2005c})
\begin{equation}\label{eq:RR-eq}
c|E_{\parallel}|\sim\frac{2ce\gamma^{4}}{3\rho_{\rm c}^{2}},
\end{equation}
yielding $\gamma=(1.5E_{\parallel}/e)^{1/4}\rho_{\rm c}^{1/2}$, the Lorentz factor corresponding to radiation reaction. For pulsars with surface magnetic field strengths $B_0 \sim10^{12}$ G and electric potentials $\Phi\sim10^{13}$ V, the $E$-field strength is $E_{\parallel}\sim10^{4}$ ${\rm statvolts/cm}$ (for young pulsars $E_{\parallel}>10^{4}$ ${\rm statvolts/cm}$) and depends on $B$ and $P$. In these strong fields, the CR spectral cutoffs are therefore around a few GeV for emitting particles with Lorentz factors of $\gamma\sim10^{7}$ \citep{Yadigaroglu1997}. These high Lorentz factors are connected to beamed radiation in the form of a cone with an opening angle $\sim1/\gamma\ll1$, implying emission tangentially to the $B$-field lines. We used this approximation to simplify the geometric models described in Section~\ref{sec:PhysPSRmodels} and Chapter~\ref{chap:OffsetPC}. Given the fact that we expect spectral cutoffs in the GeV range for typical pulsar parameters, as well as rather hard power-law low-energy tails, this process has become the standard explanation for HE pulsar spectra such as those observed by the\emph{ Fermi} LAT satellite (e.g., \citealp{Abdo2013SecondCat}). 

\subsection{Inverse Compton scattering and synchrotron self-Compton scattering} \label{subsec:ICS}

Compton scattering involves the collision between HE photons and low-energy electrons, where the photons transfer some of their momentum $p=hf/c=h/\lambda$ (with $h$ Planck's constant, $f$ the frequency and $\lambda$ the wavelength) and energy to the electrons. This transfer leads to an increase in photon wavelength, implying a lower photon energy. The inverse case of Compton scattering is ICS, where the HE electrons scatter the low-energy photons, resulting in photons with very high energies (i.e., ``boosting'' of photon energies). 

When a relativistic electron with Lorentz factor $\gamma$ upscatters a photon from a low energy to a high energy, the energy $E_{\gamma}$ of the Compton-boosted photon, with an initial energy $\epsilon$, may be approximated as \citep{Ramanamurthy1986} 
\begin{eqnarray}\label{eq:ICS-limits}
E_{\gamma}\sim\epsilon\gamma^{2}, & \gamma\epsilon\ll m_{e}c^{2} & \text{- Thomson limit} \\
E_{\gamma}\sim\gamma m_{e}c^{2}, & \gamma\epsilon\gg m_{e}c^{2} & \text{- Extreme Klein-Nishina limit}.
\end{eqnarray}
The total power lost due to ICS by an electron in an isotropic radiation field of low-energy photons, in the Thomson limit, is given by
\begin{equation}\label{eq:ICS-lossrate}
P_{\rm ICS,total}=\frac{4}{3}\sigma_{{\rm T}}(c\beta^{2}\gamma^{2})U_{{\rm rad}},
\end{equation}
which has the same form as Eq.~\eqref{eq:P-SR}, but with $U_{{\rm rad}}$ the soft-photon energy density, and $\sigma_{{\rm T}}$ the classical Thomson scattering cross section (see Section~\ref{subsec:SR}). In order to obtain the total radiated Compton spectrum, we need to integrate the production rate $dN^{\prime}(\epsilon,\gamma)/dE_{\gamma}$,
valid for a single electron, over the soft-photon energy $\epsilon$ and the electron energy $\gamma$ \citep{Blumenthal1970}:
\begin{equation}\label{eq:Compton-spectrum}
\left(\frac{dN}{dE_{\gamma}}\right)_{{\rm total}}=\iint N_{e}(\gamma)\left(\frac{dN^{\prime}(\epsilon,\gamma)}{dE_{\gamma}}\right)d\gamma d\gamma d\epsilon,
\end{equation}
with $dN_{\rm e}=N_{\rm e}(\gamma)d\gamma$ the differential number of electrons in the interval $(\gamma,\gamma+d\gamma)$. Similar to SR (see Eq.~[\ref{eq:SR-spectrum}]), if we assume that the electron spectral energy distribution is a power law, $N_{e}\sim\gamma^{-p}$ with index $p$, and a blackbody soft-photon distribution, then it follows from Eq.~\eqref{eq:Compton-spectrum} that the ICS spectrum is also power law:
\begin{equation}\label{eq:Compton-spectrum-limits}
\left(\frac{dN}{dE_{\gamma}}\right)_{{\rm total}}\propto\Biggl\{\begin{array}{ll}
E_{\gamma}^{-(p+1)/2} & \text{-- Thomson limit}\\
E_{\gamma}^{-(p+1)} & \text{-- Extreme Klein-Nishina limit},
\end{array}
\end{equation}
and is valid only in a specified energy range between the minimum and maximum cutoff energy similar to SR. In the Thomson limit, we follow a classical approach for photon energies $\gamma\epsilon\lesssim100$ keV for which $\sigma_{{\rm T}}$ is valid. However, when we consider target soft photons of higher energies, quantum effects become important and $\sigma_{{\rm T}}$ should be replaced by the Klein-Nishina cross section $\sigma_{{\rm KN}}$ \citep{Rybicki1979}. As the photon energy increases, the cross section reduces, leading to a steeper photon spectrum (reduced loss rate) in the extreme Klein-Nishina limit (with $\sigma_{{\rm T}}>\sigma_{{\rm KN}}$) and eventually a rapid spectral cutoff.

The ICS process is important for pulsars. One example is MSPs such as PSR J0437$-$4715 from which thermal and non-thermal (possibly SR) X-ray emission have been observed (see e.g., \citealp{Zavlin2002}). These energetic photons provide a background field that may be upscattered to TeV energies by relativistic particles in the magnetosphere. Another example is afforded by the pulsed very-high-energy (VHE) emission recently observed from the Crab pulsar. This has been explained using a revised OG model \citep{Hirotani2008a,Hirotani2008b} that produces IC radiation of up to $\sim$400 GeV when secondary and tertiary pairs upscatter infrared to ultraviolet photons \citep{Aleksic2012}. 

Other astrophysical examples include pulsar wind nebulae (PWNe, see e.g., \citealp{DeJager1996}) and many other VHE sources which display typical spectral components corresponding to SR and IC radiation as part of their broadband emission spectrum.

Another suggestion to explain the Crab pulsar's VHE emission is proposed by \citet{Lyutikov2012}, invoking the SSC radiation process where relativistic pairs upscatter the SR photons emitted previously by the same particle population. In the SSC process, one needs to calculate the SR from primaries and pairs at each step along all particle trajectories in the open field volume, since the SR photon density is needed to compute the SSC radiation. The SR emission is recorded at each location and photon emission direction in the inertial observers frame. Second, once the SR photon density in all directions at a certain position is determined, the SSC flux at a certain position and velocity can be calculated (see \citealt{Harding2015} and references therein). In essence, the soft-photon energy density is basically replaced by the SR photon energy density.

\subsection{Pair production}\label{subsec:PairProduct}

\begin{figure}
\begin{center}
\includegraphics{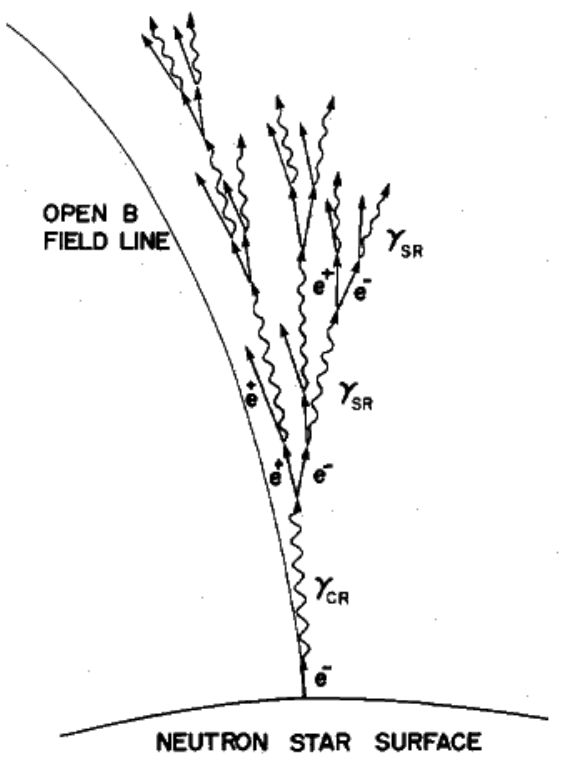}
\end{center}
\caption[Illustration of a photon pair cascade]{\label{fig:Pair-production} Schematic illustration by \citet{Daugherty1982} of a photon pair cascade emerging from the acceleration of a primary electron above the PC along the curved $B$-field of an NS. Four generations of photons, including both CR photons $\gamma_{\rm CR}$ and SR photons $\gamma_{\rm SR}$, are shown.}
\end{figure}

Pulsar magnetospheres contain of strong $B$-fields and $E$-fields, especially near the NS surface, and the latter fields can accelerate particles to relativistic energies. Efficient radiation processes (see Sections~\ref{subsec:SR} to~\ref{subsec:ICS}) and a pair creation mechanism, necessary for particle-photon cascades to ensue, also operate in these extreme environments \citep{Daugherty1982}. There are two different pair creation processes that may occur, namely single photon (magnetic) and two-photon pair production. 

\subsubsection{Magnetic (one-photon) pair creation}\label{subsub:magneticpp}

Magnetic pair production can only occur in the presence of a strong $B$-field ($B_{\perp}>10^{9}$ G, perpendicular to the photon's direction of motion) with which the photons, if they have a high enough energy, can interact to produce $e^{\pm}$-pairs. The probability that $e^{\pm}$-pairs will be produced via this process is expressed by the photon attenuation coefficient given by 
\begin{equation}\label{eq:attenuation-coeff}
\alpha^{\prime}(\chi)=\frac{1}{2}\left(\frac{\alpha_{f}}{\lambda_{\rm c}}\right)\left(\frac{B_{\perp}}{B_{{\rm cr}}}\right)T(\chi),
\end{equation}
which determines the number of pairs $n_{p}$ created for each photon that travels a path length $d$ through the $B$-field \citep{Erber1966}:
\begin{equation}\label{eq:number-of-pairs}
n_{p}=n_{\gamma}\left(1-\exp[-\alpha^{\prime}(\chi)d]\right)\simeq n_{\gamma}\alpha^{\prime}(\chi)d,
\end{equation}
with $\alpha_{f}=e^{2}/\hbar c\approx1/137$ the fine structure constant, $\chi\equiv0.5(h\nu/m_{e}c^{2})(B_{\perp}/B_{{\rm cr}})$ the Erber parameter, $\nu$ the photon frequency, $B_{{\rm cr}}=m_{e}^{2}c^{3}/eh=4.414\times10^{13}$ G the critical $B$-field value in which the electron's gyro-energy (cyclotron) equals its rest mass \citep{Daugherty1983}, $T(\chi)$ a dimensionless function, and $n_{\gamma}$ is the photon number density. \citet{Sturrock1971} approximated the threshold condition for magnetic pair production as
\begin{equation}\label{eq:pp-condition}
E_{\gamma}B\sin\theta_{\gamma B}=E_{\gamma}B_{\perp}\gtrsim10^{11.9},
\end{equation}
with $E_{\gamma}$ in units of $m_{e}c^{2}$, $\theta_{\gamma B}$ the photon propagation angle with respect to the $B$-field, and $B_{\perp}$ measured in Gauss. In the PC model (see Section~\ref{subsec:PC}) primary particles are accelerated from the PC surface along the curved field lines and CR occurs. When the emitted CR photon energy and the local $B$-field are high enough or $\theta_{\gamma B}\gg1$, magnetic pair production will occur, leading to a cascade of secondary $e^{\pm}$ pairs that will screen the $E_{\parallel}$-field ($E$-field parallel to the local $B$-field). An $E_{\parallel}$-field develops because there is a deficit of negative charges and due to the backflow of the first generation of pair $e^{+}$ to the NS surface, a space charge accumulates that counteracts any charge imbalances and screens out the accelerated $E_{\parallel}$-field, significantly so above the so-called pair-formation front (PFF). ICS photons may also be converted into $e^{\pm}$ pairs. The pair cascade is characterised by the so-called multiplicity, i.e., the number of pairs spawned by a single primary, as represented in Figure~\ref{fig:Pair-production}.

\subsubsection{Two-Photon pair creation}\label{subsub:twophotonpp}

Two-photon pair creation is due to a collision between two photons with high enough energies, where the minimum photon energy required is $E_{\gamma}=2m_{e}c^{2}\sim1$ MeV (for a head-on collision), creating an $e^{\pm}$ pair. The cross section for two-photon pair production (in a region devoid of a $B$-field) in terms of the photon energy in the centre-of-momentum frame, $\epsilon_{\rm cm}=[\epsilon_{1}\epsilon_{2}(1-\cos\theta_{12})/2]^{1/2}$, is \citep{Svensson1982} (using dimensionless energies normalised to the electron rest-mass energy)
\begin{equation}\label{eq:2pp-cross-section}
\sigma_{2\gamma}\simeq\frac{3}{8}\sigma_{T}\Biggl\{\begin{array}{ll}
(\epsilon_{\rm cm}^{2}-1)^{1/2} & \quad(\epsilon_{\rm cm}-1)\ll1\\{}
[2\ln(2\epsilon_{\rm cm})-1]/\epsilon_{\rm cm}^{6} & \quad\epsilon_{\rm cm}\gg1,
\end{array}
\end{equation}
where $\epsilon_{1}$ and $\epsilon_{2}$ refer to the energies of the photons, and $\theta_{12}$ is the angle between the photon propagation directions. Two-photon pair production can also take place in the presence of a strong $B$-field. In this case, the resulting $e^{\pm}$ pair will have non-zero velocity to ensure that the energy and parallel momentum are conserved. In a strong $B$-field, the requirement for producing a pair in the ground state, using the conservation equations, is given by \citep{Harding2006}:
\begin{equation}\label{eq:condition_2pp}
(\epsilon_{1}\sin\theta_{1}+\epsilon_{2}\sin\theta_{2})^{2}+2\epsilon_{1}\epsilon_{2}[1-\cos(\theta_{12})]>4,
\end{equation}
where $\theta_{1}$ and $\theta_{2}$ are the angles between the photon propagation directions and the $B$-field. The first term in the above equation is due to the non-conservation of perpendicular momentum, implying that pair production is possible when photons travel parallel to each other ($\theta_{12}=0$, $\theta_1=\theta_2\neq0$), an event not permitted in field-free space. 

In the high-altitude SG model, electrons are accelerated away from the NS, such that the angle of each of the photons to the $B$-field is too small to tap the perpendicular momentum of the $B$-field and no pairs are produced. However, in high-altitude OG model, particles are also accelerated downward so that these particles (and therefore their emitted CR photons) have large angles with respect to soft photons originating at the hot stellar surface. The two-photon pair creation process is therefore expected to occur. The resulting pairs play an important role in gap closure \citep{Hirotani2008a}. \citet{Burns1984} found that the one-photon pair creation process will generally dominate over the two-photon process in $B$-fields above $\approx{10^{12}}$ G, since the first is a lower-order process than the second.

%----------------------------------------------------------------------

\section{Traditional pulsar models}\label{sec:PhysPSRmodels}

\begin{figure}[t]
\begin{center}
\includegraphics[scale=0.5]{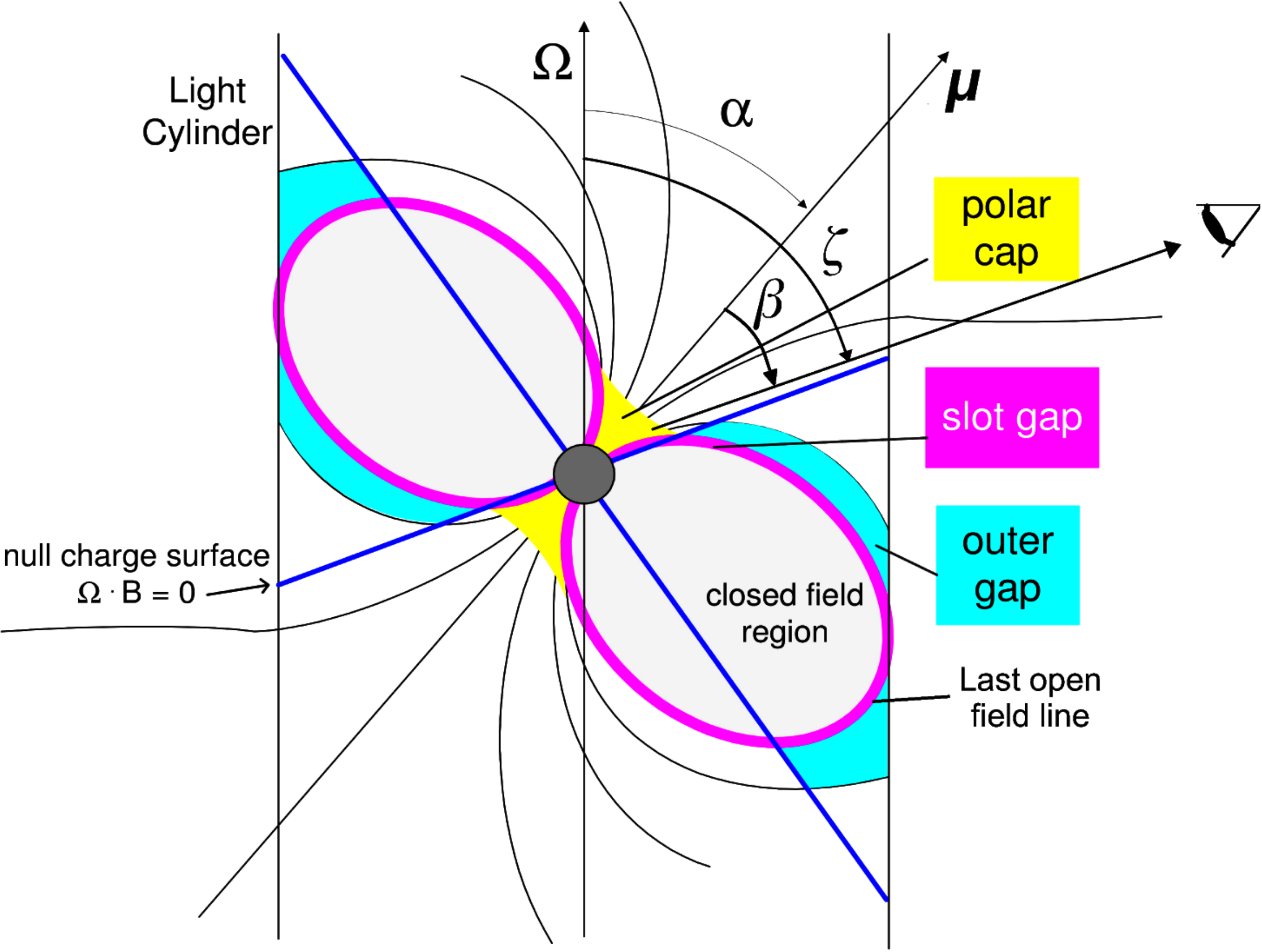}
\end{center}
\caption[Schematic view of the different traditional pulsar emission models]{\label{fig:models} Schematic view of the different traditional pulsar emission models, with $\boldsymbol{\mu}$ the magnetic axis inclined by an angle $\alpha$ with respect to the rotation axis $\boldsymbol{\Omega}$, and $\zeta$ the observer angle measured with respect to $\boldsymbol{\Omega}$.
From \citet{Harding2005a}.}
\end{figure}

Several pulsar emission models have been developed over the last forty years, including the PC, SG, and OG models. These geometries are illustrated in Figure~\ref{fig:models}. Each model differs in its assumption of the geometry and location of the acceleration region were HE radiation takes place. To simulate the HE emission from these physical models, the assumed electrodynamics and $B$-field structures are important.

\subsection{Polar cap model}\label{subsec:PC}

\begin{figure}[t]
\begin{center}
\includegraphics[scale=0.8]{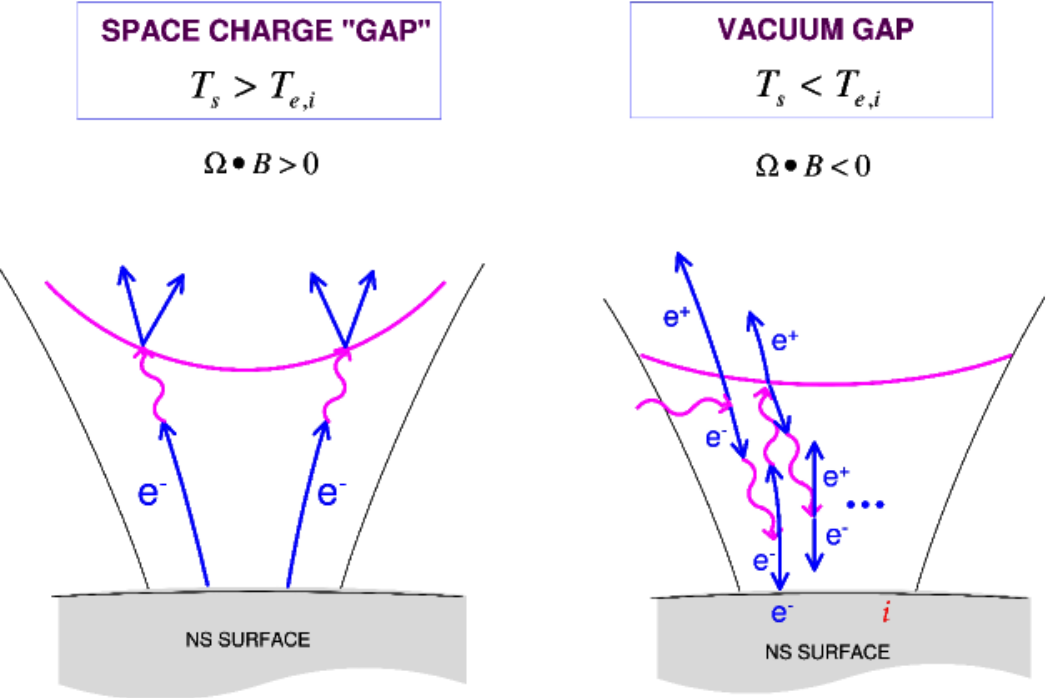}
\end{center}
\caption[Illustration of the two types of PC accelerators, including the SCLF and vacuum gap]{\label{fig:pair} Illustration of the two types of PC accelerators, including the SCLF gap on the left \citep{Daugherty1982}, and the vacuum gap on the right \citep{Ruderman1975}. If the NS surface temperature $T_{s}$ exceeds the ion or electron thermionic temperature $T_{i,e}$ then the SCLF gap will form, otherwise the vacuum gap will dominate. Figure from \citet{Harding2007a}.}
\end{figure}

In PC models \citep{Ruderman1975,Daugherty1982} HE particles ($e^{\pm}$) are assumed to originate at the NS surface layer. These are then accelerated by large, rotation-induced $E$-fields at the magnetic poles (known as the magnetic PCs) up to heights just above the PC ($h\lesssim{R}$, see Figure~\ref{fig:models}). There exist two types of polar cap accelerators: the vacuum gaps with ${\boldsymbol{\Omega}}\cdot{\boldsymbol{\rm B}}<0$ \citep{Ruderman1975,Usov1995} and space-charge-limited-flow (SCLF) gaps with ${\boldsymbol{\Omega}}\cdot{\boldsymbol{\rm B}}>0$ (\citealp{Arons1979,Harding1998}; see Figure~\ref{fig:pair}). The formation of these gaps depends primarily on the surface temperature $T_{s}$ of the NS, and the thermionic emission temperatures for the charges (electrons and ions) $T_{e,i}$. For the vacuum accelerator the surface temperature $T_{s}<T_{e,i}$, causing the charges to be trapped inside the NS surface and a full vacuum $E$-field (or potential drop) develops above the surface \citep{Usov1995}. However, at high surface temperatures, $T_{s}>T_{e,i}$ the binding energy of the charges due to lattice structures in strong $B$-fields is exceeded \citep{Medin2007} and the charges are ``boiled off'' the surface layers and flow freely along the open field lines in the SCLF regime. These two acceleration gaps differ primarily in their surface boundary conditions (at the stellar surface $r=R$) which state that for the vacuum gap the space charge $\rho(r)=0$ and $E_{\parallel}(R)\neq0$, whereas for the SCLF accelerator the full Goldreich-Julian charge can be provided, $\rho(r)=\rho_{GJ}$ (which may be modified by curvature of $B$-field lines as well as inertial frame dragging, \citealp{Muslimov1992}) and $E_{\parallel}(R)=0$ \citep{Harding2007a}. Both accelerators will be self-limited by the development of pair cascades (initiated by the conversion of radiated photons into $e^{\pm}$ pairs; see Section~\ref{subsec:PairProduct}), with the particles being accelerated to altitudes where they will reach high enough Lorentz factors to radiate $\gamma$-ray photons.

The pair production in the vacuum gap differs from that in the SCLF model. In the vacuum gap the potential breaks down when a random photon crosses the $B$-field and creates a pair. The resulting electron and positron are accelerated in opposite directions. This electron and positron can then initiate more pairs since they will radiate photons that may again be converted into pairs, causing a pair cascade and discharge of the vacuum gap. In contrast, in the SCLF model electrons and positrons are accelerated from the NS surface upwards until the radiated photons reach the pair creation threshold. A pair cascade ensues at the PFF. The $E_{\parallel}$ is screened due to the polarisation of pairs above this front, halting any further acceleration (with the relativistic charges ``coasting'' outward, potentially emitting SR). Since these accelerators can maintain a steady current, there will be an upward current of electrons ($j_{\parallel}^{-}\simeq c\rho_{GJ}$) and also a downward current of positrons ($j_{\parallel}^{+}\ll c\rho_{GJ}$), which will heat the PC. The height of the PFF determines the eventual potential of these accelerators. Simulations of time-dependent vacuum \citep{Timokhin2010} and SCLF \citep{Timokhin2013} gaps show that the pair cascades are non-steady.

\subsection{Slot gap model}\label{subsec:SG}

\begin{figure}
\begin{center}
\includegraphics[scale=0.9]{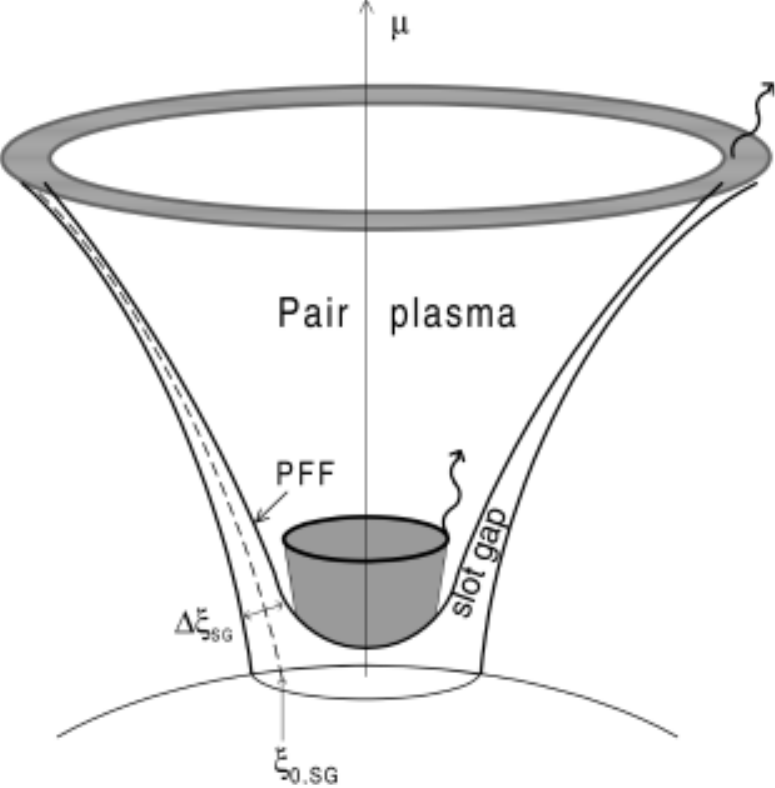}
\end{center}
\caption[A schematic view of an SG geometry]{\label{fig:SG} Schematic view of the SG, which is the region between the PFF and the outer boundary formed by the last open field lines (with $\boldsymbol{\mu}$ the magnetic axis and $\Delta\xi_{\rm SG}$ the gap width). The SG forms a hollow cone emitting HE radiation. From \citet{Muslimov2003}.}
\end{figure}

In SG models (\citealp{Arons1983}; see Figure~\ref{fig:models}) it is assumed that HE particles originate from the NS surface layer and are accelerated from the PCs along the last open field lines and up to high altitudes, comparable to the light cylinder radius $R_{{\rm LC}}=c/\Omega$ \citep{Harding2011b}. This SG model is very similar to the SCLF accelerator of the PC model, except that the SG model extends up to high altitudes. The altitude of the PFF (Figure~\ref{fig:SG}) strongly varies with magnetic colatitude across the PC due to the $B$-field geometry and the boundary conditions ($E_{\parallel}=0$ on the surface and last open $B$-field lines) assumed for the SG accelerator \citep{Harding1998}. The PFF will occur at higher and higher altitudes closer to the closed field line region boundary. This is because the mean free path for magnetic (one-photon) pair production increases as $E_{\parallel}$ decreases toward this boundary. The radiating particle therefore needs to be accelerated over a longer distance before it can radiate photons of high enough energy so that pair formation can take place. The mean free path becomes infinite at and asymptotically tangent to the last open field line. The $E_{\parallel}$ is screened above the PFF, and a narrow gap surrounded by two conducting walls will form, as represented in Figure~\ref{fig:SG} \citep{Harding2003}. The electrons accelerated in the SG will radiate CR, ICR and SR, although their Lorentz factors are constrained by the CR. We note that new solutions for the $E_{\parallel}$ were determined by \citet{Muslimov2003,Muslimov2004a}, including the general relativistic effect of inertial frame-dragging near the NS surface, enhancing this field significantly.

\subsection{Outer gap model}\label{subsec:OG}

The OG model (Figure~\ref{fig:models}) was introduced by \citet{Cheng1986}, initially assuming an inclined rotator with a charge density of $\rho_{GJ}$. They proposed that when the primary current passes through the neutral sheet (where ${\bf \bm{\Omega}}\cdot{\bf B}=0$ and thus $\rho_{GJ}=0$) the negative charges above this sheet will escape beyond the light cylinder. A vacuum gap region is then formed (in which $E_{\parallel}\neq0$). Charges will be accelerated in this gap region and will emit CR photons, the energy of which depends on the $E$-field strength. Therefore, as the vacuum region grows, $E_{\parallel}$ will increase, and hence the energy of the CR photons will increase until the photons have enough energy to produce electron and positron pairs when they collide with the background soft photons via photon-photon (or two-photon) pair production (see Section~\ref{subsec:PairProduct}, \citealp{Cheng2011}).

\begin{figure}[t]
\begin{center}
\includegraphics[scale=0.7]{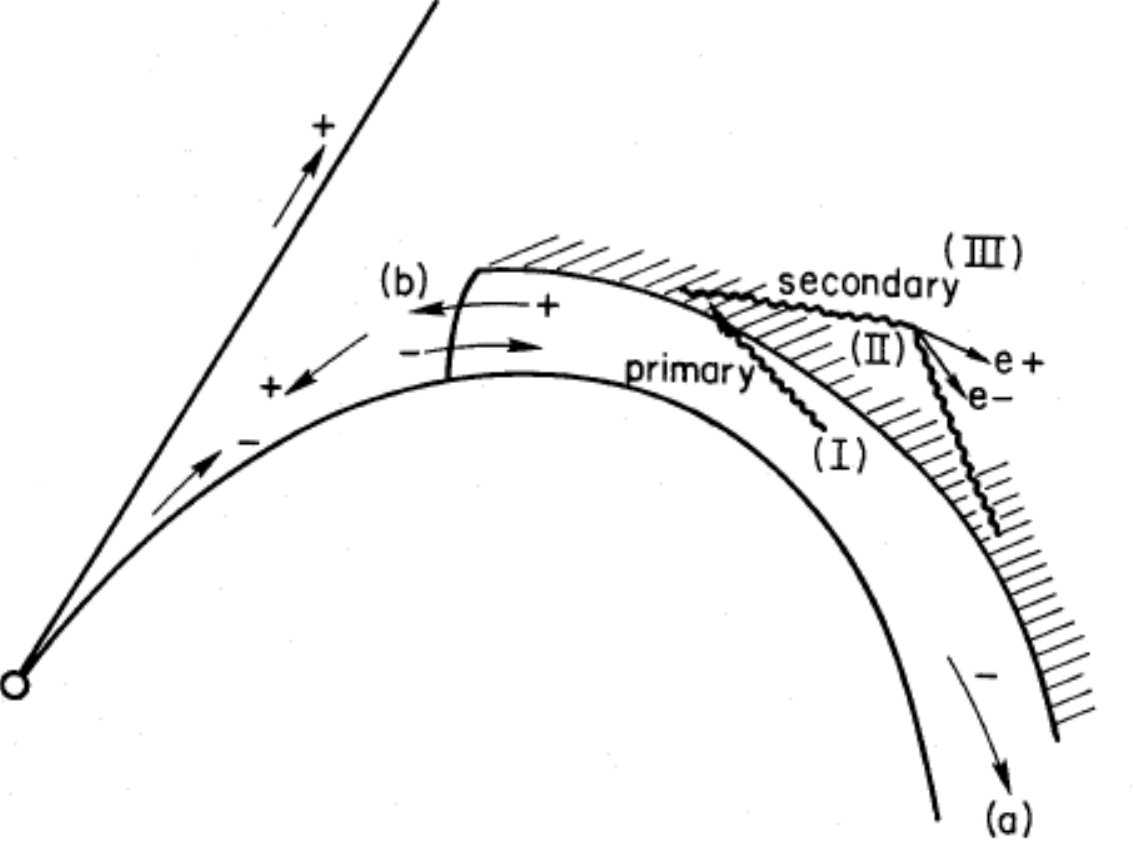}
\end{center}
\caption[Representation of particle acceleration and pair production inside an OG geometry]{\label{fig:OGmodel} Schematic representation of the location of the three regions in an OG model, including the primary (I) (where
primary particle acceleration and pair production occur), secondary (II) (where a small $E_{\parallel}$ is present and secondary pair production as well as SR occur) and tertiary (III) regions (where tertiary pair production occurs, producing soft $\gamma$-rays). From \citet{Cheng1986}.}
\end{figure}

The outer magnetosphere is conceptually divided into three regions (see Figure~\ref{fig:OGmodel}). In region I the primary electrons and positrons are accelerated in opposite directions due to the $E_{\parallel}$ present in the gap, with the acceleration limited by CR losses or ICS on infrared photons. Although some $\gamma$-rays undergo pair creation, most of them move over into region II, where the $E_{\parallel}$ is small and secondary $e^{\pm}$ pairs are produced, radiating secondary $\gamma$-rays and X-rays via SR. In region III, tertiary $e^{\pm}$ pairs are created and are responsible for the emission of softer radiation \citep{Cheng1986}. This is also the region where $\gamma$-rays are produced by ICS involving the primary pairs. The radiation in this region may furthermore interact with primary CR $\gamma$-rays to create pairs in region II. \citet{Romani1996} investigated an OG model based on CRR-limited charges (see Section~\ref{subsec:CR}) in the outer magnetosphere, showing that photon-photon pair production may limit the gap width. He also demonstrated that radiation efficiency should increase with pulsar age, and discussed spectral variations in the optical and X-ray SR spectra. More modern approaches have solved the electrodynamical equations of the OG in a 2D and 3D geometry (e.g., \citealp{Takata2004,Hirotani2006,Hirotani2008a}).

\subsection{Pair-starved polar cap model}\label{subsec:PSPC}

The PSPC model for MSPs was first introduced by \citet{Harding2005c}. They studied X-ray and $\gamma$-ray emission emitted by rotation-powered MSPs (see Section~\ref{subsec:PSRcat}). These MSPs have very low surface $B$-field strengths and short periods (compared to younger pulsars). The electrodynamics is based on a (young-pulsar) model that considers the acceleration of particles and pair production (see Section~\ref{sec:RadMechanisms}) on the open field lines above the PCs \citep{Muslimov2004b}. \citet{Harding2005c} assumed a PC geometry (see Section~\ref{subsec:PC}) and a dipole field (see Section~\ref{subsec:StatDip}). They found that most MSPs are below the CR pair death line (i.e., the $P\dot{P}$ death line for creating pairs via CR, see Section~\ref{sec:BrakingMod}), due to the low surface $B$-fields (see Eq.~[\ref{eq:pp-condition}]), implying that pairs are rather produced by ICS radiation. 
Since the pair cascade multiplicity is very low in PSPC pulsars, the accelerating $E$-field is inefficiently screened by these pairs and no PFF is formed as in the case of young pulsars. This leads to a pair-starved PC. Therefore, the primary particles and possibly a few pairs continue to accelerate to high altitudes, up to the light cylinder, over the full open volume (in contrast to the traditional PC model where particles are only accelerated up to the PFF, and then coast along the field lines after which they escape from the magnetosphere). There exists a progression between models, depending on the pair multiplicity: the SG model has very narrow gaps for pulsars with high multiplicity, but these gaps increase in thickness and the model eventually tends toward a PSPC geometry as pair creation is more and more inhibited \citep{Harding2007a}.

%----------------------------------------------------------------------
\section{Developments in the magnetic field structure calculations}\label{sec:BandEfields} 

The $B$-field is one of the basic assumptions of the geometric models (others include the gap region's location, and the emissivity $\epsilon_{\nu}$ profile in the gap). Several $B$-field structures have been studied, including the static dipole \citep{Griffiths1995}, the RVD (a rotating vacuum magnetosphere that can in principle accelerate particles but do not contain any charges or currents; \citealp{Deutsch1955}), the force-free field (FF; filled with charges and currents, but unable to accelerate particles, since the accelerating $E$-field is screened everywhere; \citealp{Contopoulos1999}), and the offset dipole (to mimic deviations from the static dipole near the stellar surface analytically; \citealp{Harding2011a,Harding2011c}). A more realistic pulsar magnetosphere, i.e., a dissipative solution \citep{Kalapotharakos2012a,Li2012,Tchekhovskoy2013,Li2014}, would be one that is intermediate between the RVD and the FF fields. The dissipative $B$-field is characterised by the plasma conductivity $\sigma_{\rm c}$ (e.g., \citealp{Lichnerowicz1967}) which can be set in order to alternate between the limiting cases of vacuum ($\sigma_{\rm c}\rightarrow{0}$) and FF ($\sigma_{\rm c}\rightarrow\infty$) magnetospheres (see \citealp{Li2012}). 

\subsection{Static dipole magnetic field}\label{subsec:StatDip}
The static dipole field has been studied since the calculations are simpler for this $B$-field. We derive its form below. In order to obtain an approximate formula for a vector potential associated with a localised current distribution, valid at distant points, a multipole expansion can be used and the potential is written in the form of a power series in $1/r$, with $r$ the radial distance to the point in question. If $r$ is very large the power series is dominated by the lowest non-vanishing contribution and the higher-order terms can be ignored. The first term (which goes like $1/r$) in the multipole expansion is called a monopole term, the second term (which goes like $1/r^{2}$) the dipole term, and the third term (which goes like $1/r^{3}$) the quadrupole, etc. \citep{Griffiths1995}. The magnetic monopole term is zero (i.e., $\nabla\cdot\boldsymbol{\mathbf{B}}=0$). The next term is the magnetic dipole. The vector potential can be written as a function of position vector ${\bf r}$: 
\begin{equation}
{\boldsymbol{\rm A}}_{{\rm dip}}({\bf r})=\frac{I}{r^{2}}\oint r^{\prime}\cos\theta^{\prime}d{\rm l^{\prime}}=\frac{I}{r^{2}}\oint(\hat{\boldsymbol{\rm r}}\cdot r^{\prime})d{\rm l^{\prime}},
\end{equation}
with $\hat{\boldsymbol{\rm r}}$ the unit radial vector and $I$ the current. By rewriting this integral, the vector potential becomes
\begin{equation}\label{eq:vector-potential}
{\boldsymbol{\rm A}}_{{\rm dip}}({\boldsymbol{\rm r}})=\frac{{\boldsymbol{\mu}}\times\hat{\boldsymbol{\rm r}}}{r^{2}},
\end{equation}
where $\boldsymbol{\mu}$ is the magnetic dipole moment (aligned with the magnetic axis, therefore sometimes defined by the same symbol) defined as
\begin{equation}
{\boldsymbol{\mu}}{\equiv}\frac{I}{c}\int{d}{\boldsymbol{\rm a}}=\frac{Ia}{c},
\end{equation}
with $\boldsymbol{\rm a}$ the enclosed area of the current loop, and $c$ the speed of light. To calculate the $B$-field of a dipole, $\boldsymbol{\mu}$ may be set at the origin, pointing in the $z$-direction. The vector potential, Eq.~\eqref{eq:vector-potential} can then be written in spherical co-ordinates,
\begin{equation}
{\boldsymbol{\rm A}}_{{\rm dip}}({\boldsymbol{\rm r}})=\frac{\mu\sin\theta}{r^{2}}\hat{\boldsymbol{\phi}},
\end{equation}
hence the dipole $B$-field is given by \citep{Griffiths1995}:
\begin{equation}\label{eq:static-dipole}
\boldsymbol{\mathbf{B}}_{{\rm static}}(\boldsymbol{\mathbf{r}})=\nabla\times\boldsymbol{\mathbf{A}}_{{\rm dip}}(\boldsymbol{\mathbf{r}})=\frac{\mu}{r^{3}}(2\cos\theta\hat{\boldsymbol{\mathbf{r}}}+\sin\theta\hat{\boldsymbol{\mathbf{\bm{\theta}}}}),
\end{equation}
in the magnetic (${\boldsymbol{\mu}}$) frame or where the inclination angle $\alpha=0$ (the angle between the rotation ${\boldsymbol{\Omega}}$ and the ${\boldsymbol{\mu}}$ axes). 

When the light curve shapes and features for the static dipole are compared to those for the other $B$-fields (see Chapter~\ref{chap:OffsetPC}), the importance of the near-$R_{{\rm LC}}$ distortions in the $B$-fields for predicted radiation characteristics can be gauged \citep{Dyks2004b}. The static (non-rotating) dipole is a special case of the retarded (rotating) dipole which we consider next.

\subsection{Retarded vacuum dipole magnetic field}\label{subsec:RVD}
The solution for a $B$-field surrounding a star rotating in vacuum was first derived by \citet{Deutsch1955}. Previous investigators (\citealp{Yadigaroglu1997}, \citealp{Arendt1998}, \citealp{Jackson1999}, \citealp{Cheng2000}, \citealp{Dyks2004b}) implemented methods that considered distortions in the $B$-field structure due to sweepback of the field lines as the NS rotates with an angular frequency $\Omega$ about the $\hat{\rm {\mathbf{z}}}$-axis. The general expression for this RVD field is given by
\begin{eqnarray}\label{eq:magnetic-moment}
{\boldsymbol{\rm B}}_{\rm ret} = -\left[\frac{\boldsymbol{\mu}(t)}{r^3}+\frac{\dot{\boldsymbol{\mu}}(t)}{cr^2}+\frac{\ddot{\boldsymbol{\mu}}(t)}{c^2r}\right]+{\hat{\boldsymbol{\rm r}}}{\hat{\boldsymbol{\rm r}}}\cdot\left[3\frac{\boldsymbol{\mu}(t)}{r^3}+3\frac{\dot{\boldsymbol{\mu}}(t)}{cr^2}+\frac{\ddot{\boldsymbol{\mu}}(t)}{c^2r}\right],
\end{eqnarray}
with
\begin{eqnarray}\label{eq:magnetic-moment}
{\boldsymbol{\mu}}(t) = \mu(\sin\alpha\cos\Omega{t}{\hat{\boldsymbol{\rm x}}}+\sin\alpha\sin\Omega{t}{\hat{\boldsymbol{\rm y}}}+\cos\alpha{\hat{\boldsymbol{\rm z}}}),
\end{eqnarray}
the magnetic moment, with $\dot{\boldsymbol{\mu}}(t)$ and $\ddot{\boldsymbol{\mu}}(t)$ its first and second time-derivatives, and ${\hat{\boldsymbol{\rm r}}}=r/{\boldsymbol{\rm r}}$ the unit radial vector. The RVD solution can be described by the following $B$-field equations in spherical co-ordinates in the laboratory frame (where $\hat{\boldsymbol{\rm z}}\parallel\boldsymbol{\Omega}$; \citealp{Dyks2004a}):
\begin{eqnarray}
B_{{\rm ret,}r} & = & \frac{2\mu}{r^{3}}\Big[\cos\alpha\cos\theta+\sin\alpha\sin\theta\big(r_{{\rm n}}\sin\lambda+\cos\lambda\big)\Big], \\
B_{{\rm ret,}\theta} & = & \frac{\mu}{r^{3}}\Big(\cos\alpha\sin\theta+\sin\alpha\cos\theta\big[-r_{{\rm n}}\sin\lambda+\big(r_{{\rm n}}^{2}-1\big)\cos\lambda\big]\Big), \\
B_{{\rm ret,}\phi} & = & -\frac{\mu}{r^{3}}\sin\alpha\Big[\big(r_{{\rm n}}^{2}-1\big)\sin\lambda+r_{{\rm n}}\cos\lambda\Big],
\end{eqnarray}
with $\lambda=r_{{\rm n}}+\phi_{\rm L}-\Omega t$, $r_{{\rm n}}={r}/{R_{{\rm LC}}}$, $\Omega$ the angular velocity, and $\phi_{\rm L}$ the phase. These equations can be rewritten to give the $B$-field components in Cartesian co-ordinates (where $\hat{\boldsymbol{\rm z}}\parallel{\boldsymbol{\Omega}}$; \citealp{Dyks2004a})
\begin{eqnarray}
B_{{\rm ret,}x} & = & \frac{\mu}{r^{5}}\Big(3xz\cos\alpha + \sin\alpha\Big\{\big[\big(3x^2-r^2\big)+3xyr_n+\big(r^2-x^2\big)r^2_n\big]\cos\big(\Omega{t}-r_n\big) \nonumber \\
                &   & +\big[3xy-\big(3x^2-r^2\big)r_n-xyr^2_n\big]\sin\big(\Omega{t}-r_n\big)\Big\} \Big), \\ 
B_{{\rm ret,}y} & = & \frac{\mu}{r^{5}}\Big(3yz\cos\alpha + \sin\alpha\Big\{\big[3xy+\big(3y^2-r^2\big)r_n-xyr^2_n\big]\cos\big(\Omega{t}-r_n\big) \nonumber \\
                &   & +\big[\big(3y^2-r^2\big)r_n-3xyr_n+\big(r^2-y^2\big)r^2_n\big]\sin\big(\Omega{t}-r_n\big)\Big\} \Big), \\
B_{{\rm ret,}z} & = & \frac{\mu}{r^{5}}\Big\{\big(3z^2-r^2\big)\cos\alpha + \sin\alpha\big[\big(3xz+3yzr_n-xzr^2_n\big)\cos\big(\Omega{t}-r_n\big) \nonumber \\
                &   & +\big(3yz-3xzr_n-yzr^2_n\big)\sin\big(\Omega{t}-r_n\big)\big]\Big\}.   
\end{eqnarray}
The above expressions are obtained when assuming the limit of the Deutsch solution where $R/R_{\rm LC}\ll{1}$, with $R$ the stellar radius. By setting $r_{\rm n}=0$, the retarded field simplifies to the non-aligned static dipole ($\alpha\neq0$). 

\begin{figure}[t]
\centering
\includegraphics[scale=0.95]{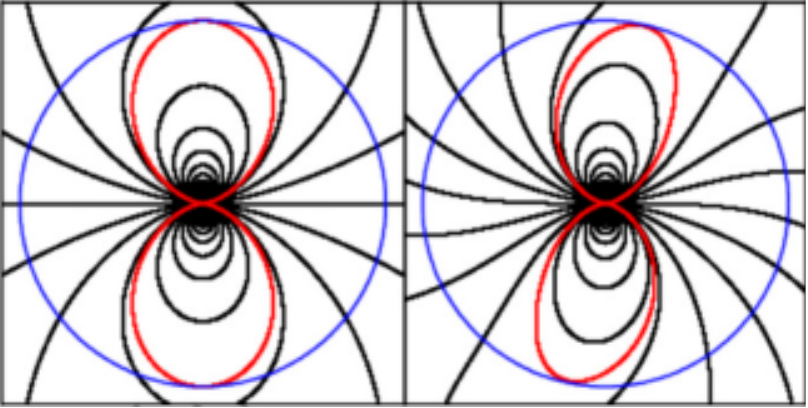}
\caption[Comparison between the static and RVD $B$-field structures]{\label{fig:SRComp_new} Illustration of the static dipole (left) and RVD (right) $B$-field structures in the equatorial plane, where $\alpha=90^\circ$. The red curve indicates the last closed field line which closes at the light cylinder, where the corotation speed is equal to \textit{c}. The sweepback of field lines is evident for the RVD case. The blue circle indicates a slice through the light cylinder. From \citet{Romani2010}.}
\end{figure}

\begin{figure}[h]
\begin{center}
\includegraphics[width=10cm]{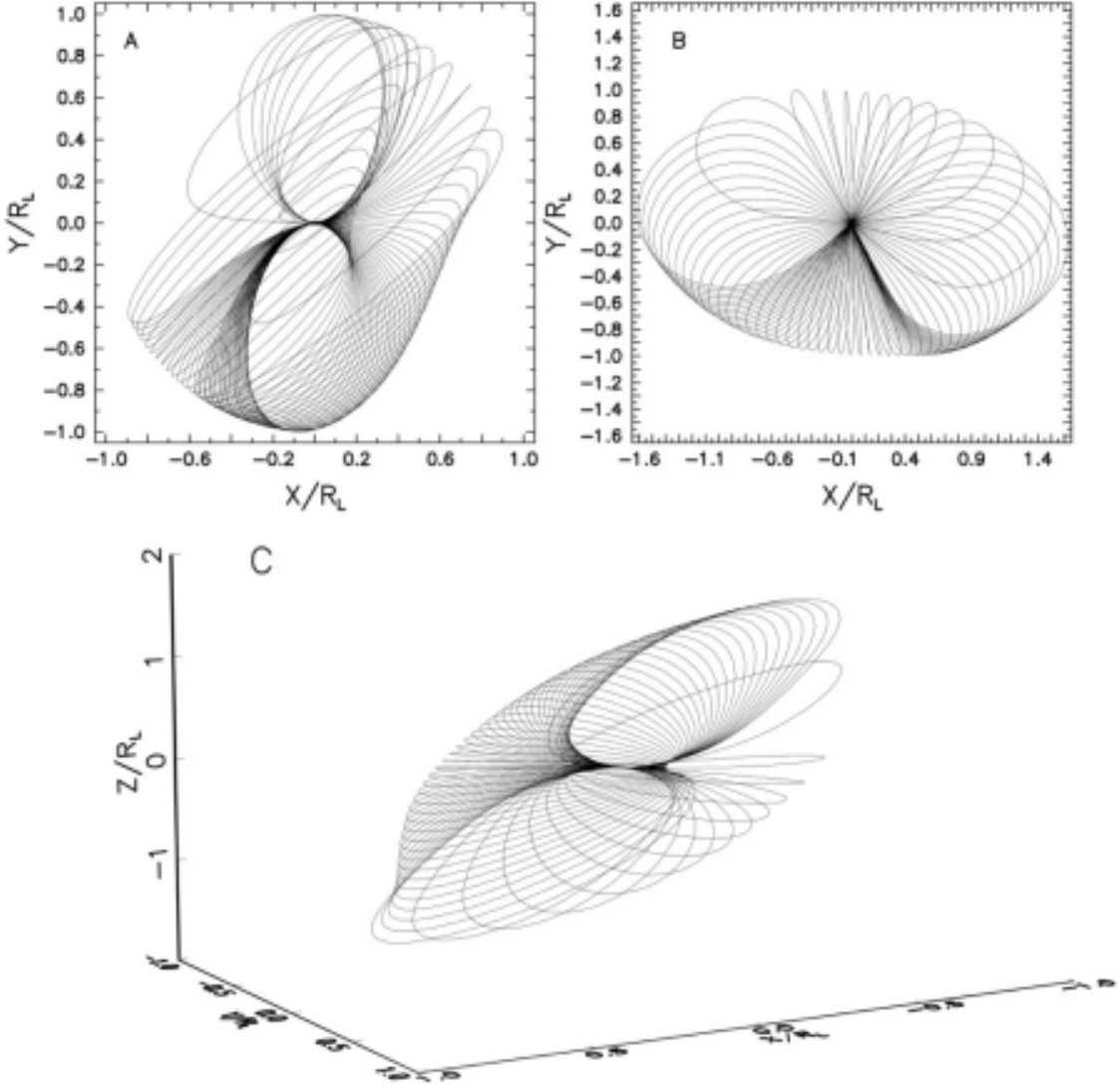}
\end{center}
\caption[An example of the distortions in an RVD $B$-field for $\alpha=65^{\circ}$]{\label{fig:RVD_example} 
Illustration of an RVD $B$-field for the last closed field lines, inclined with respect to $\boldsymbol{\Omega}$ by an angle $\alpha=65^{\circ}$. Three projections are presented: (a) looking down $\boldsymbol{\Omega}$, (b) looking down $\boldsymbol{\mu}$, and (c) a 3D view. Strong distortions are visible as the inclination becomes larger (see \citealp{Arendt1998}). From \citet{Cheng2000}.}
\end{figure}

The difference between the static dipole and RVD is illustrated in Figure~\ref{fig:SRComp_new} for $\alpha=90^\circ$, i.e., field lines in the equatorial plane. For the static dipole (left) the field lines are symmetric, whereas in the RVD case (right) the field lines are distorted due to sweepback of the field lines as the NS rotates. This has implications for the definition of the PC. These distortions in the RVD $B$-field are illustrated in Figure~\ref{fig:RVD_example} for an $\alpha=65^\circ$ as seen from different points of view.

\subsection{Offset-dipole magnetic field}\label{subsec:OffsetPC}
The offset dipole is a heuristic model of a dipolar magnetic structure that is offset from the stellar centre, leading to PCs offset from the central ${\boldsymbol{\mu}}$-axis. The $B$-field lines of an offset dipole field are azimuthally asymmetric compared to those of a pure dipole field. This leads to field lines having a smaller curvature radius $\rho_{\rm c}$ over half of the PC compared to those of the other half. Such small distortions in the $B$-field structure are due to retardation and asymmetric currents, thereby shifting the PCs by small amounts in different directions. A detailed study of the effects of this $B$-field structure on the predicted HE light curves of the Vela pulsar will be presented in Chapter~\ref{chap:OffsetPC}.

\subsection{A force-free field} \label{subsec:FFfield}

The FF $B$-field structure assumes that the entire pulsar's magnetosphere is filled with highly conductive and dense plasma so that the $E_\parallel$ is fully screened (i.e., $E_\parallel=0$; \citealt{Spitkovsky2006}). This implies that the ideal magnetohydrodynamic (MHD) condition $\mathbf{E}\cdot\mathbf{B}=0$ is valid everywhere in the magnetosphere \citep{Spitkovsky2011}. 

There are different analytic solutions that have been studied regarding the ideal MHD equation. \citet{Michel1973a} obtained a split-monopole solution (representing the magnetospheric structure farther from the NS; see \citealt{Petrova2016}), whereas \citet{Michel1973b} (see also \citealt{Michel1982}) found a solution for a corotating relativistic dipole magnetosphere with zero poloidal current, valid inside the light cylinder. Other solutions include a slightly perturbed monopole \citep{Beskin1998}, as well as an exact axisymmetric dipole with a differential rotational magnetospheric velocity distribution and general toroidal structure \citep{Petrova2016,Petrova2017}.

\begin{figure}[h]
\begin{center}
\includegraphics[width=11cm]{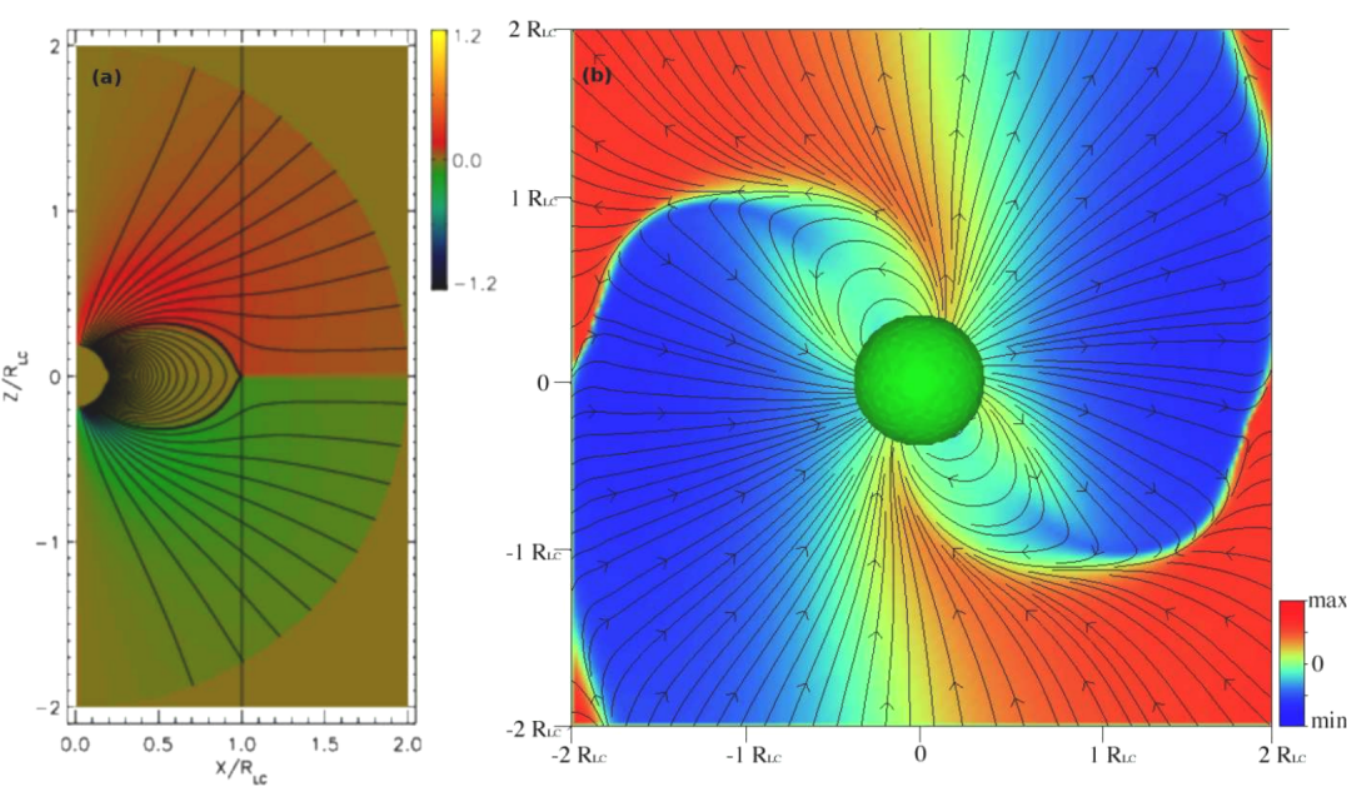}
\end{center}
\caption[Snapshots of time-dependent FF simulations of the (a) aligned and (b) oblique rotators.] {\label{fig:FF_example} 
Snapshots of time-dependent FF simulations of the (a) aligned and (b) oblique rotators. In the first instance, poloidal field lines of the steady-state solution are shown. The thick black line indicates the boundary of the light cylinder, and the colour the normalised toroidal $B$-field component. In the second panel, for an $\alpha=60^\circ$, the $B$-field lines are shown in the corotating frame where the colour indicates the field strength perpendicular to the plane (the toroidal field in the aligned rotator case). From \citet{Spitkovsky2006}.}
\end{figure}

However, the first numerical solution of the pulsar equation (valid for FF magnetospheres, relating current and magnetic flux) for a dipole $B$-field near the NS was found by \citet{Contopoulos1999}, and permits a smooth transition of the field lines at and beyond the light cylinder as well as current closure (see Figure~\ref{fig:FF_example}a). This solution is characterised by two zones where field lines are closed or open, respectively. Also, as the poloidal field lines move out to greater distances away from the NS these lines become monopolar at regions where torodial $B$-field components exist. There exist time-dependent numerical solutions as well, such as the oblique rotator (e.g., \citealt{Spitkovsky2006,Kalapotharakos2009,Contopoulos2010,Kalapotharakos2012c}; see Figure~\ref{fig:FF_example}b) with $B$-field lines similar to those of the solution by \citet{Contopoulos1999}, but the current sheet thereof has the shape of a ``ballerina skirt'' about the rotational equator. 

%----------------------------------------------------------------------

\section{Recent theoretical developments}\label{sec:bmodels}
%Sien reviews: Venter, Harding, Cerutti, Petri, etc.

\subsection{A dissipative field} \label{subsec:FIDOfield}

The dissipative $B$-field solution represents a transition from the vacuum to the FF case, and allows energy dissipation and therefore particle acceleration, which is not the case for the FF solution. Thus, the dissipative $B$-fields is characterised by a zero (in vacuum case) to infinite (FF case) macroscopic conductivity, $\sigma$ \citep{Kalapotharakos2012a,Li2012,Kalapotharakos2014}. The dissipative FF was developed in order to probe the locations where particle acceleration may take place, as well as the effect of the deviations from ideal MHD conditions on the magnetosphere structure \citep{Li2012,Kalapotharakos2012a}. In these approaches, different methods led to a non-accelerating $E$-field, e.g., a finite $\sigma$ alter the current density \citep{Kalapotharakos2012a}. For increasing $\sigma$, the regions containing large charge and current densities also increased in size and the current sheet became more pronounced, and $B$-field lines became more straight over farther distances. 

Some studies modelled energy light curves as a function of $\sigma$. \citet{Kalapotharakos2012b} modelled light curves assuming a geometric approach and found that peak widths broadened accompanied by a phase lag to the right with increasing $\sigma$. This is ascribed to the effect of the magnetospheric structure on the light curves. \cite{Kalapotharakos2014} constructed a model assuming dissipative magnetospheres, and incorporating CR and a slightly new prescription for the current density, as well as different values for the $\sigma$ based on the region in the magnetosphere. They studied the particle trajectories together with a self-consistent accelerated $E$-field, including CR energy losses. For lower $\sigma$ values emission was noticed at lower altitudes inside the light cylinder, and as $\sigma$ was increased the radiation occurred at higher altitudes near the current sheet where $E_\parallel$ is higher. The latter implies that for a decrease in $E_\parallel$ longer acceleration distances need to be followed for particles to acquire enough energy to emit CR. However, emission from the current sheet was enhanced. Also, small values of $\sigma$ were associated with broad light curves, while those corresponding to large $\sigma$'s were narrower in some instances. 

\begin{figure}
\begin{center}
\includegraphics[width=15cm]{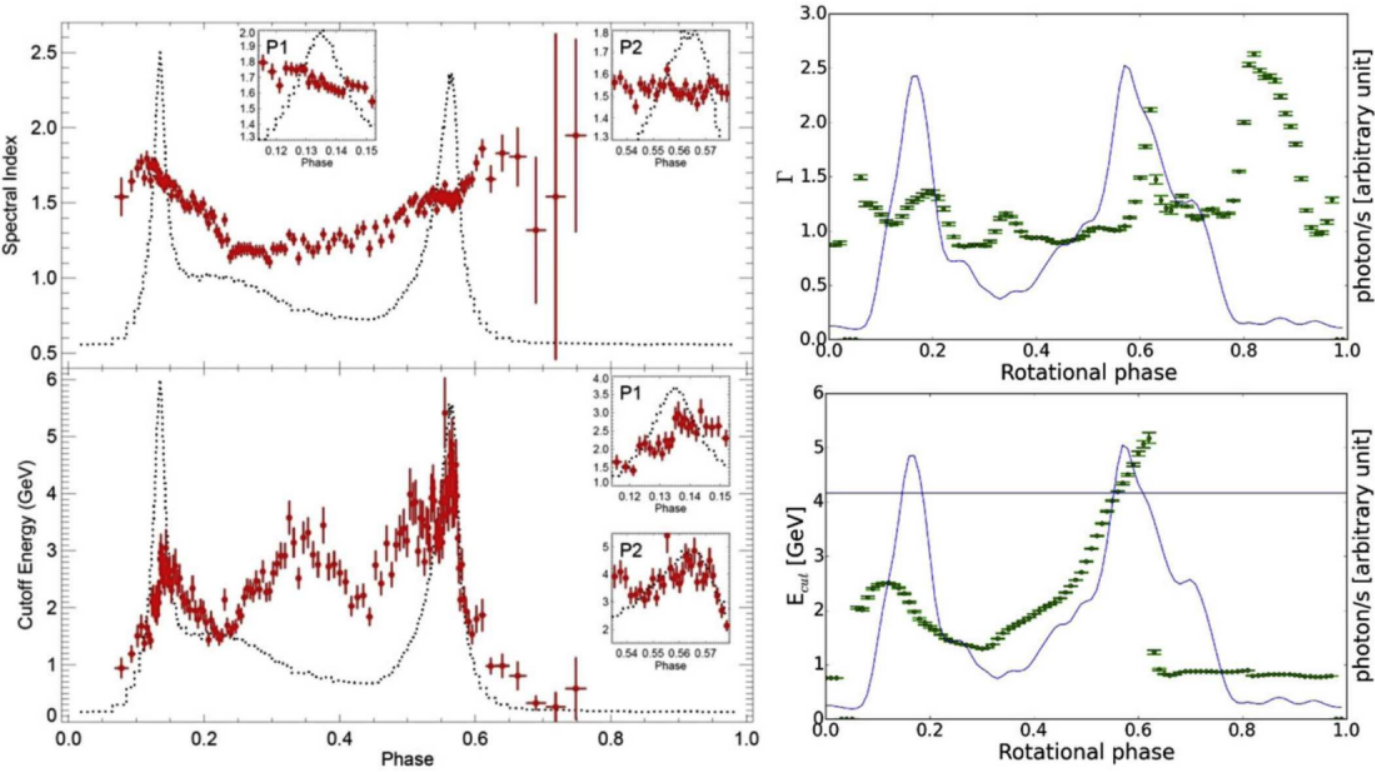}
\end{center}
\caption[The observed and predicted light curves, and phase-resolved spectra for a FIDO model]{\label{fig:FIDO_example} The observed (left) and predicted (right) light curves, and phase-resolved spectral index (top) and the cutoff energy (bottom) for the Vela pulsar using the FIDO model. From \citet{Brambilla2015}.}
\end{figure}

Another implementation of the dissipative models is the FIDO models and are defined as FF conditions existing inside the light cylinder and dissipative conditions outside (FIDO) beyond the light cylinder (into the current sheet). Thus, a large but finite $\sigma$ is chosen for the dissipative regions. \citet{Brambilla2015} calculated phase-averaged and phase-resolved spectra, using a FIDO model, for a few very luminous pulsars (including Vela; see Figure~\ref{fig:FIDO_example}) and assuming CR. They found that for a fixed $\alpha$ the spectral cutoff energy $E_{\gamma,\rm CR}$ increased for larger $\zeta$, but decreased for larger $\sigma$'s. The FIDO model also predicted that $E_{\gamma,\rm CR}$ may increase near the second light curve peak's phase. However, this is not always the case. We also demonstrate that the P1/P2 effect can be explained within an CR emission model framework using a FF $B$-field and SG current sheet model (see Chapter~\ref{chap:CREdepLCmod}).

The FIDO model still fails to replicate some trends of the light curve phenomenology as well as the phase-resolved spectral details. However, the FIDO model does provide good results regarding basic trends. Moreover, a model such as the FIDO is able to replicate the GeV light curve phenomenology relatively well and impacts future microphysical simulations. 

\subsection{Kinetic models}
%Prentjies uit Brambilla 2018, Philippov, Cerutti

The kinetic particle-in-cell (PIC) codes model the pulsar magnetosphere from first principles and resolve both the temporal and spatial scales of the problem (plasma frequency and skin depth) to avoid numerical instabilities and numerical plasma heating \citep{Brambilla2018}. However, the assumed field values of the latter codes are unrealistically low. These codes follow a two-step process: first to calculate charge and current densities, and then the fields based on these; Lastly, the process is iterative, so the influence on of the $B$-fields and $E$-fields on the charges are taken into account, and updated currents and charge densities are calculated.
Recent studies (e.g., \citealt{Philippov2014,Belyaev2015,Cerutti2017,Kalapotharakos2018}) concentrated on handling the pulsar electrodynamics self-consistently, including global current closure, the contribution of charges of different sign to the current, dissipative processes, electromagnetic emission, and the effects of pair production and general relativity (for a detailed review see \citealt{Venter2016,Cerutti2020}). 

\begin{figure}
\begin{center}
\includegraphics[width=11cm]{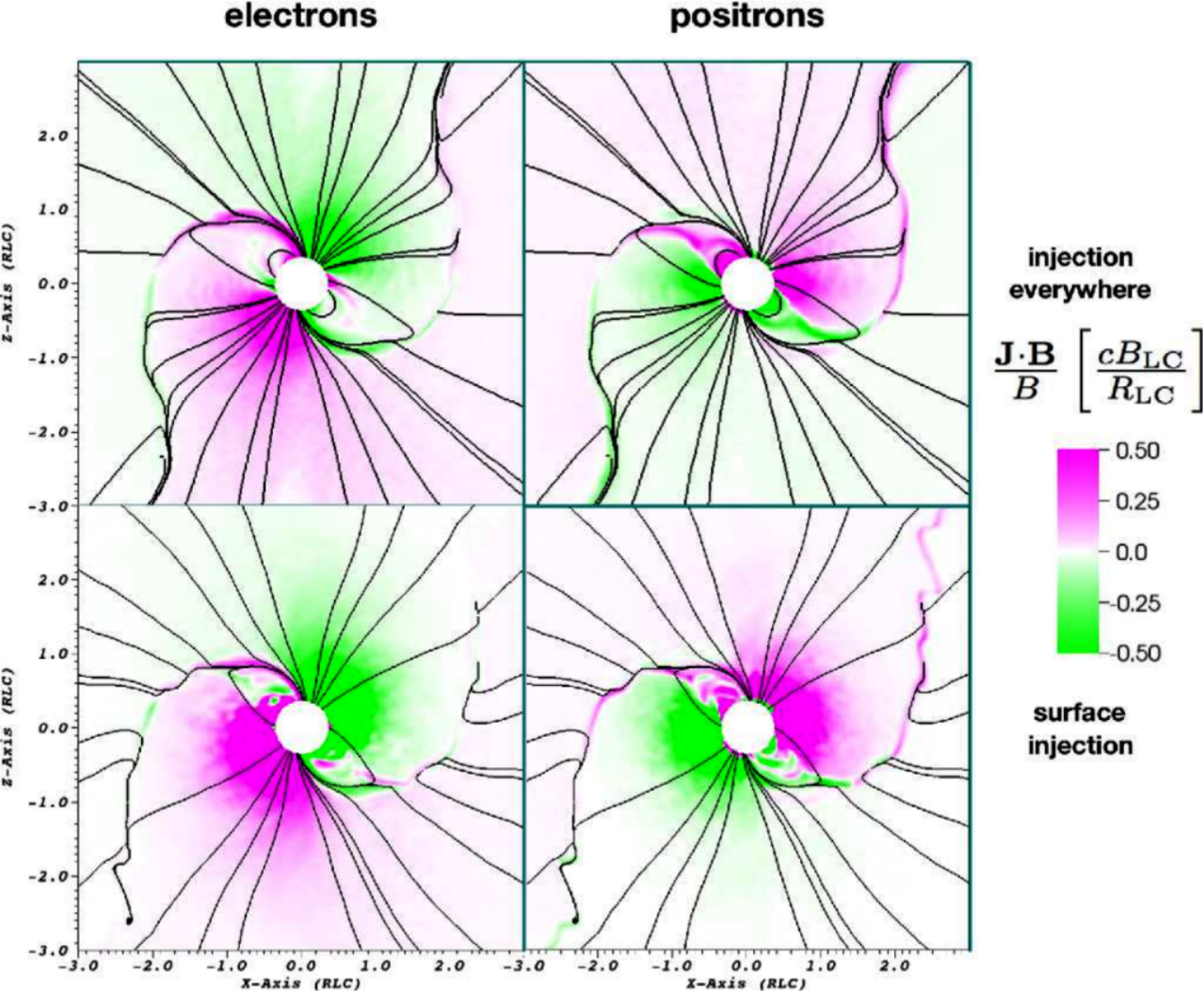}
\end{center}
\caption[The electron and positron components of the current density as predicted by the PIC model ]{\label{fig:PIC_example} The electron and positron components of the current density for magnetospheres close to being FF, as predicted by the PIC model of \citet{Brambilla2018}. One can see that the electrons and positrons both flowed out in the PC regions. The labels distinguish the cases where pair injection took place only at the surface vs. everywhere in the magnetosphere. From \citet{Brambilla2018}.}
\end{figure}

Recent reviews, see \citet{Brambilla2018} investigated current composition and flow using a new PIC code \citep{Kalapotharakos2018}, and focused on the dependence of magnetospheric properties on particle injection rate. Larger injection rates are achieved by \citet{Brambilla2018} than in previous studies. They obtained a transition of the magnetospheric solutions from a vacuum to FF solution, invoking two scenarios of particle injection, e.g., from the NS surface and everywhere in the pulsar magnetosphere. They obtained the highest dissipation for intermediate injection rates. As they increased injection rate (i.e., equivalent to a macroscopic $\sigma$ being increased), $E_\parallel$ was gradually (but not fully) screened and the FF current structure was attained. The dissipation regions also mostly moved to the current sheet. However, these two particle injection scenarios differ in particle density distribution in the sense that higher multiplicities were reached at the NS surface in the surface-injection scenario. They also studied the particle trajectories, and could probe some details of the current composition, e.g., they found that electrons and positrons both flowed out in the PC regions (which may inhibit two-photon pair production; see Figure~\ref{fig:PIC_example}), and lower-energy electrons returned to the NS surface by crossing the $B$-field lines close to the return current sheet inside the light cylinder, thus making them good candidates for emitting SR in the MeV range. 

The energetic particles flowing out along the current sheet correspond well to the FIDO model assumption \citep{Kalapotharakos2017} that invokes dissipation regions beyond the light cylinder into the current sheet. This model generally provides a good description of the \emph{Fermi} pulsar phenomenology. Thus, the latest PIC simulations are now elucidating and justifying the FIDO macroscopic assumptions and electrodynamical (or spatial accelerator) constraints derived from the GeV data when assuming CR from positrons in the current sheet \citep{Kalapotharakos2018}. 

\subsection{Striped-wind models}
%Petri, Mochol, Kirk

In these models, the current sheet (equatorial region between $B$-field lines of different polarities) is considered to be the main region for the generation of HE pulsed emission \citep{Beskin1983,Montgomery1999}. There are still many open questions regarding the basic understanding of general concepts within the framework of the current sheet models, e.g., location of the magnetic reconnection, the formation plasmoids via the tearing instability, the nature of the dissipation, and the effect of internal thermal pressure on the current sheet thickness. Several studies attempted to answer these questions (e.g., \citealt{Lyubarskii1996,Petri2011,Petri2012}) in the context of the ``striped-wind" models \citep{Coroniti1990,Michel1994} (see \citealt{Petri2016} for a more detailed review). 

\begin{figure}
\begin{center}
\includegraphics[width=15cm]{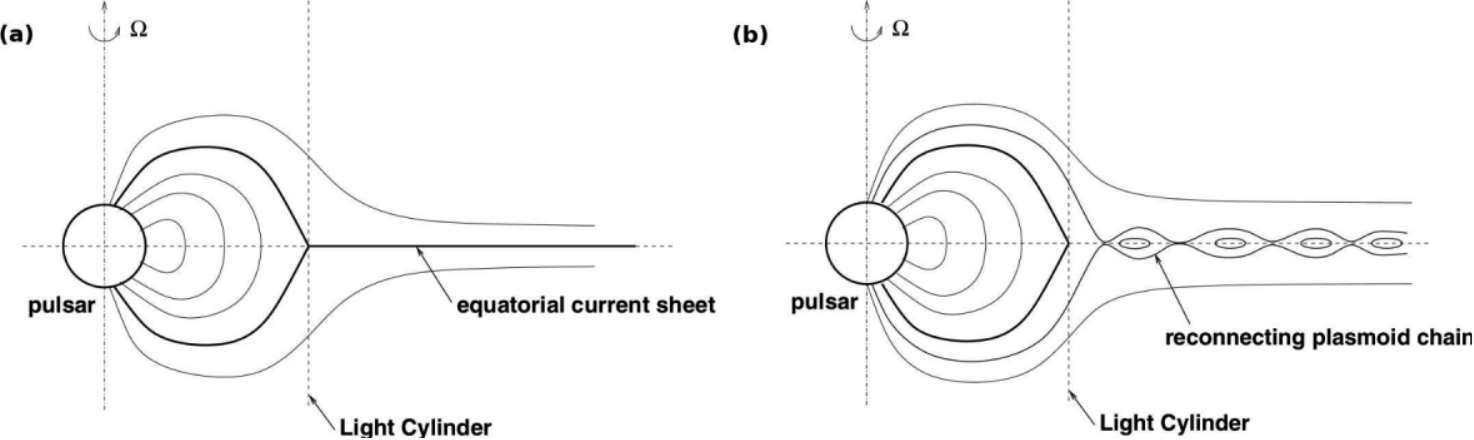}
\end{center}
\caption[Current sheet model]{\label{fig:windmodels} The (a) basic axisymmetric magnetosphere, and the (b) tearing of the equatorial current sheet (formation of plasmoids). From \citet{Uzdensky2014}.}
\end{figure}

Some studies have contrasting ideas about the current sheet. \citet{Kalapotharakos2014} notice that the physical conditions present in the current sheet greatly modifies the global magnetosphere structure. Others, such as \citet{Uzdensky2014}, argue that the current sheet is part of a rotating pulsar magnetosphere, and that the magnetic reconnection dissipates a large fraction of the pulsar spin-down power there. The numerical FF codes cannot treat the current sheet properly, therefore they developed a near-$R_{\rm LC}$ reconnection model to constrain the local plasma conditions. They stated that reconnection takes place via the formation of plasmoids (growing ``magnetic islands''; Figure~\ref{fig:windmodels}b) of different sizes. These plasmoids are continuously formed and merging with each other, and are ejected quasi-periodically. The relativistically hot reconnection layers present in pulsars undergo strong SR cooling, which leads to plasma compression. The particles can indeed radiate pulsed GeV emission by SR as well as pulsed emission in the TeV-band via ICS of ultraviolet or X-ray emission from the pulsar (see Chapter~\ref{chap:VelaTeV}).

%----------------------------------------------------------------------

\section{Summary} \label{sec:ch2concl}

This Chapter gave a broad overview of pulsar astrophysics such as the pulsar history, formation of pulsars, different pulsar classes, traditional models, etc. \citet{Harding2015} assumed a FF $B$-field structure (Section~\ref{subsec:FFfield}), in an SG and current sheet scenario, in their SSC emission model (Section~\ref{sec:RadMechanisms}) used in this study (see Chapters~\ref{chap:ch4},~\ref{chap:CREdepLCmod}, and~\ref{chap:VelaTeV}). The trends observed using the FIDO model are also seen in the GeV to TeV light curves (see Chapters~\ref{chap:CREdepLCmod} and~\ref{chap:VelaTeV}). 

In Chapter~\ref{chap:CREdepLCmod} I investigate the P1/P2 effect seen in the light curves and also explain this by investigating the phase-resolved spectra. This effect is also noted in studies regarding the dissipative models (Section~\ref{subsec:FIDOfield}). Lastly the values chosen for the two-step $E_\parallel$ used by \citet{Harding2018} are motivated by values obtained for the accelerating $E$-field in the PIC simulations, after suitable scaling of the particle energies to realistic values.
\\
\\
In the next Chapter I will discuss our study concerning the investigation of different $B$-field structures on the GeV light curves of Vela. This study deals with the offset-PC dipole field described in Section~\ref{subsec:OffsetPC}. % Background study on pulsar science
\chapter[The effect of an offset-PC $B$-field geometry on the predicted $\gamma$-ray light curves of the Vela pulsar]{The effect of an offset-PC $B$-field geometry on the predicted $\gamma$-ray light curves of the Vela pulsar\chaptermark{header}} \label{chap:OffsetPC}
\chaptermark{An offset-PC $B$-field geometry}
%Summarise article / thesis work of M.Sc.
%Focus on work not published in M thesis

%General
Recent studies using \emph{NICER} (see Section~\ref{subsec:NICER}) data point to pulsars having offset-PC $B$-field structures (\citealt{Bilous2019,Lockhart2019,Kalapotharakos2020}). These non-dipolar $B$-field geometries are also motivated by earlier observations of thermal X-ray emission, e.g., pulse profiles from MSPs such as PSR~J0437$-$4715 \citep{Bogdanov2007} and PSR~J0030+0451 \citep{Bogdanov2009}, with the $B$-fields of NSs in low mass X-ray binaries even more distorted \citep{Lamb2009}.

Within the global modelling landscape, numerical magnetic fields such as the FF solution (characterised by different PC currents than those assumed in space-charge limited flow models; \citealp{Contopoulos1999,Timokhin2006}) will undergo larger sweepback of field lines near the light cylinder, and consequently display a larger offset of the PC toward the trailing side (opposite to the rotation direction) than in the RVD field (which has offset PCs due to rotation alone; see Section~\ref{subsec:RVD}). 

We investigated the impact of different magnetospheric structures on the predicted $\gamma$-ray pulsar light curves. Using a particular implementation of an offset-PC $B$-field (see Section~\ref{subsec:OffsetPC} and Section~\ref{sec:Bfields}), we wanted to constrain the amount of offset using GeV data of the Vela pulsar. We also considered a SG (see Section~\ref{subsec:SG}) $E$-field associated with this particular $B$-field and constrain the $E$-field magnitudes when multiplying it with a factor 100. We performed geometric pulsar light curve modelling using different $B$-field structures in conjunction with geometric models. Additionally, we incorporated an SG $E$-field into our geometric modelling code. The fact that we have an $E$-field solution enables us to solve the particle transport equation on each $B$-field line for the offset-PC dipole and SG model combination for both the relatively low SG $E$-field and the increased case.

This Chapter is a summary\footnote{\bf This Chapter is based on research that was initiated during my MSc \citep{Breed2015MSc}, as well as research done during the first year of my PhD.} of work presented in \citet{Barnard2016} and will focus on the implementation of an offset-PC $B$-field and the results obtained. In this chapter we will briefly describe the offset-PC dipole $B$-field structure we considered (Section~\ref{sec:Bfields}) as well as the implementation thereof in our geometric modelling code, assuming certain geometric models (Section~\ref{sec:geometries}). This implementation involved a transformation of the $B$-field between the magnetic and rotational frames (Section~\ref{sec:transform}), as well as finding the PC rim (Section~\ref{sec:epsilon_extend}). We also describe the calculation of the associated SG $E$-field and the matching of the low-altitude and high-altitude solutions using a matching parameter (scaled radius) $\eta_{\rm c}$ in Section~\ref{sec:SGEfield}. We briefly discuss the $\chi^2$ method we applied in order to find best-fit ($\alpha$,$\zeta$) for the different model combinations (see Section~\ref{sec:chi2fit}). In Section~\ref{sec:ch3res}, we present our solution of the transport equation for the offset-PC dipole $B$-field for the usual and increased $E$-field solutions. This also includes our light curve predictions for two distinct cases: 1) the effect of lowering the minimum photon energy as well as 2) multiplying the $E$-field by a factor 100. This is followed by our best-fit ($\alpha$,$\zeta$) contours for the Vela pulsar, for both cases where $E_\gamma>$100~GeV, before we compare our results to previous multi-wavelength studies from other works in Section~\ref{sec:comparison_models}. Our conclusions follow in Section~\ref{sec:ch3concl}. For an alternative and independent implementation of an offset-dipole geometry, see \citet{Kundu2017}.

\section{Introduction}

The \textit{Fermi}'s Second Pulsar Catalogue (2PC; \citealp{Abdo2013SecondCat}) describes the properties of more than 150 pulsars in the energy range 100~MeV to 100~GeV. This catalogue includes the Vela pulsar \citep{Abdo2009Vela}, one of the brightest persistent sources in the GeV sky. A third catalogue is currently in preparation.

Despite the major advances made after nearly 50 years since the discovery of the first pulsar \citep{Hewish1968}, many questions still remain regarding the electrodynamical character of the pulsar magnetosphere, including details of the particle acceleration and pair production, current closure, and radiation of a complex multi-wavelength spectrum. Physical emission models such as the SG \citep{Muslimov2003} and OG (Section~\ref{subsec:OG}; \citealp{Romani1995}) fall short of explaining these global magnetospheric characteristics. More recent developments include the global magnetospheric properties. One example is the force-free (FF; see Section~\ref{subsec:FFfield}) inside and dissipative outside (FIDO; see Section~\ref{subsec:FIDOfield}) model \citep{Kalapotharakos2009,Kalapotharakos2014} that assumes FF electrodynamical conditions (infinite plasma conductivity $\sigma_{\rm c}\rightarrow\infty$) inside the light cylinder and dissipative conditions (finite $\sigma_{\rm c}$) outside. The wind models of, e.g., \citet{Petri2011} provide an alternative picture where dissipation takes place outside the light cylinder. There is also kinetic / particle-in-cell simulations
(PIC; \citealt{Brambilla2018,Cerutti2016,Cerutti2016current,Cerutti2020,Kalapotharakos2018,Philippov2018}). See Chapter~\ref{chap:PSRastro} for a more detailed discussion of these models.

Although much progress has been made using the physical models, geometric light curve modelling still presents a crucial avenue for probing the pulsar magnetosphere in the context of traditional pulsar models, as these emission geometries may be used to constrain the pulsar geometry (inclination angle $\alpha$ and the observer viewing angle $\zeta$ with respect to the spin axis $\boldsymbol\Omega$), as well as the $\gamma$-ray emission region's location and extent. This may provide vital insight into the boundary conditions and help constrain the accelerator geometry of next-generation full radiation models. Geometric light curve modelling has been performed by, e.g., \citet{Dyks2004b,Venter2009,Watters2009,Johnson2014,Pierbattista2015} using standard pulsar emission geometries, including a two-pole caustic (TPC, of which the SG is its physical representation; \citealp{Dyks2003}), OG, and pair-starved polar cap \citep{Harding2005c} geometry. 

A notable conclusion from the 2PC was that the spectra and light curves of both the millisecond pulsar (MSP) and young pulsar populations show remarkable similarities, pointing to a common radiation mechanism and emission geometry (tied to the $B$-field structure). The assumed $B$-field structure is essential for predicting the light curves seen by the observer using geometric models, since photons are expected to be emitted tangentially to the local $B$-field lines in the corotating pulsar frame \citep{Daugherty1982}. Even a small difference in the magnetospheric structure will therefore have an impact on the light curve predictions. For all of the above geometric models, the most commonly employed $B$-field has been the retarded vacuum dipole (RVD) solution first obtained by \citet{Deutsch1955}. However, other solutions also exist. One example is the static dipole (non-rotating) field (see Section~\ref{subsec:OffsetPC}), a special case of the RVD (rotating) field~\citep{Dyks2004a}. \citet{Bai2010b} furthermore modelled high-energy (HE) light curves in the context of OG and TPC models using an FF $B$-field geometry (assuming a plasma-filled magnetosphere), proposing a separatrix layer model close to the last open field line (tangent to the light cylinder at radius $R_{\rm LC}=c/\Omega$ where the corotation speed equals the speed of light $c$, with $\Omega$ the angular speed), which extends from the stellar surface up to and beyond the light cylinder. In addition, the annular gap model of \citet{Du2010}, which assumes a static dipole field, has been successful in reproducing the main characteristics of the $\gamma$-ray light curves of three MSPs. This model does, however, not attempt to replicate the nonzero phase offsets between the $\gamma$-ray and radio profiles. 

The $B$-field is one of the basic assumptions of the geometric models (others include the gap region's location, and the $\epsilon_{\nu}$ profile in the gap). Several $B$-field structures have been studied in this context, including the static dipole \citep{Griffiths1995}, the RVD (a rotating vacuum magnetosphere which can in principle accelerate particles but do not contain any charges or currents; \citealp{Deutsch1955}), the FF (filled with charges and currents, but unable to accelerate particles, since the accelerating $E$-field is screened everywhere; \citealp{Contopoulos1999}), and the offset-PC dipole (that analytically mimics deviations from the static dipole near the stellar surface; \citealp{Harding2011a,Harding2011c}). A more realistic pulsar magnetosphere, i.e., a dissipative solution \citep{Kalapotharakos2012a,Li2012,Tchekhovskoy2013,Li2014}, would be one that is intermediate between the RVD and the FF fields. The dissipative $B$-field is characterised by the plasma conductivity $\sigma_{\rm c}$ (e.g., \citealp{Lichnerowicz1967}) which can be chosen in order to alternate between the vacuum ($\sigma_{\rm c}\rightarrow{0}$) and FF ($\sigma_{\rm c}\rightarrow\infty$) cases (see \citealp{Li2012}). 

We studied the effect of different magnetospheric structures (static dipole, RVD, and offset-PC dipole, further discussed below) and emission geometries (TPC and OG) on pulsar visibility and $\gamma$-ray pulse shape, particularly for the case of the Vela pulsar. For the static dipole the field lines are symmetric about the $\boldsymbol{\mu}$-axis, whereas the RVD is distorted due to sweepback of the field lines as the NS rotates. This has implications for the definition of the PC (see Section~\ref{sec:epsilon_extend}).

\section{Offset-dipole $B$-field structure}\label{sec:Bfields}

\begin{figure}
\captionsetup[subfloat]{farskip=2pt,captionskip=1pt}
	\centering
	\subfloat[\label{fig:Symm_offset}]{\includegraphics[width=0.48\textwidth]{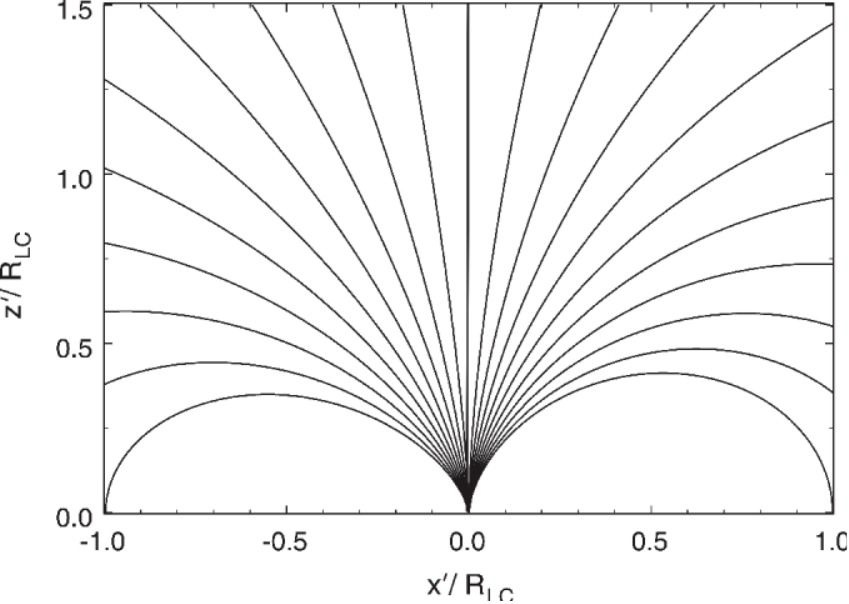}}\hfill
    \subfloat[\label{fig:Symm_offset_dir}]{\includegraphics[width=0.4\textwidth]{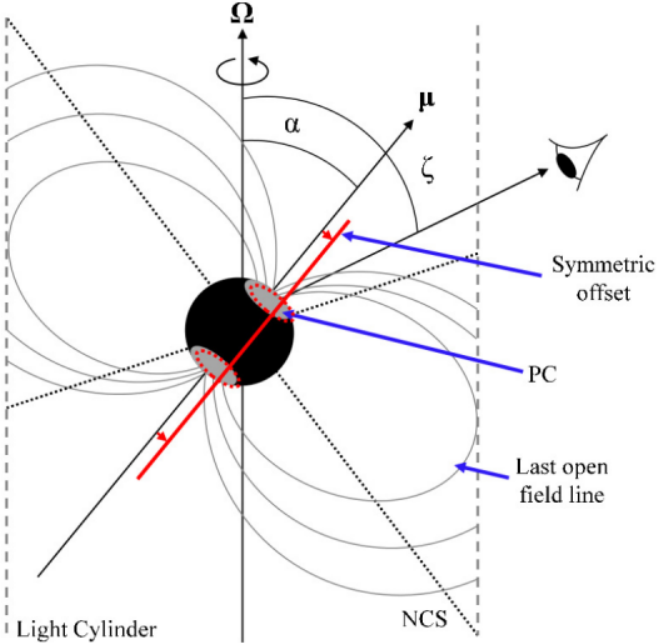}}\hfill
\caption[Symmetric offset-PC dipole \textit{B}-field]{\label{fig:Offset_Sym} In (a) we illustrate the distorted field lines of the offset dipole \textit{B}-field having symmetrically offset PCs in the $x^{\prime}-z^{\prime}$ plane, for an offset $\epsilon=0.2$ (adapted from \citealt{Harding2011c}). In (b) we represent the symmetric offset geometrically, where the PCs (grey ovals) are shifted from the magnetic axis in the same direction, resulting in symmetric PC offsets (red dashed circles). The curved arrow around $\boldsymbol{\Omega}$ indicates the direction of rotation.}
\end{figure}

\begin{figure}
\captionsetup[subfloat]{farskip=2pt,captionskip=1pt}
	\centering
	\subfloat[\label{fig:Asymm_offset}]{\includegraphics[width=0.48\textwidth]{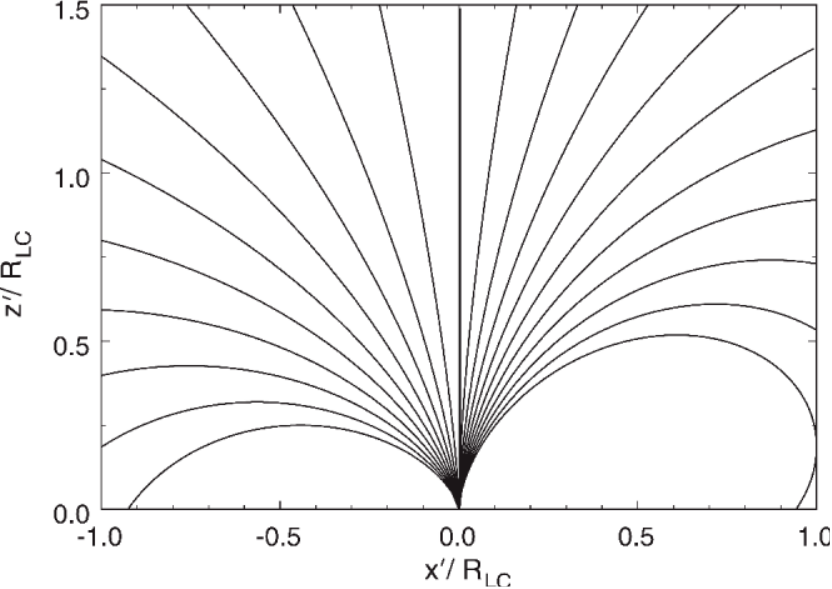}}\hfill
	\subfloat[\label{fig:Asymm_offset_dir}]{\includegraphics[width=0.4\textwidth]{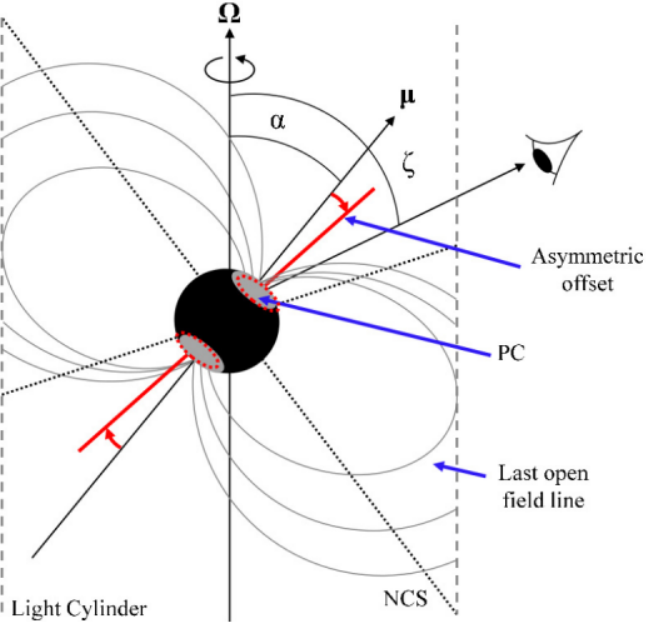}}\hfill
\caption[Asymmetric offset-PC dipole \textit{B}-field]{\label{fig:Offset_Asym} In (a) we illustrate the distorted field lines of the offset dipole \textit{B}-field having asymmetric offset PCs in the $x^{\prime}-z^{\prime}$ plane, for an offset $\epsilon=0.2$ (adapted from \citealt{Harding2011c}). In (b) we represent the offset geometrically, with the PCs (grey ovals) shifted from the magnetic axis in opposite directions, resulting in asymmetric PC offsets (red dashed circles). The curved arrow around $\boldsymbol{\Omega}$ indicates the direction of rotation.}
\end{figure}

\citet{Harding2011a,Harding2011c} considered two cases, i.e., symmetric and asymmetric PC offsets. The symmetric case involves an offset of both PCs in the same direction so that the PCs are not antipodal, and applies to NSs with some interior current distortions that produce multipolar components near the stellar surface (see Figure~\ref{fig:Offset_Sym}; \citealp{Harding2011c}). The asymmetric case is associated with asymmetric PC offsets in opposite directions and applies to PC offsets due to retardation and/or currents of the global magnetosphere (see Figure~\ref{fig:Offset_Asym}; \citealp{Harding2011c}). Both these cases were modelled by introducing an offset parameter $\epsilon$. Thus, as seen in Figure~\ref{fig:Symm_offset} the global open field lines of a centred dipole are bent toward the dipole axis on one side and bent away from the dipole axis on the other side of the PC. Therefore, one side of the PC is larger and the PC is effectively shifted from the centre of symmetry (see Figure~\ref{fig:Symm_offset_dir}). 

The general expression for a symmetric offset-PC dipole $B$-field in terms of spherical co-ordinates $(r^{\prime},\theta^{\prime},\phi^{\prime})$ in the magnetic frame (indicated by the primed co-ordinates, where $\hat{\boldsymbol{\rm z}}^{\prime}\parallel{\boldsymbol{\mu}}$) is as follows \citep{Harding2011c}
\begin{equation}\label{eq:Symm-offset}
{{\mathbf{B}}}_{\rm OPCs}^{\prime}(r^{\prime},\theta^{\prime},\phi^{\prime}) \approx \frac{\mu^\prime}{r^{\prime3}}\biggl[\cos\theta^{\prime}\hat{{\mathbf{r}}}^{\prime}+\frac{1}{2}(1+a)\sin\theta^{\prime}{\hat{\boldsymbol{\theta}}}^{\prime}-\epsilon\sin\theta^{\prime}\cos\theta^{\prime}\sin(\phi^{\prime}-\phi_{0}^\prime)\hat{\boldsymbol{\phi}}^{\prime}\biggr],
\end{equation}
where $\mu^\prime=B_{0}R^{3}$ is the magnetic moment, $R$ the stellar radius, $B_{0}$ the surface $B$-field strength at the magnetic pole, $\phi_{0}^{\prime}$ the magnetic azimuthal angle defining the plane in which the offset occurs, and $a=\epsilon\cos(\phi^{\prime}-\phi_0^\prime)$ characterises the offset direction in the $x^\prime-z^\prime$ plane. This distortion depends on parameters $\epsilon$ (related to the magnitude of the shift of the PC from the magnetic axis) and $\phi_0^\prime$ (we choose $\phi^\prime_0=0$ in what follows). If $\phi^\prime_0=0$ or $\phi^\prime_0=\pi$ the offset is in the $x^\prime$ direction (i.e., along the $x^\prime$-axis). If $\phi^\prime_0=\pi/2$ or $\phi^\prime_0=3\pi/2$ the offset is in the $y^\prime$ direction.

The general expression for an asymmetric $B$-field is as follows \citep{Harding2011c}
\begin{equation}\label{eq:Asymm-offset}
{\boldsymbol{\rm B}}_{\rm OPCa}^{\prime}(r^{\prime},\theta^{\prime},\phi^{\prime})\approx\frac{\mu^{\prime}}{r^{\prime3}}[\cos[\theta^{\prime}(1+a)]\hat{\boldsymbol{\rm r}}^{\prime}+\frac{1}{2}\sin[\theta^{\prime}(1+a)]\hat{\boldsymbol{\theta}}^{\prime}-\frac{1}{2}\epsilon(\theta^{\prime}+\sin\theta^{\prime}\cos\theta^{\prime})\sin(\phi^{\prime}-\phi_{0}^{\prime})\hat{\boldsymbol{\phi}}^{\prime}],
\end{equation}
with the distortion of the field lines also occurring in the $x^\prime-z^\prime$ plane. If we set $\epsilon=0$ both the symmetric and asymmetric cases reduce to a symmetric static dipole as given in Eq.~[\eqref{eq:static-dipole}] with the $B$-field lines being distorted in all directions. 

The distance by which the PCs are shifted on the NS surface is given by
\begin{equation}
\Delta{r_{\rm PC}}\simeq{R\theta_{\rm PC}}\left[1-\theta_{\rm PC}^\epsilon\right],
\end{equation}
where $\theta_{\rm PC}=(\Omega{R}/c)^{1/2}$ is the standard half-angle of the PC, and $\Omega$ the angular speed. This effective shift of the PCs is a fraction of $\theta_{\rm PC}$, therefore it is a larger fraction of $R$ for pulsars with shorter periods \citep{Harding2011a}. \citet{Harding2011c} found that for the RVD solution, $\epsilon=0.03-0.1$, where offsets as large as $0.1$ are associated with MSPs with large $\theta_{\rm PC}$. However, $\epsilon=0.09-0.2$ is expected for FF fields, with the larger offset values related to MSPs \citep{Bai2010b}. One of the main focus points of our study was the implementation of an offset dipole \textit{B}-field, for the symmetric case only. 

The difference between our offset-PC field and a dipole field which is offset with respect to the stellar centre can be most clearly seen by performing a multipolar expansion of these respective fields. \citet{Lowrie2011} gives the scalar potential $W$ for an equatorially offset dipole (EOD) field 
\begin{eqnarray}\label{eq:scalar-potential}
W^{\prime}(r^\prime,\theta^\prime)&=&\biggl[\frac{\mu^\prime}{r^{\prime2}}\cos\theta^\prime+\frac{\mu^\prime{d}}{r^{\prime3}}\sin\theta^\prime\cos\theta^\prime+\frac{\mu^\prime{d}}{2r^{\prime3}}\sin^{2}\theta^\prime\biggr], \nonumber
\end{eqnarray}
with $d$ being the offset parameter, and with the first few leading terms in $d/r^\prime$ listed above. From this potential, we may construct the magnetic field using $B = -\nabla W$:
\begin{eqnarray}\label{eq:offsetdipole}
{\bf B}_{\rm EOD}^{\prime}(r^\prime,\theta^\prime)&=&{\bf B}_{\rm dip}^{\prime}(r^\prime,\theta^\prime)+\biggl[\frac{3\mu^\prime{d}}{r^{\prime4}}\sin\theta^\prime\cos\theta^\prime+\frac{3\mu^\prime{d}}{2r^{\prime4}}\sin^{2}\theta^\prime\biggr]\hat{\mathbf{r^\prime}} \nonumber\\
& &-\biggl[\frac{\mu^\prime{d}}{r^{\prime4}}\cos^{2}\theta^\prime+\frac{\mu^\prime{d}}{r^{\prime4}}\sin^{2}\theta^\prime-\frac{\mu^\prime{d}}{r^{\prime4}}\sin\theta^\prime\cos\theta^\prime\biggr]\hat{\boldsymbol{\theta^\prime}}, \nonumber \\
&=& {\bf B}_{\rm dip}^{\prime}(r^\prime,\theta^\prime)+{\it O}\left(\frac{1}{r^{\prime4}}\right).
\end{eqnarray}
This means that an offset dipolar field may be expressed (to lowest order) as the sum of a centred dipole and two quadrupolar components. Conversely, our offset-PC field may be written as
\begin{eqnarray}\label{eq:Symm-offset}
{{\mathbf{B}}}_{\rm OPCs}^{\prime}(r^\prime,\theta^\prime,\phi^\prime) & \approx & {\bf B}_{\rm dip}^{\prime}(r^\prime,\theta^\prime)+{\it O}\left(\frac{\epsilon}{r^{\prime3}}\right).
\end{eqnarray}

Therefore, we can see that the EOD consists of a centred dipole plus quadropolar and other higher-order components (Eq.~[\ref{eq:offsetdipole}]), while our offset-PC model (Eq.~[\ref{eq:Symm-offset}]) consists of a centred dipole plus terms of order $a/r^{\prime3}$ or $\epsilon/r^{\prime3}$. Since $a\sim0.2$ and $\epsilon\sim0.2$, the latter terms present perturbations (e.g., poloidal and toroidal effects) to the centred dipole. \citet{Harding2011a,Harding2011c} derived these perturbed components of the distorted magnetic field while satisfying the solenoidality condition $\nabla\cdot B=0$.

In what follows, we decided to study the effect of the simpler symmetric case (which does not mimic field line sweepback of FF, RVD or dissipative magnetospheres) on predicted light curves. These complex $B$-fields usually only have numerical solutions, which are limited by the resolution of the spatial grid. Hence, it is simpler to investigate the main effects of these structures using analytical approximations such as the offset-PC dipole solution. In future, one can also include the more complex asymmetric case. 

\section{Geometric models}\label{sec:geometries}

\textit{Geometric} models assume constant emissivity $\epsilon_\nu$ in the rotational frame. We have also incorporated an SG $E$-field associated with the offset-PC dipole $B$-field (making this latter case an \textit{emission} model), which allows us to calculate the $\epsilon_{\nu}$ in the acceleration region in the corotating frame from first principles. We have only considered the TPC (assuming uniform $\epsilon_{\nu}$) and SG (assuming variable $\epsilon_{\nu}$ as modulated by the $E$-field) models for the offset-PC dipole $B$-field, since we do not have $E$-field expressions available for the OG model within the context of an offset-PC dipole $B$-field. 

The geometric TPC pulsar model was first introduced by \citet{Dyks2003}. \citet{Muslimov2003} revived the physical SG model of \citet{Arons1983}, including general relativistic (GR) corrections, and argued that the SG model may be considered a physical representation of the TPC model. This gap geometry has a large radial extent, spanning from the neutron star (NS) surface along the last closed field line up to the light cylinder. The original definition stated that the maximum radial extent reached $R_{\rm max}{\simeq}0.8{R_{\rm LC}}$ \citep{Muslimov2004a}. This was later extended to $R_{\rm max}{\simeq}1.2{R_{\rm LC}}$ for improved fits (e.g., \citealp{Venter2009,Venter2012}). Typical transverse gap extents of $1-5\%$ of the PC angle have been used \citep{Venter2009,Watters2009}. 

The OG model was introduced by \citet{Cheng1986} and elaborated by \citet{Romani1995}. They proposed that when the primary current passes through the neutral sheet or null-charge surface (NCS, with a radius of $R_{\rm NCS}$, i.e., the geometric surface across which the charge density changes sign) the negative charges above this sheet will escape beyond the light cylinder. A vacuum gap region is then formed (in which the $E$-field parallel to the local $B$-field, $E_{\parallel}\neq0$). Analogously, the geometric OG model has a radial extent spanning from the NCS to the light cylinder. We follow \citet{Venter2009} and \citet{Johnson2014} who considered a one-layer model with a transverse extent along the inner edge of the gap.

We performed geometric light curve modelling using the code first developed by \citet{Dyks2004b} which already includes the static dipole and RVD solutions. We extended this code by implementing an offset-PC dipole $B$-field (for the symmetric case), as well as the SG $E_\parallel$-field corrected for GR effects (see Section~\ref{sec:SGEfield}). We solve for the PC rim as explained in Section~\ref{sec:epsilon_extend}. The shape of PC rim depends on the $B$-field structure at the light cylinder $R_{\rm LC}$. Once the PC rim has been determined, it is divided into self-similar (interior) rings. These rings are calculated by using open-volume co-ordinates ($r_{\rm ovc}$ and $l_{\rm ovc}$). After the footpoints of the field lines on a ($r_{\rm ovc}$,$l_{\rm ovc}$) grid have been determined, particles are followed along these lines in the corotating frame and emission from them is collected in bins of pulse phase $\phi_{\rm L}$ and $\zeta$, i.e., a phase plot is formed by plotting the bin contents (divided by the solid angle subtended by each bin) for a given $\alpha$, and it is therefore a projection of the radiation beam. To simulate light curves, one chooses a phase plot corresponding to a fixed $\alpha$, then fix $\zeta$ and plot the intensity per solid angle.

The code takes into account the structure/geometry of the $B$-field (since the photons are emitted tangentially to the local field line), aberration of the photon emission direction (due to rotation, to first order in $r/R_{\rm LC}$), and time-of-flight delays (due to distinct emission radii) to obtain the caustic emission beam \citep{Morini1983,Dyks2004c}. However, \citet{Bai2010a} pointed out that previous studies assumed the RVD field to be valid in the instantaneously corotating frame, but actually it is valid in the laboratory frame (implying corrections that are of second-order in $r/R_{\rm LC}$). This implies a revised aberration formula, which we have implemented in our code.

\section{Transformation of a $B$-field from the magnetic to the rotational frame}\label{sec:transform}

We implemented an offset dipole $B$-field for the symmetric case (where the PCs of both hemispheres are offset in the same direction with respect to the magnetic ($\boldsymbol{\mu}$) axis; see Section~\ref{sec:Bfields}) in our geometric code (Section~\ref{sec:geometries}). Since the offset dipole field is given in terms of magnetic frame co-ordinates ($\hat{\boldsymbol{\rm z}}^{\prime}\parallel\boldsymbol{\mu}$; \citealp{Harding2011c}) it was necessary to transform this solution to the (corotating) rotational frame ($\hat{\boldsymbol{\rm z}}\parallel{\boldsymbol{\Omega}}$, with $\boldsymbol{\Omega}$ the rotation axis). In order to do so, we first performed transformations between the spherical and Cartesian co-ordinates and bases, and then a rotation of the co-ordinate axes to move from the magnetic frame to the rotational frame. We lastly transformed the Cartesian co-ordinates of the position vector from the magnetic to the rotational frame. For a more detailed discussion, we refer the reader to \cite{Barnard2016}.

Consider a general $B$-field specified in the magnetic frame (indicated by the primed co-ordinates), in terms of spherical co-ordinates
\begin{equation}\label{eq:rtpprime}
\boldsymbol{\mathbf{B}}^{\prime}(r^{\prime},\theta^{\prime},\phi^{\prime})=B_{r}^{\prime}(r^{\prime},\theta^{\prime},\phi^{\prime})\hat{\boldsymbol{\rm r}}^{\prime}+B_{\theta}^{\prime}(r^{\prime},\theta^{\prime},\phi^{\prime})\hat{\boldsymbol{\theta}}^{\prime}+B_{\phi}^{\prime}(r^{\prime},\theta^{\prime},\phi^{\prime})\hat{\boldsymbol{\phi}}^{\prime}.
\end{equation}
This field may then be transformed to a Cartesian basis and co-ordinate system: 
\begin{equation}\label{eq:xyzprime}
\boldsymbol{\mathbf{B}}^{\prime}(x^{\prime},y^{\prime},z^{\prime})=B_{x}^{\prime}(x^{\prime},y^{\prime},z^{\prime})\hat{\boldsymbol{\rm x}}^{\prime}+B_{y}^{\prime}(x^{\prime},y^{\prime},z^{\prime})\hat{\boldsymbol{\rm y}}^{\prime}+B_{z}^{\prime}(x^{\prime},y^{\prime},z^{\prime})\hat{\boldsymbol{\rm z}}^{\prime}.
\end{equation}
This is done using expressions that specify spherical unit vectors and co-ordinates in terms of Cartesian co-ordinates (see e.g., \citealp{Griffiths1995}). Next, one may rotate the $B$-field components (i.e., the Cartesian frame) through an angle $-\alpha$ (the angle between the $\boldsymbol{\Omega}$ and $\boldsymbol{\mu}$ axes), thereby transforming the $B$-field from the magnetic to the rotational frame (indicated by the unprimed co-ordinates). 
\begin{equation}\label{eq:xyzprime}
\boldsymbol{\mathbf{B}}(x^{\prime},y^{\prime},z^{\prime})=B_{x}(x^{\prime},y^{\prime},z^{\prime})\hat{\boldsymbol{\rm x}}^{\prime}+B_{y}(x^{\prime},y^{\prime},z^{\prime})\hat{\boldsymbol{\rm y}}^{\prime}+B_{z}(x^{\prime},y^{\prime},z^{\prime})\hat{\boldsymbol{\rm z}}^{\prime}.
\end{equation}
Lastly, we transform the magnetic co-ordinates to rotational co-ordinates: 
\begin{equation}\label{eq:xyz}
\boldsymbol{\mathbf{B}}(x,y,z)=B_{x}(x,y,z)\hat{\boldsymbol{\mathbf{x}}}+B_{y}(x,y,z)\hat{\boldsymbol{\rm{y}}}+B_{z}(x,y,z)\hat{\boldsymbol{\rm{z}}}.
\end{equation}

\section{Finding the PC rim and extending the range of $\epsilon$}\label{sec:epsilon_extend}

The object is to find the polar angle $\theta_{*}$ at each azimuthal angle $\phi$ at the footpoints of the last open $B$-field lines, lying within a bracket $\theta_{\rm min}<\theta_{*}<\theta_{\rm max}$, such that the field line is tangent to the light cylinder. The PC rim is thus defined. The magnetic structure at the light cylinder therefore determines the PC shape \citep{Dyks2004a,Dyks2004b}. 

After initial implementation of the offset-PC dipole field in the geometric code, we discovered that we could solve for the PC rim in a similar manner as for the RVD $B$-field, but only for small values of the offset parameter $\epsilon$ ($\epsilon\lesssim{0.05-0.1}$, depending on $\alpha$). We improve the range of $\epsilon$ by varying the colatitude parameters $\theta_{\rm min}$ and $\theta_{\rm max}$ which delimit a bracket (``solution space") in colatitude thought to contain the footpoint of last open field line (tangent to the light cylinder $R_{\rm LC}$). We obtain a progressively larger range of $\epsilon$ upon decreasing $\theta_{\rm min}$ and increasing $\theta_{\rm max}$. We find a maximum $\epsilon=0.18$ valid for the full range of $\alpha$. Choosing a maximal solution bracket in colatitude would in principle work, but the code would take much longer to find the PC rim compared to when a smaller bracket (that does contain the correct solution) is used. Therefore, we generalise the search for optimal $\theta_{\rm min}$ and found (by trial and error) that the following linear equation $\theta_{\rm min}=[(-31/18)\epsilon+0.6]\theta_{\rm PC}$, for a fixed $\theta_{\rm max}=2.0$, resulted in $\theta_{\rm min}$ that yielded maximum values for $\epsilon$.

\begin{figure}
    \centering
    \includegraphics[width=10cm]{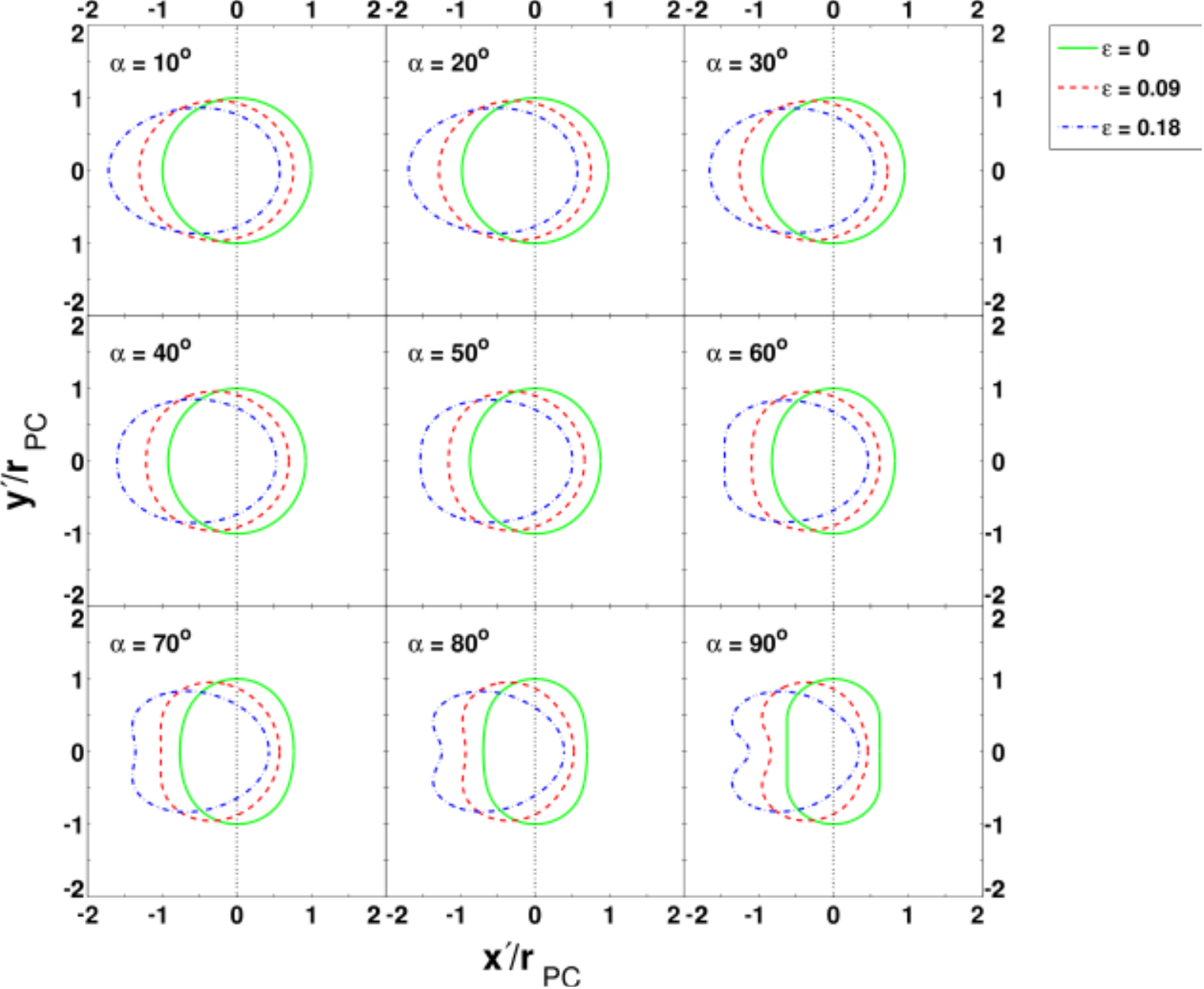}
    \caption[PC shapes of the offset-PC dipole $B$-field for a few cases of $\alpha$ and $\epsilon$ in the $x^\prime-y^\prime$ plane]{PC shapes of the offset-PC dipole $B$-field for a few cases of $\alpha$ and $\epsilon$ in the $x^\prime-y^\prime$ plane assuming that the $\mathbf{\mu}$-axis is located perpendicularly to the page at ($x^\prime,y^\prime$)=(0,0) and that $\phi^\prime$ is measured counterclockwise from the positive $x^\prime$-axis. Each PC is for a different value of $\alpha$ ranging between $10^\circ$ and $90^\circ$, with $10^\circ$ resolution. For each $\alpha$ we plot the PC shape for $\epsilon$ values of 0 (green solid circle), 0.09 (red dashed circle), and 0.18 (blue dashed–dotted circle). We note that the reference green PCs are for the static centred dipole. The horizontal line at $x^\prime$=0 (black dotted line) serves as a reference line to show the magnitude and direction of the offset as $\epsilon$ is increased. \label{fig:offsetPCs}}
\end{figure}

If the PC rims ($r_{\rm ovc}=1$) are viewed in the $x^\prime-y^\prime$ plane (in the magnetic frame) as a function of $\alpha$ and $\epsilon$ (assuming that the $\boldsymbol{\mu}$-axis is located perpendicularly to the page at $(x^\prime,y^\prime)=(0,0)$ and that $\phi^\prime$ is measured counterclockwise from the positive $x^\prime$-axis) we note that the PC shape changes considerably. As $\alpha$ and $\epsilon$ are increased the PC offset is larger in the direction of ``unfavourably curved'' $B$-field lines (i.e., $-x^\prime$-axis). For larger $\alpha$ values irrespective of $\epsilon$ the PC shape along the $x^\prime$-axis becomes narrower and irregular. This narrowing effect of the PC is also seen along the $y^\prime$-axis as $\epsilon$ increases. We illustrate the PC shape for a few cases of $\alpha$ and $\epsilon$ in Figure~\ref{fig:offsetPCs}.

\section{Incorporating an SG $E$-field}\label{sec:SGEfield}

\subsection{Approximate expressions of the associated $E$-field}\label{subsec:Efield}
It is important to take the accelerating $E$-field into account when such expressions are available, since this will modulate the emissivity $\epsilon_\nu(s)$ (as a function of arclength $s$ along the $B$-field line) in the gap as opposed to geometric models where we assume constant $\epsilon_\nu$ per unit length in the corotating frame. For the SG case, we implemented the full $E$-field in the rotational frame corrected for GR effects (e.g., \citealp{Muslimov2003,Muslimov2004a}). This solution consists of a low-altitude and high-altitude limit which we have to match on each $B$-field line. The low-altitude solution is given by (Harding, private communication)
\begin{eqnarray}\label{eq:E_low}
E_{\parallel,{\rm low}} & \approx & {-3}{\mathcal{E}_0}\nu_{\rm SG}x^a\Big\{\frac{\kappa}{\eta^4}e_{\rm 1A}\cos\alpha+\frac{1}{4}\frac{\theta_{\rm PC}^{1+a}}{\eta}\big[e_{\rm 2A}\cos\phi_{\rm PC} \nonumber \\
 & & +\frac{1}{4}\epsilon\kappa{e_{\rm 3A}}(2\cos\phi_0^\prime-\cos(2\phi_{\rm PC}-\phi_0^\prime))\big]\sin\alpha\Big\}(1-\xi_\ast^2),
\end{eqnarray}
with $\mathcal{E}_0=(\Omega{R}/c)^2(B_r/B)B_0$, $B_r$ the radial $B$-field component, $\nu_{\rm SG}\equiv({1/4})\Delta\xi_{\rm SG}^2$, and $\Delta\xi_{\rm SG}$ the colatitudinal gap width in units of dimensionless colatitude $\xi=\theta/\theta_{\rm PC}$. Also, $x=r/R_{\rm LC}$ is the normalised radial distance in units of $R_{\rm LC}$. Here, $\kappa\approx{0.15I_{45}/R_6^3}$ is a GR compactness parameter characterising the frame-dragging effect near the stellar surface \citep{Muslimov1997},~$I_{45}=I/10^{45}$ g cm$^2$, $I$ the moment of inertia, $R_6=R/10^6$ cm, $\eta=r/R$ the dimensionless radial co-ordinate in units of $R$, $e_{\rm 1A}=1+{a}(\eta^3-1)/3$, $e_{\rm 2A}=(1+3a)\eta^{(1+a)/2}-2a$ and $e_{\rm 3A}=[({5-3a})/{\eta^{(5-a)/2}}]+2a$. The magnetic azimuthal angle $\phi_{\rm PC}$ is defined for usage with the $E$-field, being $\pi$ out of phase with $\phi^\prime$ (one chooses the negative $x$-axis towards $\boldsymbol{\Omega}$ to coincide with $\phi_{\rm PC}=0$, labelling the ``favourably curved" $B$-field lines). We define $\phi^{\prime}=\arctan(y^\prime/x^\prime)$ the magnetic azimuthal angle used when transforming the $B$-field (Section~\ref{sec:transform}). Lastly, $\xi_\ast$ is the dimensionless colatitude labelling the gap field lines (defined such that $\xi_{\ast}=0$ corresponds to the field line in the middle of the gap and $\xi_{\ast}=1$ at the boundaries; \citealp{Muslimov2003}). 

We approximate the high-altitude SG $E$-field by \citep{Muslimov2004a}
\begin{eqnarray}\label{eq:E_high}
E_{\parallel,{\rm high}} & \approx & -\frac{3}{8}\Big(\frac{\Omega{R}}{c}\Big)^3\frac{B_{\rm 0}}{f(1)}\nu_{\rm SG}x^a\Big\{\Big[1+\frac{1}{3}\kappa\Big(5-\frac{8}{\eta^3_{\rm c}}\Big)+2\frac{\eta}{\eta_{\rm LC}}\Big]\cos\alpha \nonumber \\
 & & +\frac{3}{2}\theta_{\rm PC}H(1)\sin\alpha\cos\phi_{\rm PC}\Big\}(1-\xi_\ast^2),   
\end{eqnarray}
with $f(\eta)\sim{1+0.75y+0.6y^2}$ a GR correction factor of order 1 for the dipole component of the magnetic flux through the magnetic hemisphere of radius $r$ in a Schwarzchild metric. The function $H(\eta)\sim 1-0.25y-0.16y^2 -0.5(\kappa/\epsilon_{\rm g}^3)y^3(1-0.25y-0.21y^2)$ is also a GR correction factor of order 1, with $y=\epsilon_{\rm g}/\eta$, $\epsilon_{\rm g}=r_{\rm g}/R$, and $r_{\rm g}=2GM/c^2$ the gravitational or Schwarzchild radius of the NS (with $G$ the gravitational constant and $M$ the stellar mass). The factors $f(\eta)$ and $H(\eta)$ account for the static part of the curved spacetime metric and have a value of 1 in flat space \citep{Muslimov1997}. The critical scaled radius $\eta_{\rm c}=r_{\rm c}/R$ is where the high-altitude and low-altitude $E$-field solutions are matched, with $r_{\rm c}$ the critical radius, and $\eta_{\rm LC}=R_{\rm LC}/R$. This high-altitude solution (excluding the factor $x^a$) is actually valid for the SG model assuming a static (GR-corrected, non-offset) dipole field. We therefore scale the $E$-field by a factor $x^a$ to generalise this expression for the offset-PC dipole field. The general $E$-field valid from $R$ to $R_{\rm LC}$ (i.e., over the entire length of the gap) is constructed as follows (see Eq.~[59] of \citealp{Muslimov2004a})
\begin{equation}\label{eq:E_total}
E_{\parallel,{\rm SG}}{\simeq}E_{\parallel,{\rm low}}\exp[-(\eta-1)/(\eta_{\rm c}-1)]+E_{\parallel,{\rm high}}.
\end{equation}
A more detailed discussion of the electrodynamics in the SG geometry may be found in \citet{Muslimov2003} and \citet{Muslimov2004a}. In the next section, we solve for $\eta_{\rm c}(P,\dot{P},\alpha,\epsilon,\xi,\phi_{\rm PC})$ where $P$ is the period and $\dot{P}$ its time derivative.

\subsection{Determining the matching parameter $\eta_{\rm c}$}\label{subsec:matching}
\begin{figure}
\centering
\includegraphics[width=12cm]{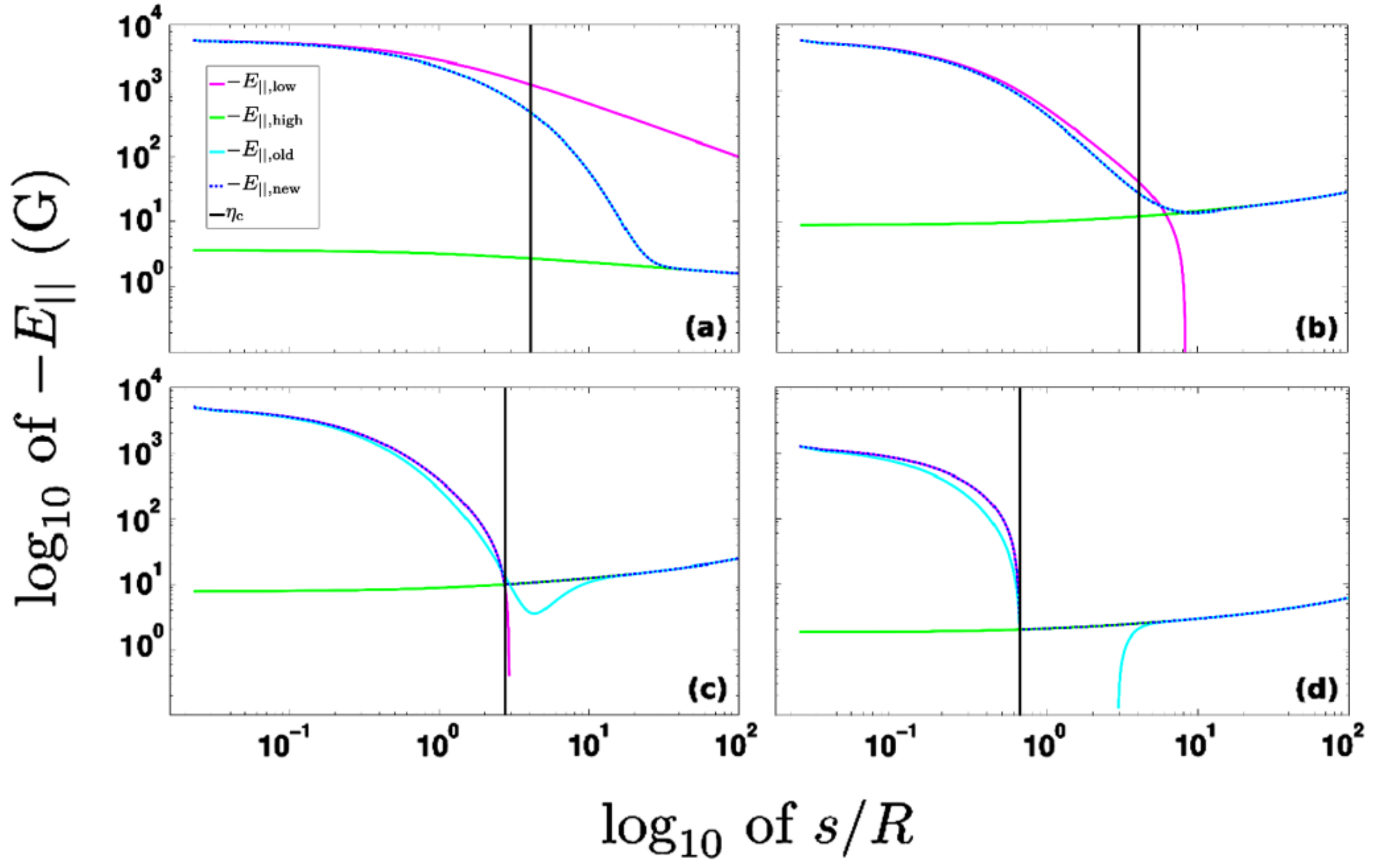}
\caption[Examples of the general SG $E$-field ($E_{\parallel,\rm new}$) we obtained by matching $E_{\parallel,\rm low}$ and $E_{\parallel,\rm high}$]{\label{fig:eta_C_examples} Examples of the general SG $E$-field ($E_{\parallel,\rm new}$, dashed dark blue line) we obtained by matching $E_{\parallel,\rm low}$ (magenta line) and $E_{\parallel,\rm high}$ (green line). We plotted the negative of the various $E$-fields as functions of the normalised $s$ along the $B$-field lines, in units of $R$. We indicated the matching parameter $\eta_{\rm c}$ (vertical black line) by using $s_{\rm c}/R\approx\eta_{\rm c}-1$ (which is valid for low altitudes). These plots were obtained for the following parameters: $P=0.0893$~s, $B_0=1.05\times{10}^{13}$~G, $R=10^6$~cm, $M=1.4M_\odot$, $\epsilon=0.18$, and $\xi=0.975$ (i.e., $\xi_\ast=0$). In (a) we chose $\alpha=90^\circ$, and $\phi_{\rm PC}=0$. Here we use $\eta_{\rm c}=5.1$ since $\eta_{\rm cut}>\eta_{\rm LC}$. In (b) we chose $\alpha=15^\circ$, $\phi_{\rm PC}=\pi$. We find a solution of $\eta_{\rm c}=5.1$. In (c) we chose $\alpha=30^\circ$, $\phi_{\rm PC}=\pi$. If $-E_{\parallel,\rm low}$ as well as $-E_{\parallel,\rm old}$ (as defined in Eq.~[\ref{eq:E_total}], light blue line) are below $-E_{\parallel,\rm high}$ beyond some radius $\eta$, we use $\eta_{\rm cut}$ (in this case $\eta_{\rm c}=\eta_{\rm cut}=3.7$) to match $E_{\parallel,\rm low}$ and $E_{\parallel,\rm high}$, resulting in $-E_{\parallel,\rm new}$ (dashed dark blue). In (d) we chose $\alpha=75^\circ$, $\phi_{\rm PC}=\pi$. For large $\alpha$ we observe that $-E_{\parallel,\rm low}$ changes sign over a small $\eta$ range. In this case we also use $\eta_{\rm c}=\eta_{\rm cut}=1.7$ to match the solutions.}
\end{figure}

At first, we matched the low-altitude and high-altitude $E$-field solutions by setting $\eta_{\rm c}=1.4$ for simplicity \citep{Breed2013}. However, we realised that $\eta_{\rm c}$ may strongly vary for the different parameters. Thus, we had to solve $\eta_{\rm c}(P,\dot{P},\alpha,\epsilon,\xi,\phi_{\rm PC})$ on each $B$-field line. In what follows we consider electrons to be the radiating particles, and our discussion will therefore generally deal with the negative of the $E$-field. Since particle orbits approximately coincide with the $B$-field lines in the corotating frame, it is important to consider the behaviour of the $E$-field as a function of $s$ rather than $\eta$.

We solved the matching parameter in the following way. First, we calculate $E_{\parallel,\rm low}$, which is independent of $\eta_{\rm c}$, along the $B$-field. If $-E_{\parallel,\rm low}<0$ for all $\eta$, it will never intersect with $E_{\parallel,\rm high}$ and we set $\eta_{\rm c}=1.1$, thereby basically using $E_{\parallel,\rm SG}\approx E_{\parallel,\rm high}$. Second, we step through $\eta_{\rm c}$ (in the range $1.1-5.1$), calculating $E_{\parallel,\rm SG}$ and $E_{\parallel,\rm high}$ as well as the ratio $S_i=S(\eta_i)=E_{\parallel,\rm SG}(\eta_i)/E_{\parallel,\rm low}(\eta_i)$ for $i=1,..,N$ at different radii $\eta_i$. If $S_i>1$ we use $1/S_i$. We next calculate a test statistic $T(\eta_{\rm c})=\sum_i^N{(S_i-1)^2}/N$ using only $E$-field values where $-E_{\parallel,\rm low}>-E_{\parallel,\rm high}$ (i.e., we basically fit $E_{\parallel,\rm SG}$ to $E_{\parallel,\rm low}$ when $-E_{\parallel,\rm low}>-E_{\parallel,\rm high}$). We then minimise $T$ to find the optimal $\eta_{\rm c}$ (similar to what was done in Figure~2 of \citealp{Venter2009}). In Figure~\ref{fig:eta_C_examples}a, the intersection radius $\eta_{\rm cut}>\eta_{\rm LC}$ (i.e., $E_{\parallel,\rm low}$ and $E_{\parallel,\rm high}$ do not intersect within the light cylinder) and therefore we impose the restriction that the solution of $\eta_{\rm c}$ should lie at or below 5.1. When $-E_{\parallel,\rm low}$ does not decrease as rapidly (e.g., as in Figure~\ref{fig:eta_C_examples}b) we find reasonable solutions. We note that $E_{\parallel,\rm SG}$ (referred to as $E_{\parallel,\rm old}$ in Figure~\ref{fig:eta_C_examples}) produces a bump when $-E_{\parallel,\rm low}$ decreases more rapidly. To circumvent this problem we test whether $-E_{\parallel,\rm SG}<-E_{\parallel,\rm high}$ and in this case we use the intersection radius $\eta_{\rm cut}$ of $E_{\parallel,\rm low}$ and $E_{\parallel,\rm high}$, rather than $\eta_{\rm c}$, to match our solutions (calling this new solution $E_{\parallel,\rm new}$; see Figure~\ref{fig:eta_C_examples}c). We lastly observe that for $\phi_{\rm PC}=\pi$ (on ``unfavourably curved'' field lines) for larger $\alpha$, $-E_{\parallel,\rm low}$ field changes sign resulting in a small $\eta_{\rm c}=\eta_{\rm cut}=1.7$ value (Figure~\ref{fig:eta_C_examples}d). We determined $\eta_{\rm c}$ (implying $E_{\parallel,\rm new}$) over the entire $\phi_{\rm L}$ and $\zeta$ range at and within the SG model boundary, as a function of $\alpha$. This is necessary for constructing phase plots, i.e., the intensity per solid angle, and light curves, i.e., a constant $\zeta$-cut (refer to as $\zeta_{\rm cut}$) on the phase plot, later on.

Since the $E$-field solutions have an $x^a=x^{\epsilon\cos(\phi^\prime-\phi^\prime_0)}=x^{-\epsilon\cos\phi_{\rm PC}}$ factor dependence, a larger (non-zero) offset results in different matching solutions vs.\ the case for $\epsilon=0$ (see Figure~3 and 4 in \citealt{Barnard2016}). In the case of $\alpha=0$, the first term $\propto\cos\alpha$ is the only contribution to the $E$-field, with the factor $x^ae_{\rm 1A}$ (with an $\epsilon$ dependence) being initially larger at low $\eta$ for $\phi_{\rm PC}=0$ than for $\phi_{\rm PC}=\pi$ ($x^a$ dominates), but rapidly decreasing with $\eta$ ($e_{\rm 1A}$ dominates), leading to a lower value of $\eta_{\rm c}$ for $\phi_{\rm PC}=0$. One should therefore note that the magnitude of one instance of the $E$-field with low $\eta_{\rm c}$ may initially be higher than another instance with high $\eta_{\rm c}$, but the first will decrease rapidly with $\eta$ and eventually become lower than the second. Therefore, there is no $\phi_{\rm PC}$-dependence for $\alpha=0$ for $\epsilon=0$, which is not the case for $\epsilon=0.18$. For a slightly larger $\alpha$ the second terms in Eq.~[\ref{eq:E_low}] and [\ref{eq:E_high}] start to contribute to the radiation. This is due to the $\sin\alpha$ term with an $\epsilon$ dependence that delivers an extra contribution which is zero in the case for $\epsilon=0$. At $\alpha=20^\circ$ the effects of the first and second terms seem to balance each other and therefore we find the same solution of $\eta_{\rm c}=5.1$ everywhere except on the SG model boundary (at $\xi\in[0.95,1.0]$) where $\eta_{\rm c}=1.1$, just as in the case of $\epsilon=0$ and $\alpha=0^\circ$. For values of $\alpha>20^\circ$ and $\epsilon=0.18$ the second term $\sim\cos\phi_{\rm PC}$ starts to dominate and thus we find solutions of $\eta_{\rm c}\sim{5.1}$ for $\phi_{\rm PC}\simeq0$ and systematically smaller solutions for $\phi_{\rm PC}\simeq\pi$ as $\alpha$ increases and the second term $\propto\sin\alpha$ becomes increasingly important (in both cases of $\epsilon$). At $\alpha=90^\circ$ we obtain the same solution as in $\epsilon=0$ case where the second term dominates (for this case $-E_{\parallel,\rm SG}<0$ for all $\eta$, since the Goldreich-Julian charge density $\rho_{\rm GJ}$ becomes positive). We note that the $\eta_{\rm c}$-distribution reflects two symmetries (one about $\phi_{\rm PC}=\pi$ and one about $\xi=0.975$, i.e., $\xi_{\ast}=0$, given our gap boundaries): that of the $\cos\phi_{\rm PC}$ term and that of the $(1-\xi_{\ast}^2)$ term in the $E_\parallel$ solutions.
After solving for $\eta_{\rm c}$, we could solve the particle transport equation along each $B$-field line (see Section~\ref{subsec:transport}).

\section{Chi-squared fitting method}\label{sec:chi2fit}

%We have implemented a chi-squared ($\chi^{2}$) method to search the multivariate solution space for optimal model parameters when we compare our predicted model light curves with \emph{Fermi} LAT data for the Vela pulsar. In this way, we are able to determine which $B$-field and geometric model combination yields the best light curve solution, how the different light curve predictions compare with each other, and which pulsar geometry ($\alpha$,$\zeta$) is optimal~\citep{Breed2013,Breed2012,Breed2014}.
We applied a standard $\chi^{2}$ statistical fitting technique to assist us to objectively find the pulsar geometry ($\alpha$,$\zeta$) which best describes the observed $\gamma$-ray light curve of the Vela pulsar. We use this $\chi^{2}$ method to determine the best-fit parameters for each of our $B$-field and geometric model combinations (spanning a large parameter space). The general expression is given by
\begin{eqnarray}\label{eq:chi2}
\chi^{2} & = & \sum_{i{\rm =1}}^{N_{{\rm bins}}}\frac{\left(Y_{{\rm d,}i}-Y_{{\rm m,}i}\right)^{2}}{\sigma_{{\rm m,}i}^{2}}\approx \sum_{i{\rm =1}}^{N_{{\rm bins}}}\frac{\left(Y_{{\rm d,}i}-Y_{{\rm m,}i}\right)^{2}}{Y_{{\rm d,}i}},
\end{eqnarray}
where $Y_{{\rm d,}i}(\phi_{{\rm L},i})$ and $Y_{{\rm m,}i}(\phi_{{\rm L},i})$ are the number of counts of the observed and modelled light curves (relative units at phase $\phi_{{\rm L},i}$), and $\sigma_{{\rm m,}i}(\phi_{{\rm L},i})$ the uncertainty of the model light curves in each phase bin $i=1,...,N_{{\rm bins}}$, with $N_{\rm bins}$ the number of bins. Since we do not know the uncertainty of the model, we approximate the model error by the data error, assuming $\sigma_{{\rm m,}i}^2(\phi_{{\rm L},i})\approx Y_{{\rm d,}i}(\phi_{{\rm L},i})$ for Poisson statistics. Since we use geometric models, with an uncertainty in the absolute {\it intensity}, we assume that the {\it shape} of the light curve is correct. The data possess a background which is also uncertain. Furthermore, \textit{Fermi} has a certain response function that influences the intrinsic shape of the light curve, which reflects the sum of counts from many pulsar rotations. Given all these uncertainties, we incorporate a free amplitude parameter $A$ to allow more freedom in terms of finding the best fit of the model light curves to the data. We normalise the model light curve to range from 0 to the maximum number of observed counts $k_{{\rm 2}}$ by using the following expression:
\begin{equation}\label{eq:chi2_norm}
Y_{{\rm m}}^{\prime}(\phi_{{\rm L},i})=\frac{Y_{{\rm m}}(\phi_{{\rm L},i})}{(k_{1}+\epsilon_{0})}A(k_{2}-{\rm BG})+{\rm BG}\approx\frac{Y_{{\rm m}}(\phi_{{\rm L},i})}{k_{1}}k_{2},
\end{equation}
with $k_{1}={\rm max}(Y_{{\rm m}}(\phi_{{\rm L},i}))$, $k_{2}={\rm max}(Y_{{\rm d}}(\phi_{{\rm L},i}))$, $\epsilon_{0}$ a small value added to ensure that we do not divide by zero, $A$ a free normalisation parameter, and BG the background level of $Y_{{\rm d}}(\phi_{{\rm L},i})$. We treat the data as being cyclic so we need to ensure that the model light curve is cyclic as well. The model light curve has to be re-binned in order to have the same number of bins in $\phi_{\rm L}$ as the data \citep{Abdo2013SecondCat}. We use a Gaussian Kernel Density Estimator function to rebin and smooth the model light curve \citep{Parzen1962}. Furthermore, we also introduce the free parameter $\Delta\phi_{\rm L}$ which represents an arbitrary phase shift of the model light curve so as to align the model and data peaks. We choose the phase shift $\Delta\phi_{\rm L}$ as a free parameter due to the uncertainty in the definition of $\phi_{\rm L}=0$ (see, e.g., \citealp{Johnson2014} who also used $A$ and $\Delta\phi_{\rm L}$). Importantly, we note that we have not changed the relative position (the radio-to-$\gamma$ phase lag $\delta$), since this is a crucial model prediction. The radio and $\gamma$-ray emission regions are tied to the same underlying $B$-field structure, and $\delta$ therefore reflects important physical conditions (or model assumptions) such as a difference in emission heights of the radio and $\gamma$-ray beams.

After preparation of the model light curve, we searched for the best-fit solution for each of our $B$-field and gap combinations over a parameter space of $\alpha\in[0^{\circ},90^{\circ}]$, $\zeta\in[0^{\circ},90^{\circ}]$ (both with $1^{\circ}$ resolution), $0.5<A<1.5$ with $0.1$ resolution, and $0<\Delta\phi_{\rm L}<1$ with $0.05$ resolution. For a chosen $B$-field and model geometry we iterate over each set of parameters and search for a local minimum $\chi^{2}$ value at a particular $\alpha$ and $\zeta$. Once we have iterated over the entire parameter space ($\alpha$,$\zeta$,$A$,$\Delta\phi_{\rm L}$), we obtain a global minimum value for $\chi^{2}$ (also called the optimal $\chi^{2}$):
\begin{eqnarray}\label{eq:chi2-opt}
\chi_{{\rm opt}}^{2} & \approx & \sum_{i=1}^{N_{{\rm bins}}}\frac{\left(Y_{{\rm d,}i}-Y_{{\rm opt,}i}\right)^{2}}{Y_{{\rm d,}i}}.
\end{eqnarray}

If faint pulsars are modelled, Poisson statistics will be sufficient to describe the observations. For the bright Vela, however we assume Gaussian statistics which yields small errors, since the emission characteristics are more significant than those of faint pulsars. However, these small errors on the data yield large values for the reduced optimal $\chi^2$ value $\chi_{{\rm opt}}^{2}/N_{\rm dof}\gg 1$. We therefore need to rescale (to compensate for the uncertainty in $\sigma_{{\rm m,}i}$) the $\chi^{2}$ values by $\chi_{{\rm opt}}^{2}$ and multiply by the number of degrees of freedom $N_{\rm dof}$ (the difference between $N_{\rm bins}$ and number of free parameters). The scaled $\chi^{2}$ is presented by \citep{Pierbattista2015}:
\begin{eqnarray}\label{eq:scaled-chi2}
\xi^{2} & = & N_{{\rm dof}}\frac{\chi^{2}}{\chi_{{\rm opt}}^{2}}.
\end{eqnarray}
From Eq.~[\ref{eq:scaled-chi2}] the $\xi^{2}$ for the optimal model are as follows
\begin{eqnarray}\label{eq:scaled-chi2-opt}
\xi_{{\rm opt}}^{2} & = & N_{{\rm dof}}\frac{\chi_{{\rm opt}}^{2}}{\chi_{{\rm opt}}^{2}}=N_{{\rm dof}},
\end{eqnarray}
with $\xi_{\rm opt}^{2}/N_{\rm dof}=\xi_{\rm opt,\nu}^{2}=1$ the reduced $\xi_{\rm opt}^{2}$. 

If one wishes to compare the optimal model to alternative models, e.g., in our case a $B$-field combined with several geometric models, confidence contours for $68\%$ ($1\sigma$), $95.4\%$ ($2\sigma$), and $99.73\%$ ($3\sigma$) can be constructed by estimating the difference in the $\xi_{{\rm opt}}^{2}$ and the $\xi^{2}$ of the alternative models:
\begin{eqnarray}\label{eq:difference-chi2}
\Delta\xi^{2} & =\xi^{2}-\xi_{{\rm opt}}^{2}=N_{{\rm dof}}\left(\chi^{2}/\chi_{{\rm opt}}^{2}-1\right).\end{eqnarray}
The confidence intervals can be estimated by reading the $\Delta\xi^{2}$ (i.e., $\Delta\xi_{1\sigma,\mu_{{\rm dof}}}^{2}$, $\Delta\xi_{2\sigma,\mu_{{\rm dof}}}^{2}$, and $\Delta\xi_{3\sigma,\mu_{{\rm dof}}}^{2}$) values from a standard $\chi^{2}$ table for the specified confidence interval at $\mu_{{\rm dof}}=2$ (corresponding to the two-dimensional ($\alpha,\zeta$) grid, \citealp{Lampton1976}). Using these values for $\Delta\xi^{2}$ and $\xi_{{\rm opt}}^{2}=N_{{\rm dof}}$, we can determine $\xi^{2}=\xi^2_{\rm opt}+\Delta\xi^2=N_{\rm dof}+\Delta\xi^2$ (i.e., $\xi_{1\sigma}^{2}$, $\xi_{2\sigma}^{2}$, and $\xi_{3\sigma}^{2}$) from Eq.~[\ref{eq:difference-chi2}], which is the value at which we plot each confidence contour. To enhance the contrast of the colours on the filled $\chi^{2}$ contours, we plot ${\rm log_{10}\xi^{2}}$ on an ($\alpha$,$\zeta$) grid, with a minimum value of ${\rm log_{10}}\xi_{{\rm opt}}^{2}={\rm log_{10}}(N_{{\rm dof}})=1.98$ (corresponding to the best-fit solution by construction, i.e., after rescaling, with $N_{\rm dof}=100-4=96$ in our study). The best-fit solution is therefore positioned at $\xi_{{\rm opt}}^{2}=96$ and enclosed by the confidence contours with values of $\xi_{1\sigma,\mu_{{\rm dof}}}^{2}=96+2.30$, $\xi_{2\sigma,\mu_{{\rm dof}}}^{2}=96+6.17$, and $\xi_{3\sigma,\mu_{{\rm dof}}}^{2}=96+11.8$ (see Eq.~[\ref{eq:difference-chi2}]; \citealp{Press1992}). We determine errors on $\alpha$ and $\zeta$ for the best-fit solution of each $B$-field and model combination using the $3\sigma$ interval connected contours. We choose errors of $1^{\circ}$ for cases when the errors were smaller than one degree (given a model resolution of $1^{\circ}$). See Section~\ref{subsec:contours_LCs} for the best-fit solutions we obtained for the offset-PC dipole $B$-field and SG model solution for two cases, i.e., the usual $E$-field, and this same $E$-field increased by a factor 100.

\begin{figure}[t]
\centering
\includegraphics[width=\textwidth]{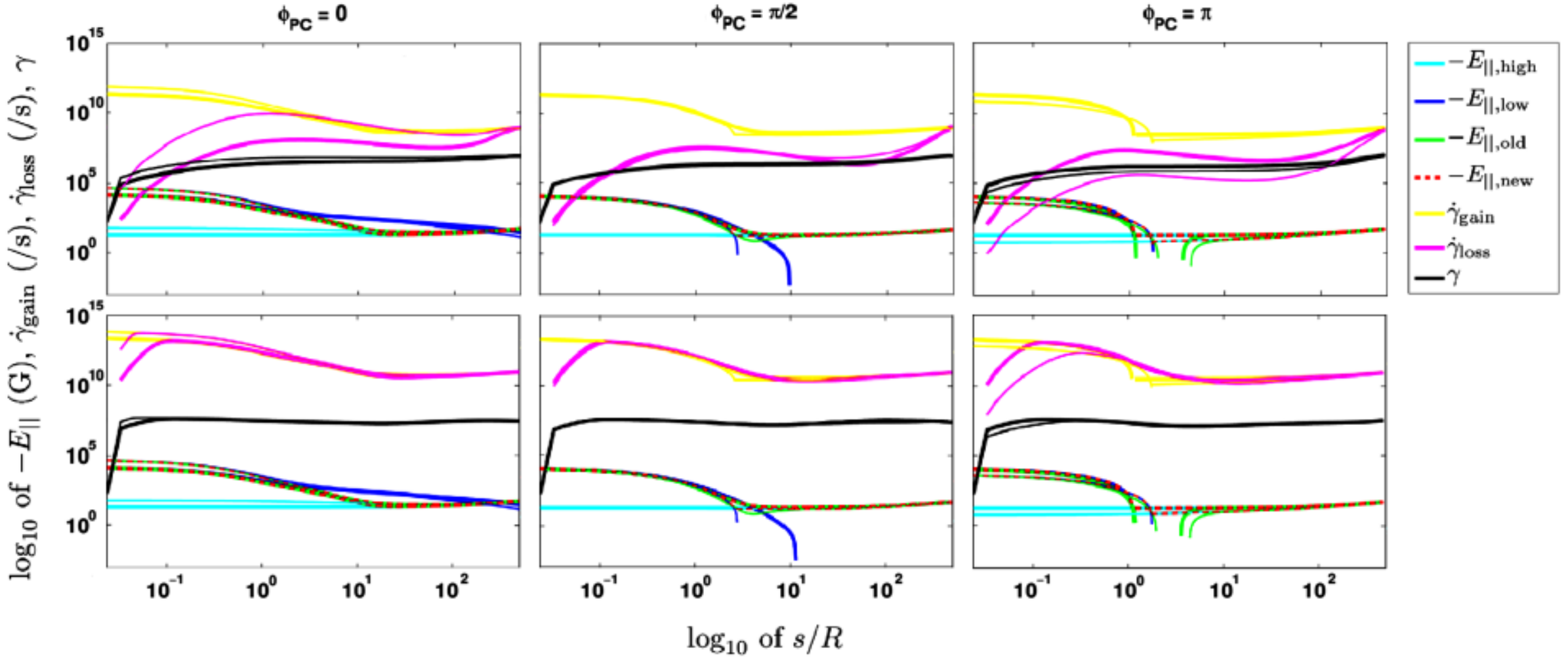}
\caption[Plot of $\log_{10}$ of $-E_{\parallel,{\rm high}}$, $-E_{\parallel,{\rm low}}$, the general $-E_{\parallel,{\rm SG}}$-field (using $\eta_{\rm c}$ as the matching parameter; $-E_{\parallel, {\rm old}}$) and a corrected $E$-field, $-E_{\parallel, {\rm new}}$, gain rate $\dot{\gamma}_{\rm gain}$, loss rate $\dot{\gamma}_{\rm loss}$, and the Lorentz factor $\gamma$ as a function of $s/R$]{\label{fig:RRLim} Plot of $\log_{10}$ of $-E_{\parallel,{\rm high}}$ (solid cyan line), $-E_{\parallel,{\rm low}}$ (solid blue line), the general $-E_{\parallel,{\rm SG}}$-field (using $\eta_{\rm c}$ as the matching parameter; $-E_{\parallel, {\rm old}}$, solid green line) and a corrected $E$-field in cases where a bump was formed using the standard matching procedure (i.e., setting $\eta_{\rm c}=\eta_{\rm cut}$; $-E_{\parallel, {\rm new}}$, dashed red line), gain rate $\dot{\gamma}_{\rm gain}$ (solid yellow line), loss rate $\dot{\gamma}_{\rm loss}$ (solid magenta line), and the Lorentz factor $\gamma$ (solid black line) as a function of $s/R$. In each case we used $\alpha=45^\circ$, $P=0.0893$~s, $B_0=1.05\times{10}^{13}$ G (corrected for GR effects), $I=0.4MR^2=1.14\times10^{45}$ g\,cm$^2$, and $\xi=0.975$ (i.e., $\xi_{\ast}=0$). On each panel we represent the curves for $\epsilon=0$ (thick lines) and $\epsilon=0.18$ (thin lines). The first row is for the `typical' $E_\parallel$-field whereas the second row represents the $E_\parallel$-field increased by a factor 100. Each column is for a different field line, i.e., the first column for ``favourably curved" field lines ($\phi_{\rm PC}=0$), the middle column for $\phi_{\rm PC}=\pi/2$, and the last panel for ``unfavourably curved" field lines ($\phi_{\rm PC}=\pi$). These choices reflect the values of $\phi_{\rm PC}$ at the stellar surface; they may change as the particle moves along the $B$-field line, since $B_{\phi}\neq 0$.}
\end{figure}

\section{Results}\label{sec:ch3res} 
\subsection{Particle transport and testing for curvature radiation reaction}\label{subsec:transport} %chapter 5

Once we solved $\eta_{\rm c}$ (see Section~\ref{subsec:matching}), we could calculate the general $E$-field ($E_{\parallel,{\rm new}}$) in order to solve the particle transport equation (in the corotating frame) to obtain the particle energy $\gamma(\eta,\phi,\xi_{\ast})$, necessary for determining the CR emissivity. By rewriting Eq.~[\ref{eq:RR-eq}] we obtain the following
\begin{equation}
\dot{\gamma}=\dot{\gamma}_{\rm gain}+\dot{\gamma}_{\rm loss}=\frac{eE_{\parallel,{\rm new}}}{m_{\rm e}c}-\frac{2e^2\gamma^4}{3\rho^2_{\rm curv}m_{\rm e}c}=\frac{1}{m_{\rm e}c^2}\left[{ceE_{\parallel,{\rm new}}}-\frac{2ce^2\gamma^4}{3\rho^2_{\rm curv}}\right],
\end{equation}
with $\dot{\gamma}_{\rm gain}$ the gain (acceleration) rate, $\dot{\gamma}_{\rm loss}$ the loss rate, $e$ the electron charge, $m_{\rm e}$ the electron mass, and $m_{\rm e}c^2$ the rest-mass energy; CRR (taking only CR losses into account) occurs when the energy gain balances the losses and $\dot{\gamma}=0$. 

In Figure~\ref{fig:RRLim} we plot the $\log_{10}$ of $-E_{\parallel,{\rm high}}$ (solid cyan line), $-E_{\parallel,{\rm low}}$ (solid blue line), the general $-E_{\parallel,{\rm SG}}$-field (using $\eta_{\rm c}$ as the matching parameter; $-E_{\parallel, {\rm old}}$, solid green line) and a corrected $E$-field in cases where a bump was formed using the standard matching procedure (see Section~\ref{subsec:matching}, i.e., setting $\eta_{\rm c}=\eta_{\rm cut}$; $-E_{\parallel, {\rm new}}$, dashed red line), $\dot{\gamma}_{\rm gain}$ (solid yellow line), $\dot{\gamma}_{\rm loss}$ (solid magenta line), and $\gamma$ (solid black line) as a function of $s/R$ along the $B$-field line. The top panels represents the usual SG $E$-field case and the bottom panels the increased $E$-field case. For each case we show $\epsilon=0$ (thick lines) and $\epsilon=0.18$ (thin lines) on the same plot. As an example we assume $\alpha=45^\circ$ for both cases in order to compare the two cases. We note that the values for $\phi_{\rm PC}$ representing each column in the figure are actually values on the stellar surface and indicate different $B$-field lines.

In the top panels of Figure~\ref{fig:RRLim} we note that $-E_{\parallel,{\rm high}}$ displays the following behaviour at low $\eta$: for $\phi_{\rm PC}=0$, $-E_{\parallel,{\rm high}}^{\epsilon\neq 0}>-E_{\parallel,{\rm high}}^{\epsilon=0}$; these are nearly equal for $\phi_{\rm PC}=\pi/2$, and $-E_{\parallel,{\rm high}}^{\epsilon\neq 0}<-E_{\parallel,{\rm high}}^{\epsilon=0}$ for $\phi_{\rm PC}=\pi$. For $\phi_{\rm PC}=\pi/2$, the first term of $-E_{\parallel,{\rm high}}$ dominates the second, and for $\phi_{\rm PC}=\pi$, the second term of $-E_{\parallel,{\rm high}}$ is always negative, but the positive first term dominates and therefore $-E_{\parallel,{\rm high}}$ does not change sign as $\eta$ increases. Similar behaviour is also seen for $-E_{\parallel,{\rm low}}$ (boosted for non-zero $\epsilon$ and $\phi_{\rm PC}=0$). For $\alpha=45^\circ$, the second term $\propto\sin\alpha$ now contributes, stopping $-E_{\parallel,{\rm low}}$ from changing sign along $\eta$ for $\phi_{\rm PC}=0$ (vs.\ the case if $\alpha=0$). The second term of $-E_{\parallel,{\rm low}}\sim x^a\cos\phi_{\rm PC}$ is comparable to the first at low $\eta$, but quickly dominates as $\eta$ increases for $\phi_{\rm PC}=0$. The second term of $-E_{\parallel,{\rm low}}$ remains positive so that we find $\eta_{\rm c}=5.1$ in this case (see Section~\ref{subsec:matching}). For $\phi_{\rm PC}=\pi/2$ we note that $-E_{\parallel,{\rm low}}^{\epsilon\neq 0}$ becomes negative with $\eta$. For $\phi_{\rm PC}=\pi$, the second term of $-E_{\parallel,{\rm low}}$ is negative, forcing this field to change sign; this change takes slightly longer to occur when $\epsilon\neq 0$. The fact that $-E_{\parallel,{\rm high}}$ is positive leads to a ``recovery" of the total $E$-field, so that it becomes positive again at larger $\eta$. The effect of matching the $E$-field is seen in the evolution of $\gamma(s)$ since $\gamma$ is determined by $E_\parallel$.

\begin{figure}
	\centering
	\includegraphics[width=\textwidth]{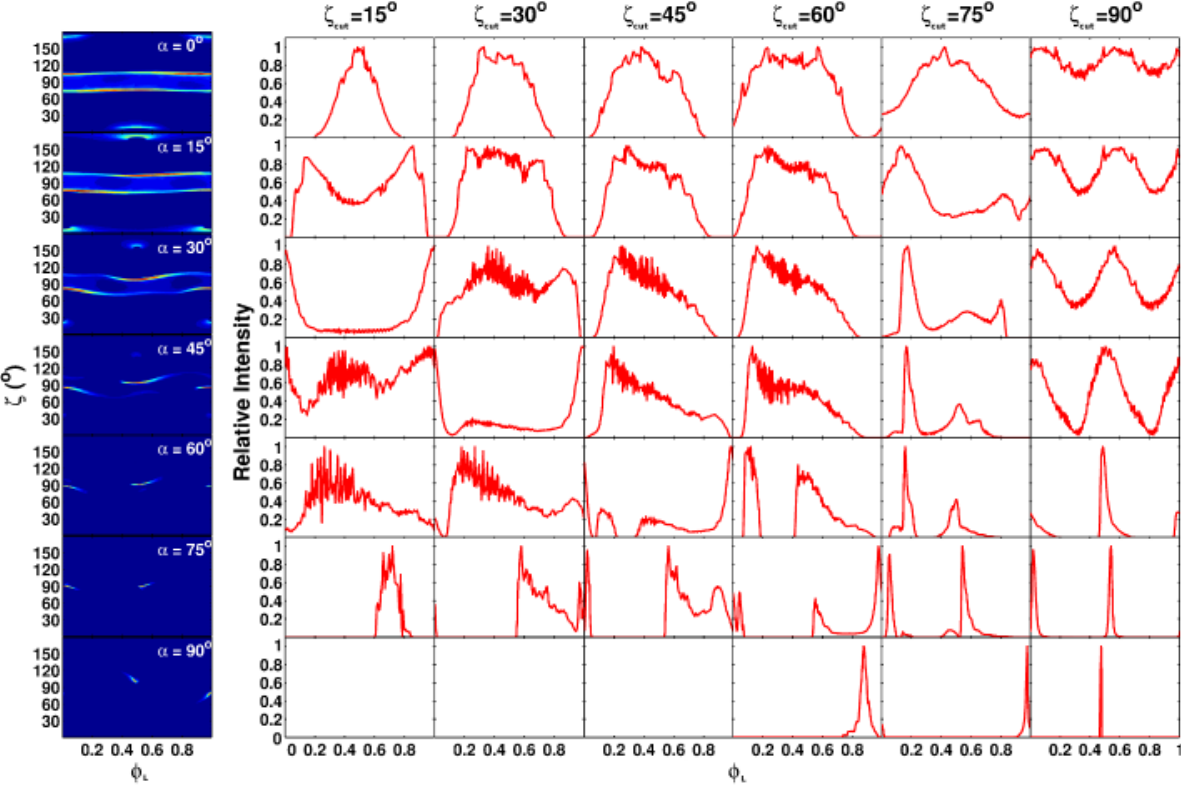}
	\caption[Phase plots and light curves for the SG model assuming an offset-PC dipole field, for $\epsilon=0.18$ and a variable $\epsilon_{\nu}$.]{\label{fig:OffsetEps018E} Phase plots (first column) and light curves (second column and onward) for the SG model assuming an offset-PC dipole field, for a fixed value of $\epsilon=0.18$ and variable $\epsilon_{\nu}$. Each phase plot is for a different $\alpha$ value ranging from $0^{\circ}$ to $90^{\circ}$ with a $15^{\circ}$ resolution, and their corresponding light curves are denoted by the solid red lines for different $\zeta_{\rm cut}$ values, ranging from $15^{\circ}$ to $90^{\circ}$, with a $15^{\circ}$ resolution.}
\end{figure}

For the usual SG $E$-field case we notice that the CRR limit is reached in some cases, but only at high altitudes (the yellow and magenta lines reach the same value): e.g., beyond $\eta\approx{R_{\rm LC}}$ for $\phi_{\rm PC}=0$ and $\alpha=45^\circ$. We note the importance of actually solving $\eta_{\rm c}(P,\dot{P},\alpha,\epsilon,\xi,\phi_{\rm PC})$ on each $B$-field line. Previously we set $\eta_{\rm c}=1.4$ for all cases and found that the particles did not attain the CRR limit \citep{Breed2013}. Only when we allowed larger values of $\eta_{\rm c}$ was $-E_{\parallel,{\rm low}}$ boosted and did we find particles reaching the CRR limit in many more cases. The relatively low SG $E$-field leads to small caustics on the phase plots constructed for photon energies $>100$~MeV (see Section~\ref{subsec:100MeV}). 
Thus, we additionally investigate the effect of \textit{increasing the $E$-field.} 

In the CRR limit we can determine the CR cutoff of the CR photon spectrum as follows, using the formula of \citet{Venter2010}
\begin{equation}\label{ECRcut}
E_{\rm CR}\sim{4}E_{\parallel,\rm 4}^{3/4}\rho_{\rm curv, 8}^{1/2} \quad {\rm GeV},
\end{equation}
with $\rho_{\rm curv, 8}\sim\rho_{\rm curv}/{10^8}$ cm the curvature radius of the $B$-field line and $E_{\parallel,4}\sim E_{\parallel}/{10^4}$ statvolt cm$^{-1}$ the $E$-field parallel to the $B$-field. As a test, we multiply the $E$-field by a factor 100. Using Eq.~[\ref{ECRcut}] we estimate the newly implied cutoff energy $E_{\rm CR}\sim4$~GeV, which is in the energy range of \textit{Fermi} ($>$100~MeV). %We also obtained a better $\chi^2$ best-fit solution for this larger $E$-field compared to the usual one, for $\epsilon=0.00$ at $\alpha=75^{+3}_{-1}$, $\zeta=51^{+2}_{-5}$, $A=1.1$, and $\Delta\phi_{\rm L}=0.55$. In Figure~\ref{fig:bestcontours} we show our significance contour ${\rm log_{10}\xi^{2}}$ left and the corresponding best-fit light curve on the right on the bottom panels. This offset-PC dipole $B$-field and SG model (using the increased $E$-field) combination therefore provides an overall optimal fit, second only to the RVD and OG model combination (see Section~\ref{sec:comparison})
We note that a higher $E$-field leads to CRR being reached at lower altitudes as seen in the bottom panels of Figure~\ref{fig:RRLim}. This also leads to extended caustic structures on these phase plots, resulting in qualitatively different light curve shapes, as noted below (see Section~\ref{subsec:100MeV}).

\subsection{Light curves in different wavebands}\label{subsec:100MeV}

\begin{figure}
\centering
\includegraphics[width=\textwidth]{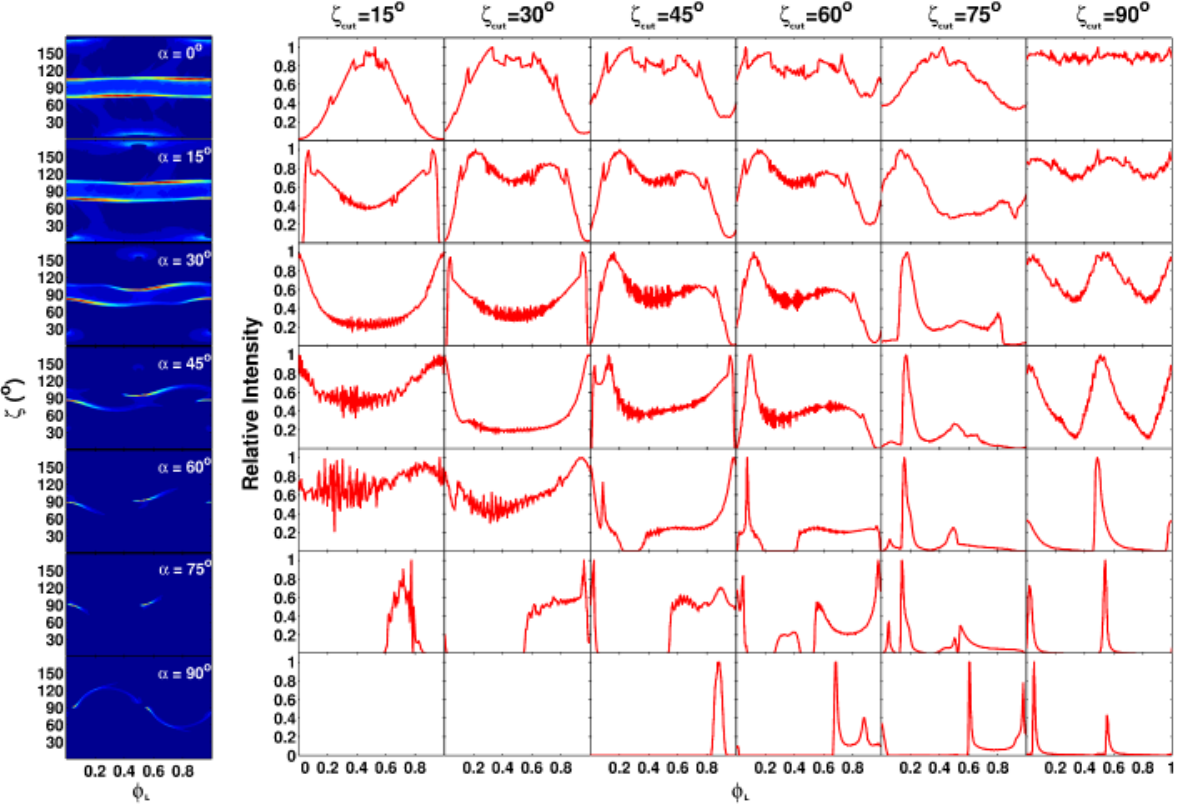}
\caption[Phase plots and light curves for the SG model assuming an offset-PC dipole field, for $\epsilon=0.18$, variable $\epsilon_{\nu}$, and a lower $E_{\gamma,\rm min}$ of 1~MeV.]{\label{fig:OffsetEps018MeV} The same as in Figure~\ref{fig:OffsetEps018E}, but for a lower $E_{\rm min}$ of 1~MeV.}
\end{figure}

In Figure~\ref{fig:OffsetEps018E} we present phase plots and light curves for the offset-PC dipole $B$-field and $\epsilon=0.18$, obtaining a variable $\epsilon_\nu(s)$ due to using an SG $E$-field solution (with CR the dominating process for emitting $\gamma$-rays; see Sections~\ref{sec:SGEfield}). The caustic structure and resulting light curves are qualitatively different for various $\alpha$ compared to the constant $\epsilon_\nu$ case (see Figure~7 in \citealt{Barnard2016}). The caustics appear smaller and less pronounced for larger $\alpha$ values (since $E_\parallel$ becomes lower as $\alpha$ increases), and extend over a smaller range in $\zeta$. If we compare Figure~\ref{fig:OffsetEps018E} with the phase plots when $\epsilon=0$ (see Figure~8 in \citealt{Barnard2016}), we note a new emission structure close to the PCs for small values of $\alpha$ and $\zeta\approx(0^\circ,180^\circ)$. This reflects the boosted $E_\parallel$-field on the ``favourably curved'' $B$-field lines (with $E_\parallel\propto{x^a}\cos\alpha$, with $a=-\epsilon{\cos\phi_{\rm PC}}$ and $\phi_{\rm PC}=0$; see Figure~\ref{fig:RRLim}). In Figure~\ref{fig:OffsetEps018E} there is also more phase space filled where the light curves generally display only one broad peak with less off-peak emission due to this non-zero $\epsilon$. As $\alpha$ and $\zeta$ increase, more peaks become visible, with emission still visible from both poles as seen for larger $\alpha$ and $\zeta$ values, e.g., $\alpha=75^{\circ}$ and $\zeta=75^{\circ}$.

If we compare Figure~\ref{fig:OffsetEps018E} with the case of constant $\epsilon_{\nu}$ (see Figure~7 in \citealt{Barnard2016}), we notice that when we take $E_{\parallel}$ into account the phase plots and light curves change considerably. For example, for $\alpha=90^{\circ}$ in the constant $\epsilon_\nu$ case a ``closed loop'' emission pattern is visible in the phase plot, which is different compared to the small ``wing-like'' emission pattern in the variable $\epsilon_\nu$ case. \textit{Therefore we see that both the $B$-field and $E$-field have an impact on the predicted light curves.} This small ``wing-like'' caustic pattern is due to the fact that we only included photons in the phase plot with energies $>100$~MeV. Given the relatively low $E$-field there are only a few photons with energies exceeding $100$~MeV. 
%To further illustrate the effect of changing $\epsilon$, we present phase plots for $\alpha=70^{\circ}$ in Figure~\ref{fig:a70z50}a and their corresponding light curves at $\zeta=50^{\circ}$ associated with the particular phase plot in Figure~\ref{fig:a70z50}b, using an offset-PC dipole field and TPC model, with $\epsilon$ ranging from 0.00 to 0.18 with intervals of 0.03, and assuming constant $\epsilon_\nu$. The caustic structure is slightly different for different values of $\epsilon$. For $\epsilon=0$ the light curve has a single peak and as $\epsilon$ increases, the peak becomes slightly narrower. Also, for larger $\epsilon$ values, the caustic structure becomes slightly broader and more pronounced. Figure~\ref{fig:a70z50}c and Figure~\ref{fig:a70z50}d represent the offset-PC dipole field and SG model with variable $\epsilon_{{\rm \nu}}$. When we compare the phase plots of Figure~\ref{fig:a70z50}a and Figure~\ref{fig:a70z50}c, the caustics are dimmer and smaller due to the low SG $E$-field, and the light curves display less off-peak emission. As $\epsilon$ increases, some small features become more pronounced. 
\begin{figure}
\centering
\includegraphics[width=\textwidth]{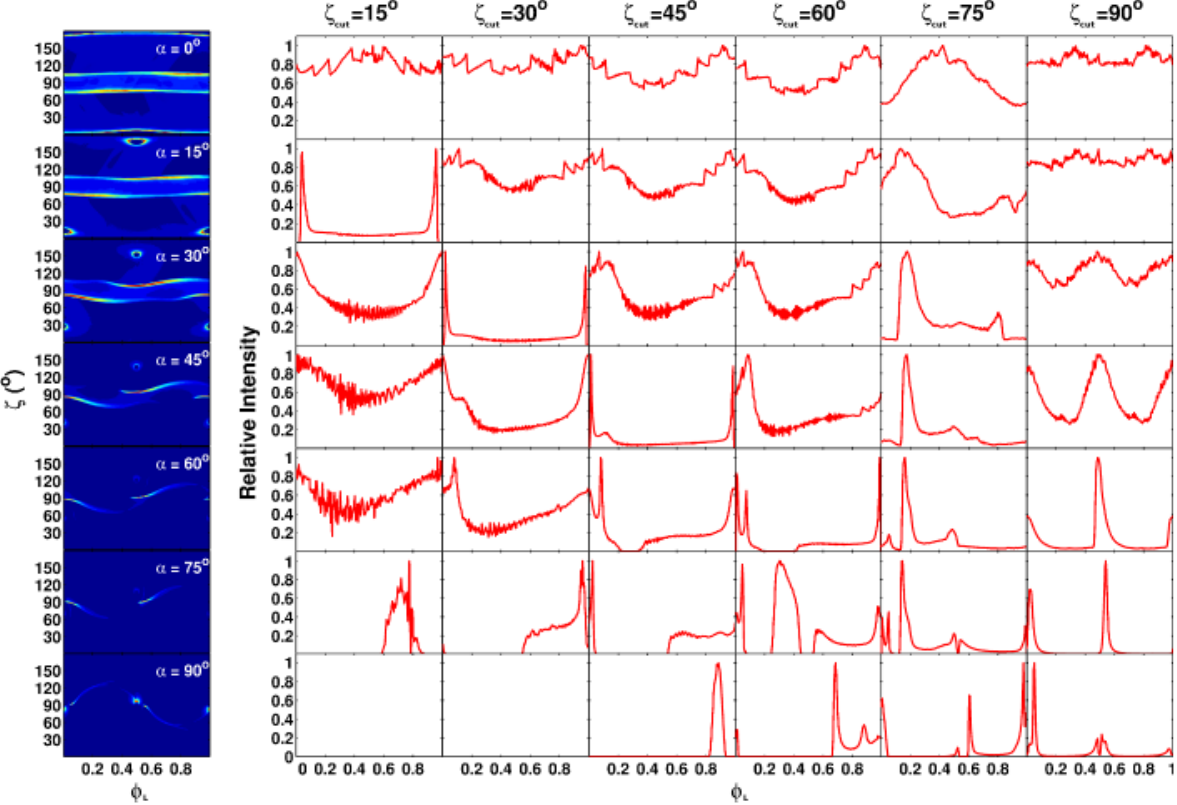}
\caption[Phase plots and light curves for the SG model assuming an offset-PC dipole field, for $\epsilon=0.18$, variable $\epsilon_{\nu}$, and a higher $E$-field.]{\label{fig:OffsetEps018GeV} The same as in Figure~\ref{fig:OffsetEps018E}, but for the case where we multiplied $E_\parallel$ by a factor 100, yielding a CR cutoff of $E_{\rm CR}\sim4$~GeV.}
\end{figure}

Since the SG $E$-field (see Section~\ref{sec:SGEfield}) is low, CRR is reached in most cases but only at high $\eta$ and small $\alpha$ (Section~\ref{subsec:transport}). This low $E$-field also causes the phase plots to display small caustics which result in ``missing structure". Therefore, we investigate the effect on the light curves of the offset-PC dipole $B$-field and SG model combination when we \textit{lower the minimum photon energy} $E_{\gamma,\rm min}$ from 100~MeV to 1~MeV above which we construct phase plots. For our given SG $E$-field with a magnitude of $E_\parallel\sim10^2$~statvolt~cm$^{-1}$ the estimated cutoff is $E_{\rm CR}\sim90$~MeV. This leads to pulsar emission being emitted in the hard X-ray waveband, and can not be compared via $\chi^2$ to \textit{Fermi} ($>$100~MeV) data for the Vela pulsar. As an illustration, we present the phase plots and light curves in Figure~\ref{fig:OffsetEps018MeV} for $\epsilon=0.18$ and $E_{\gamma,\rm min}>1$~MeV. If we compare Figure~\ref{fig:OffsetEps018MeV} with Figure~\ref{fig:OffsetEps018E} we notice that a larger region of phase space is filled by caustics, especially at larger $\alpha$, e.g., at $\alpha=90^\circ$ the visibility is enhanced. The peaks are also wider at low $\alpha$. Sometimes extra emission features appear, leading to small changes in the light curve shapes. 

We present the phase plots and light curves for this larger $E$-field (the usual one multiplied by a factor of 100) in Figure~\ref{fig:OffsetEps018GeV} for the offset-PC dipole and SG model solution with $\epsilon=0.18$. If we compare Figure~\ref{fig:OffsetEps018GeV} with Figure~\ref{fig:OffsetEps018E} we notice that more phase space is filled by caustics, especially at larger $\alpha$. At $\alpha=90^\circ$ the visibility is again enhanced. The caustic structure becomes wider and more pronounced, with extra emission features arising as seen at larger $\alpha$ and $\zeta$ values. This leads to small changes in the light curve shapes. At smaller $\alpha$ values the emission around the PC forms a circular pattern that becomes smaller as $\alpha$ increases. These rings around the PCs become visible since the low $E$-field is boosted, leading to an increase in bridge emission as well as higher signal to noise. At low $\alpha$ the background becomes feature-rich, but not at significant intensities, though. 
%When we boost the low $E$-field we find that the CRR limit is in fact reached almost immediately at lower $\eta$ for certain parameter combinations of $\alpha$, and $\phi_{\rm PC}$, as shown in Figure~\ref{fig:RRLim}.

\subsection{$\chi^2(\alpha,\zeta)$ contours and best-fit light curves}\label{subsec:contours_LCs}

In this Section, we present our best-fit solutions of the simulated light curves using the Vela data from {\it Fermi}. We plot some example contours of ${\rm log_{10}\xi^{2}}$ (colour bar) as well as the optimal ($\alpha$,$\zeta$) combination. We determine errors on $\alpha$ and $\zeta$ for the optimal solution of each $B$-field and gap model combination using a bounding box delimited by a minimum and maximum value in both $\alpha$ and $\zeta$ which surrounds the $3\sigma$ contour (see enlargement in bottom left corner of contour). %We illustrate this in Figure~\ref{fig:bestcontours}a with the white lines indicating the bounding box $[\alpha_{{\rm min}},\alpha_{{\rm max}}]$ and $[\zeta_{{\rm min}},\zeta_{{\rm max}}]$ (see enlargement in bottom left corner of panel~[a]). 
We choose errors of $1^{\circ}$ for cases when the errors were smaller than $1^\circ$ (given our chosen resolution of $1^{\circ}$) and indicate our overall best statistical fit for a certain model combination by a white star. 

\begin{figure}
\centering
\includegraphics[width=\textwidth]{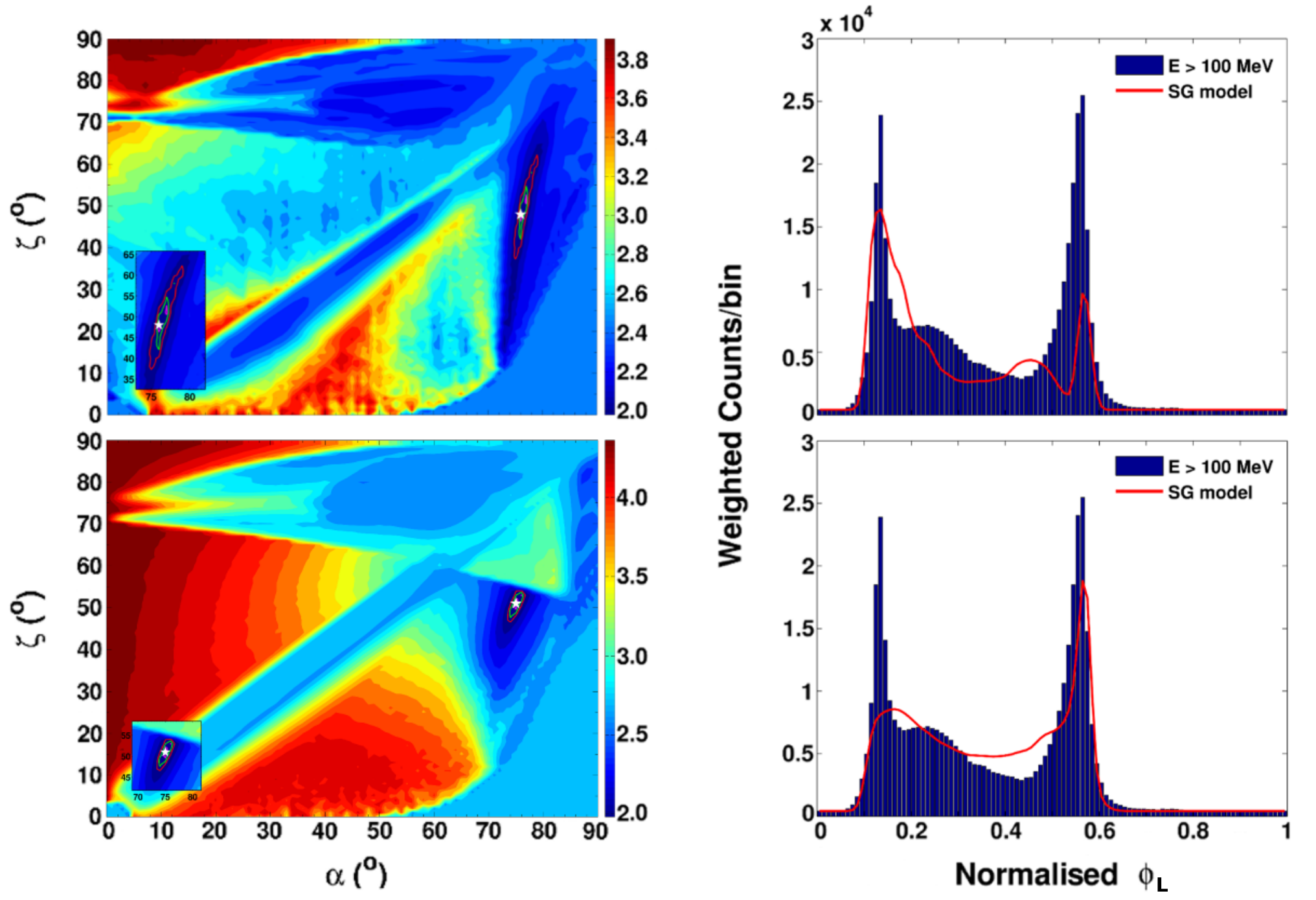}
\caption[Contour plots and best-fit light curves for each of the best-fit solutions we obtained for the offset-PC dipole $B$-field and SG model solution, for the two separate cases.]{\label{fig:bestcontours} Contour plots (left) and the corresponding best-fit light curves (right) for each of the best-fit solutions we obtained for the offset-PC dipole $B$-field and SG model solution, for the two separate cases: for the usual SG $E$-field and variable $\epsilon_{\rm \nu}$ with $\epsilon=0.15$ (top panels), and for the higher SG $E$-field and variable $\epsilon_{\rm \nu}$ with $\epsilon=0$ when we multiplied the $E_\parallel$ by a factor $100$ (bottom panels). The contours are on an ($\alpha$,$\zeta$) grid with the colour bar representing $\log_{10}\xi^2$, with 1.98 corresponding to the best-fit solution, indicated by the white star. The confidence contour for $1\sigma$ (magenta line), $2\sigma$ (green line), and $3\sigma$ (red line) are also shown with an enlargement in the bottom left corner. The blue histogram denotes the observed Vela pulsar profile (for energies $E>100$~MeV, \citealp{Abdo2013SecondCat}) and the red line the model light curve.}
\end{figure}

In Figure~\ref{fig:bestcontours} (top panel left) we present our significance contour ${\rm log_{10}\xi^{2}}$ for an SG model using an offset-PC dipole field, with $\epsilon=0.15$ and a variable $\epsilon_{\nu}$. The corresponding light curve fit of the model (solid red line) for the best-fit geometry to the Vela data (blue histogram) is also shown (Figure~\ref{fig:bestcontours} top panel, right). The observed light curve represents weighted counts per bin as function of normalised phase $\phi_{\rm L}=[0,1]$ \citep{Abdo2013SecondCat}. For this combination, we find a best-fit solution at $\alpha=76{_{-1}^{+3}}^{\circ}$, $\zeta=48{_{-11}^{+15}}^{\circ}$, $A=0.7$, and $\Delta\phi_{\rm L}=0.55$. The model light curve yields a reasonable fit to the Vela data, exhibiting distinct qualitative features including the two main peaks at the same phases, as seen in the Vela data. The peaks are lower than expected (constrained by the low level of off-peak emission, i.e., the $\chi^2$ prefers a small value for $A$), with the first peak being very broad and a small bump preceding the second peak when compared to the data. 
We also obtained a better $\chi^2$ best-fit solution for this larger $E$-field compared to the usual one, for $\epsilon=0.00$ at $\alpha=75^{+3}_{-1}$, $\zeta=51^{+2}_{-5}$, $A=1.1$, and $\Delta\phi_{\rm L}=0.55$. In Figure~\ref{fig:bestcontours} (bottom panels) we show our significance contour ${\rm log_{10}\xi^{2}}$ and its corresponding best-fit light curve. This offset-PC dipole $B$-field (with $\epsilon=0$ reduces to the static dipole field) and SG model for an increased $E$-field therefore provides an overall optimal fit, second only to the RVD and OG model combination (see Figure~\ref{fig:ModelComparisonB}).

\section{Comparison of best-fit parameters for different models}\label{sec:comparison_models}

We followed the same approach as \citet{Pierbattista2015} to compare the various optimal solutions of the different models, in two ways: (i) per $B$-field and model combination, and (ii) overall (for all $B$-field and model combinations). We determine the difference between the scaled\footnote[5]{We therefore first scale the $\chi^2$ values using the optimal value obtained for a particular $B$-field, and second we scale these using the overall optimal value irrespective of $B$-field.} $\chi^2$ of the optimal model, $\xi_{\rm opt}^{2}$ and the other models ($\xi^{2}$) using Eq.~[\ref{eq:difference-chi2}], substituting $N_{\rm dof}=96$, as summarised in Table~\ref{Summary}. The best-fit parameters for each $B$-field and geometric model combination, including the case for $100E_\parallel$, are summarised in Table~\ref{Summary}. The Table includes the different model combinations, the optimal unscaled $\chi^{2}$ value for each combination, the best-fit free parameters with $3\sigma$-errors on $\alpha$ and $\zeta$, and the comparison between models per $B$-field ($\Delta\xi^2_{\rm B}$) and overall ($\Delta\xi^2_{\rm all}$, with $\Delta\xi^{2}=0$ representing the best-fit solution for each $B$-field or the overall optimal fit; \citealp{Pierbattista2015}). We also include several multi-wavelength independent fits (all for the Vela pulsar). 

\begin{figure}
\centering
\includegraphics[width=\textwidth]{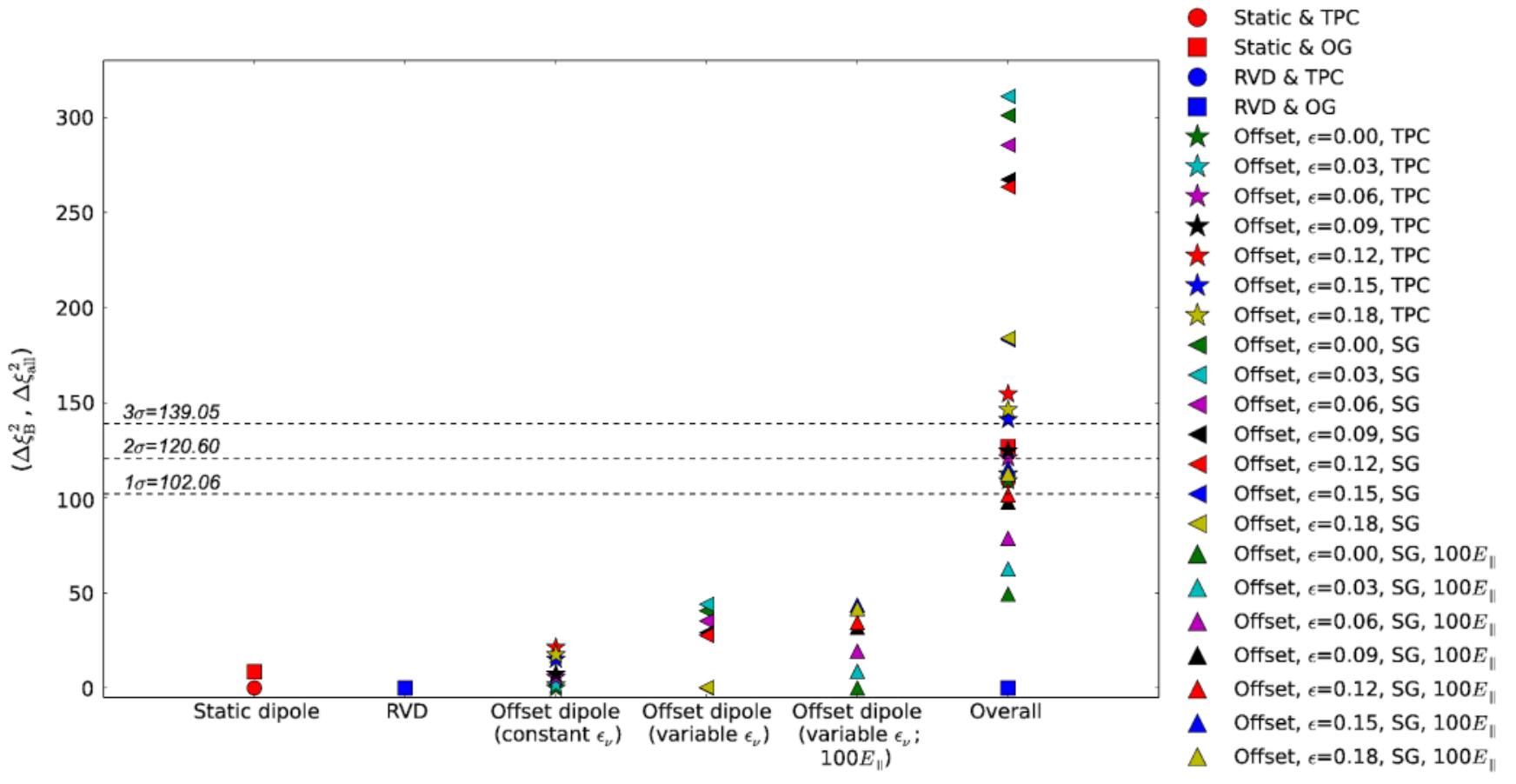}
\caption[Comparison of the relative goodness of the fit of solutions obtained for each $B$-field and geometric model combination, including the case of $100E_\parallel$]{\label{fig:ModelComparisonB} Comparison of the relative goodness of the fit of solutions obtained for each $B$-field and geometric model combination, including the case of $100E_\parallel$, as well as all combinations compared to the overall best fit, i.e., RVD $B$-field and OG model (shown on the $x$-axis). The difference between the optimum and alternative model for each $B$-field is expressed as $\Delta\xi^{2}_{\rm B}$, and for the overall fit as $\Delta\xi^{2}_{\rm all}$ (shown on the $y$-axis). The horizontal dashed lines indicate the $1\sigma$, $2\sigma$, and $3\sigma$ confidence levels. Circles and squares refer to the TPC and OG models for both the static dipole and RVD. The stars refer to the TPC (constant $\epsilon_{{\rm \nu}}$) and the left pointing triangles present the SG (variable $\epsilon_{{\rm \nu}}$) model for the offset-PC dipole field, for the different $\epsilon$ values. The upright triangles refer to our SG model and offset-PC dipole case for a larger $E$-field ($100E_\parallel$). The last column shows our overall fit comparison (see legend for symbols).}
\end{figure}

In Figure~\ref{fig:ModelComparisonB} we label the different $B$-field structures assumed in the various models as well as the overall comparison along the $x$-axis, and plot $\Delta\xi^{2}_{\rm B}$ and $\Delta\xi^{2}_{\rm all}$ on the $y$-axis. We represent the TPC geometry with a circle, the OG with a square, and for the offset-PC dipole field we represent the various $\epsilon$ values for constant $\epsilon_{\nu}$ by different coloured stars, for variable $\epsilon_{\nu}$ by different coloured left pointing triangles, and for the case of $100E_\parallel$ by different coloured upright triangles, as indicated in the legend. The dashed horizontal lines indicate our confidence levels we obtained by calculating the expected $\Delta\xi^{2}$ values using an online $\chi^{2}$ statistical calculator\footnote[6]{\url{http://easycalculation.com/statistics/critical-value-for-chi-square.php}} for $N_{\rm dof}=96$ degrees of freedom\footnote[7]{We note that \citet{Pierbattista2015} assumed that $\Delta\xi^2$ follows a $\chi^2$ distribution with $N_{\rm dof}$ degrees of freedom. We will follow this approximation here, assuming that the best-fit model provides a good fit to the observed light curves. The degrees of freedom may in reality slightly differ, however, and the matter is complicated by the fact that we want to statically compare non-nested models. A Monte Carlo approach would be preferable to find these significance levels. However, our main conclusions will not change for slight changes in these levels (which may be different for each $B$-field and model combination), and so we do not pursue this matter any further.}, i.e., using $p$-values of $p_{1\sigma}=1-0.682$, $p_{2\sigma}=1-0.954$, and $p_{3\sigma}=1-0.9973$. We found critical values of $\Delta\xi^{2}=102.06$ ($1\sigma$), $120.60$ ($2\sigma$), and $139.05$ ($3\sigma$) respectively. These confidence levels are used as indicators of when to reject or accept an alternative fit compared to the optimum fit. The last column represents fits for all models, irrespective of $B$-field.

For the static dipole field the TPC model gives the optimum fit and the OG model lies within $1\sigma$, implying that the OG geometry may provide an acceptable alternative fit to the data in this case. For the RVD field the TPC model is significantly rejected beyond the $3\sigma$ level (not shown on plot), and the OG model is preferred. We show three cases for the offset-PC dipole field, including the TPC model assuming constant $\epsilon_{\nu}$, the SG model assuming variable $\epsilon_{\nu}$, and the latter with an $E_\parallel$-field increased by a factor $100$. The optimal fits for the offset-PC dipole field and TPC model reveal that a smaller offset ($\epsilon$) is generally preferred for constant $\epsilon_{\nu}$, while a larger offset is preferred for variable $\epsilon_{\nu}$ (but not significantly), with all alternative fits falling within $1\sigma$. However, when we increase $E_\parallel$, a smaller offset is again preferred for the SG and variable $\epsilon_\nu$ case. When we compare all model and $B$-field combinations with the overall best fit (i.e., rescaling the $\chi^2$ values of all combinations using the optimal fit involving the RVD $B$-field and OG model), we notice that the static dipole and TPC model falls within $2\sigma$, whereas the static OG model lies within $3\sigma$. We also note that the usual offset-PC dipole $B$-field and TPC model combination (for all $\epsilon$ values) is above $1\sigma$ (with some fits $<2\sigma$), but the offset-PC dipole $B$-field and SG model combination (for all $\epsilon$ values) is significantly rejected ($>3\sigma$). However, the case of the offset-PC dipole field and a higher SG $E_\parallel$ for all $\epsilon$ values leads to a recovery, since all the fits fall within $2\sigma$ and delivers an overall optimal fit for $\epsilon=0$, second only to the RVD and OG model fit. 

\begin{figure}
\centering
\includegraphics[width=10cm]{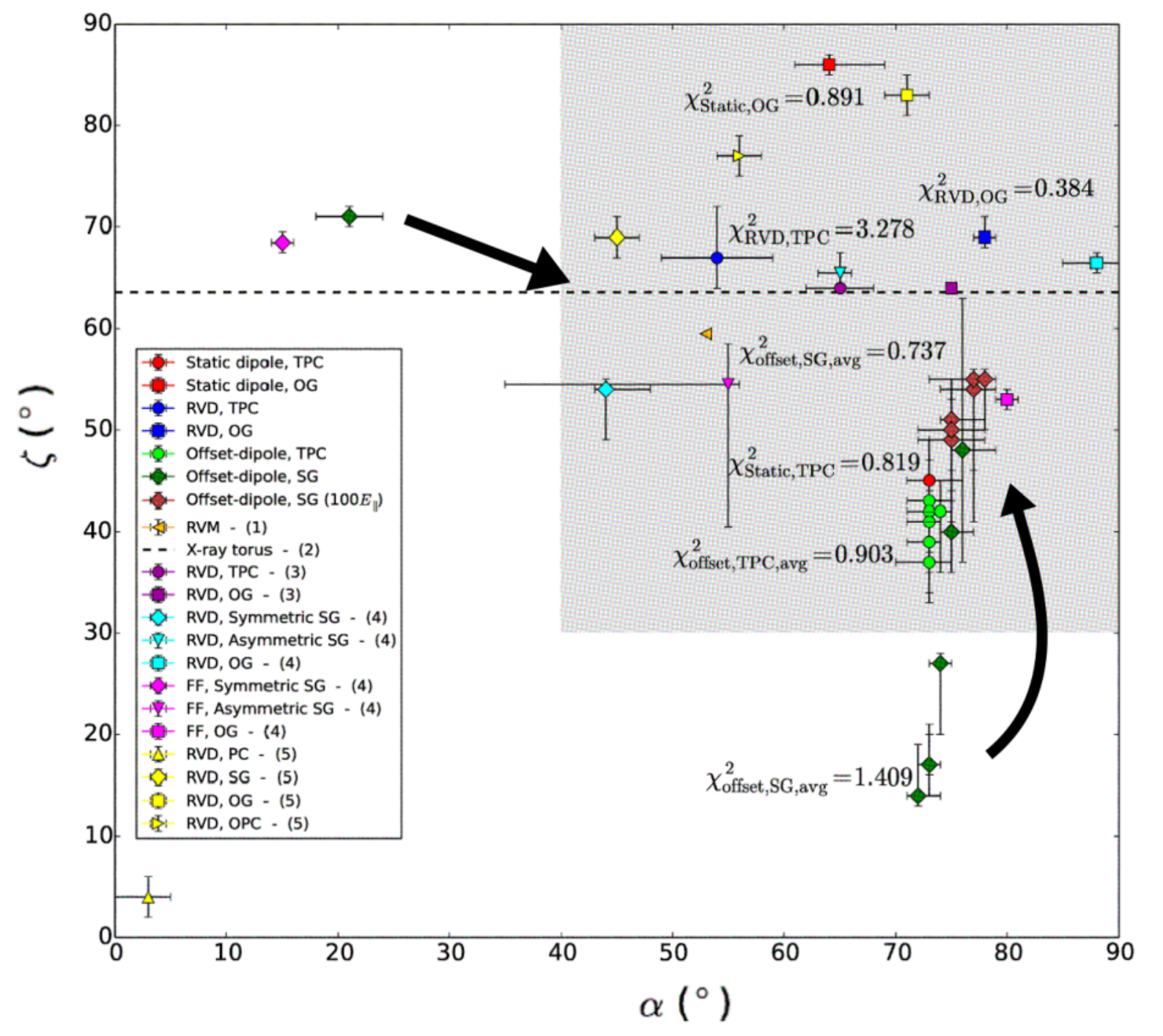}
\caption[Comparison between the best-fit $\alpha$ and $\zeta$, with errors, obtained from this and other studies.]{\label{fig:Comparison_alphazeta} Comparison between the best-fit $\alpha$ and $\zeta$, with errors, obtained from this and other multi-wavelength studies. Each marker represents a different case as summarised in Table~\ref{Summary}, with the unscaled $\chi^2$~($\times10^5$) value of our fits indicated. For the offset-PC dipole, for both the TPC and SG models we indicate the average $\chi^2$ value over the range of $\epsilon$. We also show our fits for the offset-PC dipole and SG model case with a larger $E_\parallel$. The two black arrows indicate the shift of the best fits to larger $\alpha$ and $\zeta$ if we increase our SG $E$-field by a factor 100. The shaded region contains all the fits that cluster at larger $\alpha$ and $\zeta$ values.}
\end{figure}

\begin{table}
\caption{\label{Summary} Best-fit parameters for each $B$-field and geometric model combination}
\setlength\tabcolsep{0.2cm}
\small
\begin{threeparttable}
\begin{tabular}{ccccccccccc}
\toprule
\multicolumn{3}{l}{Combinations} &  &  &  &  &  &  &  &  \\
\multicolumn{3}{l}{Models} & $\epsilon$ & $\chi^2$ & $\alpha$ & $\zeta$ & \emph{A} & $\Delta\phi_{\rm L}$ & $\Delta\xi^2_{\rm B}$ & $\Delta\xi^2_{\rm all}$ \\
\midrule
\multicolumn{3}{l}{Static dipole \textit{B}-field:} &  &  &  &  &  &  &  &  \\
\multicolumn{3}{l}{TPC} & ... & 0.819 & $73_{-2}^{+3}$ & $45_{-4}^{+4}$ & 1.3 & 0.55 &  0.00 & 108.75 \\
\multicolumn{3}{l}{OG} & ... & 0.891 & $64_{-3}^{+5}$ & $86_{-1}^{+1}$ & 1.3 & 0.05 & 8.44 & 126.75 \\
\multicolumn{3}{l}{RVD \textit{B}-field:}  &  &  &  &  &  &  &  &  \\
\multicolumn{3}{l}{TPC} & ... & 3.278 & $54_{-5}^{+5}$ & $67_{-3}^{+5}$ & 0.5 & 0.05 & 723.50 & 723.50  \\
\multicolumn{3}{l}{OG} & ... & 0.384 & $78_{-1}^{+1}$ & $69_{-1}^{+2}$ & 1.3 & 0.00 &   0.00 & 0.00 \\
\multicolumn{3}{l}{Offset-PC dipole \textit{B}-field for constant $\epsilon_{\nu}$:}  &  &  &  &  &  &  &  &  \\
\multicolumn{3}{l}{TPC} & 0.00 & 0.819 & $73_{-2}^{+3}$ & $45_{-4}^{+4}$ & 1.3 & 0.55 &  0.00 & 108.75 \\
\multicolumn{3}{l}{} & 0.03 & 0.834 & $73_{-2}^{+2}$ & $43_{-5}^{+4}$ & 1.3 & 0.55 &  1.76 & 112.50 \\
\multicolumn{3}{l}{} & 0.06 & 0.867 & $73_{-2}^{+2}$ & $42_{-5}^{+5}$ & 1.3 & 0.55 &  5.63 & 120.75 \\
\multicolumn{3}{l}{} & 0.09 & 0.882 & $73_{-2}^{+1}$ & $41_{-5}^{+3}$ & 1.3 & 0.55 &  7.39 & 124.50 \\
\multicolumn{3}{l}{} & 0.12 & 1.000 & $74_{-3}^{+1}$ & $42_{-6}^{+3}$ & 1.4 & 0.55 & 21.22 & 154.00 \\
\multicolumn{3}{l}{} & 0.15 & 0.948 & $73_{-2}^{+1}$ & $39_{-5}^{+3}$ & 1.4 & 0.55 & 15.12 & 141.00 \\
\multicolumn{3}{l}{} & 0.18 & 0.969 & $73_{-3}^{+2}$ & $37_{-4}^{+4}$ & 1.3 & 0.55 & 17.58 & 146.25 \\
\multicolumn{3}{l}{Offset-PC dipole \textit{B}-field for variable $\epsilon_{\nu}$:}  &  &  &  &  &  &  &  &  \\
\multicolumn{3}{l}{SG} & 0.00 & 1.587 & $21_{-3}^{+3}$ & $71_{-1}^{+1}$   & 0.5 & 0.85 & 40.52 & 300.75 \\
\multicolumn{3}{l}{} & 0.03 & 1.627 & $73_{-1}^{+1}$ & $17_{-3}^{+4}$   & 0.7 & 0.55 & 43.96 & 310.75 \\
\multicolumn{3}{l}{} & 0.06 & 1.525 & $72_{-1}^{+2}$ & $14_{-1}^{+5}$   & 0.5 & 0.60 & 35.18 & 285.25 \\
\multicolumn{3}{l}{} & 0.09 & 1.452 & $73_{-1}^{+1}$ & $17_{-1}^{+3}$   & 0.6 & 0.55 & 28.90 & 267.00 \\
\multicolumn{3}{l}{} & 0.12 & 1.437 & $74_{-1}^{+1}$ & $27_{-7}^{+1}$   & 0.8 & 0.55 & 27.61 & 263.25 \\
\multicolumn{3}{l}{} & 0.15 & 1.116 & $76_{-1}^{+3}$ & $48_{-11}^{+15}$ & 0.7 & 0.55 &  0.00 & 183.00 \\
\multicolumn{3}{l}{} & 0.18 & 1.119 & $75_{-1}^{+2}$ & $40_{-4}^{+6}$   & 0.5 & 0.55 &  0.26 & 183.75 \\
\multicolumn{3}{l}{Offset-PC dipole \textit{B}-field for variable $\epsilon_{\nu}$ (100$E_\gamma$):}  &  &  &  &  &  &  &  &  \\
\multicolumn{3}{l}{SG} & 0.00 & 0.581 & $75_{-1}^{+3}$ & $51_{-5}^{+2}$  & 1.1 & 0.55 &  0.00 &  49.27 \\
\multicolumn{3}{l}{} & 0.03 & 0.634 & $75_{-2}^{+2}$ & $49_{-5}^{+5}$  & 1.1 & 0.55 &  8.73 &  62.48 \\
\multicolumn{3}{l}{} & 0.06 & 0.698 & $75_{-3}^{+3}$ & $49_{-6}^{+5}$  & 1.1 & 0.55 & 19.39 &  78.61 \\
\multicolumn{3}{l}{} & 0.09 & 0.774 & $75_{-3}^{+3}$ & $50_{-9}^{+5}$  & 1.1 & 0.55 & 31.90 &  97.54 \\
\multicolumn{3}{l}{} & 0.12 & 0.789 & $77_{-3}^{+2}$ & $54_{-8}^{+2}$  & 1.1 & 0.55 & 34.42 & 101.36 \\
\multicolumn{3}{l}{} & 0.15 & 0.845 & $77_{-4}^{+2}$ & $55_{-14}^{+1}$ & 0.9 & 0.55 & 43.62 & 115.28 \\
\multicolumn{3}{l}{} & 0.18 & 0.834 & $78_{-2}^{+1}$ & $55_{-5}^{+1}$  & 0.8 & 0.55 & 41.80 & 112.51 \\
\midrule
\multicolumn{3}{l}{RVM {$^1$}} & & & 53 & 59.5 & & & & \\
\multicolumn{3}{l}{X-ray torus {$^2$}} & & &    & 63.6$^{+0.07}_{-0.05}$ & & & \\
\multicolumn{3}{l}{RVD \& TPC {$^3$}}  & & & 62--68 & 64 & & & \\
\multicolumn{3}{l}{RVD \& OG {$^3$}}   & & & 75     & 64 & & & \\
\multicolumn{3}{l}{RVD \& Symmetric SG {$^4$}} & & & 44$^{+4}_{-1}$ & 54$^{+1}_{-5}$ & & & \\
\multicolumn{3}{l}{RVD \& Asymmetric SG {$^4$}} & & & 65$^{+1}_{-2}$ & 65.5$^{+2}_{-1}$ & & & \\
\multicolumn{3}{l}{RVD \& OG {$^4$}} & & & 88$^{+2}_{-3}$ & 66.5$^{+1}_{-1}$ & & & \\
\multicolumn{3}{l}{FF \& Symmetric SG {$^4$}} & & & 15$^{+1}_{-1}$ & 68.5$^{+1}_{-1}$ & & & \\
\multicolumn{3}{l}{FF \& Asymmetric SG {$^4$}} & & & 55$^{+10}_{-20}$ & 54.5$^{+4}_{-14}$ & & & \\
\multicolumn{3}{l}{FF \& OG {$^4$}} & & & 80$^{+1}_{-1}$ & 53$^{+1}_{-1}$ & & & \\
\multicolumn{3}{l}{RVD \& PC {$^5$}} & & & 3$^{+2}_{-3}$ & 4$^{+2}_{-2}$ & & &  \\
\multicolumn{3}{l}{RVD \& SG {$^5$}} & & & 45$^{+2}_{-2}$ & 69$^{+2}_{-2}$ & & & \\
\multicolumn{3}{l}{RVD \& OG {$^5$}} & & & 71$^{+2}_{-2}$ & 83$^{+2}_{-2}$ & & & \\
\multicolumn{3}{l}{RVD \& OPC {$^5$}} & & & 56$^{+2}_{-2}$ & 77$^{+2}_{-2}$ & & & \\
\bottomrule
\end{tabular}
\begin{minipage}{\textwidth}
{\bf Note.} The table summarises the best-fit parameters $\alpha$, $\zeta$, $A$, and $\Delta\phi_{\rm L}$, for each model combination, with the errors on $\alpha$ and $\zeta$ determined by using the $3\sigma$ interval connected contours. We chose a minimum error of $1^{\circ}$ if the confidence contour yielded smaller errors. We included the unscaled $\chi^{2}$ to indicate which geometry yields the optimal fit to the Vela data (i.e., the OG model and RVD $B$-field). We included other multi-wavelength fits in the second part of the table. \\ 
{\bf References.} $^1$ \citealp{Johnston2005}; $^2$ \citealp{Ng2008}; $^3$ \citealp{Watters2009}; $^4$ \citealp{DeCesar2013}; and $^5$ \citealp{Pierbattista2015}.
\end{minipage}
\end{threeparttable}
\end{table}

Several multi-wavelength studies have been performed for Vela, using the radio, X-ray, and $\gamma$-ray data, in order to find constraints on $\alpha$ and $\zeta$. We only fit the $\gamma$-ray light curve, because we did not want to bias our results by using a geometric radio emission model \citep{DeCesar2013}. However, \citet{Johnston2005} determined the radio polarisation position angle from polarisation data by applying a rotating vector model (RVM) fit to the data finding best-fit values of $\alpha=53^{\circ}$ and $\zeta=59.5^{\circ}$, with an impact angle of $\beta=\zeta-\alpha=6.5^{\circ}$. \citet{Ng2008} applied a torus-fitting technique \citep{Ng2004} to fit the \textit{Chandra} data in order to constrain the Vela X pulsar wind nebula (PWN) geometry, deriving a value of $\zeta=63.6{_{-0.05}^{+0.07}}^{\circ}$ represented by the dashed black line in Figure~\ref{fig:Comparison_alphazeta}. \citet{Watters2009} modelled light curves using the PC, TPC, and OG geometries in conjunction with an RVD field, thereby constraining the geometrical parameters $\alpha$, $\zeta$, and also finding small $\beta$ in the case of the PC model. They found a good fit for their TPC model at $\alpha=62^{\circ}-68^{\circ}$ and $\zeta=64^{\circ}$, and for the OG geometry at $\alpha=75{}^{\circ}$ and $\zeta=64^{\circ}$. We find that our best-fit values for the RVD field, for both the TPC and OG models, are in good agreement with those found by \citet{Watters2009}. \citet{DeCesar2013} followed a similar approach to ours, but for the RVD and FF $B$-fields combined with emission geometries such as the SG (symmetric and asymmetric cases) and OG. They have different free model parameters including $\alpha$, $\zeta$, $w$ (gap width), and $R_{\rm max}$ (maximum emission radius), and determined errors on their best fits using the $3\sigma$ confidence intervals. They found best-fit solutions for the RVD and OG model at $\alpha=88{_{-3}^{+2}}^{\circ}$ and $\zeta=66.5{_{-1}^{+1}}^{\circ}$, which is within $10^{\circ}$ or less compared to our best-fit solution. Their overall best fit was for the FF $B$-field and OG geometry, with $\alpha={80^{+1}_{-1}}^{\circ}$ and $\zeta={53^{+1}_{-1}}^{\circ}$. \citet{Pierbattista2015} found a best-fit solution for Vela using the RVD field and OG model combination at $\alpha={71_{-2}^{+2}}^{\circ}$ and $\zeta=83{_{-2}^{+2}}^{\circ}$, with $\zeta$ exceeding the best-fit solution we found by nearly $15^{\circ}$. However, they fit both the $\gamma$-ray and radio light curves, which may explain this discrepancy. We summarise all these multi-wavelength fits and more in Table~\ref{Summary}.

We graphically summarise the best-fit $\alpha$ and $\zeta$, with errors, from this and other works in Figure~\ref{fig:Comparison_alphazeta}. We notice that the best fits generally prefer a large $\alpha$ or $\zeta$ or both. It is encouraging that many of the best-fit solutions lie near the $\zeta$ inferred from the PWN torus fitting \citep{Ng2008}, notably for the RVD $B$-field. A significant fraction of fits furthermore lie near the $\alpha-\zeta$ diagonal, i.e., they prefer a small impact angle, probably due to radio visibility constraints \citep{Johnson2014}. For an isotropic distribution of pulsar viewing angles, one expects $\zeta$ values to be distributed as $\sin(\zeta)$ between $\zeta=[0^\circ,90^\circ]$, i.e., large $\zeta$ values are much more likely than small $\zeta$ values, which seems to correspond to the large best-fit $\zeta$ values we obtain. There seems to be a reasonable correspondence between our results obtained for geometric models and those of other authors, but less so for the offset-PC dipole $B$-field, and in particular for the SG $E$-field case. The lone fit near $(20^\circ,70^\circ)$ may be explained by the fact that a very similar fit, but one with slightly worse $\chi^2$, is found at $(50^\circ,80^\circ)$. If we discard the non-optimal TPC / SG fits, we see that the optimal fits will cluster near the other fits at large $\alpha$ and $\zeta$. Although our best fits for the offset-PC dipole $B$-field are clustered, it seems that increasing $\epsilon$ leads to a marginal decrease in $\zeta$ for the TPC model (light green) and opposite for SG (dark green), but not significantly (see Table~\ref{Summary}). For our increased SG $E$-field case (brown) we note that the fits now cluster inside the grey area above the fits for the static dipole and TPC, and offset-PC dipole for both the TPC and SG geometries. 

\section{Conclusion}\label{sec:ch3concl}

We investigated the impact of different magnetospheric structures on predicted $\gamma$-ray pulsar light curve characteristics. We extended our code which already included the static dipole and RVD $B$-fields, by implementing an additional $B$-field, i.e., the symmetric offset-PC dipole field \citep{Harding2011a,Harding2011c} characterised by an offset $\epsilon$ of the magnetic PCs. We also included the full accelerating SG $E$-field corrected for GR effects up to high altitudes. For the offset-PC dipole field we only considered the TPC (assuming uniform $\epsilon_{\nu}$) and SG (modulating the $\epsilon_{\nu}$ using the $E$-field) models, since we do not have $E$-field expressions available for the OG model for this particular $B$-field. We matched the low-altitude and high-altitude solutions of the SG $E_\parallel$ by determining the matching parameter $\eta_{\rm c}(P,\dot{P},\alpha,\epsilon,\xi,\phi_{\rm PC})$ on each field line in multivariate space. Once we calculated the general $E$-field we could solve the particle transport equation. This yielded the particle energy $\gamma(\eta)$, necessary for determining the CR $\epsilon_\nu$ and to test whether the CRR limit is attained. For the case of a variable $\epsilon_\nu$, we found that the CRR limit is reached for many parameter combinations (of $\alpha, \epsilon$ and $\phi_{\rm PC}$; see Figure~\ref{fig:RRLim}), albeit only at large $\eta$. A notable exception occurred at large $\alpha$ where the first term of each $E$-field expression (e.g., Eq.~[\ref{eq:E_low}] and [\ref{eq:E_high}]) became lower and the second term played a larger role, leading to smaller gain rates and therefore smaller Lorentz factors $\gamma$. 

We concluded that the magnetospheric structure and emission geometry have an important effect on the predicted $\gamma$-ray pulsar light curves. \textit{However, the presence of an $E$-field may have an even greater effect than small changes in the $B$-field and emission geometries:} When we included an SG $E$-field, thereby modulating $\epsilon_\nu$, the resulting phase plots and light curves became qualitatively different compared to the geometric case.

We fit our model light curves to the observed {\it Fermi}-measured Vela light curve for each $B$-field and geometric model combination. We found that the RVD field and OG model combination fit the observed light curve the best for $(\alpha,\zeta,A,\Delta\phi_{\rm L})=({78_{-1}^{+1}}^\circ,{69_{-1}^{+2}}^\circ,1.3,0.00)$ and an unscaled $\chi^2=3.84\times{10}^4$. As seen in Figure~\ref{fig:ModelComparisonB}, for the RVD field an OG model is significantly preferred over the TPC model, given the characteristically low off-peak emission. For the other field and model combinations there was no significantly preferred model (per $B$-field), since all the alternative models may provide an acceptable alternative fit to the data, within $1\sigma$. The offset-PC dipole field for constant $\epsilon_{\nu}$ favoured smaller values of $\epsilon$, and for variable $\epsilon_{\nu}$ larger $\epsilon$ values, but not significantly so ($<1\sigma$). When comparing all cases (i.e., all $B$-fields), we noted that the offset-PC dipole field for variable $\epsilon_{\nu}$ was significantly rejected ($>3\sigma$). 

We further investigated the effect which the SG $E_\parallel$ had on our predicted light curves in two ways. First, we lowered the minimum photon energy from $E_{\gamma,\rm min}=100$~MeV to $E_{\gamma,\rm min}=1$~MeV, leading to emission in the hard X-ray waveband. We noted new caustic structures and emission features on the resulting phase plots and light curves that were absent when $E_{\gamma,\rm min}>100$~MeV. Since we wanted to compare our model light curves to \textit{Fermi} data we increased the usual low SG $E$-field by a factor of 100 (with a spectral cutoff $E_{\rm CR}\sim4$~GeV). When solving the particle transport equation, we noticed that the CRR limit is now reached in most cases at lower $\eta$. The increased $E$-field also had a great impact on the phase plots, e.g., extended caustic structures and new emission features as well as different light curve shapes emerged. We also compared the best-fit light curves for the offset-PC dipole $B$-field and $100E_\parallel$ combination for each $\epsilon$ (Figure~\ref{fig:ModelComparisonB}) and noted that a smaller $\epsilon$ was again preferred (although not significantly; $<1\sigma$). However, when we compared this case to the other $B$-field and model combinations, we found statistically better $\chi^2$ fits for all $\epsilon$ values with an optimal fit at $\alpha={75^{+3}_{-1}}^{\circ}$, $\zeta={51^{+2}_{-5}}^{\circ}$, $A=1.1$, and $\Delta\phi_{\rm L}=0.55$ for $\epsilon=0$, being second in quality only to the RVD and OG model fit. 

We graphically compared the best-fit $\alpha$ and $\zeta$, with errors, from this and other works in Figure~\ref{fig:Comparison_alphazeta}. We noted that many of the best-fit solutions cluster inside the grey area at larger $\alpha$ and $\zeta$. Some fits lie near the $\alpha-\zeta$ diagonal (possibly due to radio visibility constraints in some cases) as well as near the $\zeta$ inferred from the PWN torus fitting \citep{Ng2008}, notably for the RVD $B$-field. There was reasonable correspondence between our results obtained for geometric models and those of other independent studies. When we discarded the non-optimal TPC / SG fits, we saw that the optimal fits clustered near the other fits at large $\alpha$ and $\zeta$. For our increased SG $E$-field and offset-PC dipole combination (brown) we noted that these fits clustered at larger $\alpha$ and $\zeta$. 

There have been several indications that \textit{the SG $E$-field may be larger than initially thought.} For example, (i) population synthesis studies found that the SG $\gamma$-ray luminosity may be too low, pointing to an increased $E$-field and / or particle current through the gap \citep[e.g.,][]{Pierbattista2015}. Furthermore, if the $E$-field is too low, one is not able to reproduce the (ii) observed spectral cutoffs of a few GeV (Section~\ref{subsec:100MeV}; \citealp{Abdo2013SecondCat}). We found additional indications for an enhanced SG $E$-field. A larger $E$-field (increased by a factor of 100) led to (iii) statistically improved $\chi^2$ fits with respect to the light curves. Moreover, the inferred best-fit $\alpha$ and $\zeta$ parameters for this $E$-field (iv) clustered near the best fits of independent studies. We additionally observed that a larger SG $E$-field also (v) increased the particle energy gain rates and therefore yielded a larger particle energy $\gamma$ (giving CR that is visible in the \textit{Fermi} band) as well as leading to a CRR regime already close to the stellar surface. These evidences may point to a reconsideration of the boundary conditions assumed by \citet{Muslimov2004a} which suppressed the $E_{\parallel}$ at high altitudes. They assumed equipotentiality of the SG boundaries as well as the steady state drift of charged particles across the SG $B$-field lines, implying $E_\perp\approx0$ at high-altitudes, with the flux of charges remaining constant up to high altitudes. One possible way to bring self-consistency may be implementation of the newly developed FIDO model that includes global magnetospheric properties and calculates the $B$-field and $E$-field self-consistently.

We envision several future projects that may emanate from this study. One could continue to extend the range of $\epsilon$ for which our code finds the PC rim, since more complex field solutions, e.g., the dissipative and FF field structures, may be associated with larger PC offsets. However, the offset-PC dipole solutions have limited applicability to outer magnetosphere emission since they use the static dipole frame and do not model the field line sweep back. Therefore, it would be preferable to investigate the $B$-fields and $E$-fields of more complex $B$-field models (see Sections~\ref{sec:BandEfields} and~\ref{sec:bmodels}) and solve the transport equation to test if the particles reach the CRR limit. The effect of these new fields on the phase plots and light curves can also be studied. There is also potential for multi-wavelength studies, such as light curve modelling in the other energy bands, e.g., combining radio and $\gamma$-ray light curves (see \citealp{Seyffert2010,Seyffert2012,Pierbattista2015}). One could furthermore model energy-dependent light curves, such as those available for Vela and other bright pulsars using {\it Fermi} data (e.g., \citealp{Abdo2009Vela}). Lastly, model phase-resolved spectra can be constructed which is an important test of the $E_\parallel$-field magnitude and spatial dependence.
\\
\\
In the next Chapter we will discuss the technical details of the emission modelling code \citep{Harding2015,Harding2018} used in the rest of this study. This emission code is an extension of the geometric code discussed in Section~\ref{sec:geometries} and include a FF solution in a SG model scenario. Since we focus on the particle emission from CR (Section~\ref{subsec:CR}), we also refined the radius of curvature of the particle trajectories. % MSc work / article summarised - Offset PC
\chapter{Model description: code setup and refinements} \label{chap:ch4}

In this thesis, I modelled pulsar emission by investigating energy-dependent light curves and spectra using a full emission code. In this Chapter, I specifically discuss the technical details of this emission code, the porting and local implementation of it, and how I was able to obtain the results presented in Chapter~\ref{chap:CREdepLCmod}. Additionally, I will give a summary of \citet{Harding2018} in Chapter~\ref{chap:VelaTeV}, since those results accompany those in Chapter~\ref{chap:CREdepLCmod}.
In Section~\ref{sec:3Dcode}, we give a brief description of the pulsar emission code and the calibration thereof follows in Section~\ref{sec:callibrate}. Section~\ref{sec:cluster} gives a brief description of the parallelised version of the emission code; this is followed by Section~\ref{sec:rhocalc} where I describe a more accurate computation of the particle
trajectories and their radii of curvature $\rho_{\rm c}$. Section~\ref{sec:ch4concl} summarises the main conclusions.

\section{An SSC emission modelling code} \label{sec:3Dcode}

%Chronologies: algemene beskrywing; 2008, 2015, 2018 - som opdaterings op. Fokus uiteindelik op CR. Sit prentjie uit 2018 in. Iets oor radio-absorpsie. PC pair cascade. Al die stralingskomponente; dinamika, ens. Wat is alles in die kode? Fisika-fokus.
%Kortliks - primaries; pairs; radiation components; radio absorption; magnetic fields; electric fields; PC rim; rings; trajectories; lab frame; dynamics; etc.

\citet{Harding2008} used a 3D optical to $\gamma$-ray emission model in an SG scenario including two particle populations, i.e., primaries (leptons) being accelerated from the stellar surface up to altitudes of $R_{\rm LC}$, as well as non-accelerated electron-positron pairs radiated up to the same altitudes as the primaries. This model is based on the geometric model of \citet{Dyks2004b} that assumed an RVD solution for the $B$-field (see Section~\ref{sec:BandEfields},~\ref{sec:geometries}, and~\ref{sec:epsilon_extend}). The open volume coordinates $r_{\rm ovc}$ and $l_{\rm ovc}$ are associated with the radial (i.e., rings) and azimuthal (i.e., arc length along each ring measured in the direction of increasing $\phi_{\rm PC}$) coordinates, respectively, as mentioned in Chapter~\ref{chap:OffsetPC}.
They also explore the geometry of the radio emission in relation to that of the HE emission. They included radiation from a geometric radio emission beam model that consists of a core and a single cone \citep{Rankin1993,Gonthier2004}, in order to compute the intensity and angles of radio photons necessary for the cyclotron absorption /synchrotron emission component. The primary particles are injected at the stellar surface and are accelerated along the $B$-field lines in a SG region bounding the last open field lines via a low and high-altitude SG $E$-field, that are combined to obtain a total SG $E$-field over all altitudes (see Section~\ref{sec:SGEfield} and equations therein). In this model, the SG reaches altitudes of $R_{\rm LC}$. The electron-positron pairs are produced in PC and SG pair cascades following from the accelerated primaries. They investigated the CR from primaries, SR from both primaries and pairs that undergo cyclotron resonant absorption of radio photons and produce significant increases in pitch angle, as well as nonresonant ICS of radio photons (see Section~\ref{sec:RadMechanisms} for detailed discussions about radiation mechanisms). They followed the same particle dynamics calculation as \citet{Harding2005c} but added the CR loss and
resonant absorption terms. Due to the large acceleration and SR loss rates the ICS were neglected for the primary
particles and the pairs. This resulted in a broadband spectrum ranging from infrared to HE $\gamma$-rays. There is no SSC calculation considered in this model, however the model of \citet{Harding2015} does include this radiation component. The model of \citet{Harding2018} is based on that of \citet{Harding2015}, including extended energy rangy ranges.

Using the emission model of \citet{Harding2015}, we study the full particle acceleration and focus on the CR emission component. The model assumes a 3D FF $B$-field as the basic magnetospheric structure. This solution (formally assuming an infinite plasma conductivity, so that the $E$-field is fully screened) serves as a good approximation to the geometry of field lines implied by the dissipative models that require a high conductivity in order to match observed $\gamma$-ray light curves \citep{Kalapotharakos2012a,Kalapotharakos2014,Li2012}. This model is an extended version of the model first used by \citet{Harding2008} with the PC rim (Section~\ref{sec:epsilon_extend}) being determined in a similar fashion for this FF $B$-field structure. Both primary particles and electron-positron pairs are injected at the stellar surface. The primaries radiate CR and some of these $\gamma$-ray photons are converted into pairs in the intense $B$-fields close to the star. Using an independent code, we calculate a PC pair cascade that develops, since primaries and pairs radiate SR as well as CR that are converted into electron-positron pairs, leading to further generations of particles with lower energies. The primaries are injected with a low initial speed and are further accelerated along the $B$-field lines by a constant parallel $E$-field $E_{\parallel}$ that is parallel to the local $B$-field lines (used as a free parameter in this model) in an extended SG and current sheet scenario near the last open field lines. In this model, the SG extends beyond $R_{\rm LC}$ into the current sheet, reaching altitudes of up to $r=2R_{\rm LC}$. All calculations are performed in the laboratory frame. In order to explain the VHE emission seen from pulsars they included an SSC component (involving energetic primaries and pair SR) which could reproduce the spectral flux level as observed by MAGIC and VERITAS (see Section~\ref{sec:DevGamPSRAstro}). 

\citet{Harding2018} adapted the emission code of \citet{Harding2015} further by updating the CR component to a SC radiation component, as well as a two step accelerating $E$-field, i.e., $E_\parallel$ inside the light cylinder and $E_\parallel$ beyond the light cylinder and into the current sheet, as motivated by results of global MHD and PIC pulsar models. Our contribution was an improved calculation of the particle trajectories and their $\rho_{\rm c}$'s. They presented the broadband spectrum of Vela ranging from infrared (IR) to beyond 10 TeV energies, thereby explaining the TeV emission recently observed by \emph{H.E.S.S.} (see Chapter~\ref{chap:VelaTeV} for more details).
In Chapter~\ref{chap:CREdepLCmod} I focus on the GeV spectrum and energy-dependent CR light curves from the primary particles using the model of \citet{Harding2015}. Additionally, we included the two-step $E_\parallel$ to study the effect thereof on the light curves and spectra. In Chapter~\ref{chap:VelaTeV} I will give a short summary of my contribution to the work published in \citet{Harding2018}, which includes the refined $\rho_{\rm c}$ discussed in Section~\ref{sec:rhocalc}.

\section{Calibration of the code} \label{sec:callibrate}

Before the implementation of a refined $\rho_{\rm c}$, I first calibrated the emission code, since it is an essential component of evaluating and excluding any uncertainties in the predictions and estimates. I investigated the spectral output obtained by comparing it with the original data (from our collaborators), using the same version of the code paired with different compilers. The different choices of compilers produced variations in the model simulations. 

As a test case, I studied the variations in spectra, for the Crab pulsar, for two calculations of the observer angle $\zeta$. Using the $z$-component $(z'(s))$ of the velocity of the particle as a function of $s$, we calculated $\zeta=\arccos({z'}(s))$, and second, we used a linear calculation $\zeta = ({-\pi/2}{z'}+\pi/2)$. Later, I made a cut on $\zeta$ so that the predictions would be observer-specific (characterised by $\zeta_{\rm cut}$). Thus, different implementations of the emission direction of individual photons would influence what a particular observer at a constant $\zeta$ would see. This was to test whether the implementation of the inverse cosine had any impact on the predicted spectrum.

\begin{figure}
    \centering
    \includegraphics[width=0.6\textwidth]{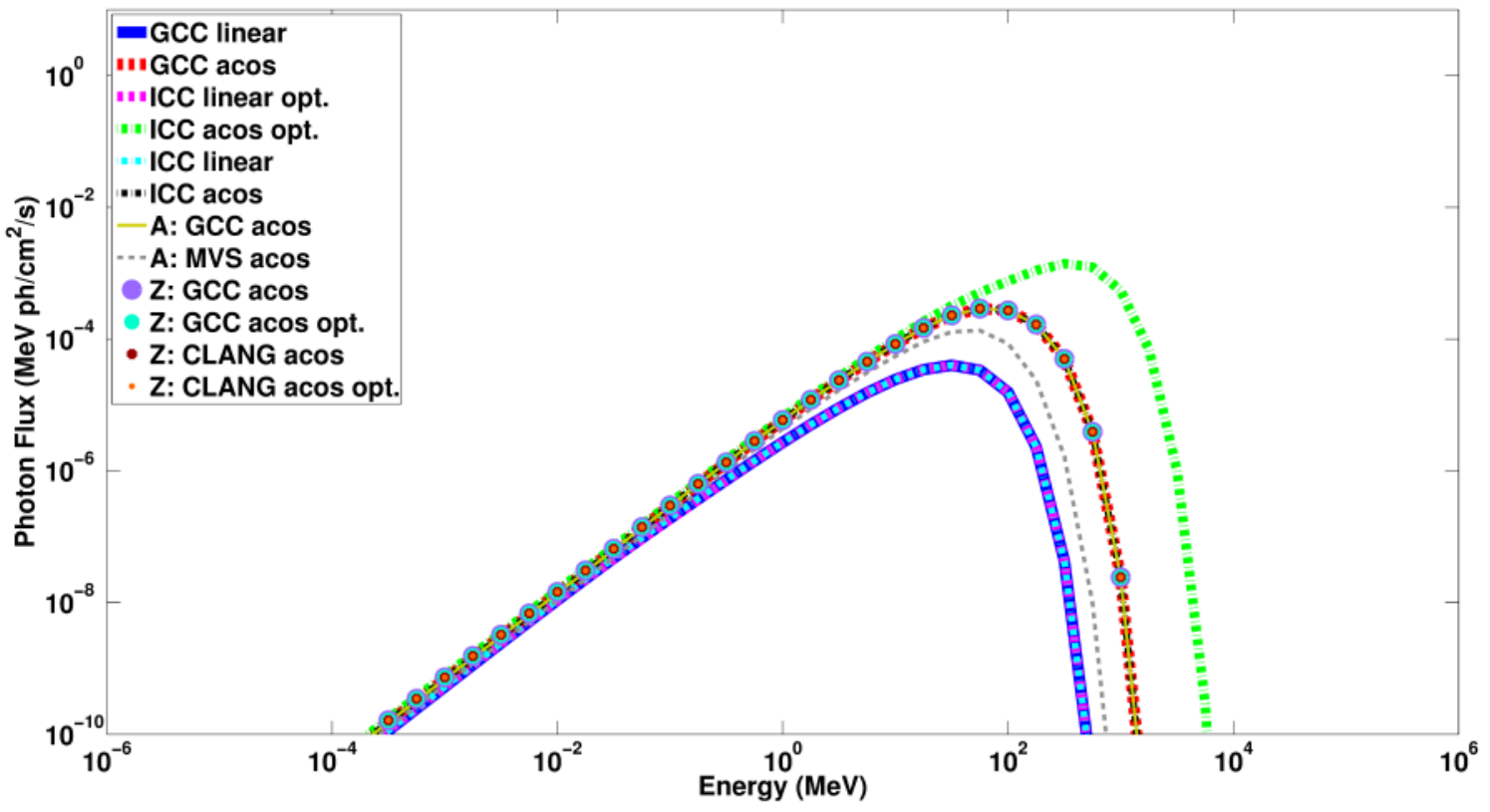}
    \caption[An example of the phase-averaged CR spectra for the Crab pulsar assuming $\alpha=45^\circ$ and $\zeta_{\rm cut}=50^\circ$]{An example of the phase-averaged CR spectra for the Crab pulsar assuming $\alpha=45^\circ$ and $\zeta_{\rm cut}=50^\circ$. Each curve represents a combination of a compiler with a different calculation of the colatitude, i.e., linear and inverse cosine. The letters A and Z refer to the results obtained by our collaborators using other compilers.} \label{fig:compilers}
\end{figure}

The impact of the compiler choice is illustrated in Figure~\ref{fig:compilers} showing the phase-averaged CR spectra for different compilers assuming $\alpha=45^\circ$, $\zeta_{\rm cut}=50^\circ$, as well as that the emission originates from primary particles only. In Figure~\ref{fig:compilers} I choose compilers such as GCC, ICC, MVS, and CLANG, for both colatitude calculations. Additionally, I considered for some compilers the optimised and non-optimised cases to further investigate the effect thereof on the calculations. The legend indicates our combination choices, where no letter denotes the simulations done by us and letters A and Z those done by two collaborators. The spectra for the two colatitude calculations, regardless of the compiler choice, are distinctively different. However, two exceptions are the MVS and the optimised ICC compilers, both for the inverse cosine case. For the linear case, the photon flux is lower than that of the inverse cosine case. Since the emission code uses the inverse cosine calculation and the output for the majority of compilers agree, I opted for the compiler that yielded results consistent with those of our collaborators, i.e., GCC. 

\begin{figure}
    \centering
    \includegraphics[width=\textwidth]{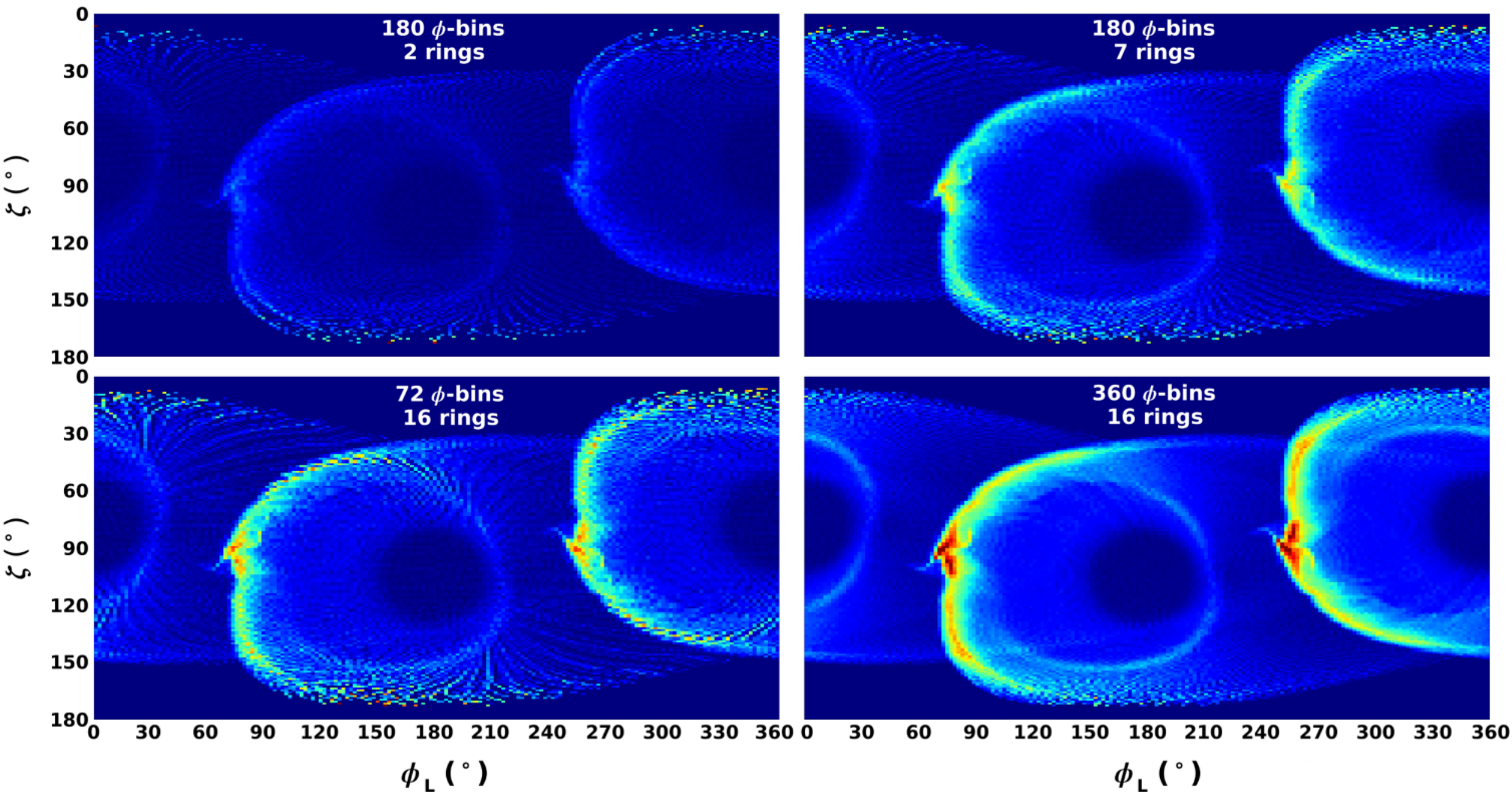}
    \caption[An example of the phase plots for the Vela pulsar assuming $\alpha=75^\circ$ and $\zeta_{\rm cut}=65^\circ$]{An example of the phase plots for the Vela pulsar assuming $\alpha=75^\circ$ and $\zeta_{\rm cut}=65^\circ$. Each phase plot is for a different resolution, i.e., number of azimuthal divisions and number of rings. Top panels are for same amount azimuthal divisions, but increasing number of rings. Bottom panels are for different azimuthal divisions but same amount of rings.} \label{fig:resolution}
\end{figure}

% DIFFERENT RESOLUTION CHOICES; PPs AND LCs
Another element of calibration is the resolution of the simulations. This is especially important when constructing results since the emission is followed along each $B$-field line. Thus the more field line footpoints there are on the PC, the more emission there is, resulting in better quality phase plots, light curves, and spectra. To investigate the effect the resolution of the model has on its accuracy, I constructed phase plots for different grid sizes of azimuthal divisions and the number of rings (self-similar to the PC rim; see Section~\ref{sec:epsilon_extend}). As an example I illustrate the phase plots for Vela assuming $\alpha=75^\circ$, $\zeta_{\rm cut}=65^\circ$, and a constant accelerating $E_\parallel$-field, i.e., $eE_\parallel/m_{\rm e}c^2=0.25$~cm$^{-1}$. In Figure~\ref{fig:resolution} I illustrate the effect on the phase plots for four different resolutions. The top panels are for a fixed amount of azimuthal divisions, but different amount of rings. The bottom panels are for a different amount of azimuthal divisions, but fixed amount of rings. If more rings are added there is more emission that accumulates to form the caustics, thus they appear brighter. This is also the case when adding more azimuthal bins, although the effect is not as prominent as in the former case. We set the resolution to 360 azimuthal bins and 7 rings for the results in the Chapter~\ref{chap:CREdepLCmod}, since the shape as well as the emission quality are reasonable for these choices.

\section{Running the code on a cluster}\label{sec:cluster} % Description of parallel code and process of running it
The emission code described in Section~\ref{sec:3Dcode} is available in both a serial and parallelised version, with a major difference in the structure of each code. The parallelised code (hereafter referred to as the parallel code) consists of several parts to ensure it is computationally effective. Thus, the radiation by particles injected on different field lines were computed in a distributed and independent fashion, and only later is the emission from all particles accumulated in a phase plot.

We calibrated the serial and parallel codes to minimise differences in the separate codes and to ensure that new implementations were not missed. We made some alterations to the parallel code in order to generate simulations on our local cluster, i.e., at the NWU, Potchefstroom. These modifications include 
\begin{itemize}
    \item The compiler needed for the code is mpiicpc, however the local cluster uses the intel/ mpi compiler. Thus, the code was slightly adapted, but the outputs were carefully calibrated against the original code.
    \item The original job script uses a Portable Batch System scheduler, i.e., computer software that performs job scheduling with the primary responsibility to allocate computational tasks on a high-performance computer. However, the local cluster uses a different job scheduler namely the Sun Grid Engine. The main difference between the two schedulers is the operation of each.
    \item We encountered memory issues caused by large dimensional arrays. We reduced the memory consumption significantly and this will be discussed below.
\end{itemize}

The parallel code uses the mpiicpc compiler on the Discover\footnote{\url{https://www.nccs.nasa.gov/}} cluster. The serial code is split up in parts and then parallelised. The $*$.sh files are needed to submit and run the job on the system. The script and code need to be tailored to each system. However, several of the parameters like $r^{\rm min}_{\rm ovc}$, $r^{\rm max}_{\rm ovc}$, the number of azimuthal divisions, and the number of rings are set in this file and override what is set in the main code. The code apportion subcalculations to the number of processors one specifies when you submit the job.

The parallelisation of the code involved splitting the serial code into different parts and then using a Makefile to compile and create the executable file. All these different parts of code, each serving a specific purpose, are called in the main code. The local cluster has two different directories. The \emph{home} directory handles smaller files such as the source code and sample job script, whereas the \emph{scratch} directory handles large amounts of data.

\begin{figure}
    \centering
    \includegraphics[width=\textwidth]{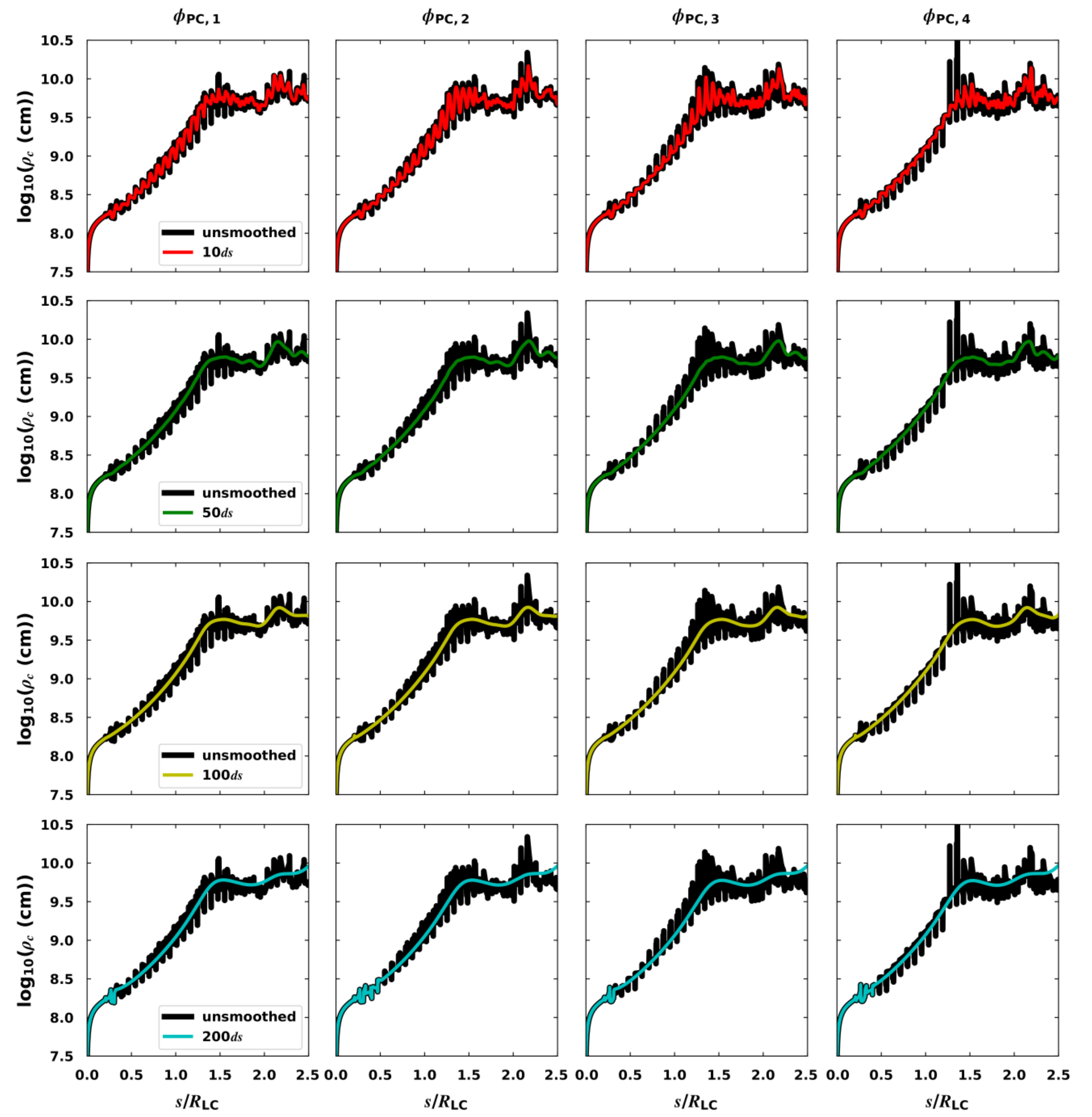}
    \caption[An example of the newly calculated $\rho_{\rm c}$ as a function of normalised arclength $s/R_{\rm LC}$ before and after smoothing and matching]{\label{fig:smoothing} An example of the newly calculated $\rho_{\rm c}$ as a function of normalised arclength $s/R_{\rm LC}$ before (solid black line) and after (coloured lines) smoothing and matching. Each column represents an arbitrary $B$-field line, and each row corresponds to different smoothing parameters, i.e., $h=[10ds,50ds,100ds,200ds]$, increasing from top to bottom as indicated in the legend.}
\end{figure}

\begin{figure}
\includegraphics[width=\textwidth]{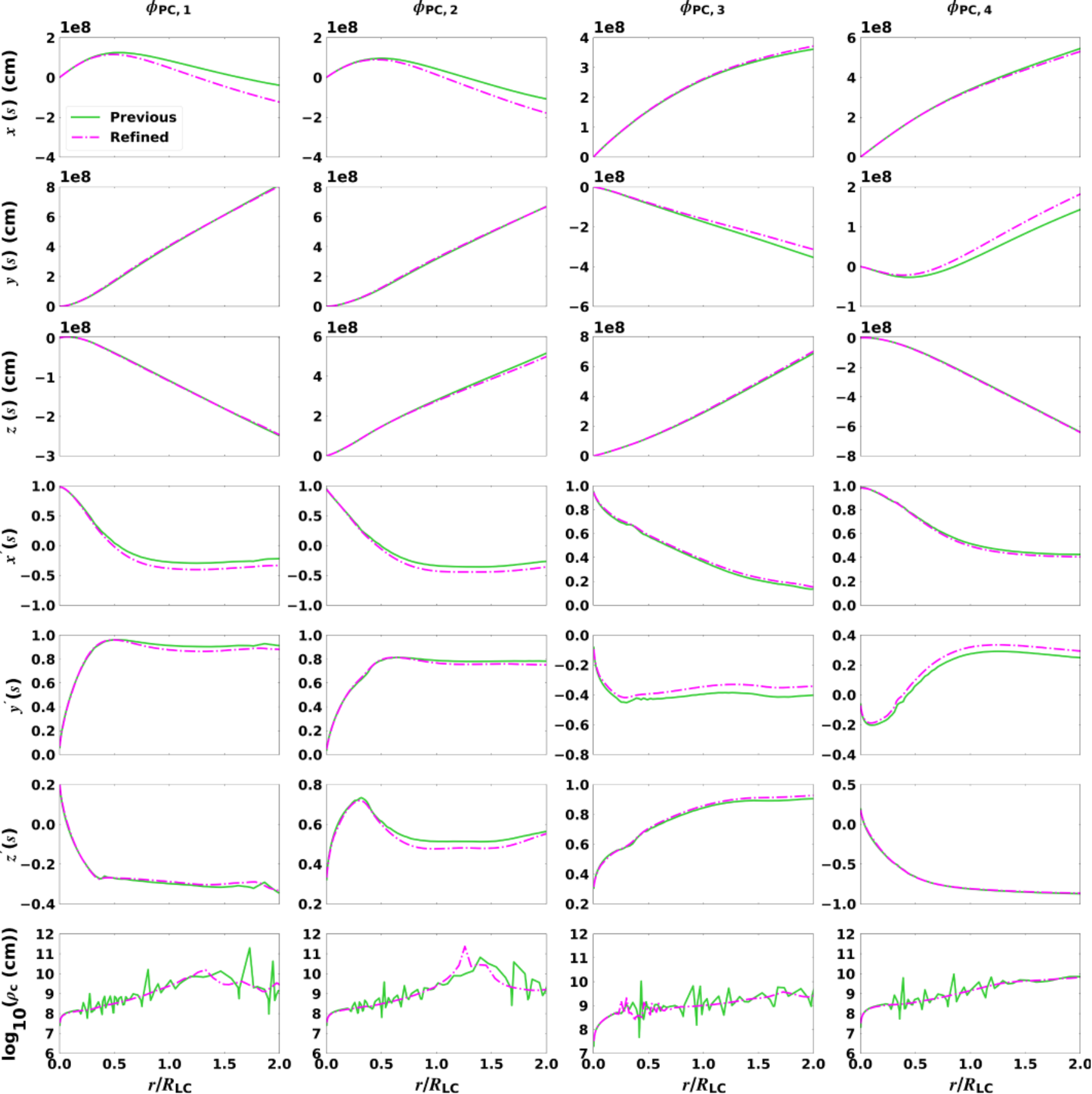}
\caption[A comparison of electron position $x(s)$,\,$y(s)$,\,$z(s)$, trajectory direction $x'(s)$,\,$y'(s)$,\,$z'(s)$ and $\log_{10}(\rho_{\rm c})$, as calculated previously and now being refined]{\label{fig:posdirrhoOvsN} A comparison of electron position $x(s)$,\,$y(s)$,\,$z(s)$, trajectory direction $x'(s)$,\,$y'(s)$,\,$z'(s)$ and $\log_{10}(\rho_{\rm c})$, as calculated previously (lime green) and now being refined (magenta), for $\alpha=75^\circ$, along four arbitrary $B$-field lines (i.e., field line footpoints with $\phi_{\rm PC,1}=45^\circ$, $\phi_{\rm PC,2}=135^\circ$, $\phi_{\rm PC,3}=225^\circ$ and $\phi_{\rm PC,4}=315^\circ$) on the outer ring of the PC.}
\end{figure} 

In the parallel code the emission of an electron is calculated and then distributed to all frequency bins, requiring write access to all frequency bins. Since it is binning emission in $E_\gamma$, $\phi_{\rm L}$, and $\zeta$ bins, huge `sky cubes' are produced from which the light curves and spectra are constructed. We take the pair spectrum as an input (calculated by another code) and then focus on the emission properties. In principle this should allow us to parallelise the code more easily, since we can distribute the individual $B$-field lines (spatial parts of the grid, i.e., stellar surface). Basically, the emission code distributes $B$-field lines among cores. Particles are transported along $B$-field lines and radiate emission that is then binned in a ``master matrix'' (therefore the need of global memory structures). We have to investigate if this can potentially cause problems. For the serial code, the `sky cubes' are filled by running field lines sequentially. On the cluster, we of course assume that lines are independent and emission from them can be superposed. 

The parallel code calculates the same broadband spectrum expected from a pulsar as the serial code. However, with the parallel code we experienced memory issues caused by seven-dimensional arrays. We reduced the memory consumption significantly by allocating the memory dynamically, however it did not fully address our problems. The dynamic allocation did speed up the code on the other hand, because the compiler did not optimise the code with the static allocation as done initially. We also encountered errors that explicitly state ``ran out of memory" as well. We solved this by distributing a number of threads over additional nodes, to get access to more memory. 

Since we were limited on time and the modifications needed for the parallel code would be time consuming we opted to use the serial code. This is acceptable since we only attempted to isolate the first to second peak (P1/P2) effect seen in the observations assuming that particles emit CR (see Chapter~\ref{chap:CREdepLCmod} for details). However, when additional radiation mechanisms are considered such as in \citet{Harding2015} and \citet{Harding2018} the parallel code is necessary since it saves computational time, and solves more complicated problems.

\section{Recalculating the curvature radius $\rho_{\rm c}$} \label{sec:rhocalc}

We refine the previous first-order calculation of $\rho_{\rm c}$ along the electron (or positron) %\footnote{We use `electron' to collectively refer to electrons and positrons.} 
trajectory, assuming that all particles injected at the footpoint of a particular $B$-field line follow the same trajectory, independent of their energy, since these are quickly accelerated to relativistic energies by the unscreened $E$-field. This independence of $\rho_{\rm c}$ on energy also reduces computational time significantly, since the calculation is done beforehand. We furthermore assume that the $B$-field is strong enough to constrain the movement of the electrons so they will move parallel to the field line in the co-rotating frame. Thus, there will be no perpendicular motion in the co-rotating frame, since the perpendicular particle energy is nearly instantly expended via SR. We thus take into account the perpendicular $\mathbf{E}\times\mathbf{B}$ drift in the lab frame.
\begin{figure*}
\centering
\includegraphics[scale=0.7]{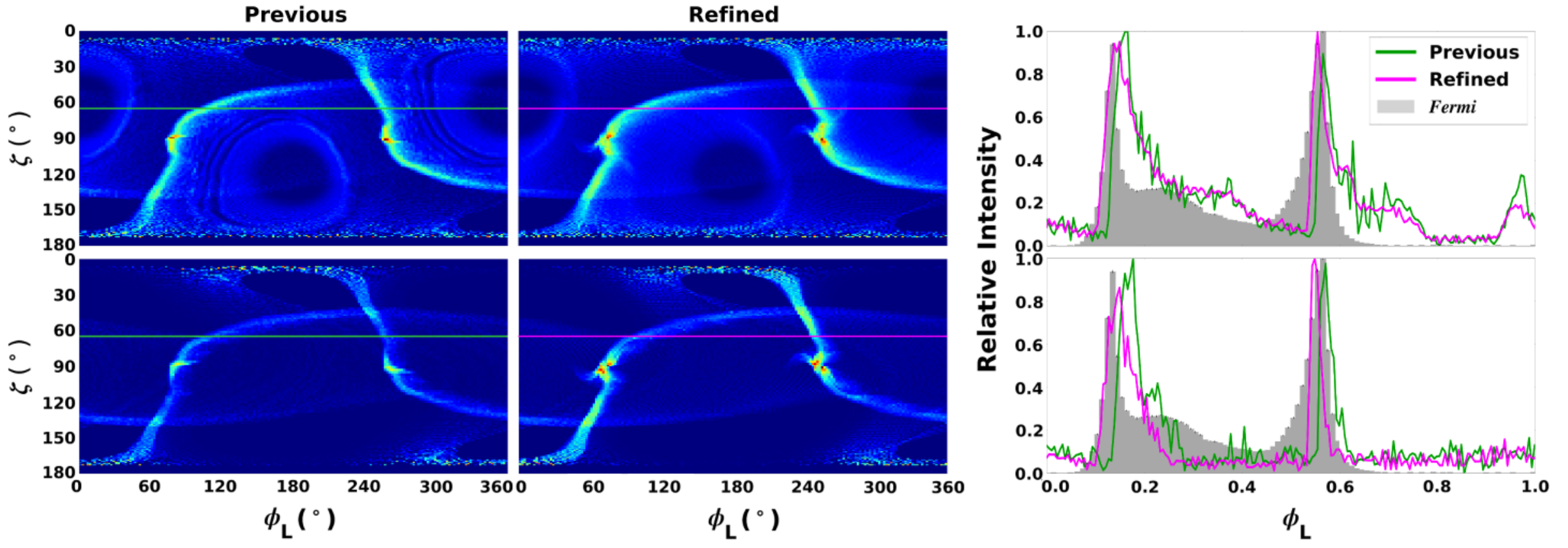}
\caption[Phase plots and pulse profiles for $\alpha=60^\circ$, $\zeta_{\rm cut}=65^\circ$, and $0.1<E_{\gamma}<50.0$~GeV]{Phase plots and pulse profiles for $\alpha=60^\circ$, $\zeta_{\rm cut}=65^\circ$, and $0.1<E_{\gamma}<50.0$~GeV. This figure serves as a comparison between phase plots for the previous (left column) and refined (centre column) $\rho_{\rm c}$ calculation, and their associated light curves (right column). The top row is for a constant $E_\parallel$ (scenario~1), and the bottom row is for a two-step $E_\parallel$ (scenario~2). We shifted the resulting $\gamma$-ray model light curves by $-0.14$ in normalised phase to fit the \emph{Fermi} LAT (\citealt{Abdo2010Vela,Abdo2013SecondCat},{\url{http://fermi.gsfc.nasa.gov/ssc/data/access/lat/2nd_PSR_catalog/}}) data points.  \label{fig:PPLCs_OvsN60}}
\end{figure*}

To calculate the electron's trajectory as well as its associated $\rho_{\rm c}$ in the lab frame, we used a small, fixed step size $ds$ (where $s$ is the arclength) along the $B$-field line. The first derivative along the trajectory (i.e., direction) is equivalent to the normalised $B$-field components as a function of $s$. Next, we smooth the directions using $s$ as the independent variable, to counteract numerical noise. 
Second, we match the unsmoothed and smoothed directions of the electron trajectory at particular $s$ values to get rid of unwanted ``tails'' at low and high altitudes, introduced by the use of a Gaussian kernel density estimator (KDE) smoothing procedure. 
Third, we use a second-order method involving interpolation by a Lagrange polynomial to obtain the second-order derivatives of the positions along the trajectory as a function of $s$ \citep{Faires2002}. This accuracy is necessary since $\rho_{\rm c}$ is a function of second-order derivatives of the electron position, and instabilities may be exacerbated if not dealt with carefully. Lastly, we match $\rho_{\rm c}$ calculated using smoothed and unsmoothed directions to get rid of ``tails'' in $\rho_{\rm c}$ at low and high altitudes, as before.

In Figure~\ref{fig:smoothing} we illustrate $\rho_{\rm c}$ (measured in cm) as a function of $s/R_{\rm LC}$, to show the impact of the smoothing parameter $h$ on the $\rho_{\rm c}$ calculation. This is shown for four arbitrary $B$-field lines with footpoints along the outer ring (rim) on the PC, as indicated by different values of the PC azimuthal coordinate $\phi_{\rm PC}$. For too small an $h$ value, e.g., 10$ds$, there is not a significant amount of smoothing and therefore the $\rho_{\rm c}$ vs.\ $s$ curve is undersmoothed, and still noisy. As we increase $h$, the $\rho_{\rm c}$ becomes smoother, however for $h=100ds$ there are some small instabilities introduced at $s/R_{\rm LC}<0.5$. This is due to the tolerance we chose when matching the unsmoothed and smoothed $\rho_{\rm c}$'s to get rid of the ``tails".

Having a pre-calculated $\rho_{\rm c}$ in hand, for a fine division in $s$ along any particular $B$-field line, I then interpolate $\rho_{\rm c}$ in my particle transport calculations to accommodate an adaptive, variable-$ds$ approach that is used to speed up the transport calculations, without losing accuracy of the trajectory. In Figure~\ref{fig:posdirrhoOvsN} the parameters describing the particle trajectory are compared for the previous and the newly calculated $\rho_{\rm c}$. These include the particle positions $x(s)$, $y(s)$, $z(s)$ in cm, dimensionless directions or spatial derivatives $x'(s)$, $y'(s)$, $z'(s)$, and $\log_{10}$ of $\rho_{\rm c}$. This comparison is shown for four arbitrary $B$-field lines with footpoints along the outer ring (rim) on the PC, as indicated by different values of $\phi_{\rm PC}$. The changes in position and direction are rather minor. However, the improved calculation smooths out some instabilities in $\rho_{\rm c}(s)$. See Appendix~\ref{chap:appRho} for a more detailed discussion and calculations. 
\begin{figure*}
\centering
\includegraphics[scale=0.7]{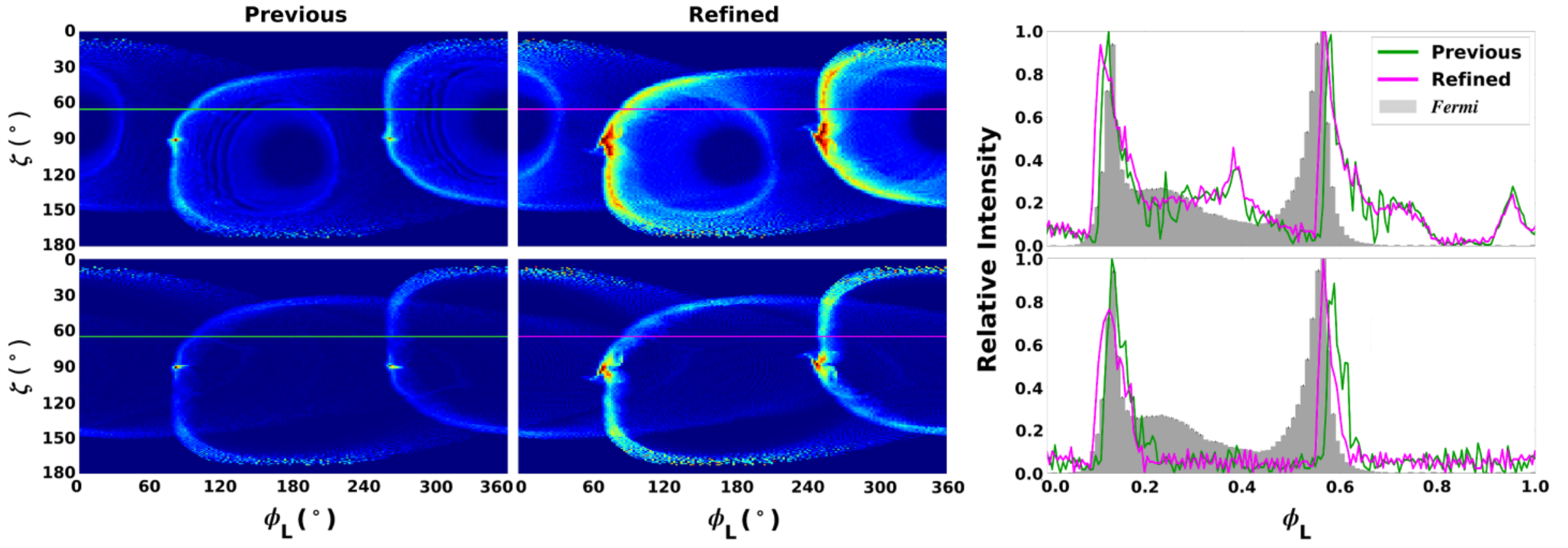}
\caption[Phase plots and pulse profiles for $\alpha=75^\circ$, $\zeta_{\rm cut}=65^\circ$, and $0.1<E_{\gamma}<50.0$~GeV]{Same as Figure~\ref{fig:PPLCs_OvsN60}, but for $\alpha=75^\circ$. \label{fig:PPLCs_OvsN75}}
\end{figure*}

We present phase plots and light curves for $\alpha=60^\circ$ and $\alpha=75^\circ$ in Figure~\ref{fig:PPLCs_OvsN60} and \ref{fig:PPLCs_OvsN75}, assuming $\zeta_{\rm cut}=65^\circ$, $r_{\rm ovc}=[0.90,0.96]$. We illustrate the effect of the previous and refined $\rho_{\rm c}$ calculation on the predicted phase plots and light curves. In Figure~\ref{fig:PPLCs_OvsN60} and \ref{fig:PPLCs_OvsN75} the phase plots and light curves associated with the previous and refined $\rho_{\rm c}$ are compared for a constant $E_\parallel$-field and a two-step constant $E_\parallel$-field (hereafter scenario~1 and scenario~2, respectively), for two different values of $\alpha$. The photon energy $E_\gamma$ extends over a wide GeV range from 100~MeV to 50~GeV. For scenario~2, I chose a constant $E_\parallel$ inside the light cylinder ($R_{\rm acc, \rm low}$; the acceleration per unit length) and a constant $E_\parallel$ outside the light cylinder and into the current sheet ($R_{\rm acc, \rm high}$). For scenario~1 (top panels) there appears inter-peak bridge emission at lower energies near the PCs (dark circles). This is not the case for scenario~2 (bottom panels), since $R_{\rm acc,\rm low}$ is too low at altitudes inside $R_{\rm LC}$, resulting in suppression of the emission as well as lowering the first peak's intensity. The caustics on the phase plots for the refined $\rho_{\rm c}$ calculation, regardless of our choice of $\alpha$, appear smoother and brighter than for the previous $\rho_{\rm c}$ calculation, although their shape is largely maintained between the two calculations. A small, additional feature becomes visible near the emission caustic (indicated by a red colour) when using the refined calculation. The caustics are also generally wider, and appear fuller (more filled out with radiation). The caustic shape furthermore depends strongly on the choice of $\alpha$. For $\alpha=60^\circ$ the caustic is more spread out in an S-curve shape, whereas for $\alpha=75^\circ$ it is rounded and concentrated around the PCs. The respective light curves for the two calculations are very similar, although they tend to be smoother for the refined calculation. The model light curves appear later in phase than the data and therefore we shifted the model with $-0.14$ in phase to fit the \emph{Fermi} data. In Figure~\ref{fig:PPLCs_OvsN60} and \ref{fig:PPLCs_OvsN75} I chose a the same resolution than what will be used in Chapter~\ref{chap:CREdepLCmod}. If I compare these to Figure~\ref{fig:resolution} I notice that our resolution choice is fairly good.

\section{Conclusion} \label{sec:ch4concl}

The full emission code is a complex code that required us to first do a calibration to understand the output. This calibration included choosing and comparing compilers used between the different systems. Another important factor was the resolution of the simulations, since higher resolution resulted in better quality light curves and spectra. Particles that are accelerated tangentially along the curved $B$-field emit CR. The curvature of the field lines is characterised by $\rho_{\rm c}$. Thus, I refined the $\rho_{\rm c}$ to improve the transport calculation. The need for this will be discussed in Chapter~\ref{chap:CREdepLCmod} and~\ref{chap:VelaTeV}. 

The serial emission code does run for solving simple problems involving single radiation mechanisms. However, for more complex problems, e.g., simulating emission from primaries and pairs for additional radiation mechanisms, the serial code needs to be parallelised. Due to the complex nature of this code and for it to work on our local cluster, future updates will be necessary. There are multiple helpful tools available that will enable us to eliminate problems, e.g., \emph{Valgrind} and \emph{blitz++} that we will investigate in future.
\\
\\
In Chapter~\ref{chap:CREdepLCmod}, we will present the results that we obtained using the serial version of the code described in Section~\ref{sec:3Dcode}. We will investigate the observed light curve trends via energy-dependent light curves and spectra in the CR regime. I will demonstrate that most of the trends seen in the energy-dependent light curves can be reproduced using our code, in a CR framework. % SSC model calibration and studying the code
\chapter[Probing the $\gamma$-ray pulsar emission mechanism]{Probing the $\gamma$-ray pulsar emission mechanism via energy-dependent light curve modelling} \label{chap:CREdepLCmod}

In this Chapter, I will be focusing on emission from the Vela pulsar in the GeV band. I use a steady-state emission model (see Section~\ref{sec:3Dcode}) to predict $E_{\gamma}$-dependent light curves and spectra that result from primary particles emitting CR; this model includes my refined calculation of $\rho_{\rm c}$ of the particle trajectory (see Section~\ref{sec:rhocalc}). In Section~\ref{subsec:revmap}, I will discuss a ``reverse mapping'' method used to isolate the spatial origin of the light curve peaks, and in Section~\ref{subsec:optparams} I performed a small parameter study to find optimal values for the model's free parameters.
In Section~\ref{sec:results}, I present sample light curves and spectra, showing the behaviour of the peaks as a function of $\rho_{\rm c}$, as applied to the Vela pulsar. For the optimal light curve and spectral fits, I study the local environment of the peaks' emission regions, finding a systematic difference in $\rho_{\rm c}$, $\gamma$, and $E_{\gamma,\rm CR}$ for the two peaks. Concluding remarks follow in Section~\ref{sec:concl}. This Chapter is a summary of an article in preparation (Barnard et al., in press) the results of which accompany those of \citet{Harding2018}. 

\section{Introduction} \label{sec:intro}
The field of pulsar science has been revolutionised by the detection of pulsed emission by ground-based telescopes. In the VHE band, MAGIC detected pulsations from the Crab pulsar at energies up to 1~TeV \citep{Ansoldi2016}, and H.E.S.S.~II detected pulsed emission from the Vela pulsar in the sub-20 GeV to 100~GeV range \citep{Abdalla2018}. New observations by H.E.S.S.\ reveal pulsed emission from Vela at a few TeV (H.E.S.S.\ Collaboration, in preparation). H.E.S.S.~II furthermore detected pulsed emission from PSR~B1706$-$44 in the sub-100~GeV energy range \citep{SpirJacob2019}. Pulsed emission from the Geminga pulsar between 15~GeV and 75~GeV at a significance of 6.3$\sigma$ was recently detected by MAGIC, although only the second light curve peak is visible at these energies. The MAGIC spectrum is an extension of the \emph{Fermi} LAT spectrum, ruling out the possibility of a sub-exponential cut-off in the same energy range at the $3.6\sigma$ level \citep{Acciari2020}. 

Interestingly, as the photon energy $E_{\gamma}$ is increased (above several GeV), the main light curve peaks of Crab, Vela and Geminga seem to remain at the same phase positions, the intensity ratio of the first to second peak (P1/P2) decreases for Vela and Geminga, the inter-peak ``bridge'' emission evolves for Vela, and the peak widths decrease for Crab \citep{Aliu2011}, Vela \citep{Abdo2010Vela} and Geminga \citep{Abdo2010Geminga}. The P1/P2 vs.\ $E_{\gamma}$ effect was also seen by \emph{Fermi} for a number of pulsars \citep{Abdo2010FirstCat,Abdo2013SecondCat}. In general, multi-wavelength pulsar light curves exhibit an intricate structure that evolves with $E_{\gamma}$ \citep[e.g.,][]{Buehler2014}, reflecting the various underlying emitting particle populations and spectral radiation components that contribute to this emission, as well as the local $B$-field geometry and $E$-field spatial distribution. 

\begin{figure}
\centering
\includegraphics[width=10cm]{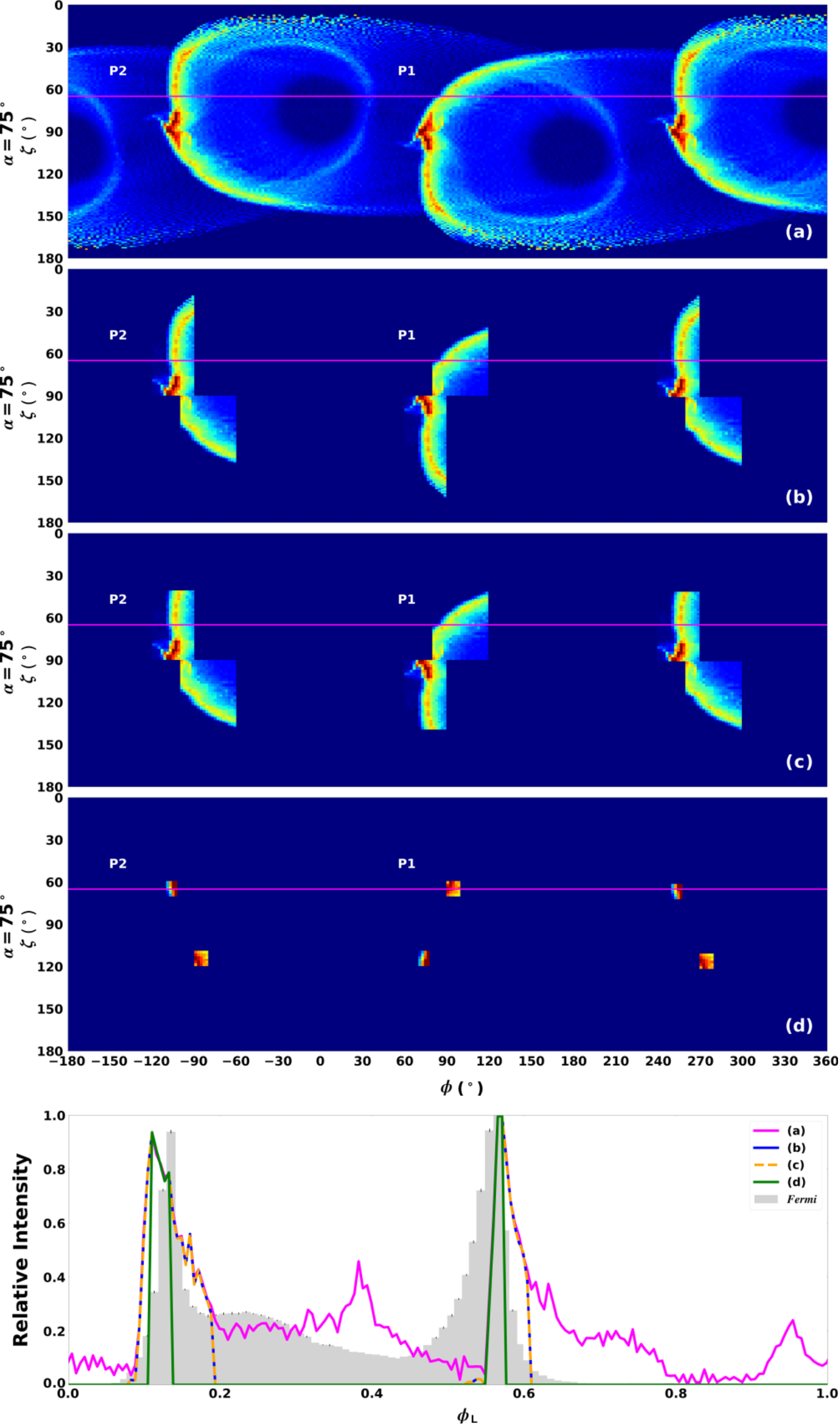}
\caption[Example phase plots with $(\phi_{\rm L},\zeta)$-``blocks''(or 2D bins) and their associated light curves for $\alpha=75^\circ$, $\zeta_{\rm cut}=65^\circ$, $R_{\rm acc}=0.25$~cm$^{-1}$, and $0.1<E_\gamma<50$~GeV]{Example phase plots with $(\phi_{\rm L},\zeta)$-``blocks''(or 2D bins) and their associated light curves for $\alpha=75^\circ$, $\zeta_{\rm cut}=65^\circ$, $R_{\rm acc}=0.25$~cm$^{-1}$, and $0.1<E_\gamma<50$~GeV. In order to indicate how we isolated the first and second light curve peaks (labelled `P1' and `P2'), we made cuts in $(\phi_{\rm L},\zeta)$ as follows: (a) no cut, (b) $\phi_{\rm L,P1}=[-100^\circ,-60^\circ]$ and $\phi_{\rm L,P2}=[-120^\circ,-90^\circ]$ for all $\zeta$,
(c) $\zeta_{\rm P1}=[90^\circ,160^\circ]$ and $\zeta_{\rm P2}=[40^\circ,90^\circ]$ for same $\phi_{\rm L}$ as in (b), and (d) $\phi_{\rm L,P1}=[-90^\circ,-81^\circ]$, $\zeta_{\rm P1}=[110^\circ,120^\circ]$, $\phi_{\rm L,P2}=[-109^\circ,-103^\circ]$, and $\zeta_{\rm P2}=[60^\circ,70^\circ]$. The light curve legend in the lower panel refers to each associated phase plot, for increasingly smaller $(\phi_{\rm L},\zeta)$ bins. The \textit{Fermi} data for Vela are indicated by a grey histogram (\citealt{Abdo2013SecondCat},\url{http://fermi.gsfc.nasa.gov/ssc/data/access/lat/2nd_PSR_catalog/}). We shifted the resulting $\gamma$-ray model light curves by $-0.14$ in normalised phase to fit the data. This reflects the degeneracy of $\phi_{\rm L}$=0 in the data (reflecting the main radio peak) and $\phi_{\rm L}$=0 (the phase of the $\boldsymbol{\mu}$-axis).} \label{fig:PPLCblocks}
\end{figure}

Some traditional physical emission models invoke CR from extended regions within the magnetosphere to explain the HE spectra and light curves. These include the SG \citep{Arons1983,Harding2003} and OG \citep{Romani1995,Cheng1986} models. However, they fall short of fully addressing global magnetospheric characteristics, e.g., the particle acceleration and pair production, current closure, and radiation of a complex multi-wavelength spectrum. 

Geometric light curve modelling \citep{Dyks2004b,Venter2009,Watters2009,Johnson2014,Pierbattista2015} presented an important interim avenue for probing the pulsar magnetosphere in the context of traditional pulsar models, focusing on the spatial rather than physical origin of HE photons. 
More recent developments include global magnetospheric models such as the FIDO model \citep{Brambilla2015,Kalapotharakos2009,Kalapotharakos2014}, equatorial current sheet models (e.g., \citealt{Bai2010b,Petri2012}), the striped-wind models (e.g., \citealt{Petri2011}), and PIC \citep{Brambilla2018,Cerutti2016current,Cerutti2016,Cerutti2020,Kalapotharakos2018,Philippov2018}. Some studies using the FIDO models assume that particles are accelerated by induced $E$-fields in dissipative magnetospheres and produces GeV emission via CR (e.g., \citealt{Kalapotharakos2014}). Conversely, in some of the wind or current-sheet models, HE emission originates beyond the light cylinder via synchrotron radiation (SR) by relativistic, hot particles that have been accelerated via magnetic reconnection inside the current sheet \citep[e.g.,][]{Petri2011,Philippov2018}. 
Given the ongoing debate between the emission mechanisms of HE emission, our motivation in this study is to explain the GeV spectrum and light curves of Vela as measured by \emph{Fermi} and H.E.S.S.\ Specifically, by modelling the $E_\gamma$-dependent light curves (and P1/P2 signature) in the CR regime of SC radiation, we hope to probe whether this effect can serve as a potential discriminator between emission mechanisms and models (see also the reviews of~\citealt{Harding2016,Venter2016,Venter2017} on using pulsar light curves to scrutinise magnetospheric structure and emission distribution).

\section{Isolating the origin of emission for each of the light curve peaks} %\label{sec:model}
\label{subsec:revmap}

Using the model described in Section~\ref{sec:3Dcode}, and for a given magnetic inclination angle $\alpha=75^\circ$, we generated phase plots (observer angle $\zeta$ vs.\ rotation phase $\phi_{\rm L}$; Figure~\ref{fig:PPLCblocks}). We inject the primaries into a roughly annular slot gap situated between $r_{\rm ovc}=0.90$ and $r_{\rm ovc}=0.96$ \citep{Dyks2004b,Harding2018}, and divide the surface projection of the slot gap situated near the rim of the PC into 7 rings, with each ring having $360$ azimuthal segments. We additionally set $ds=10^{-3}R_{\rm LC}$ with a corresponding $h=50ds$. The phase plots are emitted photon fluxes $\dot{N}_{\gamma}$ that have been normalised using the primary particle flux (the appropriate Goldreich-Julian injection rate at the stellar surface); $\dot{N}_{\gamma}$ is collected in bins of $\zeta$ and $\phi_{\rm L}$. The photon directions have been corrected for the Special Relativistic effects of rotation and time-of-flight delays. Lastly, $\dot{N}_{\gamma}$ per bin is divided by the solid angle subtended by each phase plot bin, i.e., $\delta\Omega=(\cos\zeta-\cos(\zeta+\delta\zeta))\delta\phi_{\rm L}\approx\sin\zeta{\delta\zeta}{\delta\phi_{\rm L}}$. To generate light curves, a constant-$\zeta$ cut ($\zeta_{\rm cut}$) is made through the respective phase plot (see lower panel of Figure~\ref{fig:PPLCblocks}).

As mentioned in Section~\ref{sec:intro}, the relative fading of peak~1 vs.\ peak~2 with $E_\gamma$ seems to be a common characteristic of HE light curves. We have also been able to reproduce this with the code. In order to probe the origin of this effect, it is necessary to isolate the spatial origin of each light curve peak. We start by isolating each peak on the phase plot (using increasingly smaller $(\zeta,\phi_{\rm L})$ bins) and then apply ``reverse mapping'' to uncover the emission's spatial position. This can be compared to developing a ``reverse dictionary'' that translate a chosen $(\phi_{\rm L},\zeta)$ range into a spatial range within the magnetosphere. In our code, we calculate only the emission from the northern rotational hemisphere. The contribution of the emission from the southern hemisphere is obtained taking into account the symmetry with respect to the centre of the star (i.e., $\dot{S}_{\gamma}/d\Omega(\phi_{\rm L},\zeta)=\dot{N}_{\gamma}/d\Omega(\phi_{\rm L}+180^{\circ},180^{\circ}-\zeta))$, where $S_\gamma$ and $N_\gamma$ indicate the contribution of the southern and northern rotational hemispheres, respectively. This symmetry exploitation saved computational time, and the corresponding full sky map is shown in the first panel of Figure~\ref{fig:PPLCblocks}. The implication is that one has to carefully keep track of the ($\phi_{\rm L},\zeta$) coordinates of each peak, and map them back onto the northern-hemisphere caustic (e.g., mapping P1 onto the northern-hemisphere caustic where $\zeta>90^\circ$). Using the latter, one can perform the reverse mapping to find the spatial coordinates of this emission.

\begin{figure}
\centering
\includegraphics[width=12cm]{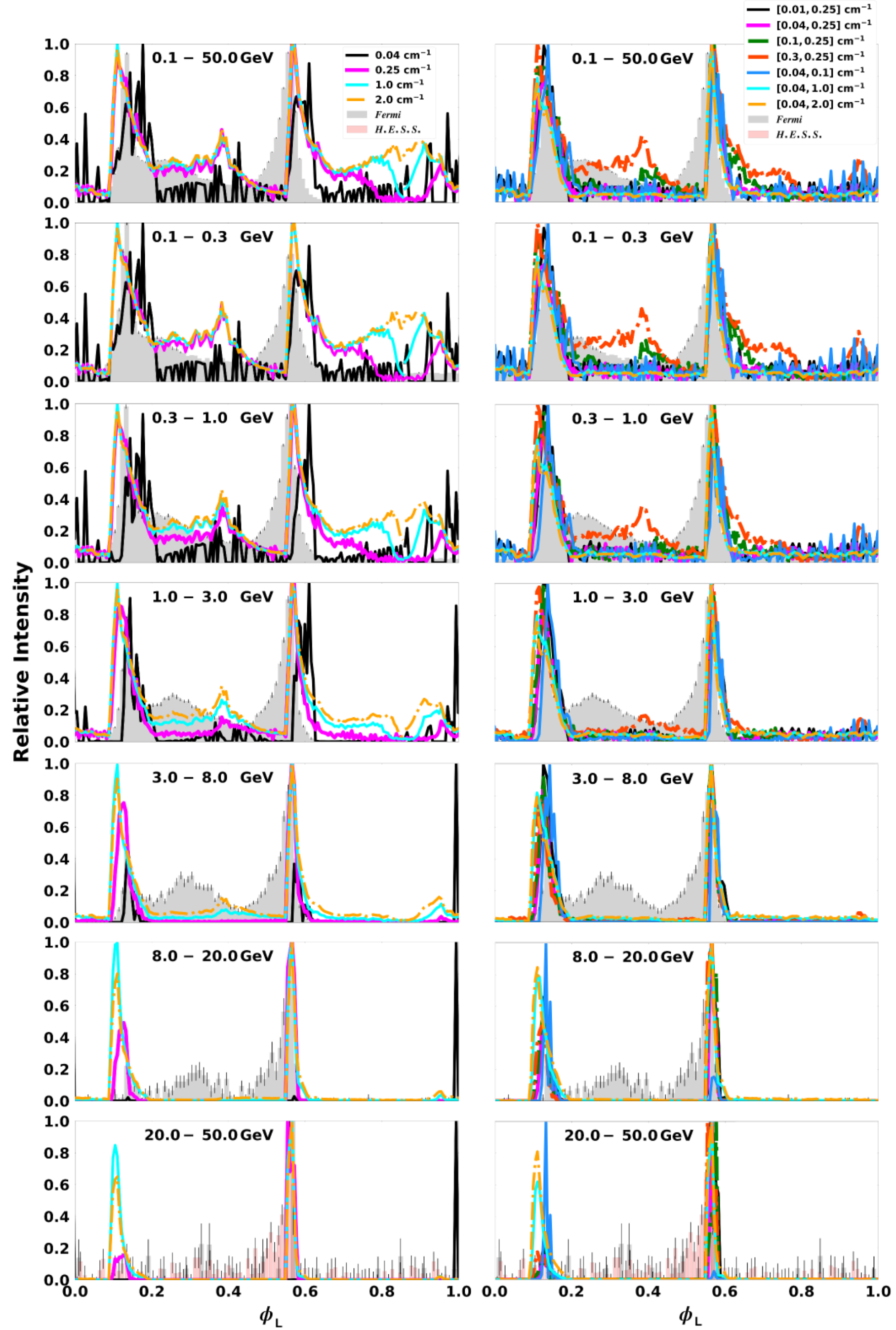}
\caption[Energy-dependent light curves for $\alpha=75^\circ$ and $\zeta_{\rm cut}=65^\circ$ for several different combinations of $R_{\rm acc}$ for both scenarios]{Energy-dependent light curves for $\alpha=75^\circ$ and $\zeta_{\rm cut}=65^\circ$ for several different combinations of $R_{\rm acc}$ for both the constant $E_\parallel$ (left column) and two-valued $E_\parallel$ (right column) case. The top panels are for the full $E_\gamma$-range, and for each panel thereafter, the minimum $E_\gamma$ is increased as indicated. We are fitting the model light curves to the \emph{Fermi} (\citealt{Abdo2010Vela,Abdo2013SecondCat},\url{http://fermi.gsfc.nasa.gov/ssc/data/access/lat/2nd_PSR_catalog/}), and H.E.S.S.\ (at $E_\gamma>20$~GeV; \citealt{Abdalla2018}) data points. We shifted the predicted light curves by $\delta=0.14$ in normalised phase. One observes that for some choices of $E_\parallel$, the P1/P2 decrease with $E_\gamma$ is more apparent than for others. \label{fig:LC_epardep}}
\end{figure}

This reverse mapping procedure is illustrated in Figure~\ref{fig:PPLCblocks} for a constant acceleration ``rate'' (acceleration per unit length) $R_{\rm acc}=eE_{\parallel}/{m_{\rm e}}c^2$~cm$^{-1}$, with $e$ the electron charge, $m_{\rm e}$ the electron mass, and $m_{\rm e}c^2$ the rest-mass energy. The first panel is for the full phase space, whereas panels (b), (c), and (d) are for different $(\phi_{\rm L},\zeta)$-``blocks'' or bins. In panel (b), we make a cut in $\phi_{\rm L}$ for both peaks but keep $\zeta$ fixed and see that only the peaks remain on the corresponding light curve (see bottom panel). If we then narrow the range in $\zeta$ for a fixed $\phi_{\rm L}$ interval (same as in (b)), we note that the light curve remains the same. Lastly, we make $\phi_{\rm L}$ and $\zeta$ small enough so that only the maximum of each peak is included in the $(\phi_{\rm L},\zeta)$ range, as seen in panel (d). These ranges in $\phi_{\rm L}$ and $\zeta$ are referred to as the ``optimal bins'' for both peaks and are necessary for constructing the phase-resolved spectra of each peak. We chose the $\zeta$-range for each peak with width of $\pm{5^\circ}$ around $\zeta_{\rm cut}=65^\circ$, to include the $\zeta$ inferred from the the pulsar wind nebula torus fit of Vela \citep{Ng2008}.

\begin{figure}
\centering
\includegraphics[width=12cm]{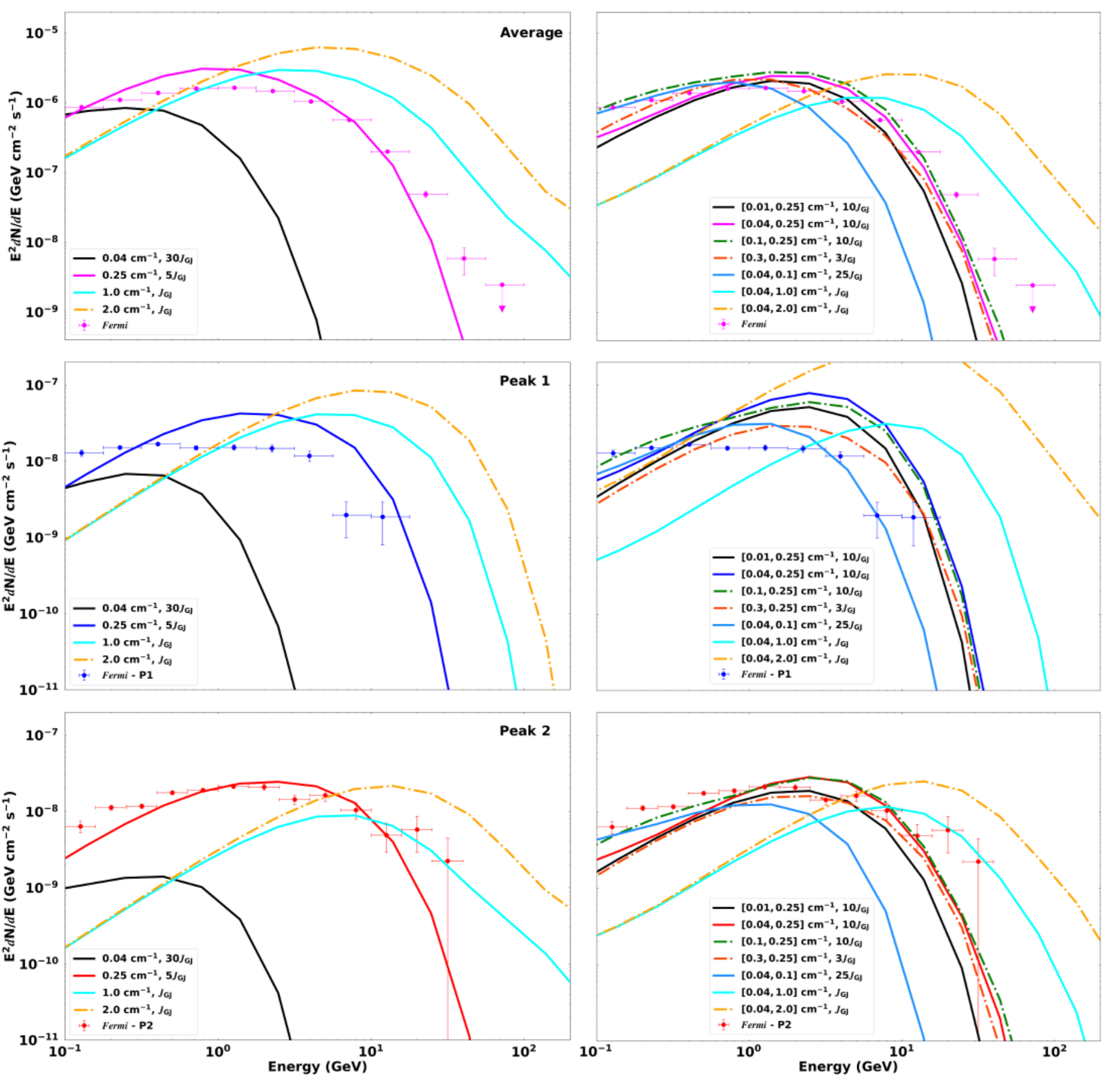}
\caption[Model phase-averaged and phase-resolved spectra associated with Figure~\ref{fig:LC_epardep}]{Model phase-averaged and phase-resolved spectra associated with Figure~\ref{fig:LC_epardep} for the same for $\alpha$, $\zeta_{\rm cut}$ and $R_{\rm acc}$-field combinations, for both scenario 1 (left column) and scenario 2 (right column). In each $E_\parallel$ case the legend indicate the chosen values for $R_{\rm acc}$, $R_{\rm acc,\rm low}$, $R_{\rm acc,\rm high}$, and the flux normalisation factor.The data points for the phase-average spectra are from \citet{Abdo2013SecondCat} (see \url{http://fermi.gsfc.nasa.gov/ssc/data/access/lat/2nd_PSR_catalog/}), and the phase-resolved spectra are updated data to those published in \citet{Abdo2010Vela} \label{fig:spec_epardep}}
\end{figure} 

\section{Results} \label{sec:results}

\subsection{Finding optimal fitting parameters} \label{subsec:optparams}
After having isolated the spatial origin of the emission of each light curve peak as described in the previous section, we first perform joint light curve and spectral fitting to find optimal model parameters; subsequently, we will consider the local environments where the respective light curve peaks originate (Section~\ref{sec:results}), given these optimal parameters. We consider two cases throughout this paper, based on either a constant or a two-step parametric accelerating $E_\parallel$-field, independent of the $\phi_{\rm PC}$, $\zeta$ and $r$. Thus, we choose (and subsequently refer to this as scenario~1 and scenario~2): (1) a constant $R_{\rm acc}$ from the stellar surface and into the current sheet (see \citealt{Harding2015}), and (2) a two-valued $R_{\rm acc}$, where $R_{\rm acc,\rm low}$ occurs inside, and $R_{\rm acc,\rm high}$ outside the light cylinder (see \citealt{Harding2018}). The two-step function for the accelerating $E_\parallel$ is motivated by global dissipative models \citep{Kalapotharakos2014,Kalapotharakos2017,Brambilla2015} and kinetic PIC models \citep{Cerutti2016,Kalapotharakos2018}, that indicate that the particle acceleration takes place primarily near the current sheet, outside the light cylinder.

I performed a preliminary parameter study to search for an optimal combination of $\alpha$, $\zeta_{\rm cut}$, and $R_{\rm acc}$ (for both scenarios, respectively), calibrated against \textit{both} the observed HE light curves \textit{and} spectra measured by \emph{Fermi} and H.E.S.S.~II. We start by fixing $\zeta_{\rm cut}$ and testing different values of $R_{\rm acc}$; later, we fix $R_{\rm acc}$ and free $\zeta_{\rm cut}$\footnote{Given how computationally expensive this exercise is, we only considered a few values of the free parameters. In future, a more robust method may be considered where parameter space of several free parameters may be searched for optimal joint light curve and spectral fits. Given the disparate nature of these data, and the complexity of such a joint fit, here we perform a pilot study to indicate the effect of the different parameters, and to find a reasonable joint fit by eye.}.

\begin{figure}
\centering
\includegraphics[width=12cm]{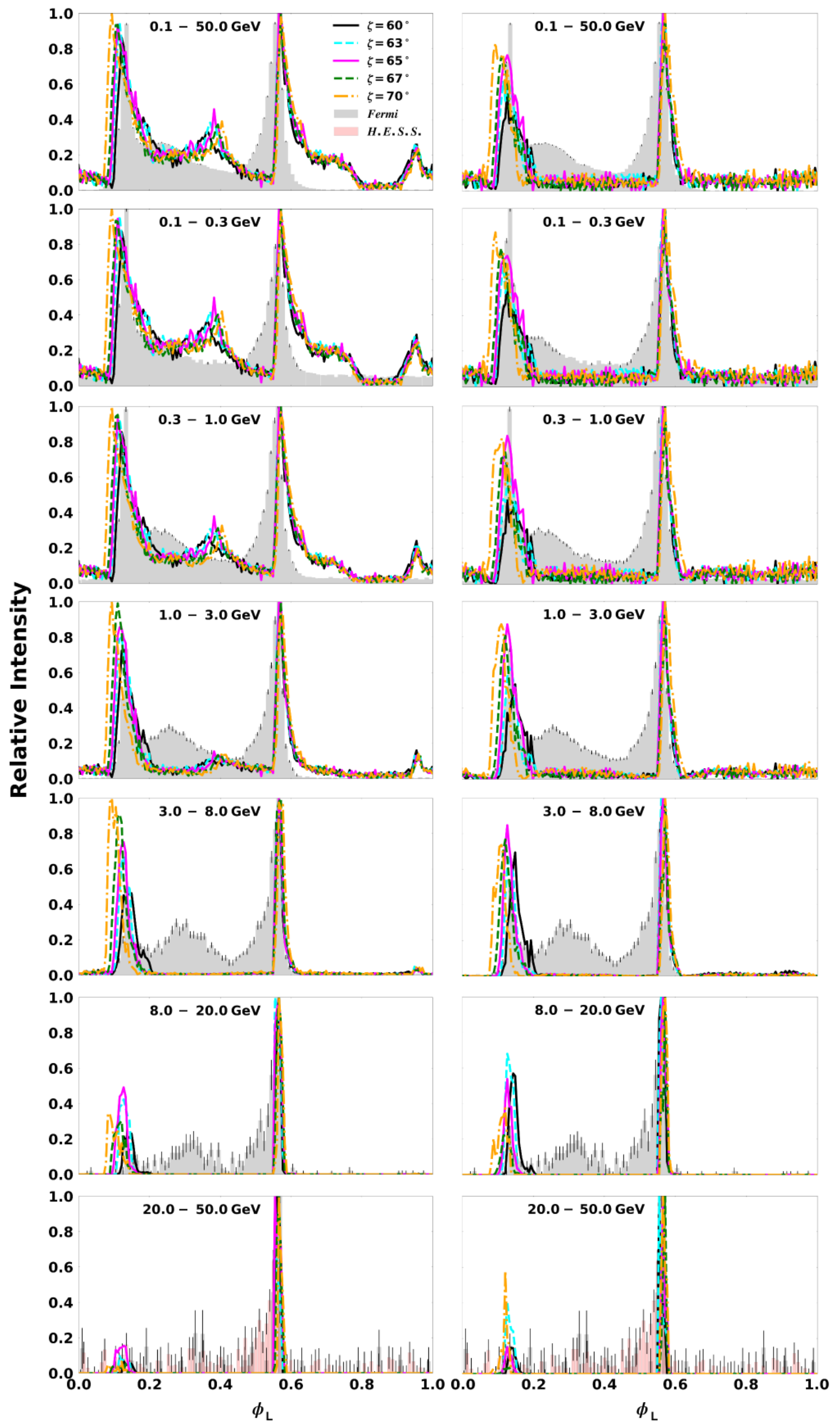}
\caption[Energy-dependent light curves for $\alpha=75^\circ$ and different $\zeta_{\rm cut}$ for the optimal values of $E_{\parallel}$-field for both scenarios]{Energy-dependent light curves for $\alpha=75^\circ$ and different $\zeta_{\rm cut}$ for the optimal values of $E_{\parallel}$-field for both scenario~1 (left) and scenario~2 (right). In each $R_{\rm acc}$ case, the legend indicates the chosen values for $\zeta_{\rm cut}$. The first row are for the full $E_\gamma$-range, and each panel thereafter is for an increase in the minimum $E_\gamma$. We are fitting the model light curves to the \emph{Fermi} (\citealt{Abdo2010Vela,Abdo2013SecondCat},\url{http://fermi.gsfc.nasa.gov/ssc/data/access/lat/2nd_PSR_catalog/}) and to the H.E.S.S.\ (at $E_\gamma>20$~GeV; \citealt{Abdalla2018}) data points, with $\delta=-0.14$. \label{fig:LC_zdep}}
\end{figure}

Figure~\ref{fig:LC_epardep} shows the $E_\gamma$-dependent light curves for scenario~1 (left column) and scenario~2 (right column). For scenario~1, we choose four arbitrary constant $R_{\rm acc}$ values, and for scenario~2, seven arbitrary $R_{\rm acc, \rm low}$ and $R_{\rm acc, \rm high}$ combinations, as indicated in the legends. We also indicate different energy ranges (with the minimum $E_\gamma$ increasing from top to bottom), with the first panel showing light curves for a full HE range $E_\gamma\in(100~{\rm MeV},50~{\rm GeV})$. The $E_\gamma$ ranges correspond to those of the \emph{Fermi} light curves in Figure~2 in \citet{Abdo2010Vela}, and \citet{Abdo2013SecondCat}, as well as $E_\gamma>$20~GeV to match the H.E.S.S.~II data \citep{Abdalla2018}. 
%In the figure legend, we indicate the combinations of $R_{\rm acc}$ for scenario~1 and $[R_{\rm acc, \rm low},R_{\rm acc, \rm high}]$ for scenario~2. 
The light curves for scenario~1 display bridge emission at $\phi_{\rm L}\geq0.25$ that diminishes as $E_\gamma$ increases. For scenario~2, bridge emission develops when $R_{\rm acc, \rm low}\geq0.10$~cm$^{-1}$ and $R_{\rm acc, \rm high}=0.25$~cm$^{-1}$. For $R_{\rm acc, \rm low}=0.3$~cm$^{-1}$ and $R_{\rm acc, \rm high}=0.25$~cm$^{-1}$ the light curve almost mimics our fit in scenario~1 for $R_{\rm acc}=0.25$~cm$^{-1}$ (since these respective values are so close). If both $R_{\rm acc, \rm low}$ and $R_{\rm acc, \rm high}$ are small, we obtain light curve shapes that are contrary to what is expected (probably because the particles do not reach radiation reaction limit), e.g., choosing $R_{\rm acc, \rm low}=0.04$~cm$^{-1}$ and $R_{\rm acc, \rm high}=0.1$~cm$^{-1}$, yields an increase of P1/P2 for at $E_\gamma>8.0$~GeV, contrary to what is observed. For combinations where $R_{\rm acc, \rm low}$ is small and $R_{\rm acc, \rm high}\geq1.0$~cm$^{-1}$, we see that P1 remains relatively high for $E_{\gamma}>8.0$~GeV, instead of following the observed trend. The optimal choice in terms of reproducing the P1/P2 effect seems to be $(R_{\rm acc, \rm low}$,$R_{\rm acc, \rm high}) = (0.04,0.25)$~cm$^{-1}$, although the bridge emission is somewhat underpredicted.

\begin{figure}
\centering
\includegraphics[width=12cm]{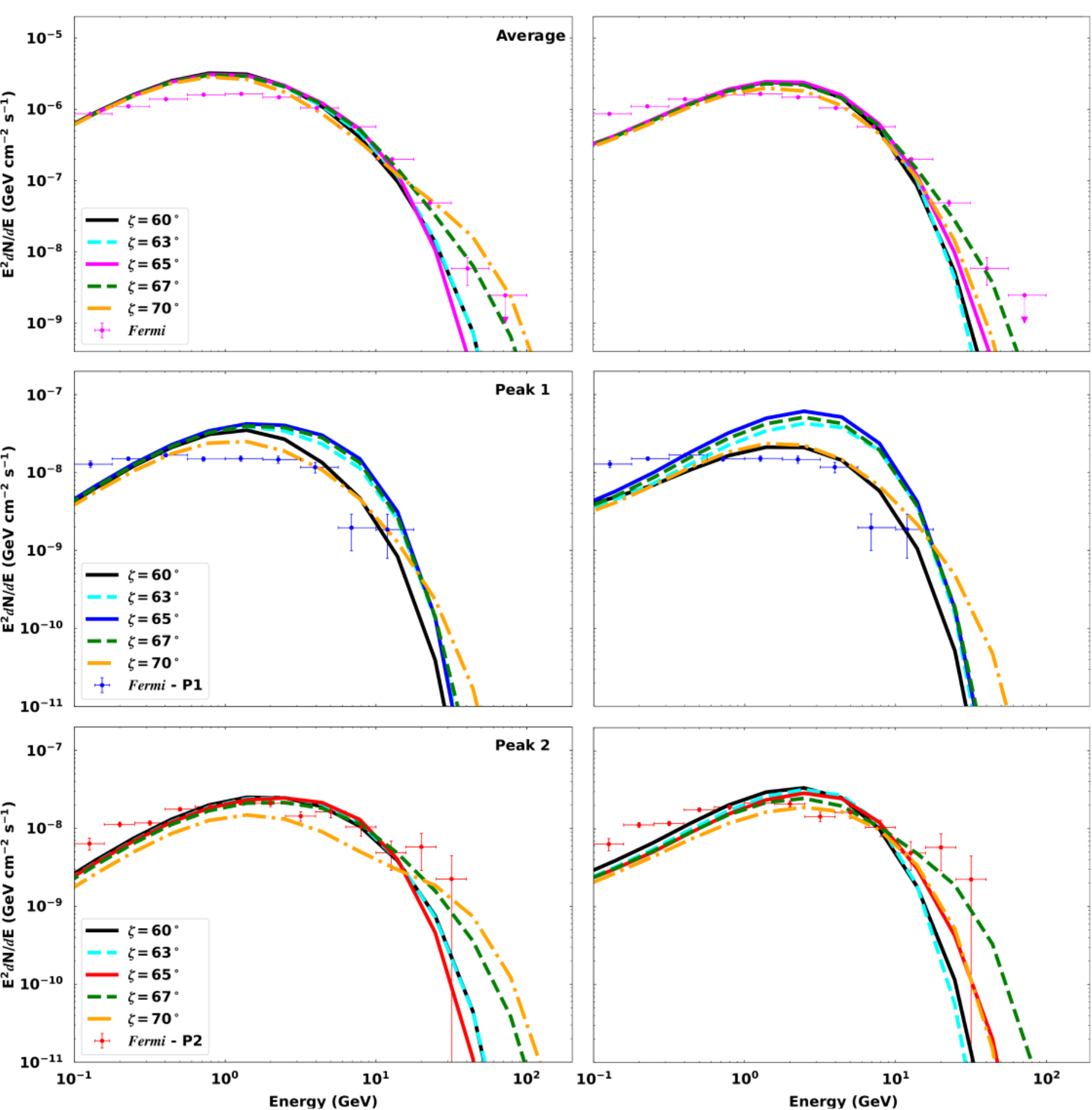}
\caption[Model phase-averaged and phase-resolved spectra associated with Figure~\ref{fig:LC_zdep}]{Model phase-averaged and phase-resolved spectra associated with Figure~\ref{fig:LC_zdep} for the same $\alpha$, $\zeta_{\rm cut}$ and optimal $R_{\rm acc}$ choices, for both scenario~1 (left) and scenario~2 (right). In each case, the legend indicates the chosen values for $\zeta_{\rm cut}$. The flux normalisation factor is $5J_{\rm GJ}$ for the first case and $10J_{\rm GJ}$ for the second. The data points for the phase-average spectra are from \citet{Abdo2013SecondCat} (see \url{http://fermi.gsfc.nasa.gov/ssc/data/access/lat/2nd_PSR_catalog/}), and the phase-resolved spectra are updated data to those published in \citet{Abdo2010Vela}. \label{fig:spec_zdep}}
\end{figure}

In both scenarios, four main trends are evident in our optimal fits to the light curves as they evolve with $E_\gamma$. First, the model peaks remain at the same phase, i.e., P1 at $\phi_{\rm L}$=[0.10,0.18] and P2 at $\phi_{\rm L}$=[0.57, 0.60], after we shifted the model in phase to fit the data. Second, the intensity ratio of P1 relative to P2 decreases as $E_{\gamma}$ increases in some cases, where the peaks are nearly equal in height at lower $E_{\gamma}$. Third, the bridge emission fades at higher energies, possibly reflecting its softer spectrum and its origin at lower altitudes, where acceleration is suppressed as compared to the current sheet environment. Lastly, the pulse width decreases with an increase in $E_\gamma$. It is encouraging that the model can broadly reproduce these observational trends. We also note that a two-step $E_\parallel$-field provides more reasonable light curve shapes, especially at lower photon energies.

The observed phase-averaged CR spectra are characterised by a power law with an (sub)exponential cutoff. In our model, this spectrum is calculated as the observed $\dot{N}_{\gamma}$ at a particular viewing angle $\zeta_{\rm cut}$, summing the fluxes (originating in different parts of the magnetosphere) over $\phi_{\rm L}$ and dividing by $2\pi{d}^2$, where $d$ (in cm) is the distance to the source. To calculate the phase-resolved spectra associated with each peak, we limit the $\phi_{\rm L}$-range to include the specific fraction of the emission we want to study. We scaled the phase-resolved flux with the ratio of the difference in each peak's $\phi_{\rm L}$-range of the \emph{Fermi} data to the model range. Figure~\ref{fig:spec_epardep} shows the phase-averaged and phase-resolved (for both P1 and P2) spectra per row. These are associated with the light curves in Figure~\ref{fig:LC_epardep}, for both scenarios and the same parameter values as in Figure~\ref{fig:LC_epardep}. The phase-resolved spectra are taken from \citet{Abdo2010Vela}, but we have removed an incorrect exposure correction that led to spectral points in the peaks being higher than the phase-averaged spectrum.
%For scenario~1 and~2 we choose the same $R_{\rm acc}$ combinations, and via the reverse mapping procedure (see Section~\ref{subsec:revmap}) we search for the correct $\phi$ range for P1 and P2 to construct the spectra (for each $R_{\rm acc}$ combination). 
Since the predicted CR $\dot{N}_\gamma$ are lower than the \emph{Fermi} data points (\citealt{Abdo2010Vela,Abdo2013SecondCat}, \url{http://fermi.gsfc.nasa.gov/ssc/data/access/lat/2nd_PSR_catalog/}), we scaled the model with a flux normalisation factor in terms of $J_{\rm GJ}$. The flux normalisation factor is a multiple of the Goldreich-Julian current density $J_{\rm GJ}=\rho_{\rm GJ}c$ (with $\rho_{\rm GJ}=-\mathbf{\Omega}\cdot\mathbf{B}/2\pi c$ the corresponding charge density; \citealt{GJ1969}). This spectrum normalisation has some freedom, since the actual multiplicity of HE particles in the pulsar magnetosphere is not absolutely certain. In the figure legend, we indicate $[R_{\rm acc},J_{\rm GJ}]$ for scenario~1 and $[R_{\rm acc,\rm low},R_{\rm acc,\rm high},J_{\rm GJ}]$ for scenario~2.

For scenario~1, at small $R_{\rm acc}$ the model does not fit the data, and the flux is too low, even with a large flux normalisation factor. This may be addressed in future by invoking SC emission, rather than pure CR \citep{Harding2018}. As $R_{\rm acc}$ increases, the model better fits the data; however, when it becomes too large, it shifts the spectra to larger $E_\gamma$'s and the spectral shape changes and deviates from the data points. This reflects the fact that a larger accelerating $E$-field is implied, leading to a larger particle energy and spectral cutoff. Also, for larger $R_{\rm acc}$ the flux normalisation factor becomes smaller. This flux factor should in principle be constant for the phase-averaged and phase-resolved spectra, but the flux level is not consistent between the different predicted spectra, e.g., at $R_{\rm acc}=0.25$ P1's model spectra overestimates the data, but not for P2 or the phase-averaged spectra. This may point to the need for a spatially-dependent normalisation of the current in future.

For scenario~2, most combinations of $R_{\rm acc,\rm low}$ and $R_{\rm acc,\rm high}$ yield a good fit to the data, except when both $R_{\rm acc,\rm low}$ and $R_{\rm acc,\rm high}$ are small, e.g., $R_{\rm acc}=[0.04,0.1]$~cm$^{-1}$, or $R_{\rm acc,\rm high}$ is high, e.g., $R_{\rm acc}=[0.04,2.0]$~cm$^{-1}$. When $R_{\rm acc,\rm low}$ is small and $R_{\rm acc,\rm high}$ is very high, the spectra extend to unreasonably high $E_{\gamma}$. For $R_{\rm acc,\rm low}=0.3$~cm$^{-1}$ and $R_{\rm acc,\rm high}=0.25$~cm$^{-1}$ the spectral fits almost mimic our fits in scenario~1 for $R_{\rm acc}=0.25$~cm$^{-1}$, although the flux normalisation is a bit lower for scenario~2. In both scenarios, $E_{\gamma,\rm CR}$ varies significantly as we change the parameters, so that for certain choices of $R_{\rm acc}$, P1 may have a larger cutoff than P2, contrary to what is observed. Thus, we settle on $R_{\rm acc}=0.25$~cm$^{-1}$ for scenario~1, and $R_{\rm acc,\rm low}=0.04$~cm$^{-1}$ and $R_{\rm acc,\rm high}=0.25$~cm$^{-1}$ for scenario~2 as optimal values for this paper.

Next, I consider the impact of different values of $\zeta_{\rm cut}$ on the predicted light curves and spectra, for the optimal values of $R_{\rm acc}$.
%For this optimal fit of $E_\parallel$ and flux factor, for both scenarios, we show light curves and spectra in
I study the energy-dependent light curves for $\alpha=75^\circ$ and $\zeta_{\rm cut}=[60^\circ,63^\circ,65^\circ,67^\circ,70^\circ]$ (Figure~\ref{fig:LC_zdep}). One notices that P1/P2 decreases with energy at different rates. For larger $\zeta_{\rm cut}$ (i.e., 67$^\circ$ and 70$^\circ$), P1 is relatively higher at lower $E_\gamma$. In scenario~2, the same happens at larger $\zeta_{\rm cut}$ but only at $E_\gamma\geq{20}$~GeV. Also, the level of bridge emission depends on the choice of $\zeta_{\rm cut}$.
Figure~\ref{fig:spec_zdep} indicates spectra for the same optimal $R_{\rm acc}$ parameters, but for different $\zeta_{\rm cut}$ values. For smaller $\zeta_{\rm cut}$, the model spectra fit the data well, but for larger $\zeta_{\rm cut}$, the model spectral cutoffs extend to higher $E_\gamma$, sometimes overshooting the data. Also, these spectra are lower in flux than those for the smaller $\zeta_{\rm cut}$ fits (we fixed the flux normalisation for all values of $\zeta_{\rm cut}$). 
In scenario~2, $E_{\gamma,\rm CR}$ varies significantly, so that for certain choices of $\zeta_{\rm cut}$, P1 has a larger cutoff than P2.
If we analyse Figure~\ref{fig:LC_zdep} and~\ref{fig:spec_zdep} concurrently, our optimal fit for both scenarios is for $\zeta_{\rm cut}=65^\circ$. 

\subsection{Optimal model fit} \label{subsec:optfit}

We perform simulations for the Vela pulsar for the following parameters\footnote{\citet{Manchester2005}.}: spin period $P=0.089$~ms, its time-derivative $\dot{P}=1.25\times10^{-13}$~s$^{-1}$, and $d=0.29$~kpc. We construct all subsequent figures, e.g., phase plots, light curves, and spectra for optimal values of $\alpha=75^\circ$, and $\zeta_{\rm cut}=65^\circ$ (we indicate spectra for $\alpha=60^\circ$ for comparison). Additional optimal values are $R_{\rm acc}=0.25$~cm$^{-1}$ and a flux normalisation factor of $5J_{\rm GJ}$ for scenario~1, and $R_{\rm acc,\rm low}=0.04$~cm$^{-1}$, $R_{\rm acc,\rm high}=0.25$~cm$^{-1}$, and $10J_{\rm GJ}$ for scenario~2. These values produce good fits to the \emph{Fermi} and H.E.S.S.\ II data. 

\begin{figure}
\centering
\includegraphics[width=16cm]{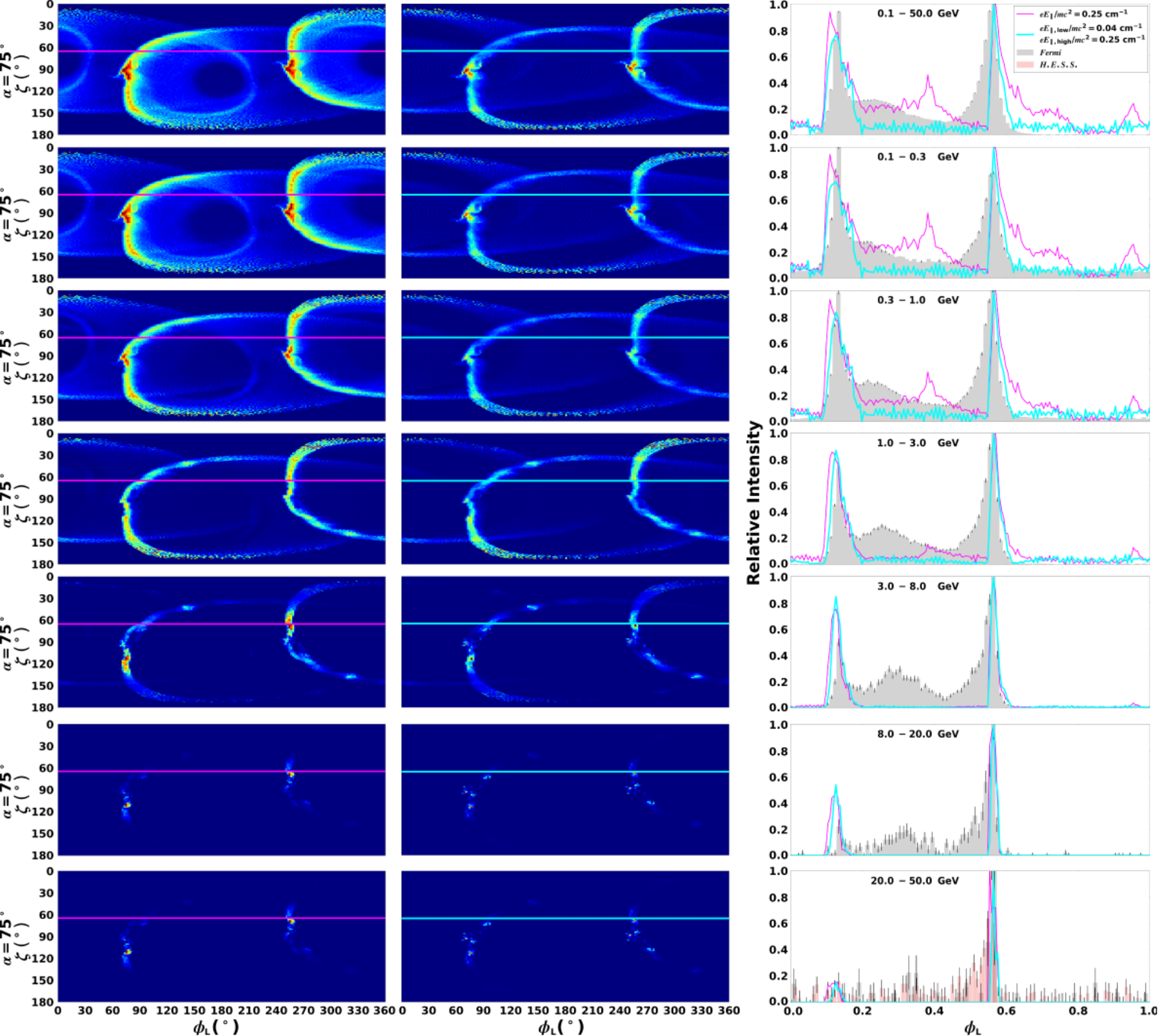}
\caption[Energy-dependent phase plots and light curves for $\alpha=75^\circ$ and $\zeta_{\rm cut}=65^\circ$ and for the optimal $R_{\rm acc}$ for both scenarios]{Energy-dependent phase plots and light curves for $\alpha=75^\circ$ and $\zeta_{\rm cut}=65^\circ$ and for the optimal $R_{\rm acc}$ for both the first (left column) and second (centre column) scenarios, plus their associated light curves (right column). The top panels are for the full $E_\gamma$-range, and each panel thereafter is for a different sub-band, as indicated by the labels in the light curve panels. Peaks were shifted by $-0.14$ to fit the \textit{Fermi} LAT and H.E.S.S.\ data. \label{fig:PPLCopt}}
\end{figure}

In Figure~\ref{fig:PPLCopt} we show the energy-dependent phase plots and accompanying light curves for our optimal fit, for both scenarios. For scenario~1 (left phase plot) the bridge and most of the off-peak emission disappears with increasing $E_\gamma$, although the light curve peak positions for both scenarios remain roughly stable. The other light curve trends mentioned in Section~\ref{subsec:optparams} are also visible here, i.e., the decrease of P1/P2 and a decrease in peak width with $E_\gamma$.

\begin{figure}
\centering
\includegraphics[width=16cm]{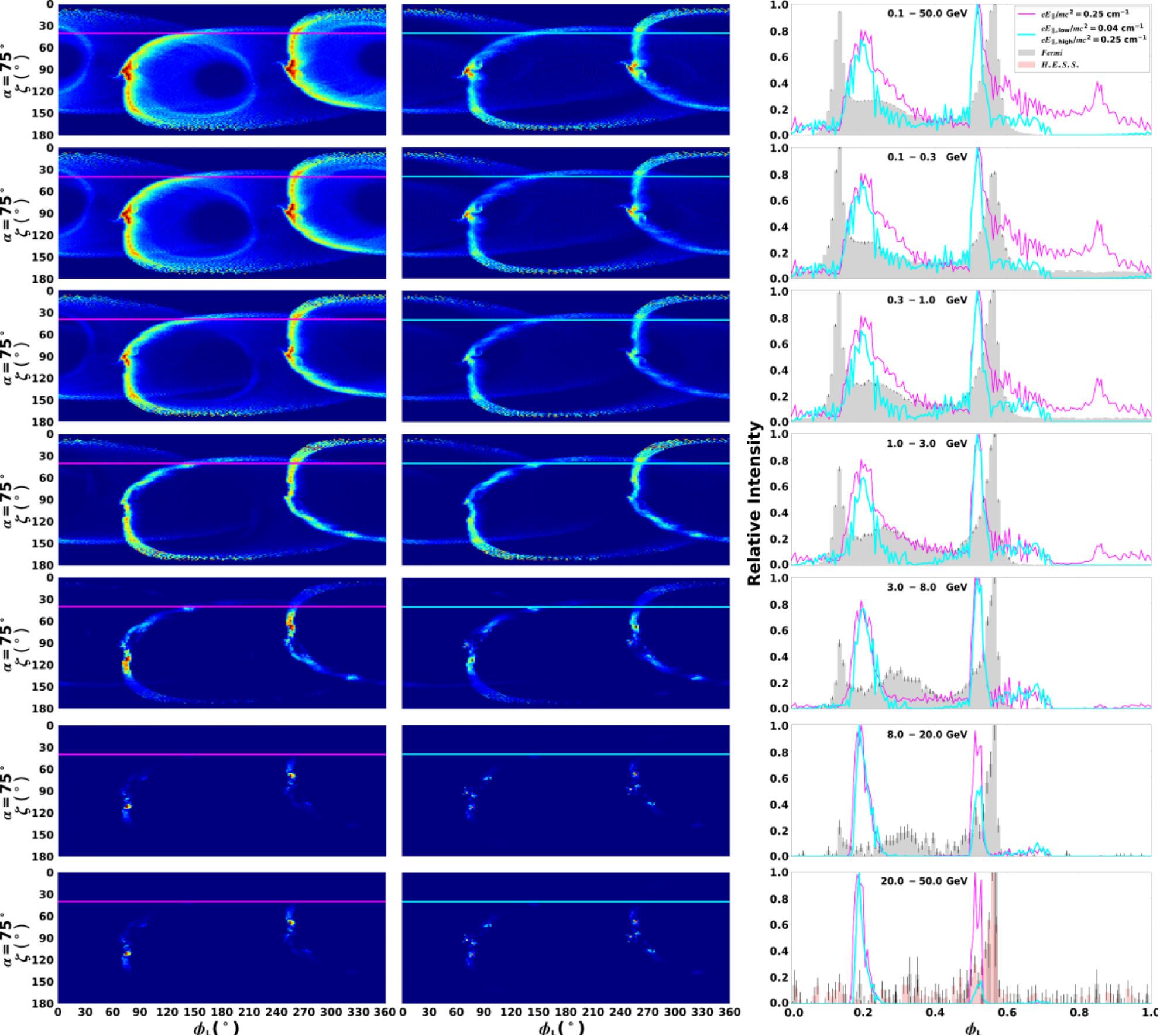}
\caption{The same as Figure~\ref{fig:PPLCopt} but for $\zeta_{\rm cut}=40^\circ$ and $\delta=-0.2$. \label{fig:PPLCrev}}
\end{figure}

To test the robustness of the P1/P2 vs.\ $E_\gamma$ effect, we studied the light curves at $\zeta_{\rm cut}=40^\circ$ to obtain a counter-example. These light curves have a different emission structure than those in Figure~\ref{fig:PPLCopt}, due to a different spatial origin of the emission. In Figure~\ref{fig:PPLCrev} the observer misses the bridge emission, since emission radiated at $\zeta_{\rm cut}=40^\circ$ is farther from the PCs than emission at $\zeta_{\rm cut}=65^\circ$. The phase plots for scenario~1 remains brighter than for the second scenario. As the energy increases, the relative flux of P1 becomes larger than that of P2. A similar study was done by \citet{Brambilla2015} assuming a FIDO model to show that the P1/P2 effect is common, but not universal, since a change in geometry can reverse the effect. Figure~\ref{fig:PPLCrev} supports this finding. We shifted the model light curves by $-0.2$ in phase to fit the \emph{Fermi} and H.E.S.S.\ data. This indicates the effect of $\zeta$ on the degeneracy of $\phi_{\rm L}$=0 in the data (reflecting the main radio peak) and $\phi_{\rm L}$=0 (the phase of the $\boldsymbol{\mu}$-axis).

\begin{figure}
\centering
\includegraphics[width=10cm]{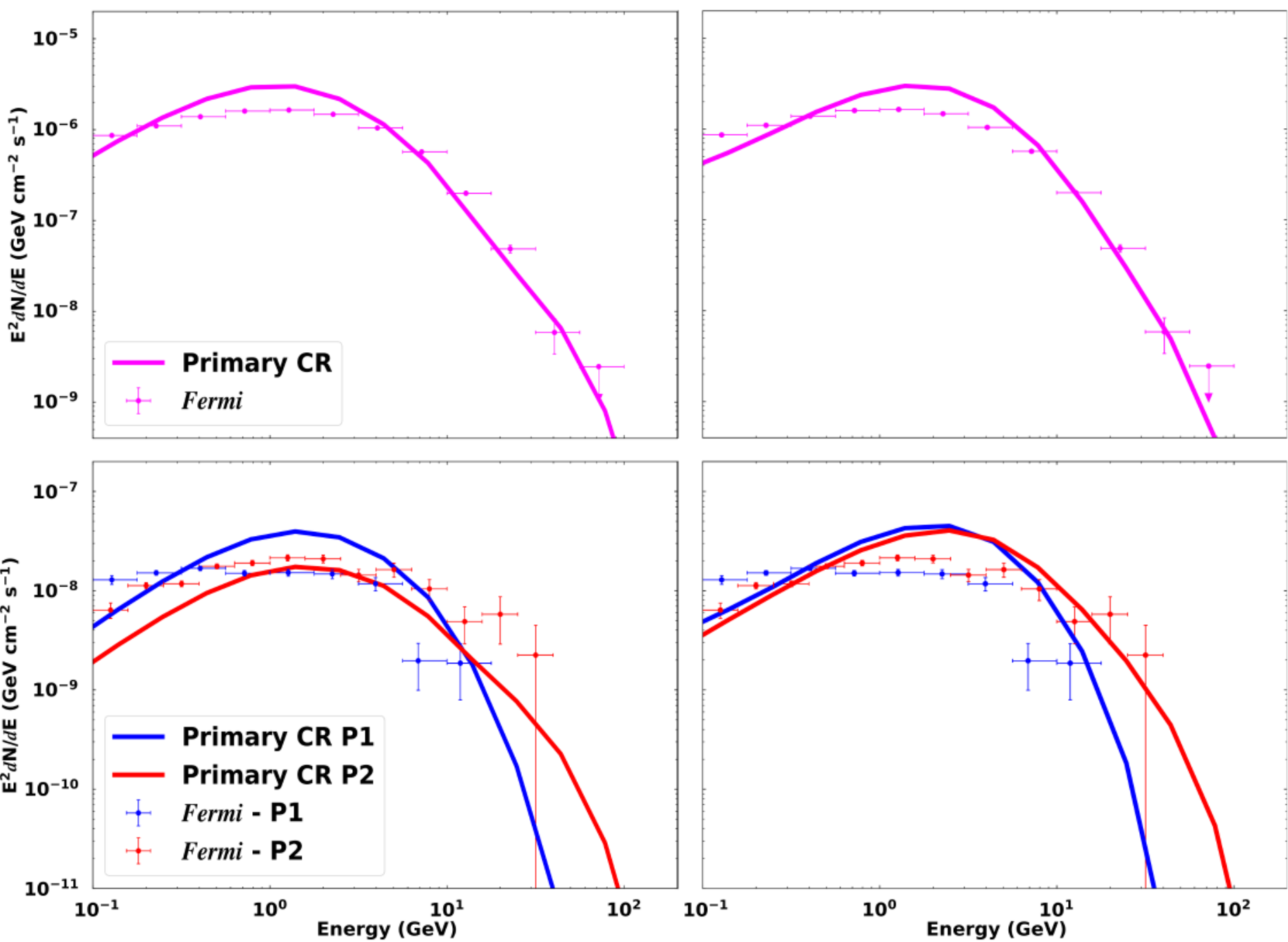}
\caption[Phase-averaged and phase-resolved spectra for the refined $\rho_{\rm c}$ calculation, for $\alpha=60^\circ$ and $\zeta_{\rm cut}=65^\circ$, for both scenarios]{Phase-averaged (top panel) and phase-resolved (bottom panel) spectra for the refined $\rho_{\rm c}$ calculation, for $\alpha=60^\circ$ and $\zeta_{\rm cut}=65^\circ$. For the first scenario (left column), the flux is normalised using $2J_{\rm GJ}$ and for the second case (right column), it is normalised using $5J_{\rm GJ}$. The data points for the phase-average spectra are from \citet{Abdo2013SecondCat} (see \url{http://fermi.gsfc.nasa.gov/ssc/data/access/lat/2nd_PSR_catalog/}), and the phase-resolved spectra are updated data to those published in \citet{Abdo2010Vela}. \label{fig:phavgspec60}}
\end{figure}

\begin{figure}
\centering
\includegraphics[width=10cm]{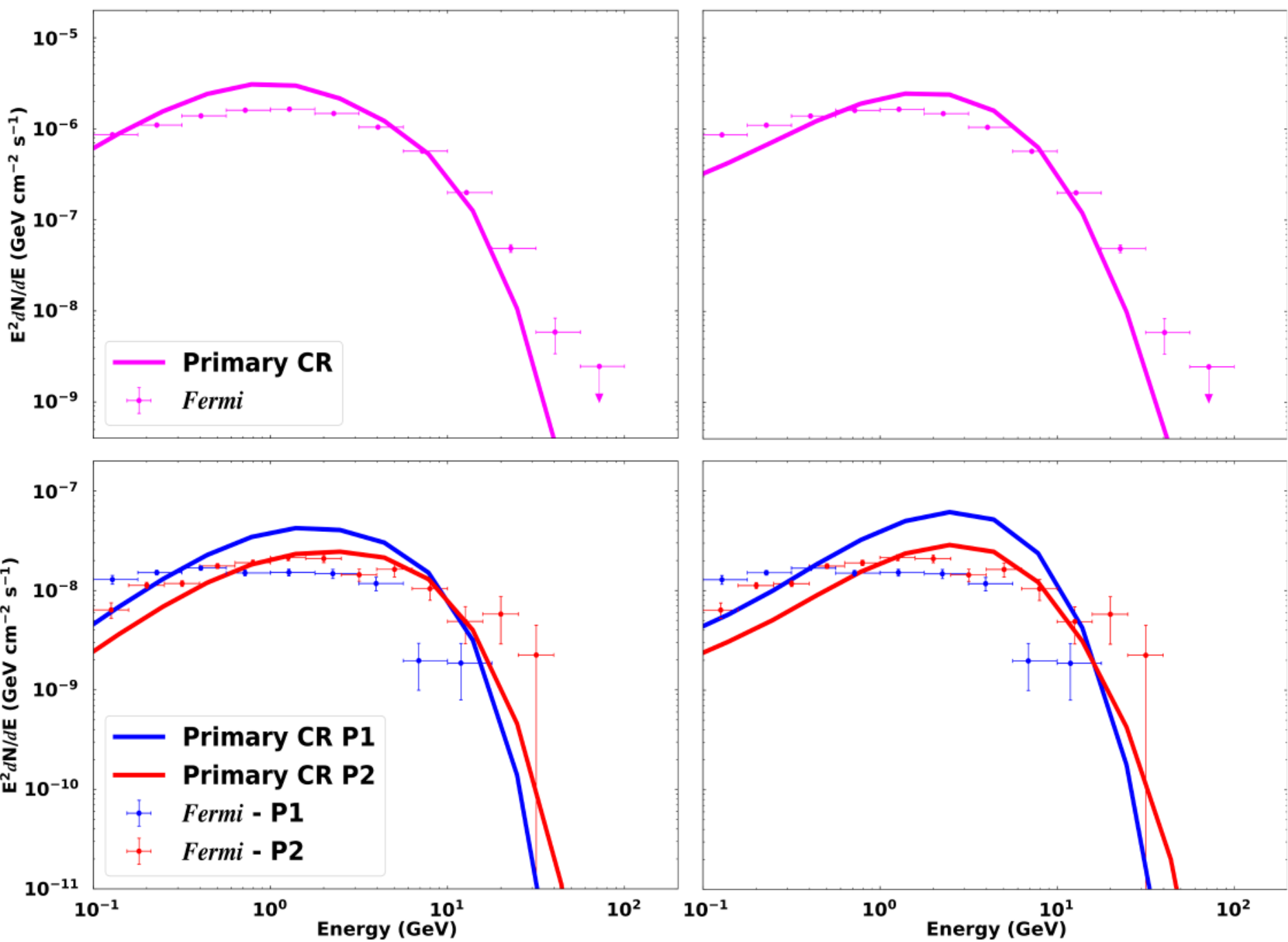}
\caption[The same as Figure~\ref{fig:phavgspec60}, but for $\alpha=75^\circ$ and $\zeta_{\rm cut}=65^\circ$]{The same as Figure~\ref{fig:phavgspec60}, but for $\alpha=75^\circ$ and $\zeta_{\rm cut}=65^\circ$. For the first scenario (left column), the flux normalisation factor is $5J_{\rm GJ}$ and for the second scenario (right column), it is $10J_{\rm GJ}$. \label{fig:phavgspec75}}
\end{figure}

In Figure~\ref{fig:phavgspec60}, the phase-averaged and phase-resolved spectra are shown for $\alpha=60^\circ$ and $\zeta_{\rm cut}=65^\circ$. The model spectra fit the \emph{Fermi} LAT points for both $R_{\rm acc}$ cases fairly well. In the first scenario, the phase-resolved spectra of P1 has a higher flux than that of P2, although the latter has a tail extending to higher $E_{\gamma}$ and a slightly larger $E_{\gamma,\rm CR}$. In Figure~\ref{fig:phavgspec75}, the phase-averaged and phase-resolved spectra are shown for the optimal parameters. The model spectra fit the data for both scenarios fairly well. In the first scenario, the phase-resolved spectra of P1 has a higher flux than that of P2, although smaller than the flux of P1 seen in Figure~\ref{fig:phavgspec60}. For P2, the high-$E_{\gamma}$ tail extends not as far in $E_\gamma$ as in Figure~\ref{fig:phavgspec60}, but $E_{\gamma,\rm CR}$ remains larger for P2, with the predicted cutoff being $E_{\gamma,\rm CR}\sim{1}$~GeV. A larger cutoff for P2 than P1 is expected for this $\zeta_{\rm cut}$ value, since the second light curve peak survives longer than P1 as $E_\gamma$ increases (see Figure~\ref{fig:LC_epardep}). This may not always be the case, as pointed out in Figure~\ref{fig:LC_zdep} where the P1 remains larger than P2 depending on the choice of $\zeta_{\rm cut}$.

\subsection{Testing the attainment of the CRR limit} \label{subsec:crr}
We solved the transport equation of a particle as it moves along a $B$-field line, focusing on CR \citep[e.g.,][]{Daugherty1982,Harding2005c}:
\begin{equation} \label{eq:transport}
{\dot{\gamma}}={\dot{\gamma}_{\rm gain}}+{\dot{\gamma}_{\rm loss}}=\frac{1}{m_{\rm e}c^2}\left[{ceE_\parallel}-\frac{2ce^2\gamma^4}{3\rho^2_c} \right],
\end{equation}
with $\dot{\gamma}$ the time-derivative of $\gamma$, $\dot{\gamma}_{\rm gain}$ the acceleration rate, and $\dot{\gamma}_{\rm loss}$ the loss rate. From Eq.~(\ref{eq:transport}), it is clear that the $\dot{\gamma}_{\rm gain}$ is dependent on $R_{\rm acc}$, and $\dot{\gamma}_{\rm loss}$ is directly proportional to $\gamma^4$ and $\rho_{\rm c}^{-2}$. Eq.~(\ref{eq:transport}) may be recast in spatial terms by dividing by $c$ (assuming relativistic outflow of particles):
\begin{equation} \label{eq:transport1}
{\frac{d\gamma}{dl}}=R_{\rm acc}-\frac{2e^2\gamma^4}{3m_{\rm e}c^2\rho^2_c}.
\end{equation}

The CR spectral energy cutoff is defined as follows \citep[e.g.,][]{Daugherty1982,Cheng1996}:
\begin{eqnarray} \label{eq:Ecut}
E_{\gamma,{\rm CR}} = \frac{3\lambdabar_{\rm c}\gamma^3}{2\rho_{\rm c}}m_{\rm e}c^2,
\end{eqnarray}
where ${\lambdabar}_{\rm c}=\hbar/(m_{e}c)$ is the Compton wavelength, and $\hbar$ the reduced Planck's constant. The curvature radiation reaction (CRR) limit is attained when the acceleration rate equals the loss rate. In this limit, the Lorentz factor is \citep[e.g.,][]{Luo2000}
\begin{equation}
    \gamma_{\rm CRR} = \left(\frac{3E_{\parallel}\rho_c^2}{2e}\right)^{1/4}.\label{Gamma_CRR}
\end{equation}
Substituting Eq.~(\ref{Gamma_CRR}) into Eq.~(\ref{eq:Ecut}), we obtain for a constant $E$-field \citep{Venter2010}
\begin{eqnarray} \label{eq:EcutCRR}
  E_{\gamma,{\rm CR}} \sim 4\left(\frac{E_{\parallel}}{10^4~{\rm statvolt\,cm^{-1}}}\right)^{3/4}\left(\frac{\rho_{c}}{10^8~{\rm cm}}\right)^{1/2}, %~{\rm GeV}.
\end{eqnarray}
measured in GeV. We generally test if the CRR limit is attained in both scenarios by plotting the $\log_{10}$ of $\rho_{\rm c}$, $\gamma$, $\dot{\gamma}_{\rm gain}$, and $\dot{\gamma}_{\rm loss})$ along the same field lines chosen in Figure~\ref{fig:posdirrhoOvsN}, checking if the acceleration and loss rates become equal at large distances. The particle dynamics depend on the $\rho_{\rm c}$, therefore an improved calculation yielding a smoother $\rho_{\rm c}$ has an impact on the particle transport and thus the energy-dependent light curves and spectra. 
\begin{figure}
\centering
\includegraphics[width=10cm]{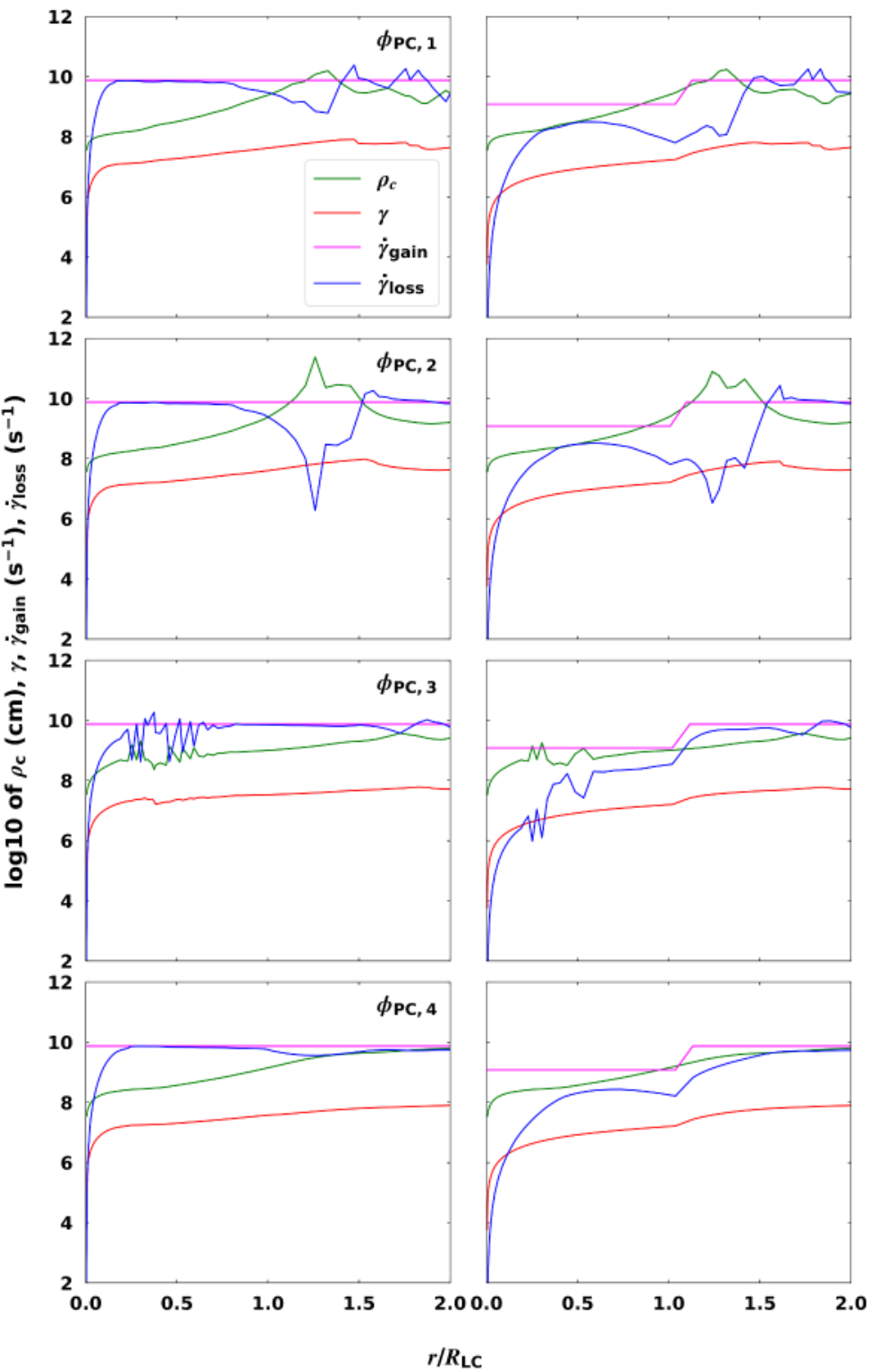}
\caption[The particle dynamics, along the same $B$-field lines as in Figure~\ref{fig:posdirrhoOvsN}, are shown for the refined $\rho_{\rm c}$ calculation, for $\alpha=75^\circ$.]{The particle dynamics, along the same $B$-field lines as in Figure~\ref{fig:posdirrhoOvsN}, are shown for the refined $\rho_{\rm c}$ calculation, for $\alpha=75^\circ$. The quantities plotted are the $\log_{10}$ of $\rho_{\rm c}$ (green), $\gamma$ (red), $\dot{\gamma}_{\rm gain}$ (magenta), and $\dot{\gamma}_{\rm loss}$ (blue), for both scenario~1 (left column) and scenario~2 (right column). \label{fig:crr}}
\end{figure}

In Figure~\ref{fig:crr}, the CRR limit is almost immediately attained in the first scenario, since the $E_\parallel$ is large enough to supply the primaries with ample energy at lower altitudes. The rapid rise of $\gamma\rightarrow \sim5\times10^7$ leads to a rapid increase in $\dot{\gamma}_{\rm loss}$, and then the CRR limit is reached around $0.2R_{\rm LC}$. However, since $\rho_{\rm c}$ oscillates or dips along some of the field lines, this limit is disturbed (since the loss rate is anti-correlated with $\rho_{\rm c}$) and in some cases only recovered later on at higher altitudes. Indeed, instabilities in $\rho_{\rm c}$ cause similar but anti-correlated oscillations in $\dot{\gamma}_{\rm loss}$. If the $E_\parallel$ is lower inside the light cylinder, as in the second scenario, the acceleration of the primaries is initially suppressed, as is $\dot{\gamma}_{\rm loss}$. However, beyond $r\sim R_{\rm LC}$, where a higher $E_\parallel$ is assumed, the particles accelerate efficiently, and the CRR limit may be reached around $r\sim 1.5R_{\rm LC}$.

\subsection{Local environment of emission regions connected to each light curve peak} \label{subsec:environment}
In order to isolate and understand the P1/P2 vs.\ $E_\gamma$ effect seen in the light curves of Vela, we investigated the values of $E_{\gamma,\rm CR}$ (Eq.~[\ref{eq:Ecut}]), $\rho_{\rm c}$, and $\gamma$ in the spatial regions where each model peak originates, for the set of optimal parameters we found as described in Section~\ref{subsec:optfit}. Thus, as explained in Section~\ref{subsec:revmap}, we performed ``reverse mapping'' and accumulated the range of values that these three quantities assume in the regions where the photons originate that make up P1 and P2. These binned quantities are presented as $E_\gamma$-dependent histograms below, where we scaled the frequency of occurrence of the quantities using the emitted $\dot{N}_{\gamma}$ to obtain a true relative probability for each chosen energy range. 

\begin{figure*}
  \subfloat[]{\includegraphics[width=8.2cm]{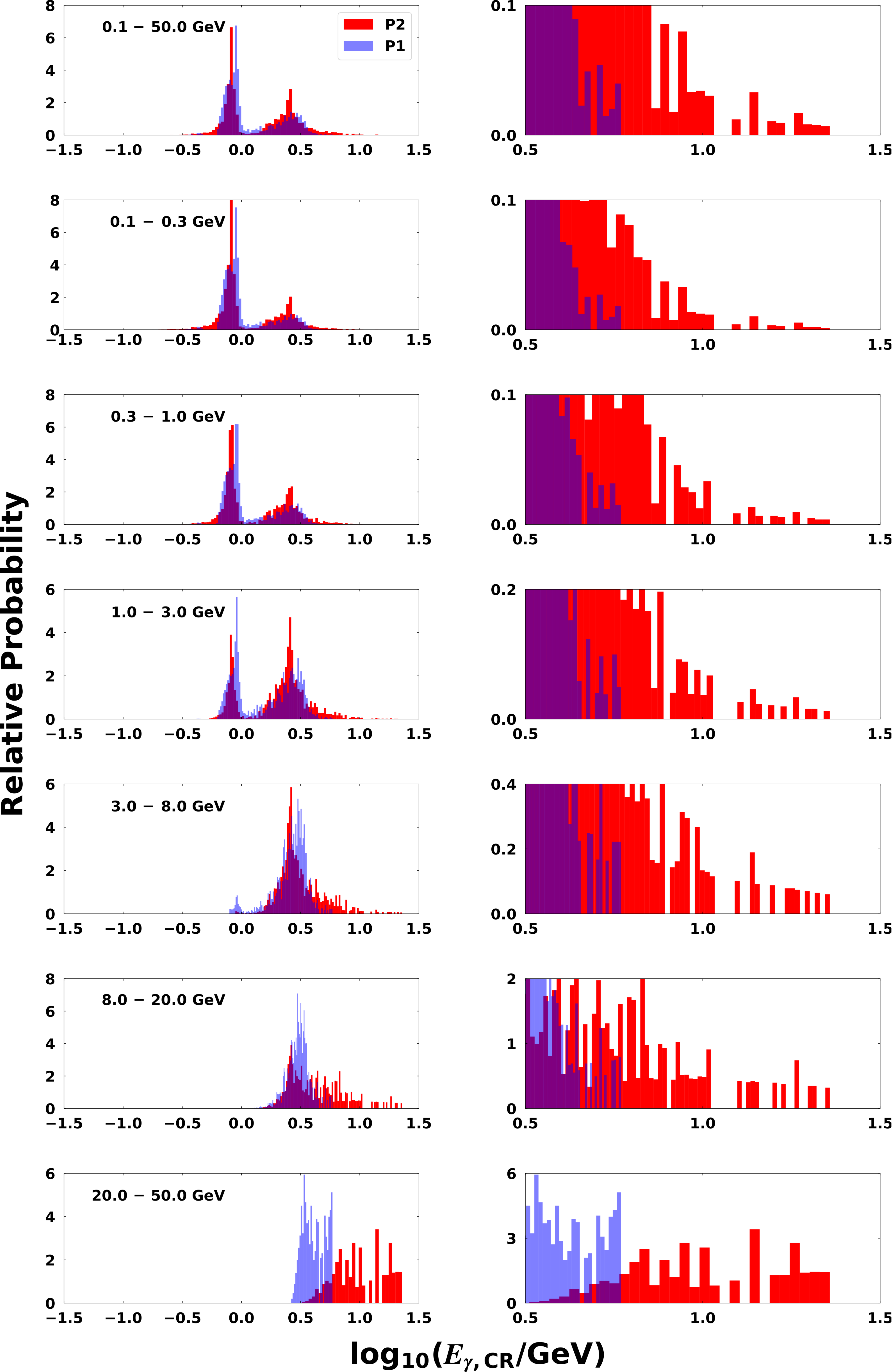}\label{fig:ecrR1}}
  \hfill
  \subfloat[]{\includegraphics[width=8.2cm]{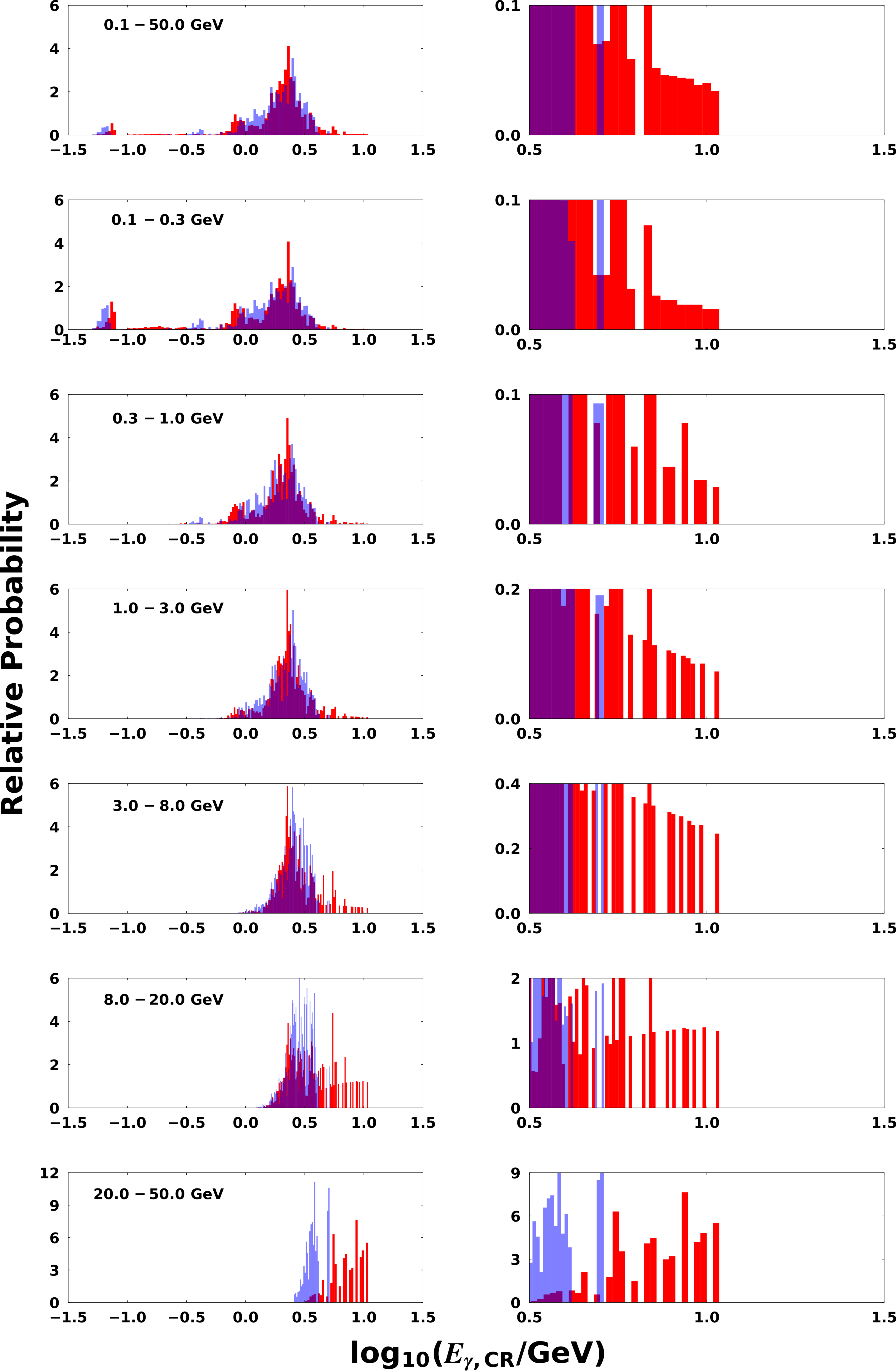}\label{fig:ecrR2}}
\caption[Energy-dependent histograms for $\log_{10}(E_{\gamma,\rm CR}/{\rm GeV})$ for P1 and P2, for both scenarios]{Energy-dependent histograms for $\log_{10}(E_{\gamma,\rm CR}/{\rm GeV})$ for P1 (blue curve) and P2 (red curve), where (a) represents scenario~1 and (b) scenario~2. The respective energy bands are indicated as labels in each panel. The second column in each case represents a zoom-in of the tails of the distributions for large values of $\log_{10}(E_{\gamma, \rm cut}/{\rm GeV})$. \label{fig:hists_ecut}}
\end{figure*}
In Figure~\ref{fig:hists_ecut} we show histograms for $\log_{10}(E_{\gamma,\rm CR}/{\rm GeV})$, for different energy ranges. This quantity is calculated using Eq.~(\ref{eq:Ecut}), specifically involving $\rho_{\rm c}$ and $\gamma$. These quantities are in principle calculated for all $E_\gamma$, but we subsequently apply cuts in $E_\gamma$ and then study the resulting distributions of the $\rho_{\rm c}$ and $\gamma$ associated with photons in a particular chosen band. In the first scenario (left column), there appears two bumps, for both peaks, at lower $E_\gamma$ (up to $\sim5$~GeV), situated around $\log_{10}(E_{\gamma,\rm CR}/{\rm GeV})\approx{-0.2}$ and ${0.4}$. The lower bump disappears with increasing $E_\gamma$. 
%This causes the second bump at $\log_{10}(E_{\gamma,\rm CR})>{0.0}$ to appear pronounced. 
In (b), we show scenario~2 where there is a small low-$E_\gamma$ bump (up to $\sim0.3$~GeV) at even smaller values of $\log_{10}(E_{\gamma,\rm CR}/{\rm GeV})\approx{-1.2}$. The existence of this bump is probably because of the lower value of $R_{\rm acc}$ inside the light cylinder that suppresses the low-altitude acceleration and emission in this scenario. Also, the lower-energy bump disappears as the $E_\gamma$ is increased, since only photons from individually-radiated spectra (that make up the cumulative spectrum seen by the observer) with higher cutoffs are then visible. The $\log_{10}(E_{\gamma,\rm CR}/{\rm GeV})$ of P2 is relatively larger than that of P1 for both scenarios, as seen in the zoom-ins. This confirms what has already been seen in the light curves in Figure~\ref{fig:PPLCopt} and spectra in Figure~\ref{fig:phavgspec75}: P2 survives with an increase in energy, since its spectral cutoff is relatively higher than that of P1. The $\log_{10}(E_{\gamma,\rm CR}/{\rm GeV})$ of P2 reaches values as high as $\sim10^{1.0}-10^{1.4}$, with larger values reached in the first scenario, given the higher $E$-field. 
\begin{figure*}
  \subfloat[]{\includegraphics[width=8.2cm]{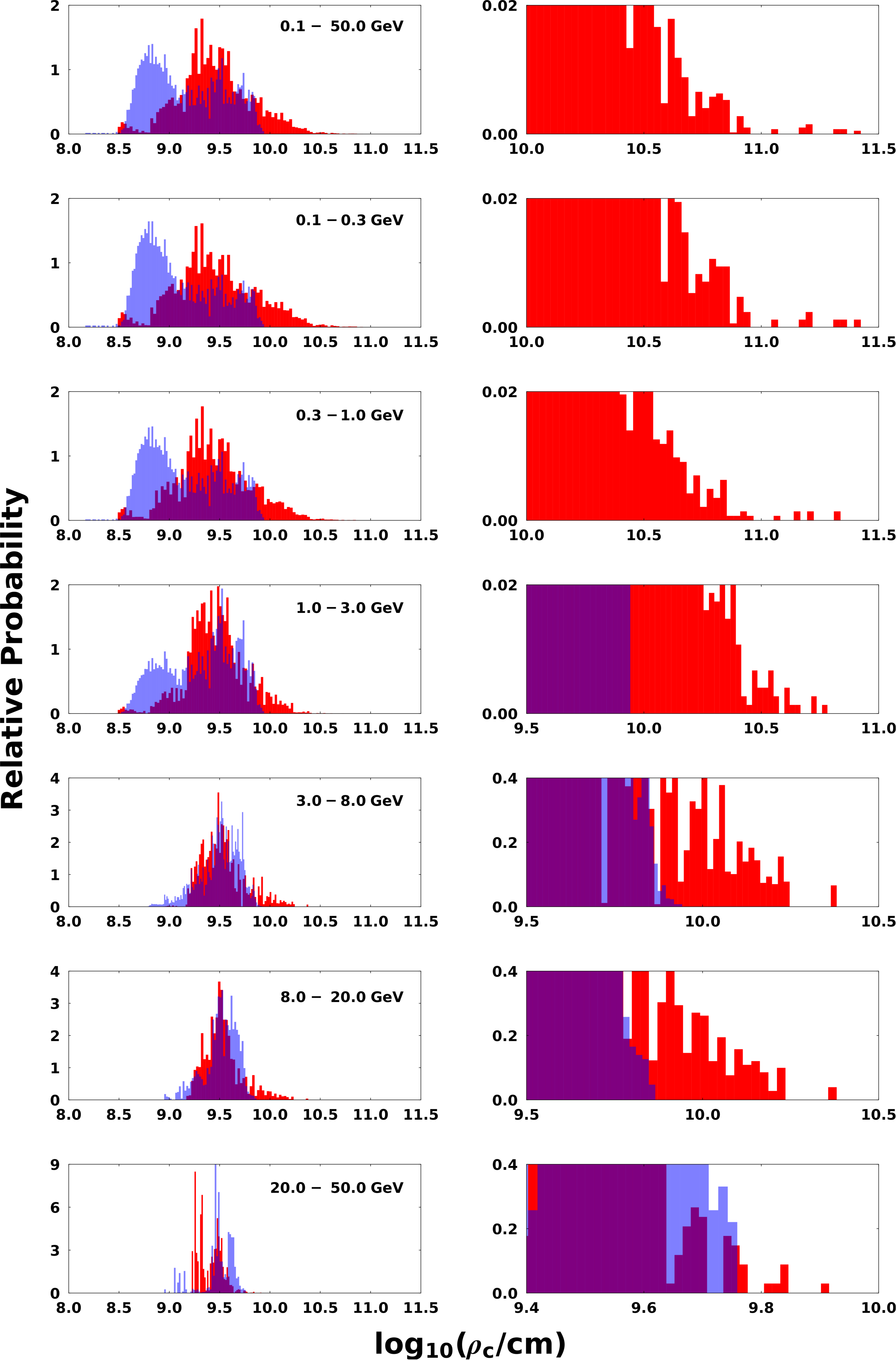}\label{fig:rhoR1}}
  \hfill
  \subfloat[]{\includegraphics[width=8.2cm]{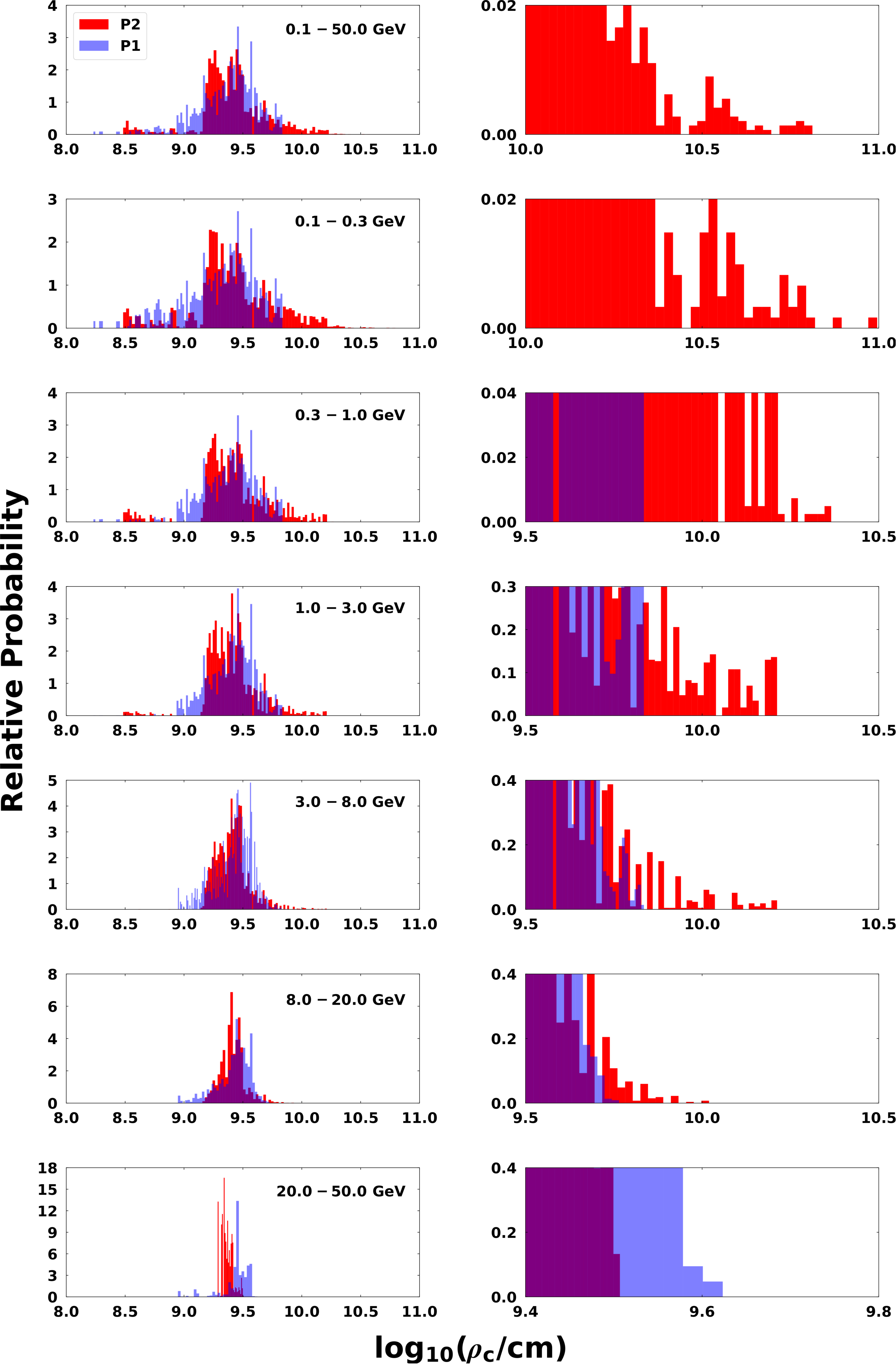}\label{fig:rhoR2}}
\caption[The same as Figure~\ref{fig:hists_ecut} but for $\log_{10}(\rho_{\rm c})$]{The same as Figure~\ref{fig:hists_ecut} but for $\log_{10}(\rho_{\rm c})$. \label{fig:hists_rho}}
\end{figure*}
In Figure~\ref{fig:hists_rho} we show histograms of the relative probability as a function of $\log_{10}\rho_{\rm c}$ for P1 (blue) and P2 (red). For the first scenario in (a), with a zoom-in of the tail of the distributions (right column), there appears a bump around $\log_{10}\rho_{\rm c}\approx{8.5}$ to~$9.0$ for P1 at lower $E_\gamma$ (up to $\sim3$~GeV), which disappears with increasing $E_\gamma$. In (b) we show scenario~2, where there is no low-$E_\gamma$ bump at smaller values of $\log_{10}\rho_{\rm c}$ as in the first scenario. This is due to the fact that in the first scenario, the accelerating $E$-field is relatively larger at lower altitudes, so that the particles can radiate in the GeV band from these lower altitudes characterised by lower values of $\log_{10}\rho_{\rm c}$. In the second scenario, however, the small value of $R_{\rm acc,\rm low}$ inside the light cylinder suppresses emission in the GeV band originating from lower altitudes, hence the missing bump. Importantly, the $\log_{10}\rho_{\rm c}$ of P2 is relatively larger than that of P1 for both scenarios, as seen in the zoom-ins, with P2's associated $\rho_{\rm c}$ reaching values as high as $\sim10^{9.8}-10^{11.5}$~cm (indicating relatively less curved orbits). The $\rho_{\rm c}$ values reached in scenario~1 for P2 are also relatively larger than those in scenario~2 for the same peak. Thus, for sustained acceleration, particles radiating at high energies are moving along slightly straighter orbits (and radiating from farther out, see Figure~\ref{fig:hists_RlcLim1}). It is only at energies above 20~GeV that the values of $\log_{10}\rho_{\rm c}$ associated with P1 becomes comparable or larger than those associated with P2.

\begin{figure*}
  \subfloat[]{\includegraphics[width=8.2cm]{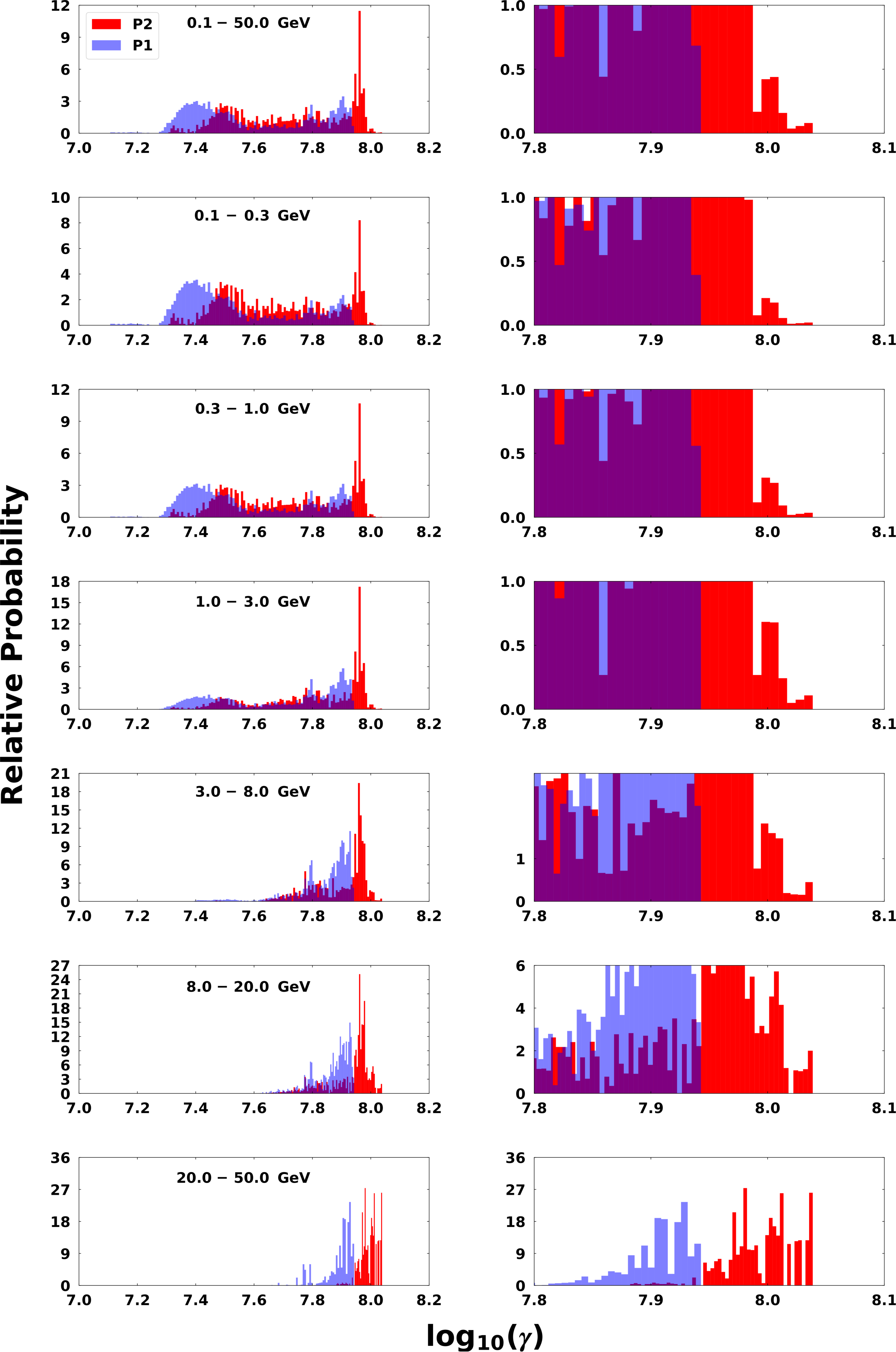}\label{fig:gamR1}}
  \hfill
  \subfloat[]{\includegraphics[width=8.2cm]{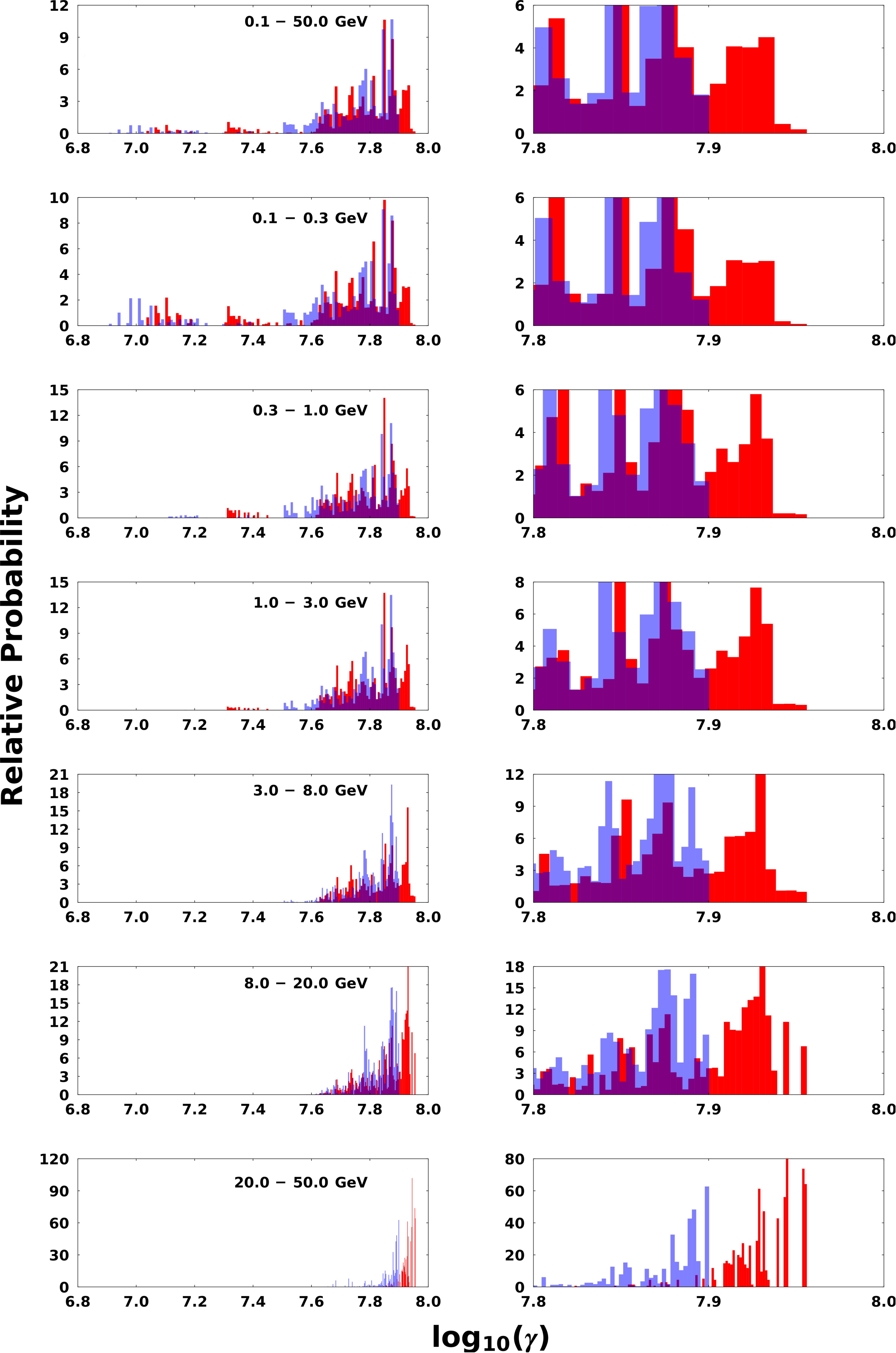}\label{fig:gamR2}}
\caption[The same as Figure~\ref{fig:hists_ecut} but for $\log_{10}(\gamma)$]{The same as Figure~\ref{fig:hists_ecut} but for $\log_{10}(\gamma)$. \label{fig:hists_gam}}
\end{figure*}
Similar to Figure~\ref{fig:hists_ecut} and Figure~\ref{fig:hists_rho}, we show histograms of $\log_{10}(\gamma)$ in Figure~\ref{fig:hists_gam} for different energy ranges. In the first scenario indicated in (a), there appears a bump around $\log_{10}(\gamma)\approx{7.3}-7.5$ for both peaks at lower $E_\gamma$ (up to $\sim3$~GeV), which disappears with increasing $E_\gamma$. In (b) we show scenario~2 where there is no low-$E_\gamma$ bump at smaller values of $\log_{10}(\gamma)$. There is also a peak in $\log_{10}(\gamma)\sim8$ for P2 in scenario~1, while $\log_{10}(\gamma)$ is relatively smaller in scenario~2, given the fact that particles experienced less acceleration in that case. The $\log_{10}(\gamma)$ of P2 is relatively larger than that of P1 as seen in the zoom-ins for both scenarios. 

\begin{figure*}
  \subfloat[]{\includegraphics[width=4.7cm]{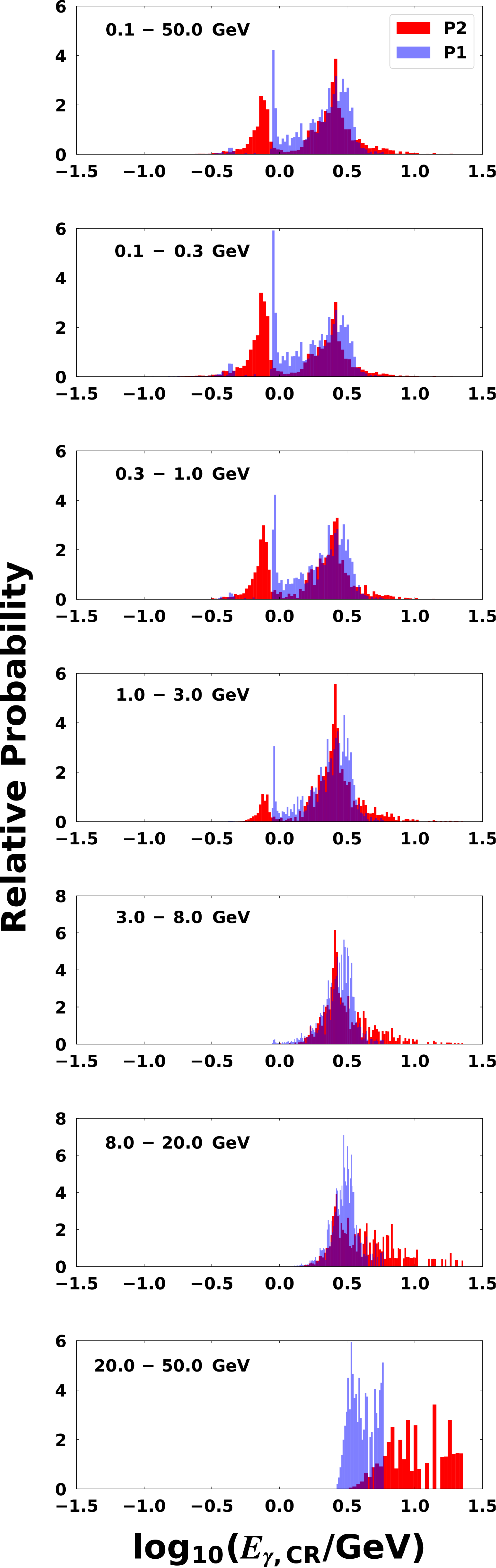}\label{fig:ecrR1lim}}
  \hfill
  \subfloat[]{\includegraphics[width=4.7cm]{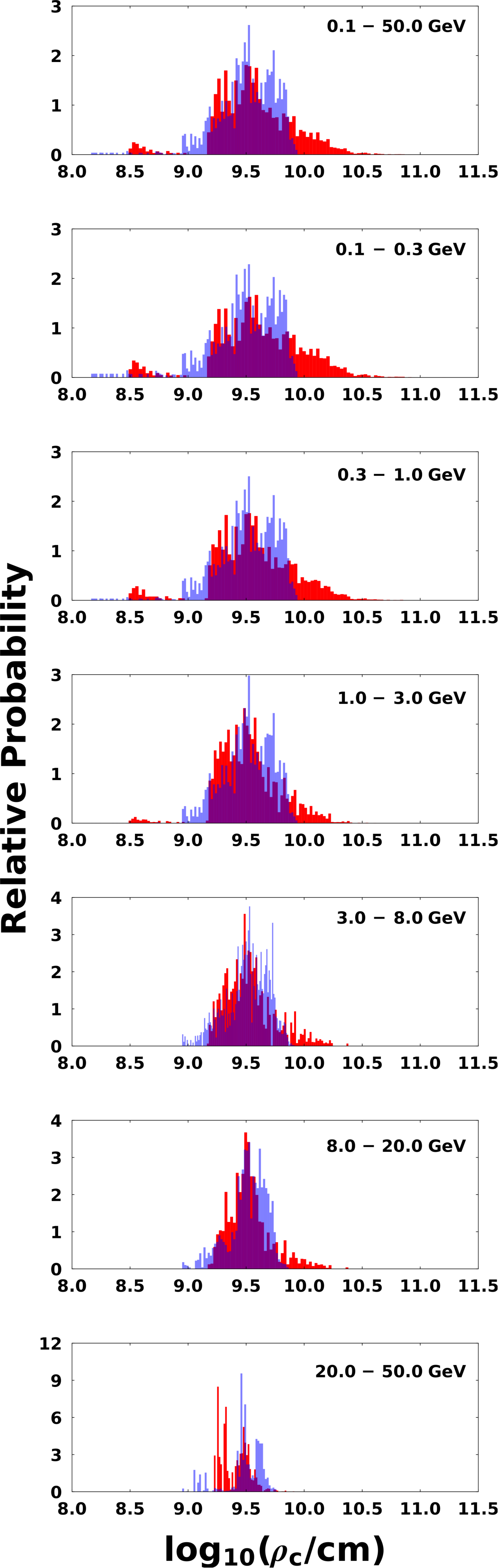}\label{fig:rhoR1lim}}
  \hfill
  \subfloat[]{\includegraphics[width=4.7cm]{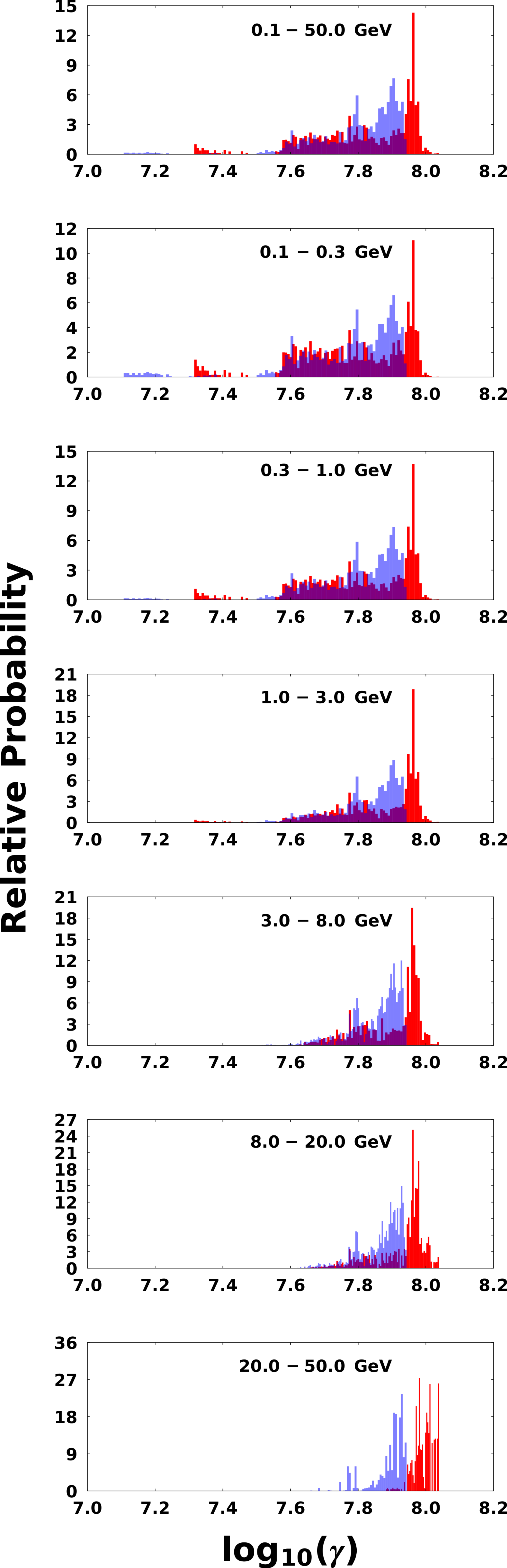}\label{fig:gamR1lim}}
\caption[Energy-dependent histograms of $\log_{10}(E_{\gamma,\rm CR}/{\rm GeV})$, $\log_{10}(\rho_{\rm c})$, and $\log_{10}(\gamma$), for P1 and P2. All three cases are for scenario~1 at altitudes equal to and beyond $R_{\rm LC}$]{Energy-dependent histograms of (a) $\log_{10}(E_{\gamma,\rm CR}/{\rm GeV})$, (b) $\log_{10}(\rho_{\rm c})$, and (c) $\log_{10}(\gamma$), for P1 (blue curve) and P2 (red curve). All three cases are for the first scenario at altitudes equal to and beyond $R_{\rm LC}$, into the current sheet. \label{fig:hists_RlcLim1}}
\end{figure*}
\begin{figure*}
  \subfloat[]{\includegraphics[width=4.7cm]{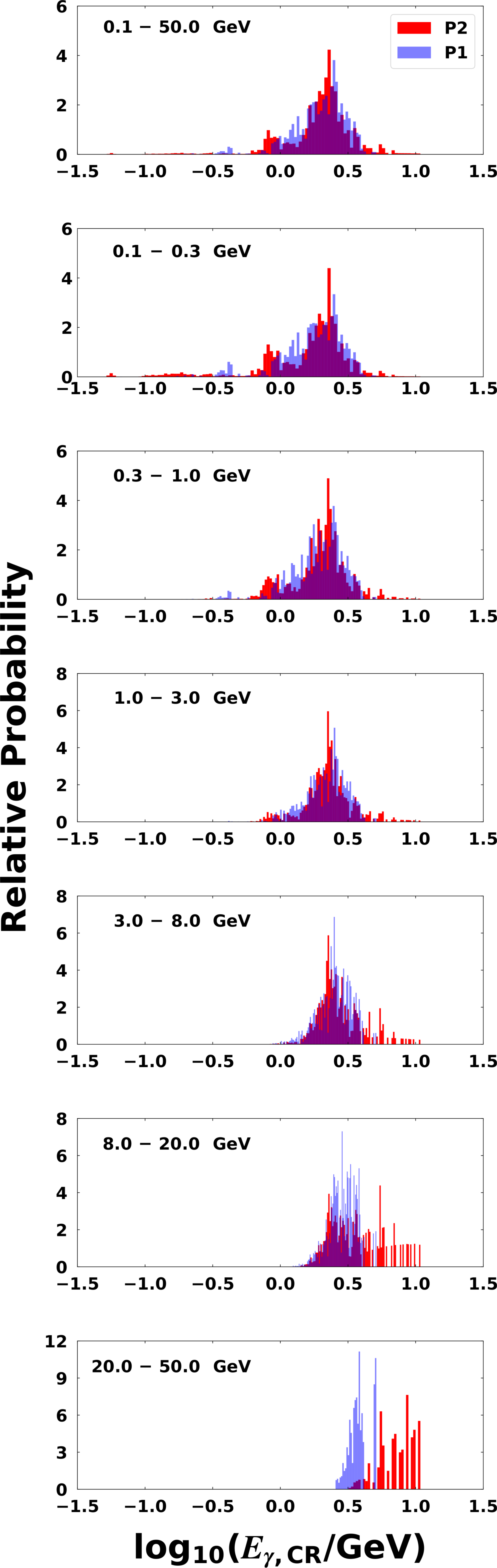}\label{fig:ecrR2lim}}
  \hfill
  \subfloat[]{\includegraphics[width=4.7cm]{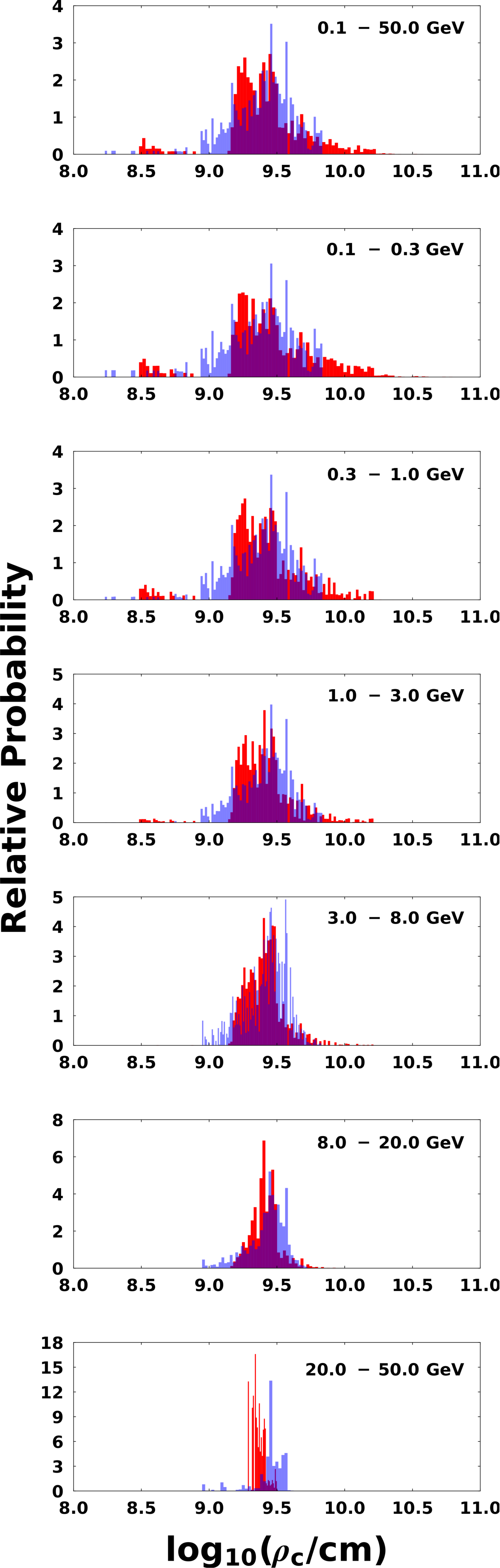}\label{fig:rhoR2lim}}
  \hfill
  \subfloat[]{\includegraphics[width=4.7cm]{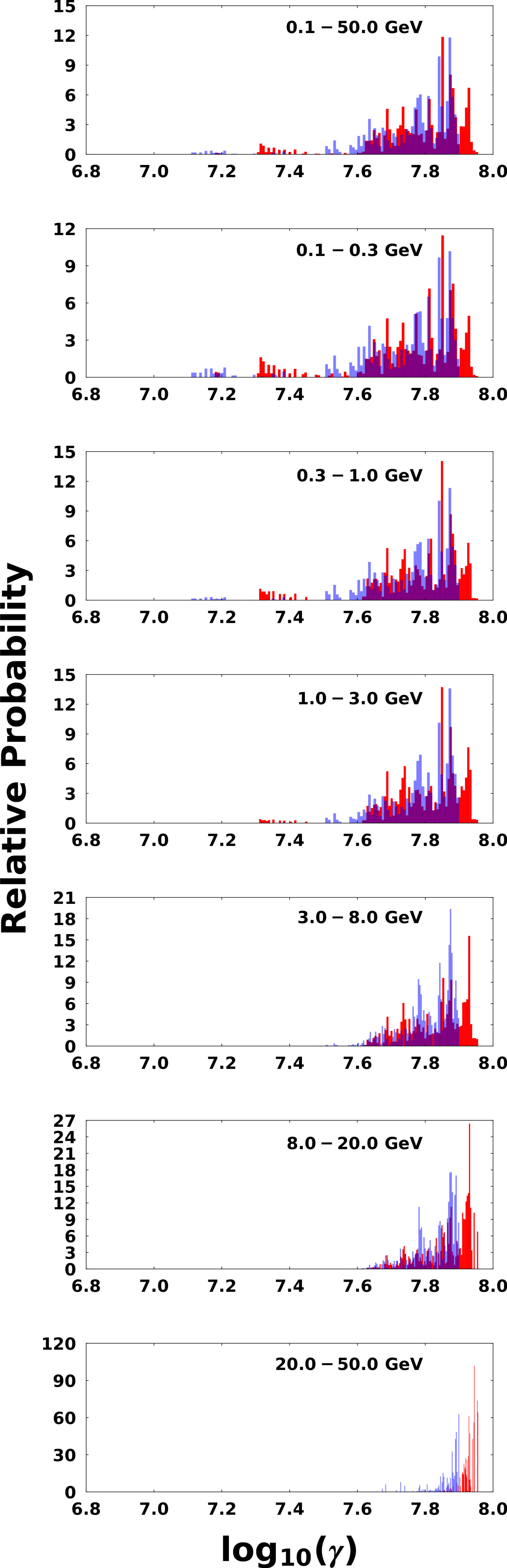}\label{fig:gamR2lim}}
\caption[The same as in Figure~\ref{fig:hists_RlcLim1} but for scenario~2]{The same as in Figure~\ref{fig:hists_RlcLim1} but for the second scenario. \label{fig:hists_RlcLim2}}
\end{figure*}
%As a comparison between the histograms (both scenarios) including the emission over all altitudes and those with emission limited to altitudes at or beyond the light cylinder, we also search for indices where $1\leq{r/R_{\rm LC}}\leq{2}$. Following the same procedure as mentioned above we search for intersecting indices between $\phi$, $\zeta$, $r/R_{\rm LC}$, and $E_\gamma$, and normalise again.
In Figure~\ref{fig:hists_RlcLim1}(a),~(b), and~(c) we limit the emission radius $r$ to altitudes at and beyond $R_{\rm LC}$ to investigate the change in the range of values for the pertinent quantities as compared to the previous cases where we considered emission from all altitudes. We show histograms for $\log_{10}$ of $E_{\gamma,\rm CR}$, $\rho_{\rm c}$, and $\gamma$, respectively, for the first scenario. At lower $E_\gamma$ (up to $\sim3$~GeV), the P1 bumps at lower values that we first noticed in Figure~\ref{fig:hists_ecut}a,~\ref{fig:hists_rho}a, and~\ref{fig:hists_gam}a are suppressed. This indicates that photons originating inside the light cylinder come from regions that are characterised by lower values of $E_{\gamma,\rm CR}$, $\rho_{\rm c}$, and $\gamma$.
%However, the P1 and P2 bump in Figure~\ref{fig:hists_gam}a disappears that were first noticed at $\log_{10}{\gamma}\sim{7.4}$. 
%Also in Figure~\ref{fig:hists_ecut}a the bump of P1 and P2 appear smaller than those over all altitudes. 
This effect of limiting the emission altitudes is not as noticeable in the second scenario in Figure~\ref{fig:hists_RlcLim2}(a),~(b), and~(c). 
For $E_{\gamma,\rm CR}$ the slight bump (including P1 and P2 emission) at low values of $E_{\gamma,\rm CR}$ disappears at lower $E_\gamma$ (up to $\sim{0.3}$~GeV).
%Although, in Figure~\ref{fig:hists_rho}b, Figure~\ref{fig:hists_gam}b, and Figure~\ref{fig:hists_ecut}b there appear lower values for P1 and then a slight bump in Figure~\ref{fig:hists_ecut}b. 
For $\rho_{\rm c}$ and $\gamma$ the difference is insignificant, given the fact that the low-altitude $E$-field already suppresses the emission.

\section{Conclusion} \label{sec:concl}
There is an ongoing debate regarding the origin of the GeV emission detected from pulsars, it being attributed either to CR or SR (or even IC; see \citealt{Lyutikov2012,Lyutikov2013}). One way in which to possibly discriminate between these options is to model the energy-dependent light curves of several bright pulsars.

We modelled $E_{\gamma}$-dependent light curves and spectra of the Vela pulsar in the HE regime assuming CR from primaries to see if we can explain the origin of the decreasing ratio of P1/P2 vs.\ $E_\gamma$, expecting that the answer may lie in a combination of the values of geometric and physical parameters associated with each peak. Since the light curves probe geometry, e.g., $\alpha$, $\zeta$ and emission gap position and extent, and the spectrum probes both the energetics and geometry, we simultaneously fit these data with our model to obtain optimal fitting parameters.

We presented a refined calculation of the $\rho_{\rm c}$ of particle trajectories, impacting the CR loss rate and leading to smoother phase plots and light curves. However, this refinement had a rather small impact, as the broad structure of caustics and light curves remained similar to what was found previously.
We assumed a FF magnetosphere (implying zero $E$-fields) as a good first approximation of the true $B$-field structure, yet considered both a constant and two-step accelerating $E$-field. We also found that the CRR limit was easily reached in the first scenario, and sometimes also in the second.
We proceeded to isolate the P1/P2 effect by selecting photons that make up these two light curve peaks, and investigating the range of associated values of $\rho_{\rm c}$, $E_{\gamma, \rm CR}$ and $\gamma$. We found that the phase-resolved spectra associated with each peak indicated a slightly larger spectral cutoff for P2, confirming that P2 survives with an increase in energy, given its larger spectral cutoff. This was also seen in energy-dependent histograms of $E_{\gamma, \rm CR}$, confirming that this quantity was systematically larger for P2. The reason for this became more clear upon discovery that both the $\rho_{\rm c}$ and $\gamma$ were systematically larger for P2, for both scenarios. If CRR is reached, one expects $E_{\gamma, \rm CR}\propto \rho_{\rm c}^{1/2}$ for a constant $E$-field, so the larger $\rho_{\rm c}$ would explain the larger spectral cutoff for P2. Conversely, even if CRR is not attained, $E_{\gamma, \rm CR}\propto \gamma^3\rho_{\rm c}^{-1}$. Given the systematic dominance of $\gamma$ for P2, and the strong dependence of the third power, the larger spectral cutoff of P2 is thus explained by the larger $\gamma$. We also found that the values of $\rho_{\rm c}$ and $\gamma$ remained larger for P2 when only considering emission beyond the light cylinder; in particular, the largest values of these quantities occurred there, pointing to dominant emission from that region to make up the GeV light curves.

We thus found reasonable fits to the energy-dependent light curves and spectra of Vela, and our model that assumes CR as the mechanism responsible for the GeV emission captures the general trends of the decrease of P1/P2 vs.\ $E_{\gamma}$, evolution / depression of the inter-peak bridge emission, plus stable peak positions and a decrease in the peak widths as $E_{\gamma}$ is increased. However, an unknown azimuthal dependence of the $E$-field as well as uncertainty in the precise spatial origin of the emission preclude a simplistic discrimination of emission mechanisms. Similar future modelling of energy-dependent light curves and spectra within a striped-wind context that assumes SR to be the relevant GeV mechanism will be necessary to see if those models can also reproduce and explain these salient features in the case of Vela and other pulsars.

We note that the drop in P1/P2 vs.\ $E_\gamma$ may not be universal, as also found by \citep{Brambilla2015}. We found a counter-example for a different choice of $\zeta_{\rm cut}$, where P1/P2 increases with $E_\gamma$. This was also the case for specific choices of the two-step acceleration $E$-field. There may also be other parameter combinations that can yield this behaviour. However, this effect seems prevalent and has been seen in both HE and VHE data of bright pulsars.
\\
\\
In the next Chapter we will discuss the study done by \citet{Harding2018} to explain the pulsed VHE emission observed by H.E.S.S.\ from the Vela pulsar, and our contribution. I will give a summary of their study, the parameter values assumed, and the implications of our improved calculation of $\rho_{\rm c}$ (see Section~\ref{sec:rhocalc}) on their results.

 % Implication 1: VHE as CR - ApJ article
\chapter{Modelling the emission from the Vela pulsar in the TeV band} \label{chap:VelaTeV}

In this Chapter, I will briefly discuss the work published in \citet{Harding2018}. In Section~\ref{sec:overviewSSC} an overview of the study is given that includes the motivation thereof, the refinements made to their SSC emission code, the assumptions made, as well as the results obtained. Section~\ref{sec:rhooutput} highlights my contribution to the study which is mainly the refined calculation of $\rho_{\rm c}$ of the particle trajectories, as well as its effect on the model light curves and spectra for different wavebands and radiation components. This is followed by our concluding remarks in Section~\ref{sec:ch6concl}.

\section{Overview}\label{sec:overviewSSC}

%Motivation
The Vela pulsar is the first pulsar detected in the VHE range by H.E.S.S.\ up to energies of a few TeV \citep{Djannati2017}. The pulsed emission at such extreme energies might be either connected to the GeV emission of \emph{Fermi}'s spectra or it could be a new radiation component. More details as to the measured spectrum will be given in a forthcoming paper by the H.E.S.S.\ Collaboration. The study by \citet{Harding2018} treated it to be a separate component that is radiated by particles that have been accelerated up to energies of several TeV. 

Some studies attempted to model the VHE pulsed emission detected from the Crab pulsar as ICS \citep{Du2012,Lyutikov2012,Harding2015}. Others modelled the emission from this same pulsar as cyclotron-self-Compton emission from electron-positron pairs in an OG model, where the pairs scatter their own SR radiation. \citet{Harding2015} modelled the emission observed from Crab, Vela and two bright MSPs (i.e., PSR B1937+21 and PSR B1821–24) in the optical-to-TeV energy range with an SSC model, where pairs from the PC scatter their own SR in the outer magnetosphere (see Section~\ref{sec:3Dcode}). Other studies have proposed that SR from particles accelerated via reconnection in the current sheet could reach TeV energies through Doppler boosting \citep{Uzdensky2014,Mochol2015}. \citet{Rudak2017} modelled Vela's measured VHE emission in an OG scenario, where primary particles are accelerated inside this gap and scatter the observed IR to optical emission, that are placed along the inner edge of the gap and believed to come from SR emitted by pairs also produced in this gap. They showed that primaries, whose CR spectrum match the \emph{Fermi} measured GeV spectrum, produce a significant ICS emission component reaching energies up to $\sim$10~TeV.

%Model refinement
A basic description of the pulsar model has been given in Chapter~\ref{chap:ch4}. For more details, see \citet{Harding2008,Harding2015,Harding2018}. 

\citet{Harding2018} extended the SSC emission code of \citet{Harding2015} 
%as described in Section~\ref{sec:3Dcode}.
To summarise, the main refinements include the following:
\begin{itemize}
    \item The spectral energy range for the radiation calculation was extended to span from IR (10$^{-3}$~eV) to VHE (100~TeV) energies. This expanded range is necessary, since it includes more soft photons whose scattering is in the Thompson limit, and for modelling the SSC emission at the highest energies.  
    \item A more accurate calculation of the particle trajectories and their $\rho_{\rm c}$'s. This in turn gives a more precise determination of the energy of the accelerated particles and of their emitted radiation spectrum. This improved trajectory calculation is discussed in Section~\ref{sec:rhocalc} and will appear in a forthcoming paper (Barnard et al. 2020, in preparation). 
    \item The accelerating $E_\parallel$ is divided into a two-step $E_\parallel$, i.e., a lower $E_\parallel$ inside the light cylinder (mostly SR emission), and a higher $E_\parallel$ at and beyond the light cylinder (into the current sheet, with the GeV emission mostly from CR). This two-step $E$-field is utilised in Chapter~\ref{chap:CREdepLCmod}. In \citet{Harding2015}, the $E_\parallel$ was set to one constant high value extending from the NS surface to 2~$R_{\rm LC}$. This change provides better agreement with recent global MHD and PIC pulsar models showing that the particle acceleration takes place primarily near the current sheet outside of the light cylinder in near-FF magnetospheres of young and middle-aged pulsars.
    \item Another update was the injection of electron–positron pairs only above the PC in regions where the global FF current density enables pair cascades. Injection was thus only done at selected azimuthal PC angles (see \citet{Harding2018} for more details).
    \item The last improvement to the model of \citet{Harding2015} is that the MeV to GeV radiation from accelerated / relativistic particles inside and outside of the light cylinder are emitted via SC radiation, although their radiation at GeV energies is mostly in the CR limit. In \citet{Harding2015}, both the accelerating primary particles (as well as the pairs) could acquire pitch angles at low altitudes through cyclotron-resonant absorption of radio photons (as in the 2018 model), but their SR and CR were treated separately.
\end{itemize}

\begin{figure}
    \centering
    \includegraphics[width=0.6\textwidth]{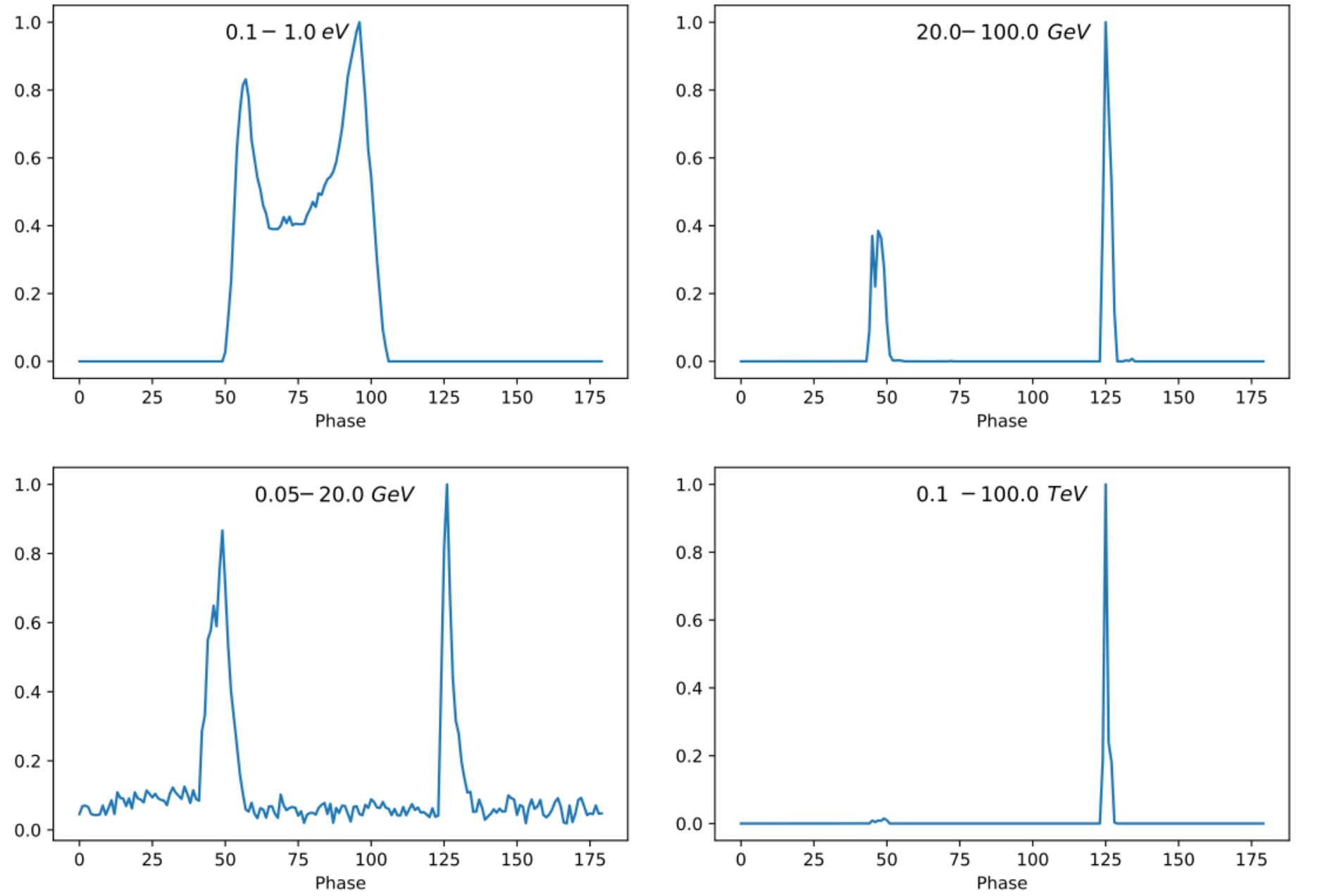}
    \caption[$E_\gamma$-dependent light curves for emission from Vela in the IR/optical band, and three different $\gamma$-ray energy bands, assuming $\alpha=75^\circ$ and $\zeta_{\rm cut}=65^\circ$.]{$E_\gamma$-dependent light curves for emission from Vela in the IR/optical band ($0.1-1$~eV), and three different $\gamma$-ray energy bands (ranging between 0.05~GeV-100~TeV), assuming $\alpha=75^\circ$ and $\zeta_{\rm cut}=65^\circ$. The phase is in degrees. One notices the relatively small peak separation in the optical; this increases with energy, and the relatively intensity of P1 decreases with an increase in energy. From \citet{Harding2018}. \label{fig:SSCLCs}}
\end{figure}

\begin{figure}
    \centering
    \includegraphics[width=0.6\textwidth]{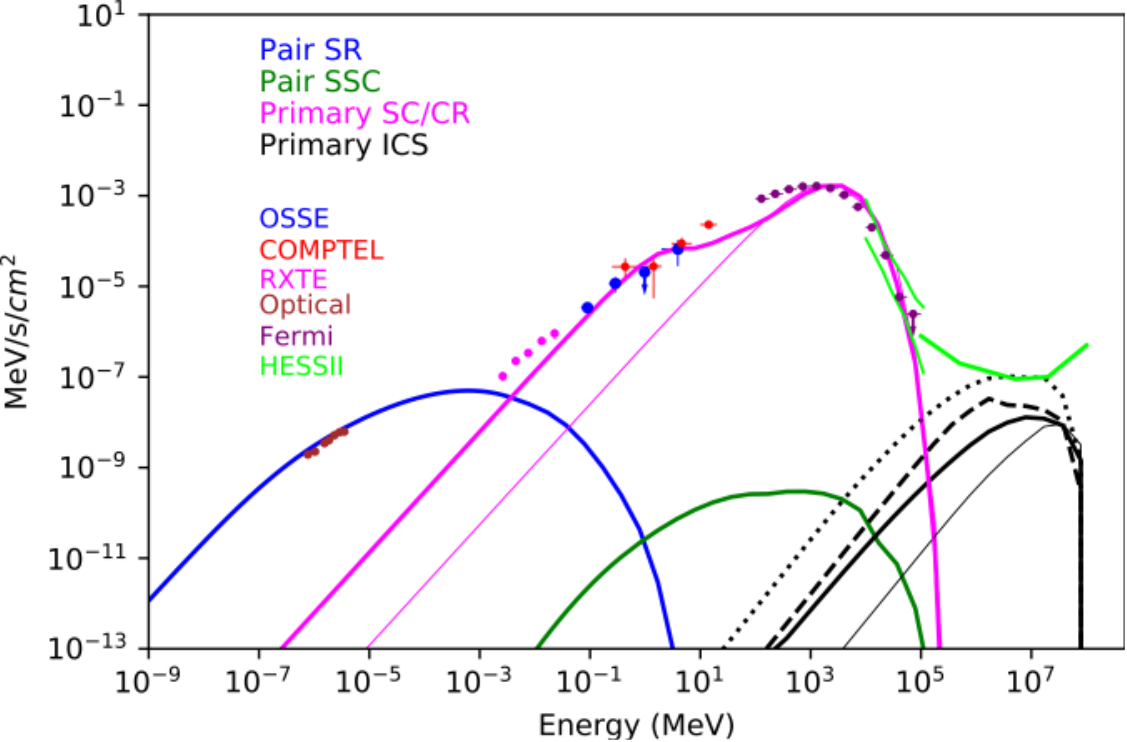}
    \caption[Model phase-averaged spectra for the Vela pulsar assuming $\alpha=75^\circ$ and $\zeta_{\rm cut}=65^\circ$]{Model phase-averaged spectra for the Vela pulsar assuming $\alpha=75^\circ$ and $\zeta_{\rm cut}=65^\circ$. The emission from primaries and pairs are included (as labelled). The solid black lines represent the ICS components from accelerated SC-emitting (thick line) or CR-emitting (thin line) primaries scattering the pair SR component (blue solid line), while the dashed and dotted black lines are the spectra from SC-emitting primaries scattering toy-model soft IR/optical photons with energy range ($0.5-4$~eV) and ($0.005-4$~eV), respectively. The model spectra are fit to the data points from \citet{Abdo2013SecondCat} (\url{http://fermi.gsfc.nasa.gov/ssc/data/access/lat/2nd_PSR_catalog/}), \citet{Shibanov2003}, and \citet{Harding2002Xray}. The H.E.S.S.\ II detection \citep{Abdalla2018} and HE sensitivity are also shown. From \citet{Harding2018}. \label{fig:SSCspec}}
\end{figure}

% Assumptions made; differences. Results (LC, spectra).
\citet{Harding2018} assumed the same parameter values as those used in Chapter~\ref{chap:CREdepLCmod}, except that their $R_{\rm acc}=0.2$ cm$^{-1}$ is slightly lower than what we used. This also impacts the spectral normalisation factor that shifts the spectral flux up or down to fit the GeV data points (see Chapter~\ref{chap:CREdepLCmod} and \citet{Harding2018} for the explanation thereof). Our normalisation factor for the two-step $E_\parallel$ is slightly larger than theirs of 10$J_{\rm GJ}$. Figure~\ref{fig:SSCLCs} and Figure~\ref{fig:SSCspec} show the first results obtained with the improved code. In Figure~\ref{fig:SSCLCs} the $E_\gamma$-dependent model light curves for Vela are illustrated, assuming $\alpha=75^\circ$ and $\zeta_{\rm cut}=65^\circ$, for the IR/optical ($0.1-1$~eV) band and three other $\gamma$-ray bands ranging between 0.05~GeV and 100~TeV. In Figure~\ref{fig:SSCspec} the phase-averaged spectrum is represented and includes all the radiation components produced by this model. The primaries and pairs respectively produce these components via emission mechanisms as discussed in Chapter~\ref{chap:PSRastro}. The TeV component is specifically produced by primary particles (that produce the GeV component via SC) that upscatter the pair SR.

% Spectra changes due to rho

\section{The effect of the refined $\rho_{\rm c}$ on the model output} \label{sec:rhooutput}

%Refined rho is needed to improve the calculation of the particle dynamics: influences CR loss term, and this couples to gamma. This influences radiation mechanisms - depend on gamma. Dynamical equations solved step by step, so using refined gamma will improve radiation calculations (from primary particles - prim SR, prim IC, SSC, etc.). This improves spectral calculations. Specifically CR (depends on both gamma and rho).

The equations of motion for the Lorentz factor, $\gamma$, and perpendicular momentum, $p_\perp$ (in units of $mc$), of a particle as it
moves along a $B$-field line are as follow \citep{Harding2005c}
\begin{eqnarray}\label{eq:ssctransport}
\frac{d\gamma}{dt}=\frac{eE_\parallel}{mc}-\frac{2e^4}{3m^3c^5}B^2p_\perp^2- \frac{2e^2\gamma^4}{3\rho_{\rm c}^2}+\left(\frac{d\gamma}{dt}\right)^{\rm abs}-\left(\frac{d\gamma}{dt}\right)^{\rm SSC},
\end{eqnarray}
\begin{eqnarray} \label{eq:momperp}
\frac{dp_\perp}{dt}=-\frac{3c}{2r}p_\perp-\frac{2e^4}{3m^3c^5}B^2\frac{p_\perp^3}{\gamma}+\left(\frac{dp_\perp(\gamma)}{dt}\right)^{\rm abs}.
\end{eqnarray}
In Eq.~(\ref{eq:ssctransport}), the different terms on the right-hand side are the CR acceleration rate, SR losses, CR losses, cyclotron/synchrotron absorption, and IC losses. In Eq.~(\ref{eq:momperp}), the various terms on the right-hand side are adiabatic momentum change along the field line, SR losses, and cyclotron/synchrotron resonant absorption. The SSC losses are negligible for $p_\perp$.

The refined calculation of $\rho_{\rm c}$ is needed to improve the calculation of the particle dynamics that form the basis to describe the emission from the particles. The particles can emit radiation via CR, SC, SSC and ICS with a detailed calculation of these processes and their emission directions described in \citet{Harding2008,Harding2015,Torres2018}. The CR loss term (see Eq.~(\ref{eq:ssctransport}) is strongly dependent on $\gamma$ and $\rho_{\rm c}$. This loss term influences the particle dynamics due to the coupling between this term and the particle $\gamma$ (i.e., the $\gamma(s)$ at the next step in arclength $s$ along the $B$-field line is determined by the losses suffered by the particle at the previous step, in particular the CR loss term). This in turn influences the other radiation mechanisms' loss terms, as they also depend on $\gamma(s)$. The dynamical equations are solved step by step for each radiation mechanism, so using a refined $\gamma(s)$ will improve the radiation calculations, i.e., from primary particles that emit SR, IC, SSC, etc. This also improves spectral calculations, specifically the CR component which depends on both $\gamma$ and $\rho_{\rm c}$.

%Second, moving ``off'' field lines: thus impacts emission directions (both prim and pair). This means, updated caustics and LCs. As a second-order effect: different trajectory will influence pair dynamics (probe slightly different B-field and radio photon distribution).}
%a second-order effect
As seen in Figure~\ref{fig:posdirrhoOvsN} the particle trajectories and $\rho_{\rm c}$ calculated previously and now via a refined calculation deviate slightly from each other. This implies that particles may move ``off'' the $B$-field lines in the old calculation (i.e., the adaptive step length sometimes sampled a particular $B$-line crudely, especially at high altitudes, and this led to particles numerically `jumping' to an adjacent field line). Thus, the new calculation impacts the emission directions (both of the primaries and the pairs), since the new trajectories that are calculated using fixed but small steps in arclength are slightly different. Therefore, a more accurate calculation of the particle trajectories and their $\rho_{\rm c}$'s leads to refined phase plots, i.e., the caustic structure is more pronounced, although the structure remains roughly the same. The respective light curves are also somewhat smoother and occur at slightly different phase, as shown in Figure~\ref{fig:PPLCs_OvsN75} in Chapter~\ref{chap:ch4}. Also, as the $E_\gamma$ increases the light curve trends investigated in Chapter~\ref{chap:CREdepLCmod}, in the CR regime, are still present at TeV energies. A different trajectory will influence the pair dynamics as well to some extent, since it probes a slightly different $B$-field and radio photon distribution. It was thus necessary to refine the trajectory and $\rho_{\rm c}$ calculation, given the interconnectedness and dependence of the particle dynamics and emission calculations on these basic spatial calculations.

\section{Conclusion} \label{sec:ch6concl}

%Motivation - TeV SSC, Important to refine model: there exist a number of assumptions (radio emission height, pair spectra parameters, etc.). If one can refine calculation, then data can more directly be used to constrain the unknown assumptions (not due to unreliable method). CTA
This study's motivation was to explain the TeV emission observed from the Vela pulsar by H.E.S.S.\ This observation provided a fundamental lower limit to the relativistic particles of a few TeV. They claimed that in order to obtain this measurement of several TeV, accelerated particles are needed that obtain such high energies. These particles will radiate CR at GeV energies, and also give rise to SC, SSC and ICS emission, where the primary ICS (i.e., SSC by primaries involving the pair SR) component is close to the H.E.S.S.\ sensitivity. Thus, the TeV spectrum can be explained within this framework that invokes CR to explain the GeV emission. This also supports our argument and findings in Chapter~\ref{chap:CREdepLCmod}.

It is important to keep on refining the model calculations, e.g., of particle trajectories. If one can refine these calculations to get a more accurate model, then data can more directly be used to constrain the unknown assumptions (radio emission height, pair spectra parameters, etc.), since discrepancies with respect to the data will more likely be due to model assumptions, and not the model implementation. This is especially important since upcoming ground-based telescopes such as CTA should be able to detect more VHE pulsars with its higher sensitivity than current telescopes.
\\
\\
In Chapter~\ref{chap:conclusion} we will give a summary of the aims, the process of achieving them, and the findings from the study as a whole. I will lastly also give a future outlook on what can still be done.

 % Implication 2: VHE emission as ICS - VHE 2018 article
\chapter{Conclusions} \label{chap:conclusion}
%BIG PICTURE OF WHOLE PhD

Pulsars are dense NSs that spin rapidly and contain strong $B$-fields, $E$-fields, and gravitational fields (e.g., \citealt{Abdo2010FirstCat}). These compact stars emit pulsed emission across the entire electromagnetic spectrum, injecting HE particles into the local environment. The \emph{Fermi} LAT has revolutionised the field of pulsar science with the number of $\gamma$-ray pulsar detections increasing from 7 pre-\emph{Fermi} to 117 (in 2013) to over 250. The ongoing detections of pulsed VHE emission from pulsars by ground-based telescopes paved the way to explore a new region of this $\gamma$-ray energy range. Light curve profiles in different energy bands and spectral properties from these pulsars have been studied in great detail. Therefore, energy-dependent light curve and spectral modelling served as useful tools to constrain the $B$-field structure, pulsar geometry ($\alpha$ and $\zeta$), constrain the GeV emission region's location and extent (i.e., within and beyond the light cylinder), and this may also help to discriminate between different emission mechanisms (e.g., \citealt{Dyks2004b,Venter2009,Watters2009,Johnson2014,Pierbattista2015}). In this study, I focused on HE and VHE $\gamma$-ray emission from the Vela pulsar, one of the brightest sources detected by \emph{Fermi}. 

\section{Significant contributions and results}
\label{contrib}
%Restate the thesis research topic and questions - summarise the main points and highlights from the body paragraphs.
%Show how you have addressed your aims and objectives - Explain the contribution the study makes.
%State the importance of the research and results.
%Explain the limitations of the study.

\subsection{Offset-dipole studies} \label{subsec:offsetConcl1}

The first aim of this study was to investigate the impact of different assumed magnetospheric structures on the predicted HE $\gamma$-ray pulsar light curve features of the Vela pulsar as observed by \emph{Fermi}. I adopted a geometric pulsar modelling code \citep{Dyks2004b} that already included the static dipole and RVD $B$-fields. I implemented a symmetric offset-PC dipole field \citep{Harding2011a,Harding2011c} characterised by an offset $\epsilon$ from the magnetic PCs. This included transforming the $B$-field from the co-rotating to the lab frame. For each $B$-field, I considered both the TPC and OG models, assuming uniform $\epsilon_{\nu}$. Additionally, I implemented the full accelerating SG $E$-field corrected for GR effects up to high altitudes, which modulated $\epsilon_{\nu}$. For the offset-PC dipole field I only considered the TPC and SG models, since there are no OG $E$-field expressions available for this particular $B$-field solution. I obtained a general SG $E_\parallel$-field by matching the low-altitude and high-altitude solutions of this $E_\parallel$-field by determining the matching parameter $\eta_{\rm c}(P,\dot{P},\alpha,\epsilon,\xi,\phi_{\rm PC})$ on each field line in multivariate space. 

After the general SG $E$-field was calculated I could solve the particle transport equation, which yielded the particle energy $\gamma(\eta)$, necessary for determining the CR $\epsilon_\nu$ and to test whether the CRR limit was attained. I found that the CRR limit was reached (see Figure~\ref{fig:RRLim}), albeit only at large $\eta$. Thus, the SG $E$-field is relatively low, therefore the particle energy only becomes large enough to yield significant CR at large altitudes above the stellar surface and thus particles do not always attain the CRR limit. Given this low SG $E$-field, I further investigated the effect that the SG $E_\parallel$ had on the predicted light curves in two ways. First, the minimum photon energy was lowered from $E_{\gamma,\rm min}=100$~MeV to $E_{\gamma,\rm min}=1$~MeV, leading to radiation in the hard X-ray waveband. Second, since I wanted to compare our model light curves to \textit{Fermi} data the usual low SG $E$-field was increased by a factor of 100 (with a spectral cutoff now at $E_{\gamma,\rm CR}\sim4$~GeV). 

We fit our model light curves to the observed \emph{Fermi}-measured Vela light curve for each $B$-field and geometric model combination. We found that our overall optimal light curve fit to the data was for the RVD field and OG model as seen in Figure~\ref{fig:ModelComparisonB}. For the other $B$-field and model combinations there were no significantly preferred model (per $B$-field), since all the alternative models may provide an acceptable alternative fit to the data, within $1\sigma$. The offset-PC dipole field preferred smaller values of PC offsets when assuming constant $\epsilon_{\nu}$, and larger values for variable $\epsilon_{\nu}$, but not significantly so ($<1\sigma$). When comparing all cases (i.e., all $B$-field and model combinations), we found that the offset-PC dipole field for variable $\epsilon_{\nu}$ was significantly rejected. When I lowered $E_\gamma$ we noted new caustic structures and emission features on the resulting phase plots and light curves that were absent when $E_{\gamma,\rm min}>100$~MeV. When I increased the SG $E$-field by a factor of 100, we found improved phase plots and light curve fits, e.g., extended caustic structures and new emission features as well as different light curve shapes emerged, as well as CRR being now reached in most cases at lower $\eta$. I also compared the best-fit light curves for the offset-PC dipole $B$-field and $100E_\parallel$ combination for each $\epsilon$~(Figure~\ref{fig:ModelComparisonB}) and noted that a smaller $\epsilon$ was again preferred (although not significantly; $<1\sigma$). In particular, when I compared this case to the other $B$-field and model combinations, we found statistically better fits for all $\epsilon$ values with this combination being second in quality only to the RVD and OG model fit. 

I compared the best-fit $\alpha$ and $\zeta$, with errors, from this and other independent studies ( Figure~\ref{fig:Comparison_alphazeta}) and noted that many of the best-fit solutions cluster inside the grey area at larger $\alpha$ and $\zeta$. Some fits lie near the $\alpha-\zeta$ diagonal (possibly due to radio visibility constraints in some cases) as well as near the $\zeta$ inferred from the PWN torus fitting \citep{Ng2008}, notably for the RVD $B$-field. Therefore, there was reasonable correspondence between my results and those of other studies. However, when I discarded the non-optimal TPC / SG fits, I saw that the optimal fits clustered near the other fits at large $\alpha$ and $\zeta$. For the increased SG $E$-field and offset-PC dipole combination, I noted that these fits also clustered at larger $\alpha$ and $\zeta$.

\subsection{Energy-dependent CR light curves and spectral modelling} \label{subsec:edepConcl1}

The second aim of this study was to investigate the HE light curve trends as a function of $E_\gamma$ as well as the phase-resolved spectra so as to contribute to the debate of whether CR or SR is responsible for the HE emission. As a first approach, I modelled the $E_{\gamma}$-dependent light curves and phase-resolved spectra of the Vela pulsar in the HE regime assuming CR from primary particles in order to explain the origin of the decreasing ratio of P1/P2 vs.\ $E_\gamma$ as seen in the observations, expecting that it is due to the combination of the values of geometric and physical parameters associated with each peak. Since the light curves probe geometry, e.g., $\alpha$, $\zeta$ and emission gap position and extent, and the spectrum probes both the energetics and geometry, I simultaneously fit our model to the \emph{Fermi} and H.E.S.S.\ data points to obtain optimal fitting parameters.

We used a full emission code \citep{Harding2015}, but only assumed CR, and implemented a refined calculation of the $\rho_{\rm c}$ of particle trajectories, impacting the CR loss rate and leading to smoother phase plots and light curves. However, this improved $\rho_{\rm c}$ had a rather small effect on the spatial distribution of the emissivity, as the broad structure of caustics and light curves remained similar to what was found previously. I assumed an SG and current sheet model in an FF magnetosphere (implying an $E$-field that is fully screened) as a good first approximation of the true $B$-field structure, yet I considered both a constant and two-step accelerating $E$-field. I solved the particle transport for both cases and found that the CRR limit was easily reached in the first scenario, but not always in the second. In order to isolate the P1/P2 effect, I selected photons that make up these two light curve peaks, and investigated the range of associated values of $\rho_{\rm c}$, $E_{\gamma,\rm CR}$ and $\gamma$ for each peak. I found that the phase-resolved spectra associated with each peak indicated a slightly larger spectral cutoff for P2, confirming that P2 survives longer with an increase in $E_\gamma$, given its larger spectral cutoff. This behaviour is also seen in the spectral observations by \emph{Fermi} and other ground-based Cherenkov detectors. This was also seen in energy-dependent histograms of $E_{\gamma, \rm CR}$, confirming that this quantity was systematically larger for P2. The reason for this became more clear upon discovery that both the $\rho_{\rm c}$ and $\gamma$ were systematically larger for P2, for both $E$-field scenarios. If CRR is reached, one expects $E_{\gamma, \rm CR}\propto \rho_{\rm c}^{1/2}$ for a constant $E$-field, so the larger $\rho_{\rm c}$ would explain the larger spectral cutoff for P2. Conversely, even if CRR is not attained, $E_{\gamma, \rm CR}\propto \gamma^3\rho_{\rm c}^{-1}$. Given the systematic dominance of $\gamma$ for P2, and the strong dependence of the third power, the larger spectral cutoff of P2 is thus explained by the larger $\gamma$. I also found that the values of $\rho_{\rm c}$ and $\gamma$ remained larger for P2 when only considering emission beyond the light cylinder into the current sheet. In particular, the largest values of these quantities occurred there, pointing to dominant emission from that region to make up the GeV light curves.

\subsection{SSC modelling} \label{subsec:sscConcl1}

The aim of the study done by \citet{Harding2018} was to explain the pulsed emission observed by H.E.S.S.\ from the Vela pulsar up to energies of a few TeV \citep{Djannati2017}. The pulsed emission at such extreme energies might be either connected to the GeV emission of \emph{Fermi}'s spectra or it could be a new radiation component. This study by \citet{Harding2018} viewed it to be a separate emission component, i.e., SSC emission, that is radiated by particles that have been accelerated up to energies of several TeV.

\citet{Rudak2017} modelled Vela's measured VHE emission in an OG scenario, where primary particles are accelerated inside this gap and scatter the observed IR to optical emission that are placed along the inner edge of the gap and believed to come from SR emitted by pairs also produced in this gap. They showed that primaries, whose CR spectrum match the \emph{Fermi} measured GeV spectrum, produce a significant ICS emission component reaching energies up to $\sim$10 TeV. 

We contributed to the study done by \citet{Harding2018}. \citet{Harding2018} expanded the SSC emission code of \citet{Harding2015} and included the following refinements: (1) extention of the spectral energy range (spanning from IR to VHE energies) in order to include more soft photons, the scattering of which is in the Thompson limit, and for modelling the SSC emission at the VHE energies, (2) a more accurate calculation of the particle trajectories and their radii of curvature, giving a more precise determination of the accelerating particle energy and of their emitted radiation spectrum, (3) a two-step accelerating $E_\parallel$, i.e., a lower $E_\parallel$ inside the light cylinder, and a higher $E_\parallel$ at and beyond the light cylinder (this is in agreement with recent global MHD and PIC pulsar models), (4) the injection spectrum of electron–positron pairs only above the PC in regions, (5) the MeV to GeV radiation from accelerated/relativistic particles inside and outside of the light cylinder are emitted via SC radiation, although their radiation at GeV energies is mostly in the CR limit.

Our contribution to the above mentioned study was the refined calculation of the particle trajectories and $\rho_{\rm c}$ as noted in point 2, and was needed to improve the calculation of the particle dynamics that was necessary to describe the emission from the particles. 

\section{Research implications}\label{sec:implications}

\subsection{Offset-dipole studies} \label{subsec:offsetConcl2}

I conclude that the magnetospheric structure and emission geometry have an important effect on the predicted $\gamma$-ray pulsar light curves. \textit{However, the presence of an $E$-field may have an even greater impact than small changes in the $B$-field structure and emission geometries:} When we included an SG $E$-field, thereby modulating $\epsilon_\nu$, the resulting phase plots and light curves became qualitatively different compared to the geometric case. 

There have been several indications that \textit{the SG $E$-field may be larger than initially thought.} For example, 
\begin{itemize}
    \item Population synthesis studies found that the SG $\gamma$-ray luminosity may be too low, and therefore an increased $E$-field and / or particle current through the gap are necessary (e.g., \citealt{Pierbattista2015}). 
    \item If the $E_\parallel$-field is too low the observed spectral cutoffs of a few GeV are not obtained (Section~\ref{subsec:100MeV}; \citealp{Abdo2013SecondCat}). 
    \item We found additional indications for an enhanced SG $E$-field. An increased $E$-field (multiplied by a factor of 100) led to statistically improved fits with respect to the light curves. This also yielded the second-best fit, next to the RVD and OG combination.
    \item Moreover, the inferred best-fit $\alpha$ and $\zeta$ parameters for this increased $E$-field clustered near the best fits of independent studies.
    \item We observed that a larger SG $E$-field also increased the particle energy gain rates and therefore yielded a larger particle energy $\gamma$, leading to particles reaching the CRR regime at altitudes close to the stellar surface.
\end{itemize}
The above mentioned arguments may point to a reconsideration of the boundary conditions assumed by \citet{Muslimov2004a} that suppressed the $E_{\parallel}$ at high altitudes, or of the current distribution in the gap. One possible way to bring self-consistency between $B$-field and $E$-field calculations may be to implement the newly developed FIDO model or PIC models that include global magnetospheric properties.

\subsection{Energy-dependent CR light curves and spectral modelling} \label{subsec:edepConcl2}

We found reasonable fits to the $E_{\gamma}$-dependent light curves and spectra of Vela as measured by \emph{Fermi} and H.E.S.S.\, and our model that assumes CR as the mechanism responsible for the GeV emission captures the four general observed trends: (i) the decrease of P1/P2 vs.\ $E_{\gamma}$, (ii) evolution / depression of the inter-peak bridge emission, (iii) peak positions remaining at constant phases, and (iv) a decrease in the peak widths with an increase in $E_{\gamma}$. However, an unknown azimuthal dependence of the $E$-field as well as uncertainty in the precise spatial origin of the emission preclude a simplistic discrimination of the emission mechanism responsible for the GeV emission inside and beyond the light cylinder. Similar future modelling of energy-dependent light curves and spectra within a striped-wind context that assumes SR to be the relevant GeV mechanism will be necessary to see if those models can also reproduce and explain these prominent features in the case of Vela and other pulsars. We note that the drop in P1/P2 vs.\ $E_\gamma$ may not be universal, as a study by \citet{Brambilla2015} points out. We found a counter-example for a different choice of $\zeta_{\rm cut}$ (where P1/P2 increases with $E_\gamma$) as well as for specific choices of the two-step acceleration $E$-field. There may also be other parameter combinations that can yield this behaviour. However, this effect seems prevalent and has been seen in both HE and VHE data of bright pulsars.

\subsection{SSC modelling} \label{subsec:sscConcl2}

\citet{Harding2018} modellled the emission from particles that emit radiation via CR, SC, SSC and ICS. Upon solving the particle dynamics the refined $\rho_{\rm c}$ significantly impacted the calculation of $\gamma(s)$. The CR loss term strongly depends on $\gamma(s)$ and $\rho_{\rm c}$ and in return influences the particle dynamics due to the coupling between this term and the particle $\gamma(s)$. An improved $\gamma(s)$ would in turn influence other emission mechanisms' loss terms, as they also depend on $\gamma(s)$. Therefore, if the dynamical equations are solved for each emission mechanism (using a refined $\gamma(s)$) the radiation calculations (from primary particles) would also be more accurate. This also improved spectral calculations, specifically the CR component.

I found that the inclusion of the refined $\rho_{\rm c}$ and more accurate particle trajectories led to refined phase plots, i.e., the caustic structure was more pronounced, although the shape remained roughly the same. Also, the respective light curves were also slightly smoother and occurred at a different phase (but not significantly different). As $E_\gamma$ was increased the light curve trends remained at TeV energies. 

\section{Future prospects}\label{sec:future}
%Lay out questions for further research
%Explain how to address the limitations of the study

There are several future projects that may emerge from this study. The projects related to the study of the impact of the $B$-field structure on the light curves include extending the range of $\epsilon$ for which our code finds the PC rim, since more complex field solutions, such as the dissipative and FF field structures, may be associated with larger PC offsets. However, the offset-PC dipole solutions may be more applicable to MSPs discovered by \emph{NICER}. It would still be preferable to investigate the self-consistent $B$-fields and $E$-fields of the dissipative models and solve the transport equation to test if the particles reach the CRR limit. The effect of these new fields on the phase plots and light curves can also be studied. 

The VHE pulsed emission from pulsars depends strongly on the electrodynamics and the magnetospheric structure. Therefore, one could extend the SSC emission code by implementing more realistic $B$-field structures and their associated $E$-field distributions, thereby constraining the magnetospheric physics. One could use global dissipative magnetosphere models that represent solutions where a finite conductivity is specified on $B$-field lines above the stellar surface. These dissipative solutions fill the gap between the RVD (assuming zero conductivity) and the FF (assuming infinite conductivity) solutions (see \citealt{Kalapotharakos2012b,Li2012}). In future, an SC radiation mechanism can be incorporated as done by \citet{Harding2018}. This mechanism seems to be able to produce spectra that are relatively higher at lower MeV energies, and that may provide better fits to the light curve and spectral data. Study the effects of different (azimuthally-dependent) injection spectra on TeV component.

There is also potential for multi-wavelength studies, such as light curve modelling in other energy bands, e.g., combining radio and $\gamma$-ray light curves (see \citealp{Seyffert2010,Seyffert2012,Pierbattista2015}). One could furthermore continue to model energy-dependent light curves, such as those available for other bright pulsars using data from {\it Fermi} and ground-based telescopes (e.g., \citealp{Abdo2010Crab,Abdo2010Geminga}). Lastly, model phase-resolved spectra can be constructed, which will be an important test of the $E_\parallel$-field magnitude and its spatial dependence. This multi-wavelength study would assist in studying the evolution of P1/P2 with $E_\gamma$ in more depth and could shed some more light on the underlying emission geometry and radiation mechanisms responsible for the HE and VHE emission. For example, modelling of the VHE pulsed emission could scrutinise the general emission framework of any particular model, as well as constraining particle energetics. Once the model has been thus tested, we will apply it to future detections of VHE pulsars (expected from the list of potential VHE sources released by \textit{Fermi}; \citealp{Ackermann2013}). 

Lastly, the application of the code to jointly fit light curve, spectral, and polarisation data may yield further constraints. Attempting to reproduce HE phase-resolved spectra, i.e., behaviour of spectral index / cutoff with phase, may yield important constraints.
\begin{appendices}
\chapter{Refined calculation of the curvature radius $\rho_{\rm c}$} \label{chap:appRho}
In this Appendix, I describe the updated procedure to calculate the radius of curvature of particle trajectories in the lab frame.

We refine the previous first-order calculation of $\rho_{\rm c}$ along the electron\footnote{We use ``electron'' to collectively refer to electrons and positrons.} trajectory in the lab frame, assuming that all particles injected at the footpoint of a particular $B$-field line on the stellar surface follow the same trajectory, independent of their energy, since these are quickly accelerated to relativistic energies by the unscreened $E$-field. We furthermore assume that the $B$-field is strong enough to constrain the movement of the electrons so they will move parallel to the field line in the co-rotating frame. Thus, we do not consider any perpendicular motion in this frame, since the perpendicular particle energy is nearly instantly expended via SR. We thus take into account the perpendicular $\mathbf{E}\times\mathbf{B}$ drift in the lab frame in which our new calculation for $\rho_{\rm c}$ takes place.

To calculate the electron's trajectory as well as its associated $\rho_{\rm c}$, we first use a small, fixed step size $ds$ (arclength interval) along the $B$-field line in the lab frame.
% //Step length dsp_iof_e
The particle is injected at the stellar surface, and we trace its motion using this step length.
%Start at stellar surface and use (xprimary, yprimary, zprimary) position in IOF: 
%rp1 = sqrt(xprimary*xprimary + yprimary*yprimary + zprimary*zprimary);
%rp_eq1 = sqrt(xprimary*xprimary + yprimary*yprimary);
Since we are using a numerical solution of the FF $B$-field at a particular magnetic inclination angle $\alpha$, the three Cartesian components of the local first-order derivative of the position are available at any specified position (albeit that they may have to be interpolated, given the set resolution of the numerical solution of the FF field). One can thus use this information to map out a particle's trajectory for a given step length $ds$. 
The particle positions ($x$,$y$,$z$) as well as the local first derivatives ($x^\prime$,$y^\prime$,$z^\prime$) along the trajectory (i.e., the normalised velocity or the direction of motion), which is equivalent to the normalised $B$-field components, as a function of the cumulative arclength $s$ are used to compute both the full particle trajectory and its $\rho_{\rm c}(s)$.

The calculation involves three positions (previous, current, and next, denoted by indices $i-1$, $i$ and $i+1$, respectively). At injection, let the particle position be $(x_0,y_0,z_0)$. Viewing this position as being at the `previous' step, let us denote this as $(x_{i-1},y_{i-1},z_{i-1})$.
%Start with (xprimary,yprimary,zprimary) to have initial position 
%  pos_1[1] = xprimary;
%  pos_1[2] = yprimary;
%  pos_1[3] = zprimary;
%Create table of arclength along electron trajectory in IOF tot_le_tab[jj]
%The first-order derivatives are also available: $(x_0^\prime,x_0^\prime,z_0^\prime)$. As one moves one step length $ds$ along the $B$-field line, the cumulative arclength $s$ is updated, and the first-order derivative at that position is avalable: $(x_1^\prime,y_1^\prime,z_1^\prime)$
%$x_{\rm i-1}(s),x_{\rm i}(s),x_{\rm i+1}(s)$ refers to the previous, current and next positions respectively for the $x$ dimension (similar for the $y$ and $z$ dimensions)
%First time update (approximate) tangent dir vectors at next position
%    cr_inertial_iof(&pos_1[1], &pos_1[2], &pos_1[3], &t_e, &dir_1[1], &dir_1[2], &dir_1[3], &en_fac1, &xprimary, &yprimary, &zprimary, &tot_le, &dl_e_dummy);
The first-order derivatives at this position is also available: $(x_{i-1}^\prime,y_{i-1}^\prime,z_{i-1}^\prime)$. We next step along the field line, updating the arclength $s$.
%Accumulate arclengths at the various positions where the first entry should be zero to correspond to trace_primary functionality
%    tot_le_tab[0] = tot_le - dsp_iof_e;
The position was then updated according to the Euler method:   
%Move to next position using Euler method.
\begin{equation}
    x_i = x_{i-1} + x^\prime_{i-1}\cdot ds,
\end{equation}
and similar for the other two coordinates.
%    pos[ii] = pos_1[ii] + dsp_iof_e * dir_1[ii];
%Update (approximate) tangent vectors at next position
%    cr_inertial_iof(&pos[1], &pos[2], &pos[3], &t_e, &dir[1], &dir[2], &dir[3], &en_fac1, &xprimary, &yprimary, &zprimary, &tot_le, &dl_e_dummy);
%Save positions and directions (normalized B-field components)
%    pos_unsmoothed[i][ip] = pos[i+1];
%    dir_unsmoothed[i][ip] = dir[i+1];
%    pos_tab_unsmoothed[i][ip] = pos[i+1];
%    dir_tab_unsmoothed[i][ip] = dir[i+1];
The current position and derivative $(x_{i}^\prime,y_{i}^\prime,z_{i}^\prime)$ were then saved. We similarly moved to the next position
\begin{equation}
    x_{i+1} = x_{i} + x^\prime_{i}\cdot ds,
\end{equation}
also for $y$ and $z$.
%Move to next position using Euler method.
%    pos1[ii] = pos[ii] + dsp_iof_e * dir[ii];
%Update (approximate) tangent vectors at next position
%    cr_inertial_iof(&pos1[1], &pos1[2], &pos1[3], &t_e, &dir1[1], &dir1[2], &dir1[3], &en_fac1, &xprimary, &yprimary, &zprimary, &tot_le, &dl_e_dummy);
%Thus obtained we obtain: 
%Previous, current, and next positions along the $B$-field are given respectively by pos_1, pos, and pos1 for each spatial coordinate ($x$,$y$,$z$). $x_{\rm i-1}(s),x_{\rm i}(s),x_{\rm i+1}(s)$ refers to the previous, current and next positions respectively for the $x$ dimension (similar for the $y$ and $z$ dimensions)
%Previous, current, and next first derivatives (directions of $B$-field) are given by dir_1, dir, and dir1 respectively for 
%each each spatial coordiante ($x$,$y$,$z$). $x^\prime_{\rm i-1}(s),x^\prime_{\rm i}(s),x^\prime_{\rm i+1}(s)$ refers to the previous, current and next directions respectively for the $x$ dimension (similar for the $y$ and $z$ dimensions).
We thus have position and local direction components at three adjacent points with which we start the process.

We step along the particle trajectory (this stepping procedure is repeated until some large radius is reached), so that the $x_{i+1}$ becomes the current position, and similar for the first-order derivative.
%Step along trajectory; save unsmoothed positions and directions.
%  while ((rp_eq1 < 1.25*maxrh*Rlc) && (rp1 < 1.25*maxrd*Rlc)) {
%    ip++;
%    Update values of pos and dir
%    for (ii = 1; ii <= 3; ii++) {
%        //Update previous position with current value
%        pos_1[ii] = pos[ii];
%        //Update tangent vectors at previous position with current value
 %       dir_1[ii] = dir[ii];
 %       //Update current position with old "next" value
 %       pos[ii] = pos1[ii];
 %       //Update tangent vectors at current position with old "next" value
 %       dir[ii] = dir1[ii];
 %       //Give a step along the trajectory
%        pos1[ii] = pos[ii] + dsp_iof_e * dir[ii];
%    }
%    Save tot_le 
%    tot_le_tab[ip] = tot_le_tab[ip-1] + dsp_iof_e;//Save left boundary of interval
%    for (i=0; i<=2; i++) {
%        pos_unsmoothed[i][ip] = pos[i+1];
%        dir_unsmoothed[i][ip] = dir[i+1];
%        pos_tab_unsmoothed[i][ip] = pos[i+1];
%        dir_tab_unsmoothed[i][ip] = dir[i+1];        
%    }
%//Move to next position
%    cr_inertial_iof(&pos1[1], &pos1[2], &pos1[3], &t_e, &dir1[1], &dir1[2], &dir1[3], &en_fac1,&xprimary, &yprimary, &zprimary, &tot_le, &dl_e_dummy);
%//Update radial position (to test while loop condition)
%    rp1 = sqrt(pos1[1]*pos1[1] + pos1[2]*pos1[2] + pos1[3]*pos1[3]);
%    rp_eq1 = sqrt(pos1[1]*pos1[1] + pos1[2]*pos1[2]);
%  } //end of while loop over primary electron trajectory

First, at the current position, we smooth the three spatial coordinates ($x$,$y$,$z$) using $s$ as the independent variable to counteract numerical noise or uncertainties that may be present in the numerical calculation of the global $B$-field structure (and also taking into account the spatial grid on which this $B$-field was calculated). The smoothing is performed using a Gaussian Kernel Density Estimator (KDE; \citealt{Parzen1962}) smoothing procedure. We choose the smoothing parameter $h$ as a fraction of $R_{\rm LC}$; this needs to be adapted when increasing or decreasing the step size. The smoothing parameter used in the KDE procedure sets the level of smoothing (i.e., the spatial range in $s$ over which smoothing occurs), and needs to be connected to $ds$ to avoid under- or over-smoothing. After some testing, we set $h=50 ds$.

%SMOOTHING PROCEDURE
%//size_old, x_old[], y_old[], size_new, x_new[], y_new[], hh
%//ip+1, tot_le_tab,pos_unsmoothed[i],ip+1, tot_le_tab, pos_smoothed[i],h_smooth;

%\begin{eqnarry}
%A & = & \left(\frac{x_old[jj]-x_new[kk]}{hh}\right)^2, \\
%T & = & \sum{y_old[jj]e{^{-0.5A}}}, \\
%B & = & \sum{e{^{-0.5A}}}, \\
%y_new[kk] & = & T/B,
%\end{eqnarry}

%Smooth positions
%  for (i = 0; i < 3; i++) {
%    Smoothing(ip+1,tot_le_tab,pos_unsmoothed[i],ip+1,tot_le_tab,pos_smoothed[i],h_smooth);  //size_old,x_old,y_old,size_new,x_new,y_new,h
%   }

Second, we noticed that the our use of a KDE smoothing procedure on the position coordinates introduced some artificial ``tails'' at low and high altitudes, thus, the procedure is failing at the edges of the position range. We thus piecewise match (using some small tolerance on the allowed fraction that the smoothed and unsmoothed positions may differ) the unsmoothed and smoothed spatial positions of the electron trajectory at particular $s$ values to get rid of these unwanted ``tails'' and to end up with the most satisfactory set of positions that constitute three smooth but realistic functions of arclength (i.e., a combination of the smoothed and unsmoothed positions as functions of $s$). 
%Use to a certain tolerance tol = 1.0e-3 to match pos, dir, and rho_c. Match the smoothed and unsmoothed positions, since smoothing introduces an unwanted low-altitude tail.
%Search for index where unsmoothed and smoothed pos intersect.
%  for (ii = 0; ii < 3; ii++) {
%    i = 0;
%    while(fabs((pos_smoothed[ii][i] - pos_unsmoothed[ii][i])/pos_smoothed[ii][i]) > tol){
%      i++;
%    }
%  for (jj = 0; jj < i; jj++) {
%       pos_tab_smoothed[ii][jj] = pos_unsmoothed[ii][jj];
%    }
%  for (jj = i; jj < ip; jj++) {
%      pos_tab_smoothed[ii][jj] = pos_smoothed[ii][jj];
%    }
%  }
%Match positions again - from the high-altitude end:
%  for (ii = 0; ii < 3; ii++) {
%    i = ip;
%Search for index where unsmoothed and smoothed pos intersect.    
%    while(fabs((pos_smoothed[ii][i] - pos_unsmoothed[ii][i])/pos_smoothed[ii][i]) > tol){
%        i--;
%      }
%    for (jj = i; jj < ip; jj++) {
%      pos_tab_smoothed[ii][jj] = pos_unsmoothed[ii][jj];
%    }
%  }

Third, we also smooth and then piecewise match the unsmoothed and smoothed directions of the electron trajectory at particular $s$ values to get rid of these unwanted ``tails'', as was done with the position coordinates.
%and to end up with the most satisfactory set of directions that constitute three smooth but realistic functions of arclength (i.e., a combination of the smoothed and unsmoothed directions as functions of $s$).
%Smooth directions
%   for (i = 0; i < 3; i++) {
%     Smoothing(ip+1,tot_le_tab,dir_unsmoothed[i],ip+1,tot_le_tab,dir_smoothed[i],h_smooth);  //size_old,x_old,y_old,size_new,x_new,y_new,h
%   }
%Matching the smoothed and unsmoothed dir, since smoothing introduces an unwanted low-altitude tail
%  for (ii = 0; ii < 3; ii++) {
 %   i = 0;
 %   while(fabs((dir_smoothed[ii][i] - dir_unsmoothed[ii][i])/dir_smoothed[ii][i]) > tol){
 %     i++;
 %   }
 %   for (jj = 0; jj < i; jj++) {
 %     dir_tab_smoothed[ii][jj] = dir_unsmoothed[ii][jj];
 %   }
 %   for (jj = i; jj < ip; jj++) {
 %     dir_tab_smoothed[ii][jj] = dir_smoothed[ii][jj];
 %   }
  %}
%Match directions again - from the high-altitude end:
%  for (ii = 0; ii < 3; ii++) {
%    i = ip;
%    while(fabs((dir_smoothed[ii][i] - dir_unsmoothed[ii][i])/dir_smoothed[ii][i]) > tol){
%        i--;
%      }
%    for (jj = i; jj < ip; jj++) {
%      dir_tab_smoothed[ii][jj] = dir_unsmoothed[ii][jj];
%    }
%  }

Fourth, we use a second-order method involving interpolation by a Lagrange polynomial to obtain the second-order derivatives of the positions along the trajectory as a function of $s$, based on the (smoothed and matched) first derivatives \citep{Faires2002}:
\begin{equation}
x^{\prime\prime}(s) = \frac{\left(-3x^\prime_{i-1} + 4x^\prime_{i} - x^\prime_{i+1}\right)}{2ds},
\end{equation}
and similar for $y$ and $z$. 
This increased accuracy is necessary since $\rho_{\rm c}$ is a function of second-order derivatives (acceleration) of the electron position, and instabilities may be exacerbated if not dealt with carefully. 
%Calculate second-order derivatives using Lagrange polynomials.
%We compute $\rho_{\rm c}$ Lagrange we use the three-point endpoint formula for $f(x_0)$ from \citet{Faires2002}:
We do this both for the smoothed and unsmoothed first-order derivatives of the position.
Fifth, with the second-order derivatives in hand, we calculate two instances of $\rho_{\rm c}$, one involving the unsmoothed (`us') and one involving the smoothed (`s') accelerations:
%Second-order derivative of unsmoothed directions ($x^\prime_{\rm us}$) is given by df2_LPus_dir and from smoothed directions ($x^\prime_{\rm sm}$) by df2_LPsm_dir
%Compute $\rho_{\rm c}$ Lagrange using formula for f'(x0), but using directions instead of f(x), and s instead of x.
%Curvature from unsmoothed directions is given by K3us_dir and from smoothed directions by K3sm_dir.
%    df2_LPus_dir[ii] = (-3.0*dir_unsmoothed[ii][jj-1] + 4.0*dir_unsmoothed[ii][jj] - dir_unsmoothed[ii][jj+1])/(2.0*dsp_iof_e);
%    df2_LPsm_dir[ii] = (-3.0*dir_smoothed[ii][jj-1] + 4.0*dir_smoothed[ii][jj] - dir_smoothed[ii][jj+1])/(2.0*dsp_iof_e);
%\\
%\\
%    K3us_dir += df2_LPus_dir[ii]*df2_LPus_dir[ii];
%    K3sm_dir += df2_LPsm_dir[ii]*df2_LPsm_dir[ii];
%    K3us_dir = sqrt(K3us_dir);
%    K3sm_dir = sqrt(K3sm_dir);
%    rho_c_Lagrange_us_dir[jj] = 1.0/K3us_dir;
%    rho_c_Lagrange_sm_dir[jj] = 1.0/K3sm_dir;
\begin{equation}
\rho_{\rm c,us}(s) = \frac{1}{\sqrt{x^{\prime\prime}_{\rm us}(s)^2+y^{\prime\prime}_{\rm us}(s)^2+z^{\prime\prime}_{\rm us}(s)^2}}, \\
\end{equation}
\begin{equation}
\rho_{\rm c,s}(s) = \frac{1}{\sqrt{x^{\prime\prime}_{\rm s}(s)^2+y^{\prime\prime}_{\rm s}(s)^2+z^{\prime\prime}_{\rm s}(s)^2}}. \\
\end{equation}

%Set first value equal to second.
%  rho_c_Lagrange_us_dir[0] = rho_c_Lagrange_us_dir[1];
%  rho_c_Lagrange_sm_dir[0] = rho_c_Lagrange_sm_dir[1];
Finally, we piecewise match these two results for $\rho_{\rm c}(s)$ to get rid of ``tails'' in $\rho_{\rm c}$ at low and high altitudes, as before. 
%Match rho_c to get rid of ''smoothing tails''.
%  i = 0;
%  while(fabs(rho_c_Lagrange_sm_dir[i] - rho_c_Lagrange_us_dir[i])/rho_c_Lagrange_sm_dir[i] > tol){
%    i++;
%    }
%  for (jj = 0; jj < i; jj++) {
%     rho_c_tab_smoothed[jj] = rho_c_Lagrange_us_dir[jj];
%  }
%  for (jj = i; jj < ip-1; jj++) {
%    rho_c_tab_smoothed[jj] = rho_c_Lagrange_sm_dir[jj];
%  }

%Match rho_c again - from the high-altitude end:
%    i = ip;%
%    while(fabs((rho_c_Lagrange_sm_dir[i] - rho_c_Lagrange_us_dir[i])/rho_c_Lagrange_sm_dir[i]) > tol){
%        i--;
%      }
%    for (jj = i; jj < ip; jj++) {
%      rho_c_tab_smoothed[jj] = rho_c_Lagrange_us_dir[jj];
%    }
Having a pre-calculated $\rho_{\rm c}$ in hand, as well as a particle trajectory (particle positions), for a fine division in arclength along any particular $B$-field line, we then interpolate $\rho_{\rm c}$ in our particle transport calculations to accommodate an adaptive, variable-$ds$ approach that is used to speed up the transport calculations, without losing accuracy of the trajectory. 
\end{appendices}
\bibliography{PhDthesis_MBarnard}

\end{document}